\documentclass[11pt, a4paper]{article}
\pdfoutput=1
\usepackage[utf8]{inputenc}
\usepackage{jheppub,amsmath,amssymb,slashed,url,bm,textgreek,upgreek,hyperref}
\usepackage{jheppub}  
\usepackage{tikz,lipsum,lmodern}
\usepackage[most]{tcolorbox}
\usepackage{amssymb} 
\usepackage{amsmath}
\usepackage{mathtools}
\usepackage{amsfonts}    
\usepackage{dsfont}
\usepackage{pdfpages}
\usepackage{verbatim}
\usepackage{tensor}
\usepackage{mathrsfs}
\usepackage{textgreek} 
\usepackage[mathscr]{euscript}
\usepackage[normalem]{ulem}
\usepackage{tikz}
\usepackage{makecell}
\usepackage[verbose]{placeins}
\usepackage{subcaption}
\usetikzlibrary{3d, arrows.meta, decorations.pathreplacing, decorations.markings,calc,shapes.misc,decorations.pathmorphing,patterns.meta, math}
\usepackage{pgfplots}
\pgfplotsset{compat=1.18}

\newcommand{\beq}{\begin{equation}}
\newcommand{\eeq}{\end{equation}}
\newcommand{\no}{\nonumber\\} 
\newcommand{\bea}{\begin{eqnarray}}
\newcommand{\ea}{\end{eqnarray}}
\newcommand{\barr}{\begin{array}}
\newcommand{\earr}{\end{array}}

\def\be{\begin{equation}}
\def\ee{\end{equation}}
\def\ba#1\ea{\begin{align}#1\end{align}}
\def\bg#1\eg{\begin{gather}#1\end{gather}}
\def\bm#1\em{\begin{multline}#1\end{multline}}
\def\bmd#1\emd{\begin{multlined}#1\end{multlined}}

\def\vp{\chi}

\def\la{\label}

\def\er{\eqref}

\def\pa{\partial}

\def\wg{\wedge}

\def\no{\nonumber}

\def\({\left(}
\def\){\right)}
\def\[{\left[}
\def\]{\right]}
\def\<{\langle}
\def\>{\rangle}

\def\bea{\begin{eqnarray}}
\def\eea{\end{eqnarray}}

\newcommand{\tr}{\operatorname{tr}}

\newcommand{\cO}{{\mathcal O}}
\newcommand{\zb}{{\bar z}}

\def\nn{\nonumber}

\begin{document}

\global\long\def\aad{(a\tilde{a}+a^{\dagger}\tilde{a}^{\dagger})}%

\global\long\def\ad{{\rm ad}}%

\global\long\def\bij{\langle ij\rangle}%

\global\long\def\df{\coloneqq}%

\global\long\def\bs{b_{\alpha}^{*}}%

\global\long\def\bra{\langle}%

\global\long\def\dd{{\rm d}}%

\global\long\def\dg{{\rm {\rm \dot{\gamma}}}}%

\global\long\def\ddt{\frac{{\rm d^{2}}}{{\rm d}t^{2}}}%

\global\long\def\ddg{\nabla_{\dot{\gamma}}}%

\global\long\def\del{\mathcal{\delta}}%

\global\long\def\Del{\Delta}%

\global\long\def\dtau{\frac{\dd^{2}}{\dd\tau^{2}}}%

\global\long\def\ul{U(\Lambda)}%

\global\long\def\udl{U^{\dagger}(\Lambda)}%

\global\long\def\dl{D(\Lambda)}%

\global\long\def\da{\dagger}%

\global\long\def\id{{\rm id}}%

\global\long\def\ml{\mathcal{L}}%

\global\long\def\mm{\mathcal{\mathcal{M}}}%

\global\long\def\mf{\mathcal{\mathcal{F}}}%

\global\long\def\ket{\rangle}%

\global\long\def\kpp{k^{\prime}}%

\global\long\def\lr{\leftrightarrow}%

\global\long\def\lf{\leftrightarrow}%

\global\long\def\ma{\mathcal{A}}%

\global\long\def\mb{\mathcal{B}}%

\global\long\def\md{\mathcal{D}}%

\global\long\def\mbr{\mathbb{R}}%

\global\long\def\mbz{\mathbb{Z}}%

\global\long\def\mh{\mathcal{\mathcal{H}}}%

\global\long\def\mi{\mathcal{\mathcal{I}}}%

\global\long\def\ms{\mathcal{\mathcal{\mathcal{S}}}}%

\global\long\def\mg{\mathcal{\mathcal{G}}}%

\global\long\def\mfa{\mathcal{\mathfrak{a}}}%

\global\long\def\mfb{\mathcal{\mathfrak{b}}}%

\global\long\def\mfb{\mathcal{\mathfrak{b}}}%

\global\long\def\mfg{\mathcal{\mathfrak{g}}}%

\global\long\def\mj{\mathcal{\mathcal{J}}}%

\global\long\def\mk{\mathcal{K}}%

\global\long\def\mmp{\mathcal{\mathcal{P}}}%

\global\long\def\mn{\mathcal{\mathcal{\mathcal{N}}}}%

\global\long\def\mq{\mathcal{\mathcal{Q}}}%

\global\long\def\mo{\mathcal{O}}%

\global\long\def\qq{\mathcal{\mathcal{\mathcal{\quad}}}}%

\global\long\def\ww{\wedge}%

\global\long\def\ka{\kappa}%

\global\long\def\nn{\nabla}%

\global\long\def\nb{\overline{\nabla}}%

\global\long\def\pathint{\langle x_{f},t_{f}|x_{i},t_{i}\rangle}%

\global\long\def\ppp{p^{\prime}}%

\global\long\def\qpp{q^{\prime}}%

\global\long\def\we{\wedge}%

\global\long\def\pp{\prime}%

\global\long\def\sq{\square}%

\global\long\def\vp{\varphi}%

\global\long\def\ti{\widetilde{}}%

\global\long\def\wg{\widetilde{g}}%

\global\long\def\te{\theta}%

\global\long\def\tr{{\rm Tr}}%

\global\long\def\ta{{\rm \widetilde{\alpha}}}%

\global\long\def\sh{{\rm {\rm sh}}}%

\global\long\def\ch{{\rm ch}}%

\global\long\def\Si{{\rm {\rm \Sigma}}}%

\global\long\def\sch{{\rm {\rm Sch}}}%

\global\long\def\vol{{\rm {\rm {\rm Vol}}}}%

\global\long\def\reg{{\rm {\rm reg}}}%

\global\long\def\zb{{\rm {\rm |0(\beta)\ket}}}%

\title{Unveiling horizons in quantum critical collapse
\vspace{-0.8cm}}

\author[a]{Marija Toma\v{s}evi\'c}
\author[b]{and Chih-Hung Wu}

\affiliation[a]{Institute for Theoretical Physics, University of Amsterdam, Science Park 904, 1090 GL Amsterdam, The Netherlands}
\affiliation[b]{Department of Physics, University of Washington, Seattle, WA 98195, USA}

\emailAdd{m.tomasevic@uva.nl}
\emailAdd{chwu29@uw.edu}

\abstract{Critical gravitational collapse offers a unique window into regimes of arbitrarily high curvature, culminating in a naked singularity arising from smooth initial data---thus providing a dynamical counterexample to weak cosmic censorship. Near the critical regime, quantum effects from the collapsing matter are expected to intervene before full quantum gravity resolves the singularity. Despite its fundamental significance, a self-consistent treatment has so far remained elusive. In this work, we perform a one-loop semiclassical analysis using the robust anomaly-based method in the canonical setup of Einstein gravity minimally coupled to a free, massless scalar field. Focusing on explicitly solvable near-critical solutions in both $2+1$ and $3+1$ dimensions, we analytically solve the semiclassical Einstein equations and obtain controlled, quantitative results for several long-standing questions within the dominant $s$-wave sector. We find that regularity uniquely selects a Boulware-like quantum state, encoding genuine vacuum polarization effects from the collapsing matter. Remarkably, the resulting quantum corrections manifest as a growing mode. Horizon-tracing analyses, incorporating both classical and quantum modes, reveal the emergence of a finite mass gap, signaling a phase transition from classical Type II to quantum-modified Type I behavior, thereby providing a quantum enforcement of the weak cosmic censorship. The most nontrivial aspect of our analysis involves dealing with non-conformal matter fields in explicitly time-dependent critical spacetimes. Along the way, we uncover intriguing and previously underexplored features of quantum field theory in curved spacetime.}

\maketitle

%%%%%%%%%%%%%%%%%%%%%%%%%%%%%%%%%%%%%%%%%%%%%%%%%%
\newpage

\section{Introduction}
\label{sec:intro}

Singularities are ubiquitous in solutions of Einstein's General Relativity, signaling a breakdown in predictability and highlighting the limits of our current best understanding of gravity. Among the most troubling possibilities is that such singularities could be visible to distant observers: \textit{a naked singularity}. This raises profound questions not only about determinism in classical physics but also about whether we would be forced to confront quantum gravity directly. In response, Penrose proposed the \textit{weak cosmic censorship conjecture}~\cite{Penrose:1969pc}, suggesting that gravitational collapse generically produces event horizons that cloak singularities from external observers. Hawking distilled this expectation into a simple remark: ``\textit{Nature abhors a naked singularity},'' reflecting the hope that the laws of physics protect observers from directly witnessing spacetime pathologies~\cite{Hawking:1994ss}.\footnote{However, studies on naked singularities have shown that they can, in principle, be observationally distinguished from black holes~\cite{Virbhadra:1998dy, Virbhadra:2002ju, Virbhadra:2007kw, Virbhadra:2022iiy}.}

For probing the formation of singularities and the limits of weak cosmic censorship, the study of gravitational collapse under scalar fields has long served as a fruitful arena~\cite{Giamb__2024}. Early explorations in the late 1960s, motivated by the idea of boson stars, laid the foundation for using scalar fields as a tractable model for dynamical spacetime phenomena~\cite{Liebling:2012fv}. A major turning point came with the seminal works of Christodoulou~\cite{Christodoulou:1986zr, Christodoulou:1991yfa, Christodoulou:1993, Christodoulou:1994hg, Christodoulou_1999on, Christodoulou:2008nj}, who initiated a rigorous program analyzing the $3+1$-dimensional \textit{spherically symmetric collapse of a free, massless scalar field}, where the dynamics is described by the wave equation~\cite{Christodoulou_1999on}. Christodoulou proved that sufficiently small initial data evolve into a globally regular spacetime that is asymptotically flat and null geodesically complete~\cite{Christodoulou:1986zr}, giving strong evidence in favor of cosmic censorship in certain regimes. However, while analyzing sufficient conditions on initial data leading to trapped surface formation~\cite{Christodoulou:1991yfa, Christodoulou:1993, Christodoulou:1994hg, Christodoulou_1999on},\footnote{See also~\cite{Dafermos:2003qz} for the regular extension to the past null infinity.} he also discovered analytical examples of a naked singularity forming from smooth initial data, despite being unstable in the larger space of bounded variation~\cite{Christodoulou:1994hg}. Nevertheless, it shows that cosmic censorship can fail in this model under non-generic conditions.

On the other hand, the dynamical picture was not fully understood. This tension between stability and instability is mirrored in the groundbreaking numerical work of Choptuik, who uncovered a different route to naked singularities at the threshold of black hole formation~\cite{PhysRevLett.70.9}. By finely tuning a one-parameter family of initial data for a collapsing massless scalar field, Choptuik observed the emergence of a discretely self-similar solution characterized by \textit{echoing}, a periodic self-similarity in logarithmic spacetime scales. Intuitively, it means the solution repeats its structure at progressively smaller scales with a fixed period. This leads to a critical exponent that governs the scaling of black hole mass near the threshold of its formation. 

Moreover, it displays universality features where the boundary between black hole formation and dispersion is governed by a codimension-one critical surface in phase space. The end states depend only on the positions of the data with respect to the critical submanifold, strongly reminiscent of critical phenomena in statistical physics. The resulting spacetime, now known as a \textit{critical solution} or a \textit{choptuon}, exhibits a point-like naked singularity at its center and a Cauchy horizon forming along its future outgoing null cone~\cite{Gundlach:1995kd, Gundlach:1996eg, Gundlach_2003, Martin-Garcia:2003xgm, Frolov:2003dk}.

Choptuik's discovery presents an immediate counterexample to cosmic censorship. Its universality features, extending beyond Christodoulou's unstable bounded variation solutions,\footnote{One may view the choptuon as the numerical counterpart of Christodoulou’s counterexample. Intriguingly, while they both describe naked singularities with some overlapping features, the two solutions cannot coincide since Christodoulou’s solution is continuously self-similar, whereas the choptuon exhibits discrete self-similarity. We emphasize that the existence of a choptuon comes from numerical inference (see, however, a computer-assisted proof~\cite{Reiterer:2012hnr} establishing the existence of a real-analytic solution interpreted as Choptuik's solution). An analytic or numerical connection between the two cases remains to be understood~\cite{Gundlach:2002sx, Giamb__2024}.} raise intricate questions about what constitutes ``generic" initial data, a notion central to the interpretation of the cosmic censorship, since genericity critically depends on which quantities are held fixed and which are allowed to vary~\cite{Giamb__2024}. Nevertheless, Hawking eventually conceded and refined cosmic censorship to exclude these ``technicalities."

The research paradigm is now referred to as \textit{critical gravitational collapse} and will be the focus of this work, where we provide precise characterizations in Section~\ref{sec:criticalcollapse}. A particularly relevant universal feature is the threshold of black hole formation in a one-parameter family of initial data labeled by $p$. In the so-called Type II collapse, arbitrarily small black holes form near the threshold, exhibiting a characteristic scaling behavior for the mass:\footnote{Type I, by contrast, involves a finite mass gap. The black hole mass serves as an ``order parameter."}
\be
M \propto (p-p^\ast)^\delta,
\ee
in the supercritical regime $p> p^\ast$ with $p^\ast$ denoting the critical point. The critical exponent $\delta$ is universal in the sense that it is independent of the specific one-parameter family $p$, depending only on the type of collapsing matter. Choptuik's critical solution with a free massless scalar field is precisely of Type II with $\delta=0.37$. Furthermore, the near-critical spacetime has some defining properties in the large-curvature region. It is regular and approaches a discretely or continuously self-similar solution. When one perturbs slightly away from the critical point, there exists exactly one growing mode associated with black hole formation. This means that, despite being finely tuned, it is the most stable solution in the sense that it is only unstable to a single mode~\cite{Evans:1994pj, Maison:1995cc, Gundlach:1995kd,  Gundlach:1996eg, Gundlach:1996vv, Hod:1996az, Martin-Garcia:1998zqj}. Since Choptuik's discovery, this framework has been generalized to a wide variety of matter models in spherical symmetry~\cite{Bartnik:1988am, PhysRevLett.64.2844, PhysRevLett.66.1659, Evans:1994pj, Maison:1995cc, Gundlach:1995kd, Hamade:1995jx, Eardley:1995ns, Hirschmann:1995pr, Gundlach:1996vv, Hara:1996mc, Hod:1996ar,vanPutten:1996mt, Liebling:1996dx, Gundlach:1996eg, Gundlach:1996je, Choptuik:1996yg, Choptuik:1997rt, Brady:1997fj, Rein:1998uf, Neilsen:1998qc, Liebling:1998xu, Bizon:1998kq, Bizon:1998ix, Liebling:1999ke, Choptuik:1999gh, Garfinkle:1999zy, Husain:2000vm, Pretorius:2000yu, Hawley:2000dt, Olabarrieta:2001wy, Martin-Garcia:2001jtc, Birukou:2002kk, Millward:2002pk, Garfinkle:2003jf, Ventrella:2003fu, Rinne:2014kka, Rocha:2018lmv}, revealing extremely rich phase structures depending on the number of dimensions and the type of matter content.

However, this cannot be the whole story. Type II collapse involves ever-increasing curvature at progressively smaller scales in the self-similar regime, where quantum effects are expected to become significant. This provides a natural setting for exploring quantum corrections from collapsing matter before full quantum gravity becomes necessary to resolve the singularity. While the classical theory predicts the formation of a naked singularity, the key open question is whether quantum effects intervene to prevent it, saving cosmic censorship at the end of the day. Despite the fundamental importance of this issue, it has received only sporadic attention in the literature~\cite{Strominger:1993tt, Kiem:1994sh, Zhou:1995zj, Peleg:1996ce, Bose:1996pi, Ayal:1997ab, Chiba, Brady:1998fh, Berczi:2020nqy, Guenther:2020kro, Berczi:2021hdh, Hoelbling:2021axl, Varnhorst:2023dew, Moitra:2022umq, Husain:2008tc, Ziprick:2009nd, Benitez:2020szx, Benitez:2021zjs}, and a clear, conclusive treatment of quantum effects in critical collapse remains elusive. The phase structure of the semiclassical system near the threshold is expected to be even richer than in the classical theory, but it is still poorly understood.

There remain several open conceptual and technical points in the existing literature. At the end of Section~\ref{sec:onelooptheory} as well as Appendix~\ref{sec:appendixD}, we will discuss why some seemingly promising approaches face important limitations in the present setting, and we will explain how our one-loop formalism addresses these issues. Here we provide a brief summary and highlight the key distinctions.

\subsubsection*{Challenges of modeling quantum effects in critical collapse}

The foremost fact is that most quantum analyses of critical collapse have been confined to specific models with simplifying assumptions that do not capture essential aspects of the problem. Crucial quantum corrections can be effectively encoded in simple two-dimensional dilaton gravity models~\cite{Strominger:1993tt, Zhou:1995zj, Bose:1996pi, Peleg:1996ce, Chiba, Moitra:2022umq}, as we will also show in Section~\ref{sec:onelooptheory}. A common strategy for incorporating quantum corrections, particularly successful in black hole spacetimes, is to rely on the trace anomaly, which is one-loop exact and ultraviolet in origin~\cite{Christensen}. However, trace anomaly accounting for a minimally coupled massless scalar, which is also conformal in two dimensions, does not capture the higher-dimensional counterpart.\footnote{In fact, even the gravitational sector is not fully suitable for studies~\cite{Strominger:1993tt, Zhou:1995zj, Bose:1996pi, Peleg:1996ce} based on the CGHS model~\cite{Callan:1992rs}, since it does not descend from Einstein gravity in higher dimensions.} One could instead consider the trace anomaly in $3+1$ dimensions, but this applies only to conformally coupled matter~\cite{Brady:1998fh}. As we will explain in Section~\ref{sec:onelooptheory} and Appendix~\ref{sec:appendixD}, such treatments face additional subtleties in critical spacetimes and, as usually implemented, do not straightforwardly apply. 

Regardless, these approaches do not directly address two particularly relevant scenarios: the collapse of a free, massless scalar field in $3+1$ dimensions, and the particularly intriguing $2+1$-dimensional case explored in Section~\ref{sec:quantum2+1}, where the trace anomaly remains unknown in both settings. Moreover, different models adopt renormalization prescriptions that are not always directly comparable and may involve additional assumptions (for example, in some cases leading to violations of conservation laws~\cite{Ayal:1997ab}), which can complicate the interpretation and cross-comparison of results. An alternative is to appeal to numerical simulations with quantum sources. However, such approaches often adopt particular quantum states to facilitate computations, whose physical selection criteria may not be uniquely fixed within the framework~\cite{Berczi:2020nqy, Guenther:2020kro, Berczi:2021hdh, Hoelbling:2021axl, Varnhorst:2023dew, Zahn:2025tnh}.

Second, in many such models, the semiclassical source is not restricted to the backreaction only from the collapsing matter itself. Instead, they typically introduce external sources into the quantum expectation value of the stress-energy tensor $\langle T_{\mu \nu} \rangle$, effectively introducing additional incoming or outgoing energy. These sources do not originate from vacuum polarization of the collapsing field but rather behave analogously to Hawking fluxes in black hole spacetimes. However, such fluxes are generally expected to be subleading for the near-threshold dynamics governing critical collapse, which concerns the mass scaling of the first-formed or earliest marginally trapped surface.\footnote{See, however, a series of works initiated by Vaz and Witten~\cite{Vaz:1993eg, Vaz:1995rn, Vaz:1995ig, Vaz:1996kh, Barve:1998ad, Vaz:1998gd, Barve:1998tv} that studied the Hawking effects in the presence of a naked singularity in certain two-dimensional dilaton gravity and self-similar collapse models, in the same spirit of defending cosmic censorship through quantum theory. Their results are also qualitatively similar in the sense that quantum effects must be large when there is a naked singularity, and exotic behaviors occur in contrast to those of black holes.} This issue is fundamentally tied to the choice of quantum state. In Section~\ref{sec:onelooptheory}, we will argue that the physically appropriate state is one that reflects only the vacuum polarization of the collapsing scalar field, without introducing externally prescribed fluxes at null infinities. Such a state must be asymptotically Minkowskian, similar to the Boulware state in a black hole spacetime~\cite{Boulware1975}, ensuring that $\langle T_{\mu \nu} \rangle$ vanishes near both past and future null infinities. We will demonstrate in Sections~\ref{sec:quantum2+1} and \ref{sec:collapse3+1} that such a state arises naturally and is fixed unambiguously within our formalism, given physical regularity criteria.
 
The third issue concerns the nature of critical behavior in quantum settings: whether the collapse remains Type II, undergoes a phase transition to Type I with a mass gap, or exhibits qualitatively new features remains under debate. Some studies argue for the emergence of a mass gap~\cite{Bose:1996pi, Peleg:1996ce, Brady:1998fh}, while others find no such evidence~\cite{Strominger:1993tt, Zhou:1995zj, Ayal:1997ab}, and yet others report only minor deviations from the classical picture~\cite{Berczi:2020nqy, Guenther:2020kro, Berczi:2021hdh, Hoelbling:2021axl, Varnhorst:2023dew}. Likewise, whether self-similarity is preserved under quantum effects remains open, with studies supporting it~\cite{Ayal:1997ab, Brady:1998fh, Berczi:2020nqy, Guenther:2020kro, Berczi:2021hdh, Hoelbling:2021axl, Varnhorst:2023dew, Moitra:2022umq} and against it~\cite{Strominger:1993tt, Zhou:1995zj, Bose:1996pi, Peleg:1996ce}. The behavior of the quantum Lyapunov exponent is similarly inconclusive: in some models, it appears to be kinematically determined~\cite{Strominger:1993tt, Kiem:1994sh, Brady:1998fh}, while in others it is sensitive to the details of quantum backreaction or specific choices of model parameters, which suggests that its universality may depend on the regime and assumptions. Conclusions ranging from claims that quantum effects are so large that critical phenomena may cease to exist altogether~\cite{Strominger:1993tt, Zhou:1995zj, Chiba}, inducing a phase transition~\cite{Bose:1996pi, Peleg:1996ce, Ayal:1997ab, Brady:1998fh}, or are merely small corrections in the near-critical regime~\cite{Berczi:2020nqy, Guenther:2020kro, Berczi:2021hdh, Hoelbling:2021axl, Varnhorst:2023dew}. 
Moreover, due to the lack of exactly solvable models, many of these analyses rely on numerical methods, where the extracted critical exponent could be sensitive to the numerical scheme. This diversity of conclusions is plausibly related to the aforementioned differences in setups and assumptions.\footnote{On the other hand, loop quantum gravity-inspired models, which aim to incorporate genuine quantum gravity effects, also yield divergent conclusions about self-similarity and mass gap formation~\cite{Husain:2008tc, Ziprick:2009nd, Benitez:2020szx, Benitez:2021zjs}. }

We therefore conclude that a fully self-consistent semiclassical treatment of critical collapse remains to be developed. This gap is particularly significant given that quantum effects generically violate the classical energy conditions underpinning many foundational results in general relativity, including theorems on geodesic completeness, horizon formation, and singularity development.\footnote{Several recent studies have argued for a semiclassical extension of classical theorems in General Relativity \cite{Bousso:2014sda, Bousso:2015mna, Bousso:2019var, Bousso:2019bkg, Shahbazi-Moghaddam:2022hbw}. Notably, in \cite{Wall:2010jtc, Bousso:2022cun, Bousso:2022tdb, Bousso:2025xyc} they show that a singularity is still present within a trapped region, even with quantum effects included. This result emphasizes that perturbative quantum effects are generically not enough to resolve a singularity.}

\subsubsection*{Main methods, results, and their implications}

Motivated by this gap, we develop an anomaly-based semiclassical framework, formulated in the path integral approach and tailored to dynamical self-similar critical spacetimes, that enables explicit analytic control of one-loop backreaction in tractable models of scalar-field collapse in Einstein gravity. This framework captures the dominant $s$-wave sector via spherical dimensional reduction, and we use it to obtain explicit results that address several of the questions outlined above. The most nontrivial features we incorporate are \textit{non-conformal matter} and the \textit{time-dependent} critical spacetime background. We obtain a generic $1+1$-dimensional dilaton gravity from a higher-dimensional Einstein-scalar system that features a non-minimally dilaton-coupled matter sector, which has been the subject of extensive study~\cite{Mukhanov_1994, Bousso:1997cg, Mikovic_1998, Balbinot_1999, Balbinot_1999_2, Buric:1998xv, Kummer_1999, Kummer_1999-1, Gusev:1999cv, Balbinot:2000at, Balbinot_2001, Balbinot:2002bz, Grumiller:2002nm, Fabbri:2003vy, Hofmann:2004kk, Hofmann:2005yv}, owing to its strong physical motivation from higher-dimensional origins. Classically, this theory is equivalent to the $s$-wave sector of the Einstein-scalar system. However, at the quantum level, previous studies have yielded confusing or even unphysical results when applied to black hole spacetimes. In a recent work~\cite{Wu:2023uyb}, we constructed a consistent one-loop quantum effective theory for this system, recovering physical results in the black hole background and laying the groundwork for a reliable semiclassical treatment of critical collapse. 

To outline the calculations carried out in this paper, the key observation is that the two-dimensional matter theory is no longer conformally invariant due to the dilaton coupling, though it remains invariant under local Weyl rescalings. From a one-loop heat kernel analysis, one can extract its state-independent trace anomaly, which in turn allows us to construct an anomaly-induced one-loop effective action. However, unlike the minimally coupled scalar field, where the one-loop theory is uniquely given by the non-local Polyakov action~\cite{Polyakov:1981rd}, the effective action in our case is highly non-unique. It contains additional non-local, Weyl-invariant terms that cannot be written in closed form. This inherent issue has contributed to confusion in previous applications to black hole spacetimes. Importantly, this feature reflects a degree of freedom in specifying the quantum state of the matter field and is therefore sensitive to boundary conditions~\cite{Karakhanian:1994gs, Jackiw:1995qh, Navarro-Salas:1995lmi}.

As briefly alluded to earlier, there exists a natural choice of quantum state in critical spacetimes: one that exhibits no quantum fluxes near null infinity and captures only the vacuum polarization of the collapsing matter. We will demonstrate through specific examples in $2+1$ and $3+1$ dimensions, that such a state automatically arises from having a regular $\langle T_{\mu \nu} \rangle$ in the defined domain of critical spacetimes, closely analogous to how classical perturbation modes are determined. As explicit demonstrations of the formalism, we focus on two known closed-form solutions of the Einstein-scalar system that serve as candidate critical spacetimes: the Garfinkle solution in $2+1$ dimensions~\cite{Garfinkle:2000br} and the Roberts solution in $3+1$ dimensions~\cite{Roberts:1989sk}. While their classical properties are well understood, we will explain why these solutions do not, in fact, meet the strict criteria for critical collapse, but how they nevertheless remain closely connected to true critical systems~\cite{Garfinkle:2002vn, Jalmuzna:2015hoa, Brady:1994aq, Brady:1994xfa, Frolov:1997uu, Frolov:1998tq, Frolov:1999fv}. We will see that the general lessons extracted from the examples apply to general self-similar critical collapse systems.

Remarkably, the closed-form nature of these spacetimes enables us to carry out the one-loop quantum corrections analytically, yielding explicit expressions for both $\langle T_{\mu \nu} \rangle$ and the semiclassical geometries. Perhaps most surprisingly, the resulting stress-energy tensor exhibits a universal growing behavior that can be interpreted as a quantum instability near criticality, characterized by a Lyapunov exponent determined purely kinematically. This opens a path toward understanding universal features of quantum corrections in higher-dimensional scalar collapse, where such exponents are expected to depend only on the spacetime dimensionality. By studying the interplay between this quantum growing mode and the classical perturbations near the black hole threshold, we can trace the dynamics of apparent horizon formation numerically using a quasi-local mass function. Our perturbative results suggest the emergence of a finite mass gap, indicative of \textit{a phase transition from the classical Type II to a quantum-modified Type I behavior.} In a sense, this is expected: quantum effects introduce a \textit{dynamically relevant} scale, a hallmark of classical Type I collapse. However, it remains open whether it corresponds precisely to the same Type I behavior observed in classical systems without studying a potential metastable soliton phase, which we will comment in Section~\ref{sec:discussion}. Nevertheless, the calculation provides a concrete mechanism by which quantum effects can trigger apparent horizon formation, shielding the naked singularities predicted by classical theory and offering a potential quantum resolution to cosmic censorship. It therefore marks an important step toward a more complete understanding of quantum critical collapse.

Before closing, we emphasize the scope and limitations of our analysis, and briefly place our results in the context of complementary approaches in the literature. 

In particular, we stress that a number of complementary first-principles approaches---including treatments based on canonical quantization and alternative renormalization prescriptions outlined earlier~\cite{Strominger:1993tt, Kiem:1994sh, Zhou:1995zj, Peleg:1996ce, Bose:1996pi, Ayal:1997ab, Chiba, Brady:1998fh, Berczi:2020nqy, Guenther:2020kro, Berczi:2021hdh, Hoelbling:2021axl, Varnhorst:2023dew, Moitra:2022umq, Husain:2008tc, Ziprick:2009nd, Benitez:2020szx, Benitez:2021zjs, Zahn:2025tnh} (in particular, the rigorous analyses of~\cite{Brady:1998fh, Zahn:2025tnh}; see Appendix~\ref{sec:appendixD} for a detailed comparison and discussion)---are pursued under different approximations and address closely related questions from different angles. Our approach is formulated in the path integral framework, and its main practical advantage is that the relevant quantum state is fixed concretely by physical regularity criteria, which in turn enables explicit analytic control of semiclassical backreaction in near-critical spacetimes.\footnote{After the publication of our manuscript,~\cite{Zahn:2025tnh} presented a complementary analysis based on Hadamard renormalization for quantum fields on self-similar Roberts and Hayward backgrounds. Their results support our finding that the quantum Lyapunov exponent is kinematically fixed as $\omega_q=D-2$, while also emphasizing subtleties associated with spherical dimensional reduction, the dimensional reduction anomaly, and the use of the two-dimensional trace anomaly in reconstructing stress-energy components. Because the two treatments adopt different assumptions and approximations, some quantitative expressions in specific spacetimes differ. We clarify these points when introducing our one-loop formalism in Section~\ref{sec:onelooptheory}, and provide a more detailed comparison of assumptions and regimes of applicability in Appendix~\ref{sec:appendixD}.} 

At the same time, we work within the $s$-wave sector, and our main qualitative conclusions, most notably the emergence of a universal growing quantum mode and the associated mass gap, should be understood as statements within this controlled setup. A complete treatment of higher angular modes is beyond our present scope (but has been addressed in some recent works~\cite{Berczi:2020nqy, Guenther:2020kro, Berczi:2021hdh, Hoelbling:2021axl, Varnhorst:2023dew}); nevertheless, we provide a systematic discussion of plausible ways higher-$l$ sectors could affect the results, and we explain why---under the same regularity and boundary conditions that restrict admissible initial data in the gauge-invariant linear analysis---higher-$l$ modes are expected to remain subleading in Section~\ref{sec:discussion}.

A companion Letter highlighting the general lessons and phenomenological implications appears simultaneously~\cite{Tomasevic:2025kqy}.

The structure of the paper is as follows. In Section~\ref{sec:criticalcollapse}, we provide a precise characterization of classical critical collapse, which sets the stage for the semiclassical analysis. In Section~\ref{sec:onelooptheory}, we formulate the quantum one-loop theory relevant to scalar field collapse. This framework is then applied to the $2+1$- and $3+1$-dimensional cases in Sections~\ref{sec:quantum2+1} and \ref{sec:collapse3+1}, respectively. We finish by describing a few subtle issues and outlining future directions in Sections~\ref{sec:discussion} and \ref{sec:outlook}. In Appendix~\ref{sec:horizonGarfinkle}, we present numerical studies for the $n=4$ Garfinkle spacetime and discuss properties for general $n$. In Appendix~\ref{sec:WeylGarfinkle}, we discuss a new $1+1$-dimensional solution exhibiting rich global structures qualitatively similar to a critical spacetime, and use it as an example to illustrate nontrivial properties of quantum field theory in curved background that were not known in general. In Appendix~\ref{sec:Hayward}, we examine an extreme example of critical spacetime to highlight how global causal structure, in addition to self-similarity, influences the choice of quantum state. In Appendix~\ref{sec:appendixD}, we offer remarks on canonical quantization approaches to critical collapse and compare them with our path integral framework.

\section{Critical gravitational collapse: overview and global structure}
\label{sec:criticalcollapse}

In this section, we provide a self-contained overview of classical critical gravitational collapse, with an emphasis on its defining criteria, universal features, self-similarity, and global causal structure. These classical aspects form the foundation upon which our semiclassical considerations will build. Readers already familiar with classical critical collapse may skip this section and proceed to Section~\ref{sec:onelooptheory} on the quantum formulation.

\subsubsection*{Defining criteria of critical collapse}

In Choptuik’s canonical setup~\cite{PhysRevLett.70.9}, one considers the $s$-wave sector of $3+1$-dimensional Einstein-scalar system, evolving with incoming, spherically symmetric scalar wave packets. These wave packets are energy pulses with compact support that are turned on at some point along the past null infinity. For a sufficiently weak incoming wave, the pulse simply reflects off the origin and disperses as an outgoing wave. However, the behavior changes as the initial amplitude $p$ increases. Above a threshold $p>p^\ast$, the incoming wave crosses its own Schwarzschild radius before reaching the origin and forms a black hole. The spacetime exhibits discrete self-similarity characterized by an echoing period $\Delta \approx 3.44$, manifesting as a periodic structure in logarithmic proper time. This corresponds to scale invariance under rescalings $\tau \to \tau e^{- \Delta}$, where $\tau=t^\ast-t$ is the proper time to the accumulation point $t^\ast$, associated with the formation of the critical black hole.

Choptuik systematically explored several one-parameter families of initial data by evolving the system with many different values of the parameter $p$. The black hole formed from \textit{marginally supercritical} data exhibits a universal scaling law for its mass $M \propto (p-p^\ast)^\delta$ with $\delta \approx 0.37$, independent of the choice of initial data family. The limiting case $p \to p^\ast$ is a zero-mass black hole, which can be viewed as a collapsing ball of scalar field energy where inward collapse is exactly balanced by radiative loss. This limiting solution represents a \textit{naked singularity}, visible from future null infinity.

\begin{figure}
    \centering
    \includegraphics[width=0.4\linewidth]{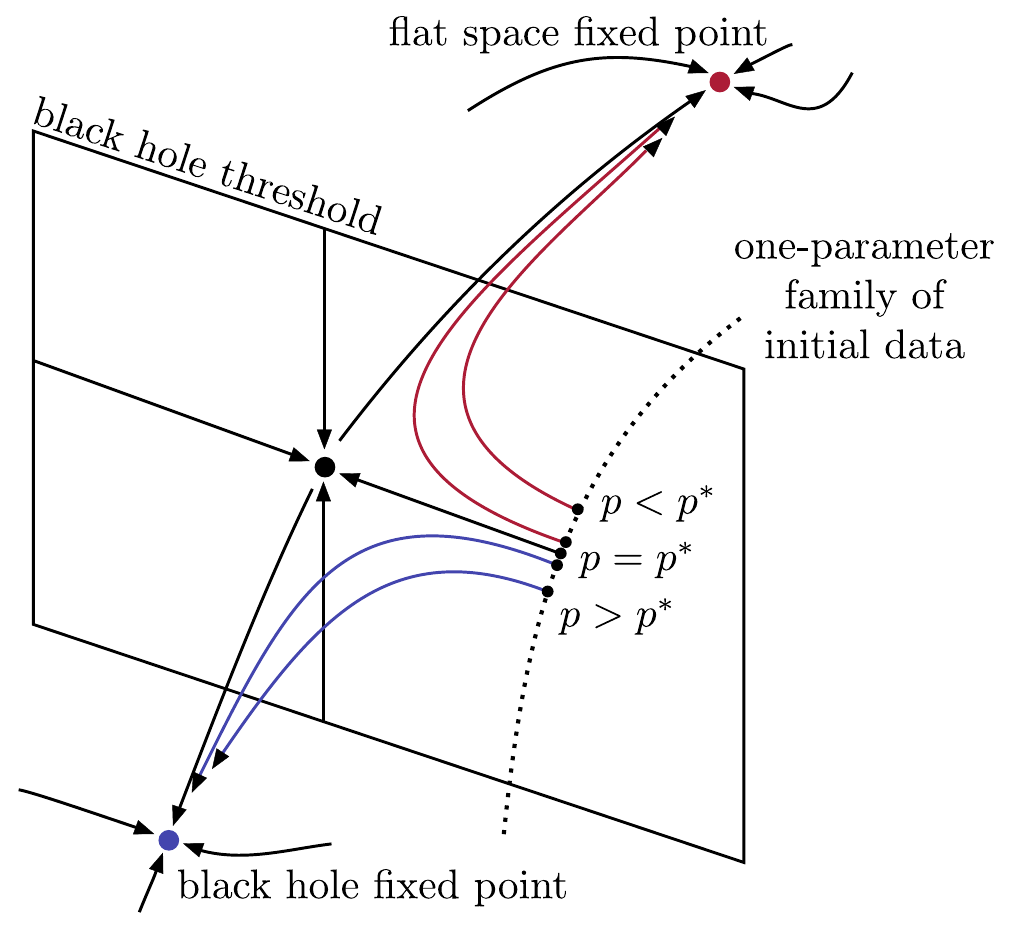}
    \includegraphics[width=0.48\linewidth]{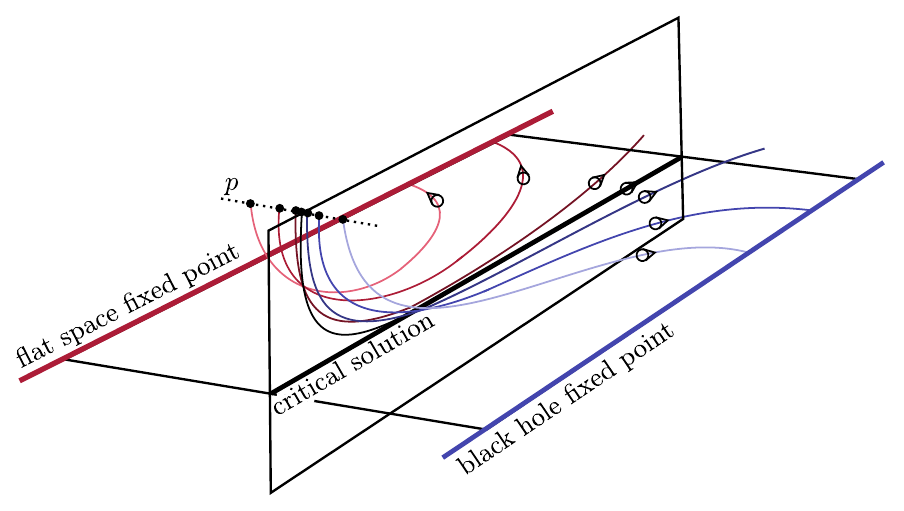}
    \caption{In the left panel, we depict the generic phase space structure near the threshold of black hole formation. Each point represents an initial data set, with arrows indicating the corresponding solution curves. A one-parameter family of initial data, labeled by $p$, intersects the threshold and evolves toward the critical spacetime at $p=p^\ast$. The right panel offers a different perspective, where the direction perpendicular to the threshold represents the global scale. In this view, the scale-invariant Type II critical spacetime appears as a straight solid line. Only precisely fine-tuned initial data will asymptotically approach the critical solution with decreasing scale. Nearby data points, although initially drawn toward the critical solution, eventually deviate, leading either to black hole formation or to dispersion.}
    \label{fig:critical1}
\end{figure}

This corresponds to Type II critical phenomena, whose defining criteria are outlined below. Consider a one-parameter family of smooth initial data in Einstein gravity, labeled by a parameter $p$, interpolating between configurations that lead either to black hole formation or to dispersion. There exists a critical value $p^\ast$ at the threshold of black hole formation, around which \textit{critical phenomena} emerge:
\begin{itemize}
    \item For marginally supercritical $p > p^\ast$,  a black hole forms with mass scaling as
    \be
    M \propto (p-p^\ast)^\delta,
    \ee
    where $\delta$ is the critical exponent.
    \item The exponent $\delta$ is universal with respect to the choice of initial data family. That is, it depends only on the type of collapsing matter, not on how the parameter $p$ is embedded in the space of initial data.
    \item Near-critical spacetimes evolve toward a self-similar solution, known as the critical solution, which is also universal and characterized by either continuous self-similarity (CSS) or discrete self-similarity (DSS).
\end{itemize}
See Figure~\ref{fig:critical1} for an intuitive illustration of the phase space structure. These features have been confirmed in a wide range of matter models under spherical symmetry, and in some axisymmetric cases as well, with extensions incorporating electric charge, angular momentum, and varying spacetime dimensionality~\cite{Bartnik:1988am, PhysRevLett.64.2844, PhysRevLett.66.1659, Evans:1994pj, Maison:1995cc, Gundlach:1995kd, Hamade:1995jx, Eardley:1995ns, Hirschmann:1995pr, Gundlach:1996vv, Hara:1996mc, Hod:1996ar,vanPutten:1996mt, Liebling:1996dx, Gundlach:1996eg, Gundlach:1996je, Choptuik:1996yg, Choptuik:1997rt, Brady:1997fj, Rein:1998uf, Neilsen:1998qc, Liebling:1998xu, Bizon:1998kq, Bizon:1998ix, Liebling:1999ke, Choptuik:1999gh, Garfinkle:1999zy, Husain:2000vm, Pretorius:2000yu, Hawley:2000dt, Olabarrieta:2001wy, Martin-Garcia:2001jtc, Birukou:2002kk, Millward:2002pk, Garfinkle:2003jf, Ventrella:2003fu, Rinne:2014kka, PhysRevLett.70.2980, Rostworowski:2025hvj, Garfinkle:1998tt, Choptuik:2003ac, Baumgarte:2018fev, Gundlach:2024eds}. Nevertheless, it remains an open question how generic critical phenomena truly are across different matter models and beyond spherical symmetry~\cite{Gundlach:2025yje}.

In this work, we focus on Einstein gravity minimally coupled to a free, massless scalar field $f$ under spherical symmetry with $D=d+1$ spacetime dimensions
\be \label{eq:Einsteinscalar}
S=\int d^D x \sqrt{-g} \bigg[\frac{1}{16 \pi G_N}(R-2 \Lambda)-\frac{1}{2}(\nabla f)^2 \bigg], 
\ee
\be
G_{\mu \nu}+\Lambda g_{\mu \nu}=8 \pi G_N T_{\mu \nu}, \quad T_{\mu \nu}=\nabla_\mu f \nabla_\nu f-\frac{1}{2}g_{\mu \nu} (\nabla f)^2, \quad \Box f=0,
\ee
\be
ds^2=g_{ab}(x^c)dx^a dx^b+r^2(x^c)d \Omega^2_{D-2}.
\ee
The action, equations of motion, and spherically symmetric metric ansatz together define what we refer to as the \textit{Einstein–scalar system}. Self-similarity will later be introduced as an additional symmetry imposed on the system. The main motivation for studying critical collapse lies in the fact that self-similar critical solutions provide a mechanism for reaching arbitrarily large spacetime curvature, leading, in the critical limit, to a naked singularity.

\FloatBarrier

Let us now describe the general characteristics of critical collapse in the dynamical picture. Critical collapse admits a natural formulation in terms of an infinite-dimensional dynamical system, where points in phase space correspond to initial data sets, comprising the spatial three-metric, extrinsic curvature, and matter fields that satisfy the Einstein constraint equations. The evolution of these data under a suitable gauge defines trajectories through this phase space. The solution curves of the dynamical system can be foliated by specific Cauchy surfaces of constant time $t$.

For a massless scalar field in spherical symmetry, the end states are limited to either black hole formation or complete dispersion. Consequently, the phase space divides into two distinct regions, separated by a \textit{critical surface} of codimension-one. Trajectories lying on this surface evolve toward an attracting fixed point, called the critical point, which corresponds to a self-similar solution. The attractor nature of the critical solution can be understood in its linear perturbation spectrum: the tangent space to the critical surface is spanned by an infinite number of decaying perturbation modes, yet there is a single growing mode transverse to the surface that drives the evolution away from criticality. This structure underlies the universality and scaling behavior observed near the black hole threshold~\cite{Evans:1994pj, Maison:1995cc, Gundlach:1995kd, Gundlach:1996eg, Gundlach:1996vv, Hod:1996az, Martin-Garcia:1998zqj}, and motivates an additional defining criterion:
\begin{itemize}
\item The critical solution possesses exactly one growing mode transverse to the critical surface; all other perturbation modes decay.
\end{itemize}

Martín-García and Gundlach performed a detailed analysis of non-spherical perturbations around the scalar field critical solution by solving a linear eigenvalue problem under regularity assumptions~\cite{Martin-Garcia:1998zqj}. They found that the only growing perturbation mode is the known spherically symmetric one; all non-spherical modes decay. This strongly suggests that the critical solution acts as a codimension-one attractor not only in the space of spherically symmetric data but also in a finite neighborhood of general initial data, modulo linearization stability.\footnote{In particular, angular momentum enters only at second order when perturbing the scalar field around spherical symmetry. All such angular perturbations are shown to decay, and a critical exponent $\mu \simeq 0.76$ is derived for the angular momentum in the massless scalar field case~\cite{Garfinkle:1998tt}. However, see~\cite{Choptuik:2003ac, Baumgarte:2018fev, Gundlach:2024eds} for indications of an additional non-spherical $l=2$ growing mode, which may suggest a subtle tension.}

\begin{figure}
    \centering
    \includegraphics[width=0.9\linewidth]{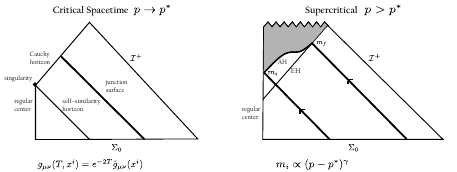}
    \caption{Penrose diagrams of the exact critical spacetime and the supercritical regime featuring black hole formation are shown, with $\Sigma_0$ denoting the initial data surface. In the left panel, the critical solution exhibits a naked singularity, with a Cauchy horizon (CH) emanating along its future light cone. Beyond this CH, the spacetime admits no unique continuation. The geometry is self-similar in the interior region, obeying $g_{\mu \nu}(T,x^i) =e^{-2T}\tilde{g}_{\mu \nu}(x^i)$, with the self-similarity horizon (SSH) lying along the past light cone of the singularity. This region is matched to an asymptotically flat exterior across the junction surface. In the right panel, the supercritical regime depicts matter collapsing to form a black hole. The event horizon (EH) is a null surface, determined only when the infalling matter has stopped, with the final mass $m_f$ independent of $p$. In contrast, the apparent horizon (AH), upon its initial formation, exhibits the characteristic Choptuik scaling relation $m_i \propto (p - p^\ast)^\gamma$.}
    \label{critsupercrit}
\end{figure}

\subsubsection*{Self-similarity and universal features}

The critical solutions often exhibit additional symmetries and can be classified into two main types, known as Type I (with a mass gap) and Type II (no mass gap).\footnote{There is also a ``Type III" critical collapse observed in the Einstein–Yang–Mills system~\cite{Choptuik:1999gh, Rinne:2014kka}, where both subcritical and supercritical branches evolve into black holes with different masses and scaling behaviors.} 

\paragraph{Type II collapse.} Here, the critical solution is self-similar, and black holes of arbitrarily small mass form near the threshold. We distinguish between the scale invariance of the field equations, which holds in the absence of dimensionful parameters, and the self-similarity of the solution, which refers to the spacetime itself admitting either continuous or discrete scale symmetry. Specifically, the critical solution may exhibit either CSS or DSS, where the latter is often referred to as scale-periodicity. 

For instance, the critical solution for a spherically symmetric perfect fluid exhibits CSS~\cite{Evans:1994pj, Maison:1995cc, Neilsen:1998qc}. A CSS spacetime admits a
homothetic Killing vector field $\xi$, defined by
\be
\mathcal{L}_\xi g_{\mu \nu}= 2 g_{\mu \nu},
\ee
where the constant $2$ is a convention that normalizes $\xi$. By picking coordinates $x^\mu=(T, x^i)$ adapted to the symmetry, so that
\be
\xi = -\frac{\pa}{\pa T}
\ee
with the minus sign yet another convention assuming that smaller scales are in the future. The integral curves of $\xi$ therefore provide a preferred fibration of the spacetime, with the metric taking the form
\be \label{eq:CSSmetric}
g_{\mu \nu} (T, x^i)=\ell^2 e^{-2 T} \tilde{g}_{\mu \nu} (x^i).
\ee
Here, $T$ can be interpreted as the negative logarithm of a spacetime scale, and $x^i$ as angles around the singular spacetime point $T = \infty$. The scaling parameter $\ell$ in \eqref{eq:CSSmetric} is an arbitrary length that fixes the overall size of the spacetime, rendering the coordinates $(T, x^i)$ dimensionless. Since the classical field equations are themselves scale‐free, no particular choice of $\ell$ is preferred, and physics must be invariant under this constant rescaling. This implies, however, that the physical radius acquires its dimension through $r_{\text{phys}} = \ell r$. It is important to highlight this point when incorporating one-loop quantum effects associated with a new scale $\hbar$, since all perturbations must then be expressed in terms of the dimensionless ratio $\frac{\hbar}{\ell^{D-2}}$ in geometrized units.

\begin{figure}[t!]
    \centering
    \includegraphics[width=0.35\linewidth]{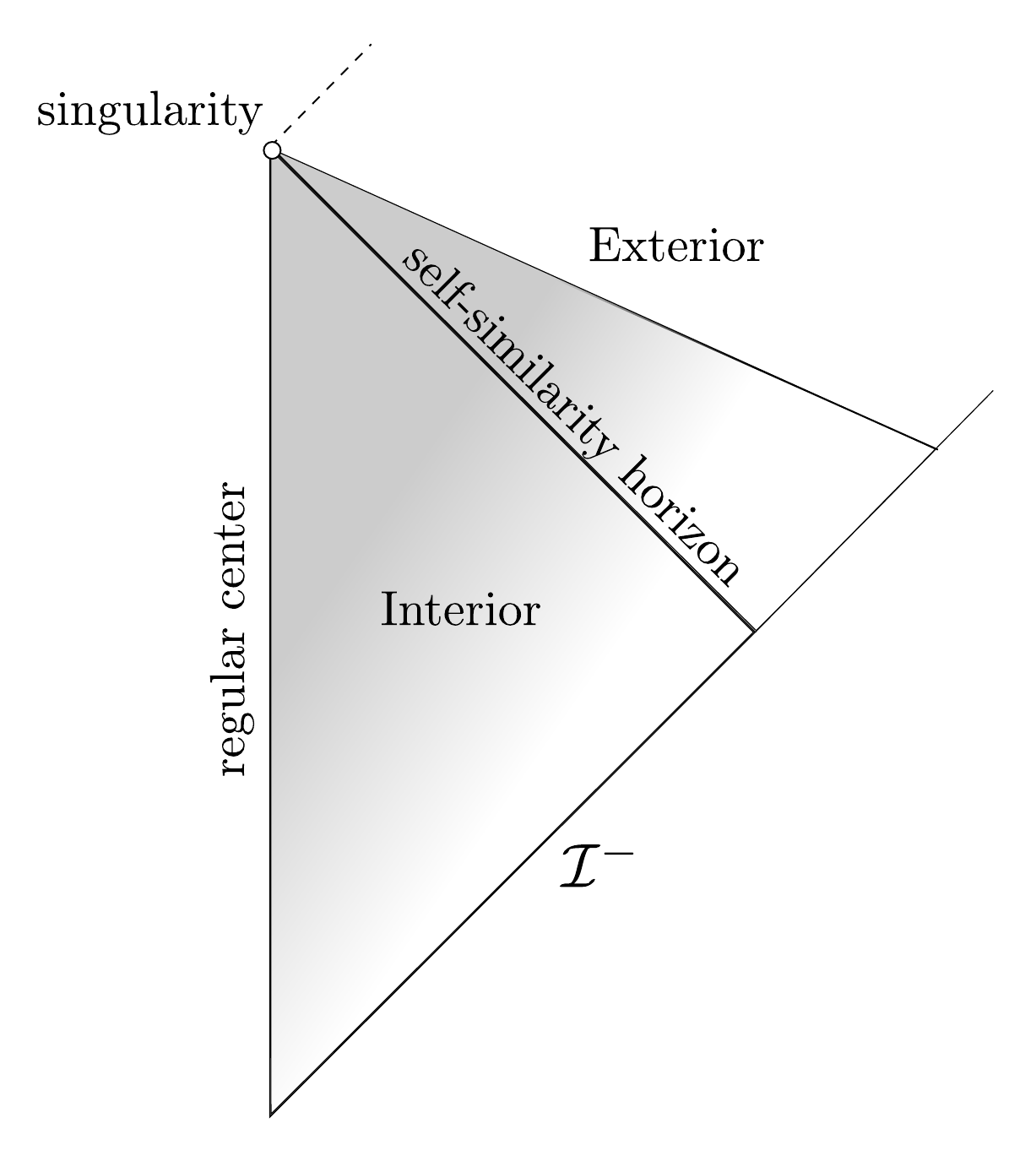}
    \includegraphics[width=0.35\linewidth]{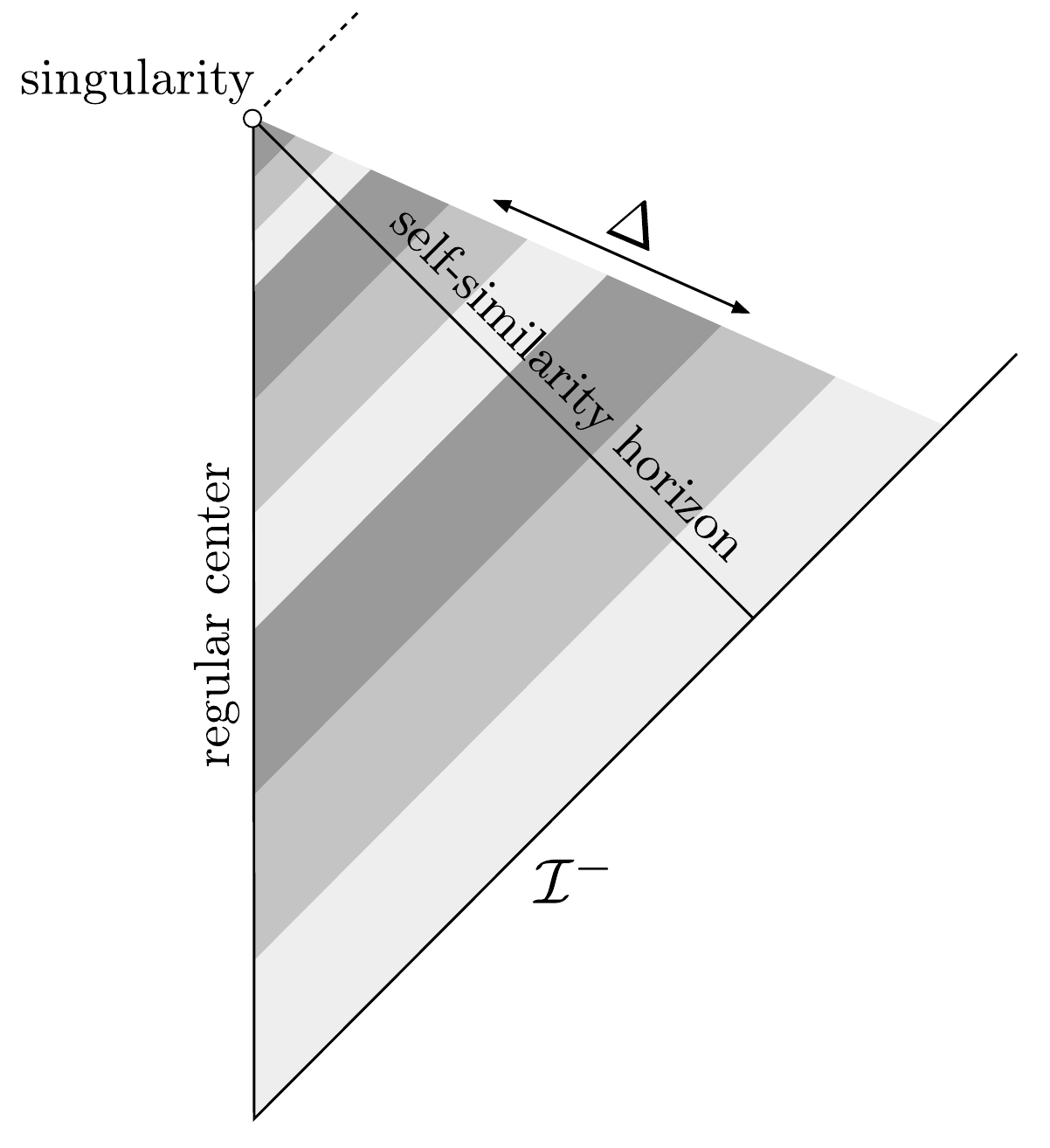}
    \caption{Penrose diagrams of continuously self-similar (CSS, left panel) and discretely self-similar (DSS, right panel) critical spacetimes, each featuring a naked singularity as $T \to \infty$. In a self-similar spacetime, the geometry repeats itself under rescalings of spacetime coordinates. A CSS spacetime possesses a homothetic Killing vector, with the metric $g_{\mu \nu}(T,x^i)$ varying continuously in $T$. In contrast, a DSS spacetime exhibits periodicity in $T$, with a fixed period $\Delta$ that depends on the specific matter model and is typically determined numerically.}
    \label{fig:cssdss}
\end{figure}

On the other hand, systems such as the spherically symmetric massless scalar field and axisymmetric gravitational waves~\cite{PhysRevLett.70.2980, Rostworowski:2025hvj} typically yield DSS.\footnote{A notable exception occurs is the massless scalar field in $2+1$ dimensions, which exhibits CSS instead~\cite{Pretorius:2000yu, Husain:2000vm}; this case will be discussed further in Section~\ref{sec:quantum2+1}.} In adapted coordinates, $g_{\mu \nu} (T, x^i)$ is periodic in $T$ with a period $\Delta$, reflecting a discrete conformal isometry~\cite{Gundlach:1995kd}. See Figure~\ref{fig:cssdss} for Penrose diagrams that distinguish CSS and DSS critical spacetimes.

We emphasize that in the ADM formalism, the lapse and shift functions can be (non-uniquely) chosen so that the coordinates become adapted to self-similarity once the solution develops it~\cite{Garfinkle:1996kt, Garfinkle:1998df, Garfinkle:1999cm}. Then $T$ can be understood both as a time coordinate where constant $T$ surfaces are Cauchy surfaces, and as the logarithm of the overall scale at fixed spatial position $x^i$. 

\FloatBarrier

To understand scaling behavior near criticality, let us consider linear perturbation theory, assuming a CSS critical solution for simplicity. Let $Z$ represent a set of scale-invariant variables, such as $\tilde{g}_{\mu \nu}$ and suitably rescaled matter variables. These variables can typically be identified through dimensional analysis. The critical solution corresponds to $Z(T, x)=Z^\ast(x)$, where linear perturbations around it evolve exponentially in $T$. To linear order, the perturbed solution takes the form
\be
Z(x, T) \simeq Z^\ast (x)+ \sum_{i=1}^\infty C_i(p) e^{\omega_i T} Z_i (x),
\ee
where $C_i(p)$ are amplitudes depending on the initial data, and $\omega_i$ are the corresponding Lyapunov exponents. By definition, there is exactly one exponent with a positive real part (and, in fact, shall be purely real), and we simply denote it by $\omega$.\footnote{In Section~\ref{sec:collapse3+1}, we will analyze the Roberts solution~\cite{Roberts:1989sk, Brady:1994aq, Brady:1994xfa} as the candidate critical solution in $3+1$ dimensions. This solution is CSS and has a \textit{continuous and complex} spectrum of perturbation modes~\cite{Frolov:1997uu, Frolov:1998tq}, which suggests that it cannot be the true critical solution but is nevertheless connected to the true DSS critical spacetime~\cite{Frolov:1999fv}.} When $T \to \infty$, all other perturbations vanish. We always consider such a limit at moderately large but finite $T$, retaining only the single growing unstable mode. This structure leads to the universal scaling law for the black hole mass near criticality:
\be
M \propto e^{-T^\ast} \propto (p-p^\ast)^{1/\omega},
\ee
so that the critical exponent is given by $\delta=1/\omega$. On the other hand, for marginally subcritical data $p<p^\ast$, it is the maximum value of curvature (Ricci scalar), visible from future infinity, that has the scaling behavior
\be
|R|_{\text{max}} \sim (p^\ast-p)^{-2 \gamma}.
\ee
In   $d+1$ dimensions where $d \geq 3$, we have the relations
\be \label{eq:dimensionalrelation}
\delta = \gamma(d-2),
\ee
which essentially comes from dimensional analysis. This relation immediately highlights the $2+1$-dimensional case with $d=2$ as a special case, which we will discuss in Section~\ref{sec:quantum2+1}.

We refer the reader to the review~\cite{Gundlach:2025yje} for a detailed discussion of subtleties including coordinate choices, boundary conditions, phase space variables, and ambiguities in the slicing of Cauchy surfaces. For our purposes, the questions are clear and well-defined when we consider specific examples in Sections~\ref{sec:quantum2+1} and~\ref{sec:collapse3+1}. However, one important clarification concerns the slicing dependence of the black hole mass. Since the apparent horizon is sensitive to the choice of slicing, we typically refer to the black hole mass as that of the first-formed or earliest apparent horizon along certain spacelike or null slices. Additional matter may fall in afterward and increase the final mass. Fine-tuning near the critical point controls the initial mass $m_i \propto (p-p_\ast)^\delta$, while the final mass $m_f$ becomes insensitive to $p$. The key point is that the apparent horizon associated with $m_i$ depends only \textit{weakly} on the slicing.

\paragraph{Type I collapse.} The phase space picture remains the same, yet the nature of the critical solution differs. The critical solution is stationary or time-periodic, rather than self-similar or scale-periodic. It has a finite mass and can be interpreted as a metastable star. The relevant dimensionful quantity that exhibits scaling is not the black hole mass, but rather the lifetime $t_p$ of the intermediate state approximated by the critical solution:
\be \label{eq:lifetime}
t_p = -\frac{1}{\omega} \ln{|p-p^\ast|}+\text{const}.
\ee
Universality in this context means that the black hole mass near the threshold is independent of the initial data family; it is always some fixed fraction of the mass of the stationary critical solution.

Type I phenomena typically arise when a length scale present in the field equations becomes dynamically relevant. Although this scale does not necessarily determine the exact mass of the critical solution, and there may exist a family of critical solutions depending on initial conditions, it explicitly breaks the scale invariance that defines Type II behavior. Conversely, Type II phenomena occur when the field equations either lack any intrinsic scale or when such a scale becomes irrelevant in the near-critical regime. Many systems, such as a massive
scalar field, could exhibit both Type I and Type II behavior in different regions of the initial data space~\cite{Brady:1997fj, Hawley:2000dt, PhysRevLett.66.1659}.

The distinction between Type I and Type II collapse is clearly analogous to phase transitions in statistical mechanics~\cite{Gundlach:2025yje}, when the black hole mass is viewed as an order parameter. Type I collapse, featuring a finite mass gap near the threshold, is akin to a first-order phase transition, where the order parameter changes discontinuously. In contrast, Type II collapse exhibits a continuous scaling of the black hole mass down to zero, reminiscent of a second-order (critical) phase transition, where the order parameter vanishes smoothly and scale invariance emerges. This analogy helps frame gravitational critical phenomena within a broader universality class, where fine-tuning, self-similarity, and critical exponents mirror the behavior of thermodynamic systems near their critical points.

\paragraph{Universality of self-similar solutions.} The massless scalar field system is inherently scale-free, but realistic matter models often introduce dimensionful parameters, and the field equations then do not allow for exactly self-similar solutions. Nevertheless, even in the presence of such parameters, the equations may still admit approximately self-similar solutions at sufficiently small spacetime scales, where the dimensionful parameters become dynamically irrelevant. This behavior can be systematically explored via an expansion in powers of a small parameter $\ell e^{-T}$, where $\ell$ is a characteristic length scale. The zeroth-order solution corresponds to the exactly self-similar critical solution of the $\ell=0$ system, and crucially, both the critical exponent and the echoing period are determined entirely by this limit, rendering them independent of $\ell$~\cite{Gundlach:1996vv} (see also~\cite{Dafermos:2004ws, Dafermos:2004wr, Langfelder:2004sk, Christodoulou_1999on, Landsman:2022hrn, Gunzig:2000ce}). We will see that this point plays a crucial role in resolving tensions surrounding scalar field collapse in $2+1$ dimensions in Section~\ref{sec:quantum2+1}, where a cosmological constant enters.

This framework enables predictive power even in systems where the field equations are not strictly scale-free, see for example~\cite{Gundlach:1996je, Gundlach:1996vv}. A simple case is when any scalar field potential $V(f)$ becomes dynamically irrelevant compared to the
kinetic term $(\nabla f)^2$ in the self-similar regime. In such situations, all scalar field models with potentials flow into the same universality class as the free massless scalar field. Remarkably, even disparate matter models can fall into the same universality class as the massless scalar case~\cite{Bizon:1998ix}. This indicates that studying the Einstein-scalar system is representative enough to capture the essential features of critical phenomena.

As a final remark, self-similarity has emerged in numerous contexts beyond gravitational collapse. These include the asymptotic structure of spatially homogeneous cosmological models~\cite{Wainwright_Ellis_1997} and inflationary scenarios~\cite{Coley:1996cy, Coley2003}, where it plays a role in the cosmic no-hair theorem~\cite{Wald:1983ky}, providing evidence that cosmological spacetimes are asymptotically attracted to exact self-similar solutions. Notably, self-similarity also appears in the formation of primordial black holes~\cite{Zeldovich:1967lct, Carr:1974nx}, which is a potential real-world astrophysical realization of both self-similarity and critical collapse~\cite{Niemeyer:1997mt, Yokoyama:1998xd, Niemeyer:1999ak, Jedamzik:1999am, Green:1999xm,  IHawke_2002, Musco:2004ak, Polnarev:2006aa, Musco:2008hv, Kuhnel:2015vtw}.

Nevertheless, despite its ubiquity, the origin of self-similarity remains deeply mysterious~\cite{Barenblatt1972, Carr:1998at, Carr:2005uf}, and appears to be generically linked to the presence of naked singularities~\cite{Eardley:1986en, Ori:1987hg, Waugh:1988ud, Ori:1989ps, Waugh:1989qcs, Zannias1991, Henriksen1991, Lake1991, Foglizzo:1993unt, Gregory:1993vy, Christodoulou:1994hg, Brady:1994aq, Carr:1999qr, Carr:2002me}. In critical collapse, it is not fully understood why particular matter models give rise to either CSS or DSS in their critical solutions, which also concerns their stability properties~\cite{christodoulou1999, Harada:2001hk, Harada:2003jg, Nolan:2000hi, Nolan:2002hr}.

The analysis surrounding the similarity properties further highlights the massless scalar case as particularly nontrivial, as it corresponds to a stiff fluid~\cite{Evans:1994pj, Wainwright_Ellis_1997, Neilsen:1998qc, Neilsen:1999we, Brady:2002iz, Harada:2003jg}. In this case, the similarity equations become formally singular, rendering the analysis less straightforward. Furthermore, given the difficulty of obtaining analytic results for DSS spacetimes, spherically symmetric solutions with CSS could offer more tractable analytic insights into critical behavior. In Sections~\ref{sec:quantum2+1} and \ref{sec:collapse3+1}, we will likewise focus on exact CSS-type solutions, where quantum effects can be precisely extracted.

\subsubsection*{Global structure}

We now turn to the global structure of the critical solution. Given that the metric of the critical spacetime takes the form $g_{\mu \nu} (T, x^i)=e^{-2 T} \tilde{g}_{\mu \nu} (x^i)$ in adapted coordinates, the conformal factor leads to a curvature singularity at $T \to \infty$, where curvature invariants generically diverge as $e^{4 T}$. This singularity is reached in finite proper time from regular points in the past. 

In spherically symmetric settings, the four-dimensional spacetime is the product of a two-dimensional Lorentzian manifold with coordinates $(T, x)$ and a round two-sphere. Then $x$ labels spatial locations, with one value corresponding to the regular center. The limit $T \to \infty$ for all values of $x$ corresponds to a single spacetime point at the center, the \textit{accumulation point}, where the curvature diverges.  Another value of $x$ corresponds to the past light cone of this point, typically a \textit{self-similarity horizon}, where the critical solution remains regular.\footnote{A self-similar hypersurface can be continuously deformed into another, but this is not the case for a null hypersurface, where the homothetic vector becomes null. Hence, we call the self-similar null hypersurface the \textit{self-similarity horizon.} A detailed classification of the self-similarity horizon provides a natural way of constructing Penrose diagrams in critical spacetimes, see~\cite{Martin-Garcia:2003xgm, Gundlach_2003}.} The critical solution can often be continued beyond this past light cone to the future light cone of the accumulation point, which is a Cauchy horizon. A further continuation beyond the Cauchy horizon is not unique, while the known DSS-compatible extensions either contain a regular center or a timelike central singularity~\cite{Gundlach_2003, Martin-Garcia:2003xgm, Carr:2002me}. However, this extended region lies outside the domain of dependence relevant to critical collapse and will not be our focus here.

\begin{figure}[t!]
    \centering
    \includegraphics[width=0.4\linewidth]{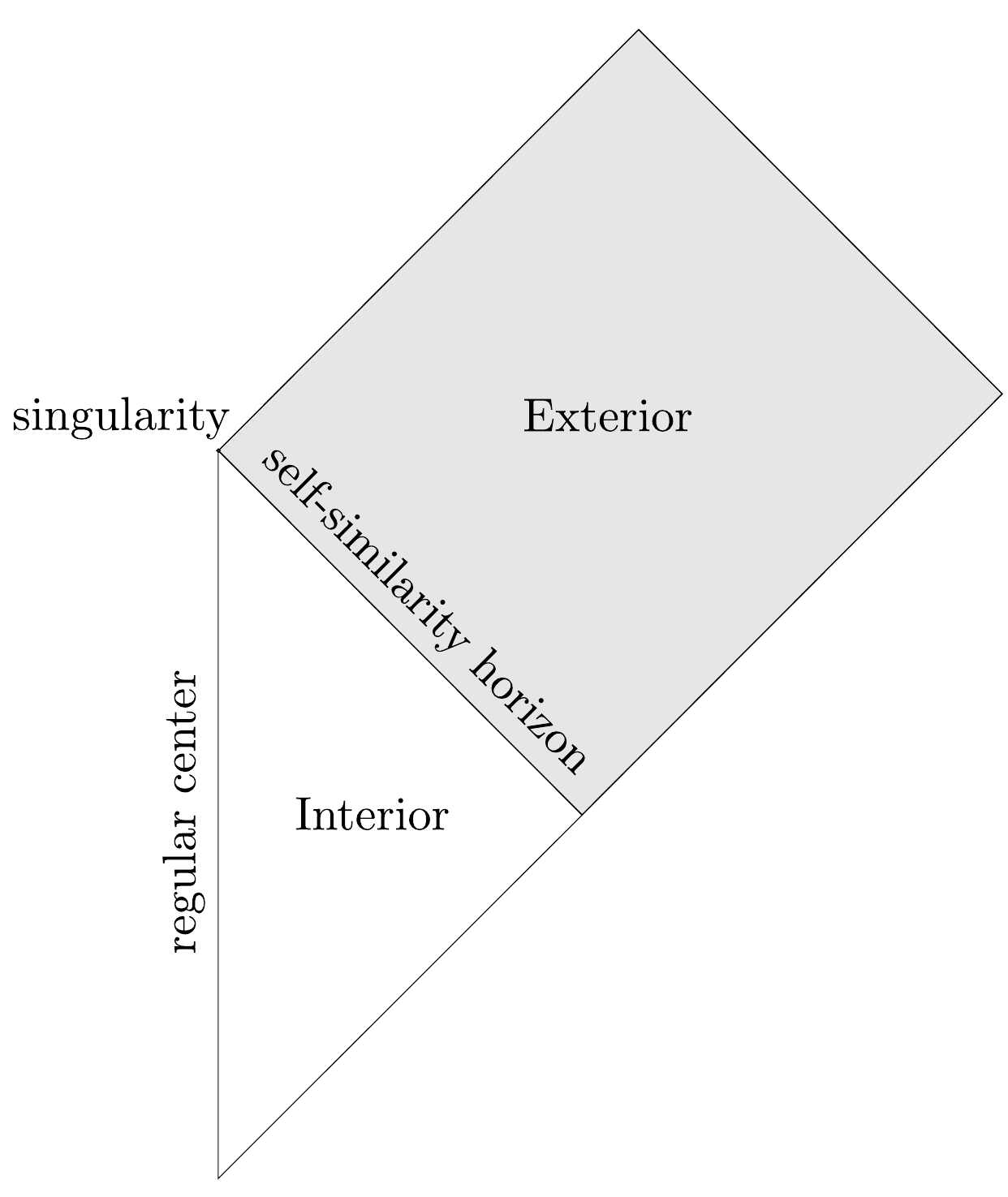}
    \includegraphics[width=0.4\linewidth]{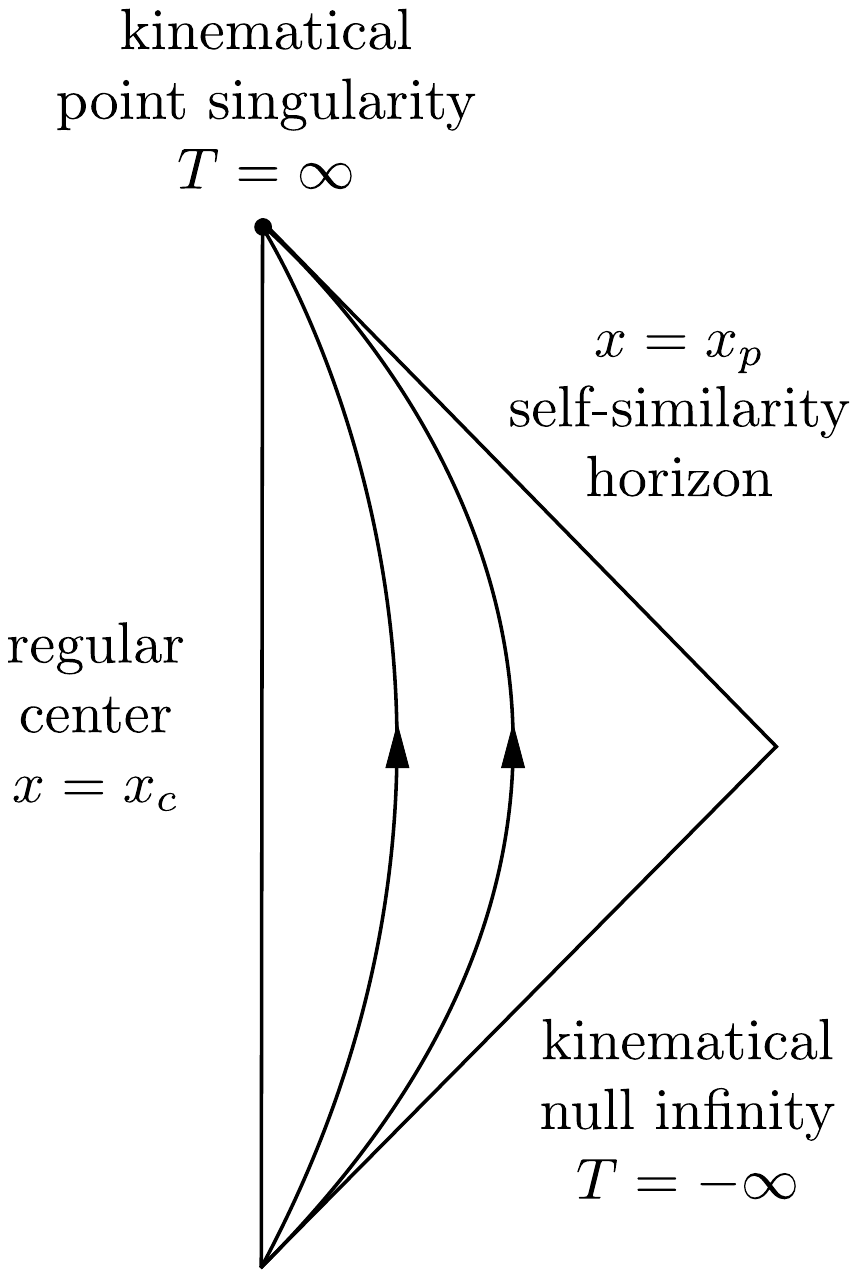}
    \caption{A typical naked singularity problem consists of two regions separated by the self-similarity horizon: the interior fill-in region and the exterior-naked singularity region, as illustrated in the left panel. The right panel shows the global structure of the interior region, modeled by a CSS critical spacetime, which is the focus of the present work.}
    \label{fig:global}
\end{figure}

The past light cone of the singularity divides the spacetime into two distinct regions, which naturally correspond to two components of the naked singularity problem~\cite{Christodoulou:1994hg, Brady:1994aq, Wang:1996xh, Jalmuzna:2015hoa, Rodnianski:2019ylb, Shlapentokh-Rothman:2022byc, Cicortas:2024hpk}. The first is the region exterior to the past light cone, extending all the way to future null infinity. This portion of spacetime, which we call the \textit{exterior-naked singularity region} following~\cite{Cicortas:2024hpk}, ends on an incomplete null boundary due to the presence of the future light cone of the singularity. This region yields the ``nakedness" property, the visibility of the singularity from infinity. The second component is the \textit{interior fill-in region}~\cite{Cicortas:2024hpk}, which provides the dynamical origin of the singularity from smooth initial data; see Figure~\ref{fig:global}. A consistent spacetime must smoothly join the interior and exterior regions across the past light cone, which typically involves a cut-and-paste procedure that glues together distinct geometries.\footnote{While this matching is a well-established technique in classical gravity, its validity becomes less clear in the semiclassical context: quantum effects may modify the geometries on both sides in ways that cause the junction conditions to fail unless explicitly verified. Ensuring smoothness across the junction in a semiclassical spacetime requires a complete computation of the renormalized quantum stress-energy tensor in both regions. In our analysis of quantum critical collapse in Sections~\ref{sec:quantum2+1} and \ref{sec:collapse3+1}, we focus primarily on the horizon formation problem within the interior fill-in region. A detailed treatment of the exterior region and the full global structure is left for a future study.}

\FloatBarrier

To understand the geometric structure more precisely, we distinguish between \textit{kinematical} and \textit{dynamical} consequences of self-similarity combined with spherical symmetry~\cite{Martin-Garcia:2003xgm, Gundlach_2003}. In this context, kinematical features arise purely from the conformal factor $e^{-2T}$, whereas dynamical features depend on the detailed structure of $\tilde{g}_{\mu \nu}(x)$. Of course, the dynamical picture involves time evolution, yet kinematical effects here refer to consequences that follow directly from the imposed symmetry, independent of the equations of motion. In this language, $T \to \infty$ is clearly a kinematical singularity. Additional dynamical singularities may occur along isolated CSS or DSS lines for certain values of $x$, and the behavior in the simultaneous limit $T \to \infty$ may depend on the path taken in spacetime.

Meanwhile, the surfaces $T= \pm \infty$ represent kinematical boundaries. Self-similar curves terminate at finite affine parameter as $T \to \infty$ since the affine parameter scales as $e^{-T}$, while they extend infinitely as $T \to -\infty$. Thus, $T =-\infty$ typically corresponds to a kinematical infinity, often associated with divergent areal radius. Dynamical boundaries can also arise independently, as singularities or infinities may be induced by $\tilde{g}_{\mu \nu}(x)$. For example, a curvature singularity localized at a particular value of $x$ naturally marks a boundary of the spacetime; conversely, if the curvature remains finite, one might expect a regular extension beyond that point. See~\cite{Martin-Garcia:2003xgm, Gundlach_2003} for an in-depth classification of possible global causal structures in self-similar spacetimes.

\section{Trace anomaly and quantum one-loop effective theory} \label{sec:onelooptheory}

In this section, we discuss how the $s$-wave sector of a higher-dimensional, spherically symmetric Einstein-scalar system can be described by a general two-dimensional non-minimally coupled dilaton gravity theory. We then introduce the corresponding quantum one-loop effective theory. The procedure we describe constitutes a precise and rigorous quantum field theory in curved background formulation without any ad hoc input. Clearly, the system involves a non-conformal matter theory evolving in a critical spacetime without an exact timelike Killing symmetry, thereby extending beyond the well-understood regime of the formulation.

Our goal is to solve the semiclassical Einstein equation
\be \label{eq:semiclassicalEinstein}
G_{\mu \nu}= T_{\mu \nu}+ \frac{\hbar}{\ell^{D-2}} \langle T_{\mu \nu} \rangle,
\ee
where we have adopted the geometrized units with $c=1$ and $8 \pi G_N=1$, but explicitly retain $\hbar$ to track quantum effects, which are encoded in $\langle T_{\mu \nu} \rangle$ itself. We reserve Latin indices $a, b, \dots$ for two-dimensional quantities, and Greek indices $\mu, \nu, \dots$ for higher-dimensional ones. Dimensionality will be indicated explicitly with subscripts or superscripts whenever ambiguity arises. The purpose of this section is to lay out the formulation of computing the renormalized quantum expectation value of the stress-energy tensor $\langle T_{\mu \nu} \rangle$ in the Einstein-scalar system. We aim to present the construction in a comprehensive manner, taking care of various subtleties that may arise.

\subsubsection*{The two-dimensional dilaton gravity model}

As discussed in Section~\ref{sec:criticalcollapse} from the linear perturbation perspective, there are strong justifications for focusing on the quantum effects of the $s$-wave mode in our analysis. Further remarks on the role of higher angular modes will be provided in Section~\ref{sec:discussion}. The $s$-wave sector can be obtained by performing a spherical dimensional reduction of the higher-dimensional Einstein-scalar system, resulting in an effective two-dimensional theory with a dilaton field $\phi$ that encodes the areal radius of the transverse sphere. 

Consider a $D=d+1$-dimensional spacetime, factorized into a $(D-n)$-dimensional spacetime with metric $g_{ab}$ and dilaton $\phi$, and an $n$-dimensional internal space. We adopt the ansatz~\cite{Kummer_1999, Kummer_1999-1, Grumiller:2002nm}
\be \label{eq:sphericalansatz}
ds^2_{(D)}= g_{ab} dx^a dx^b+ \ell^2 e^{-\frac{4 \phi}{n}}d \Omega^2_{n}=g_{ab} dx^a dx^b+\ell^2e^{-\frac{4 \phi}{D-2}} d \Omega^2_{D-2},
\ee
where we always take $D-n=2$ with $a,b=0,1$. Here, $d \Omega^2_{D-2}$ denotes the line element on a unit sphere $S^{D-2}$. The exponential parametrization of the dilaton ensures positivity of the areal radius, and the normalization is fixed such that $\sqrt{-g} \propto e^{-2 \phi}$, regardless of the dimensionality of the internal space.

Upon dimensional reduction, the two-dimensional Lagrangian densities for the gravity and matter sectors obtained from the Einstein-scalar system \eqref{eq:Einsteinscalar} become 
\be \label{eq:gravityaction}
\mathcal{L}_{\text{grav}} = \frac{\Omega_{D-2} \ell^{D-2}}{16 \pi G_N^{(D)} } \sqrt{-g_{(2)}}e^{-2 \phi} \bigg(R^{(2)}+\frac{4(D-3)}{D-2} (\nabla \phi)^2 -\frac{1}{ \ell^2}(D-2)(D-3) e^{\frac{4 \phi}{D-2}}\bigg),
\ee
\be \label{eq:matteraction}
\mathcal{L}_{\text{matter}}= -\frac{\Omega_{D-2} \ell^{D-2}}{2}\sqrt{-g_{(2)}} e^{-2 \phi}  (\nabla f)^2.
\ee
Here $\Omega_{D-2}$ represents the volume of the unit sphere $S^{D-2}$. We emphasize that this theory is distinct from other well-known two-dimensional dilaton gravity models, such as JT~\cite{Jackiw:1984je, Teitelboim:1983ux, Achucarro:1993fd}, CGHS~\cite{Callan:1992rs}, and the low-energy effective string theory~\cite{Callan:1985ia, Mandal:1991tz, Witten:1991yr}; rather, it results from a direct and systematic spherical dimensional reduction that captures the $s$-wave sector of the original theory.\footnote{Note that the dilaton kinetic term carries a positive sign, which might raise concerns about ghost-like behavior. However, this does not imply a pathology: the two-dimensional theory is not automatically in the Einstein frame, and such terms emerge naturally from dimensional reduction.} A noteworthy feature is that the matter sector becomes non-minimally coupled to the dilaton $\phi$, a structure that will play a central role in the quantum theory.

The basic principle of dimensional reduction is that any higher-dimensional solution $g_{\mu \nu}$ of the original equations of motion must yield a corresponding lower-dimensional solution of the form
\be
g_{ab}(x^a), \quad \phi(x^a),
\ee
in the reduced theory. However, this statement holds only at the classical level. Quantum mechanically, additional contributions arise in the effective action that are not captured by dimensional reduction. This discrepancy---known as the \textit{dimensional reduction anomaly}~\cite{Frolov_1999, Cognola:2000xp, Cognola:2000wd, Balbinot_2001}---originates from the fact that renormalization and dimensional reduction generally do not commute. While the unrenormalized $D$-dimensional effective action can be expanded as a sum over lower-dimensional modes (e.g., through spherical harmonics), this decomposition breaks down after renormalization as it may violate the na\"ive mode-by-mode correspondence. 

Importantly, such mismatch can appear in both the ultraviolet (UV) and infrared (IR) regimes. The UV sector contains indispensable, local, and state-independent contributions that are insensitive to angular decomposition (i.e., ultimately reconstructible from local symmetry-fixed functionals, despite angular structure in intermediate mode sums), while the IR sector involves non-local, state-dependent effects where angular dependence re-emerges. In the UV, the anomaly can be understood from the fact that the number and types of local counterterms required to regularize divergences differ across spacetime dimensions; consequently, extra terms are needed for each mode to restore matching if one insists on performing the angular decomposition before renormalization. By properly incorporating the dilaton coupling for the $s$-wave projection in the two-dimensional theory, as in~\eqref{eq:matteraction}, together with the conservation laws and the additional dilaton anomaly terms that we will introduce in~\eqref{eq:traceanomaly}, the UV structure can be fully accounted for---modulo the usual freedom in finite local covariant counterterms (scheme dependence)---without loss of nontrivial UV information from higher dimensions. In contrast, the IR sector, which does allow angular decomposition, may generate additional Weyl-invariant non-local terms even in the $s$-wave that we will discuss in detail below. These terms, while physically meaningful, should not be regarded as part of the anomaly itself but rather as reflecting an enhanced state-dependent IR physics \cite{Wu:2023uyb}.

To focus on the $s$-wave contribution within the semiclassical framework, we therefore first perform the dimensional reduction and only then quantize the resulting two-dimensional matter theory. This procedure isolates the $s$-wave sector in a controlled semiclassical framework, allowing us to analyze the leading IR growing mode together with the universal UV structure encoded by the dilaton-coupled trace anomaly. Under the same regularity and boundary conditions that restrict admissible initial data in the gauge-invariant linear analysis (see Section~\ref{sec:criticalcollapse}), higher-$l$ perturbations are expected to remain decaying and therefore not to introduce additional IR instabilities at leading order. At the same time, we emphasize that higher angular modes can be subtle in the quantum theory---through state dependence, cumulative mode sums/resummation, or nonlinear mode coupling---and we discuss these plausible scenarios and the resulting scope of our claims systematically in Section~\ref{sec:discussion}.

\subsubsection*{Trace anomaly and one-loop effective action}

Now we are ready to quantize the matter theory. The model given in \eqref{eq:matteraction} has received significant attention due to its physical relevance as an effective description of higher-dimensional dynamics~\cite{Mukhanov_1994, Bousso:1997cg, Mikovic_1998, Balbinot_1999, Balbinot_1999_2, Buric:1998xv, Kummer_1999, Kummer_1999-1, Gusev:1999cv, Balbinot:2000at, Balbinot_2001, Balbinot:2002bz, Grumiller:2002nm, Fabbri:2003vy, Hofmann:2004kk, Hofmann:2005yv, Wu:2023uyb}, and it has been applied to various scenarios~\cite{Bousso:1997wi, Nojiri:1998ww, Bytsenko:1998md, Buric:1998xv, Buric:2000cj, Bousso:1998bn, Bousso:1999ms, Niemeyer:2000nq, vanNieuwenhuizen:1999nu, Nojiri:1999br, Kadoyoshi:1997wt, Gates:1998py, Nojiri:1998wb, Nojiri:1999pm, Ho:2017joh, Navarro-Salas:2024ogp}. A standard route for quantization is via the heat kernel analysis in the Euclidean path integral. Assuming suitable regularity for the quantum fields, the effective action can be expanded in an asymptotic series at small proper time, corresponding to the UV regime, referred to as the Schwinger-DeWitt or Seeley-DeWitt expansion~\cite{DeWitt:1964mxt, Vassilevich:2003xt, Shapiro:2008sf, parker_toms_2009}. The expansion involves an infinite series of local curvature and potential terms in the second-order elliptic operator of Laplace type, yet non-locality appears in the resummed effective action. The expansion facilitates the analysis of the UV divergences of the theory, where the conformal anomaly arises robustly from the regularization in the UV, and is independent of the regularization scheme.

Conformal anomaly originates from the fact that the classical action is invariant under conformal transformation, while the path integral measure is typically not. As a result, the quantum expectation value of the stress-energy tensor acquires a non-vanishing trace. In the literature, the terms “trace anomaly” and “conformal anomaly” are often used interchangeably. However, a key feature for the two-dimensional matter theory \eqref{eq:matteraction} is that it is not conformally invariant under infinitesimal conformal transformations due to the dilaton coupling, but is still classically Weyl-invariant under local Weyl rescalings \cite{Fabbri:2003vy}.

This feature enables a well-defined computation of the anomaly using heat kernel methods, for which the trace of the quantum stress-energy tensor is given by~\cite{Mukhanov_1994, Bousso:1997cg, Mikovic_1998, Elizalde, Ichinose, Dowker_1998, katanaev1997generalized, Nojiri:1999vv}\footnote{Note that the prefactor $\hbar/24 \pi$ is universal and does not depend on the coefficient $\Omega_{D-2} \ell^{D-2}$ in \eqref{eq:matteraction}: in the path integral any such overall coefficient can only produce a constant Jacobian, which drops out upon variation. However, if we include $N$ massless scalar fields, we need to insert an overall $N$ in \eqref{eq:traceanomaly} since the path integral then sums identically over each field.}
\be \label{eq:traceanomaly}
\langle T^a{}_a \rangle= \frac{\hbar}{24 \pi} (R-6 (\nabla \phi)^2+6 \Box \phi).
\ee
The first term, proportional to the Ricci scalar $R$, reproduces the familiar two-dimensional conformal anomaly for a free massless scalar field, central to Hawking radiation calculations in black hole spacetimes~\cite{Christensen}. The additional dilaton-dependent terms reflect the non-conformal coupling and are essential for accurately capturing the $s$-wave quantum effects of the higher-dimensional theory. These properties allow for full analytic control while revealing subtle quantum effects absent in purely conformal models, and are clearly much less understood.\footnote {See~\cite{Casarin:2018odz, Ferrero:2023unz} for recent developments on the fate of trace anomalies in generic non-conformal matter theories.}

The trace anomaly \eqref{eq:traceanomaly} is derived from analyzing the UV behavior of the heat kernel and is known to be one-loop exact. It is a purely local, geometric quantity; hence, crucially, it is independent of the choice of quantum state.\footnote{This property fails for generic non-conformal matter, such as a massive scalar field or the inclusion of higher angular momentum modes ($l \neq 0$) from dimensional reductions. In such cases, even classical Weyl invariance is broken, and the trace becomes state-dependent, making the subject extremely delicate.} Unfortunately, these favorable properties do not extend to the full stress-energy tensor $\langle T_{ab} \rangle$. An exception is the case of a free massless scalar field that is conformal, where we could rely on the conservation law
\be
\nabla^a \langle T_{ab} \rangle=0.
\ee
By the canonical definition 
\be \label{eq:defin}
\langle T_{ab} \rangle \equiv \frac{-2}{\sqrt{-g}}\frac{\delta \Gamma_{\text{one-loop}}}{\delta g^{ab}},
\ee
together with the conformal anomaly one can fix the following one-loop non-local action uniquely 
\be \label{eq:Polyakov}
\Gamma_\text{\text{one-loop}}=-\frac{\hbar}{96 \pi} \int d^2 x \sqrt{-g} R \frac{1}{\Box} R,
\ee
known as the Polyakov action~\cite{Polyakov:1981rd}. This approach also agrees with analysis directly from the heat kernel~\cite{Vassilevich:2003xt, Barvinsky:1990up, Barvinsky:1994ic}. In the effective action, the choice of quantum state corresponds to boundary conditions imposed on the Green's function in the non-local operator $\Box^{-1}$. Thus, while the effective action is formally unique, the derived expectation value $\langle T_{ab} \rangle$ still encodes quantum state dependence through its non-local structure.

With dilaton-coupling, we must also take into account the diffeomorphism transformation of the dilaton field, and the conservation law is modified to be~\cite{Balbinot_1999, Kummer_1999}
\be
\nabla^a \langle T_{ab} \rangle-\frac{1}{\sqrt{-g}}  \frac{\delta \Gamma_{\text{one-loop}}}{\delta \phi} \nabla_b \phi=0.
\ee
This modified conservation law can also be understood as the dimensionally reduced form of the higher-dimensional conservation law $\nabla^\mu \langle T^{(D)}_{\mu \nu} \rangle=0$ under spherical dimensional reduction \eqref{eq:sphericalansatz}. Because of the additional functional degree of freedom associated with $\phi$, the quantum effective action can no longer be uniquely fixed by the trace anomaly and conservation law alone. Additional input, such as the quantum state, symmetry assumptions, or renormalization conditions, would be necessary to determine the full expression.

Unless the finite part of the heat kernel can be extracted by carefully accounting for both UV and IR divergences, the one-loop action, and hence the stress-energy tensor cannot be obtained in closed form. This issue is particularly relevant in our setup. While the Seeley-DeWitt expansion provides a systematic asymptotic series that captures the UV structure of the heat kernel, it offers no control in the IR regime. To overcome this, a variety of methods have been developed to extend the analysis into the IR, including both perturbative and non-perturbative techniques~\cite{Barvinsky:1987uw, Barvinsky:1990up, Barvinsky:1990uq, Barvinsky:1993en, Barvinsky:1994ic, Barvinsky:2002uf, Barvinsky:2003rx, Barvinsky:2004he}. 

However, for \eqref{eq:matteraction}, the heat kernel suffers from additional IR divergences, and it remains an open question whether the theory can be renormalized with a finite number of counterterms and resummed into a finite effective action~\cite{Hofmann:2004kk, Hofmann:2005yv}. Truncating the covariant curvature expansion na\"ively may yield unphysical results, such as black hole anti-evaporation, where the black hole appears to absorb energy from the vacuum; or divergences in $\langle T_{\mu \nu} \rangle$ even in regular quantum states on well-behaved spacetimes such as Schwarzschild~\cite{Buric:1998xv, Balbinot_1999, Balbinot_1999_2, Buric:2000cj, Balbinot:2002bz, Hofmann:2004kk, Hofmann:2005yv}. These pathologies stem from both the IR sensitivity and the enhanced state dependence introduced by the dilaton coupling.

Nevertheless, physically meaningful semiclassical predictions, such as Hawking radiation, are expected to be robust and largely insensitive to higher-curvature corrections in the expansion. These subtleties have been summarized and recently revisited in~\cite{Wu:2023uyb}, where it was emphasized that much of the ambiguity ultimately traces back to the choice of quantum state, a complication made sharper by the presence of the dilaton.\footnote{Notably, anti-evaporation occurs for a near-Nariai black hole, which is simply a Schwarzschild black hole in de Sitter background that is close to extremality~\cite{Bousso:1997wi, Nojiri:1998ue, Nojiri:1998ph}. In this limit, the calculation appears to become insensitive to the dilaton coupling. This apparent puzzle is resolved in \cite{Shi:2026wnk}.}  We will illustrate these issues and briefly compare them with the canonical quantization approach, which leads to Wald’s axiomatic formulation for identifying physically reasonable expectation values of the stress-energy tensor~\cite{Wald:1977up, Wald1978, Wald:1978ce}.

The trace anomaly \eqref{eq:traceanomaly} continues to provide valuable guidance. One can solve the defining equation \eqref{eq:defin}, viewed as a functional differential equation, to obtain a particular solution for the one-loop effective action up to Weyl-invariant terms and local counterterms~\cite{Mukhanov_1994, Balbinot_1999}
\be \label{eq:fulloneloop}
\Gamma_{\text{one-loop}}=\Gamma_{\text{anom}}+ \text{Weyl-invariant terms} + \text{Local counterterms}
\ee
\be \label{eq:anomalyinduced}
\Gamma_{\text{anom}}=-\frac{\hbar}{96 \pi} \int d^2 x \sqrt{-g} \bigg(  R \frac{1}{\Box} R -12 (\nabla \phi)^2 \frac{1}{\Box} R +12 \phi R \bigg).
\ee
We refer to the particular solution $\Gamma_{\text{anom}}$ as the \textit{anomaly-induced effective action}. We recognize the first piece in $\Gamma_{\text{anom}}$ as precisely the Polyakov action \eqref{eq:Polyakov}, while the remaining dilaton-dependent terms arise due to dilaton-coupling. The most general solution as $\Gamma_{\text{one-loop}}$ may contain additional Weyl-invariant terms (due to the IR sensitivity we described above), which cannot be determined in closed form yet. The only information we know is that they must be Weyl-invariant, otherwise it would be inconsistent with trace anomaly \eqref{eq:traceanomaly}.\footnote{A similar structure appears in the Riegert non-local action~\cite{Riegert:1984kt}, which is the anomaly-induced effective action from the four-dimensional trace anomaly. But there are several clear reasons that we should not adopt it. First, it applies only to conformally coupled matter fields in four dimensions. Second, even with the conservation law, the stress-energy tensor remains significantly underdetermined due to the larger number of independent components and its sensitivity to the quantum state. Furthermore, the construction would compel us to include quantum higher-$l$ modes in the IR, an otherwise dynamically irrelevant complication, while the physically relevant four-dimensional UV structure is already captured by~\eqref{eq:traceanomaly}.} We keep local counterterms in the full one-loop action that are state-independent, which will play a role in the discussion below.

In curved spacetime, the absence of global Poincar\'e symmetry eliminates any canonical choice of vacuum. One can no longer define the vacuum uniquely as in flat space, and instead, infinitely many inequivalent vacua may be defined depending on the coordinate chart or observer. This leads to ambiguities that are both physical and unavoidable.

To proceed, we have to make it very precise the nature of $\Gamma_{\text{one-loop}}$ \eqref{eq:fulloneloop} within the framework of quantum field theory in curved spacetime~\cite{Karakhanian:1994gs, Jackiw:1995qh, Navarro-Salas:1995lmi, Fabbri:2005}. It is precisely the ignorance of Weyl-invariant terms and the sensitivity to the choice of quantum state that causes confusion when applying this model~\cite{Wu:2023uyb}. In a curved background, we can no longer take the normal-ordered $\langle :T_{ab}: \rangle$ as our definition of stress-energy tensor in the semiclassical Einstein equation since it fails to be generally covariant. To restore covariance, we must supplement it with a local, geometric term we denote as $\langle T^{\text{geo}}_{ab} \rangle$. The full covariant stress-energy tensor is
\be 
\langle T_{ab} \rangle=\langle :T_{ab}: \rangle+\langle T^{\text{geo}}_{ab} \rangle.
\ee
Each term individually breaks general covariance, but their sum respects it. By construction, $\langle T^{\text{geo}}_{ab} \rangle$ must consist of only local geometric quantities and be state-independent, for the simple reason that it must vanish when we go back to flat space, where the Minkowski vacuum is unambiguously defined. This is manifested in the effective action formalism as the freedom of local counterterms in the regularization scheme.

The $\langle:T_{ab} :\rangle$ represents the state-dependent part of the covariant stress-energy tensor, and the undetermined Weyl-invariant terms must contribute to this part. Since the trace anomaly is local and state-independent, captured entirely by $\langle T^{\text{geo}}_{ab} \rangle$, and therefore cannot be affected by the Weyl-invariant sector. This strongly suggests that the Weyl-invariant contributions must be \textit{non-local} and \textit{state-dependent}. Such terms can naturally arise from the non-local structure of the full heat kernel expansion beyond the leading UV divergences. For example, we may encounter $(\nabla \phi)^2 \frac{1}{\Box} (\nabla \phi)^2$, which is both Weyl-invariant and non-local~\cite{Wu:2023uyb, Mukhanov_1994, Hofmann:2004kk, Hofmann:2005yv}. In contrast, local Weyl-invariant terms, such as $(\nabla \phi)^2$ or $\Box \phi$, would contribute to the trace on-shell and must therefore be absorbed into $\langle T^{\text{geo}}_{ab} \rangle$ rather than the state-dependent sector.

A tricky part concerns the first two non-local terms in $\Gamma_{\text{anom}}$: $ R \frac{1}{\Box} R$ and $(\nabla \phi)^2 \frac{1}{\Box} R$. They are not Weyl-invariant as they are designed to reproduce the trace anomaly, yet they are non-local, indicating they encode state-dependent information.  Indeed, one can formally decompose these terms into two non-covariant pieces, each of which contributes separately to the state-dependent part $\langle :T_{ab}: \rangle$ and to the local state-independent piece $\langle T^{\text{geo}}_{ab} \rangle$~\cite{Karakhanian:1994gs, Jackiw:1995qh, Navarro-Salas:1995lmi, Wu:2023uyb}. Therefore, these are precisely the terms that straddle both contributions in the covariant stress-energy tensor.

\subsubsection*{Canonical quantization and Wald's axioms}

In canonical quantization, a similar story unfolds. When renormalizing the stress-energy tensor, one standard method is to analyze the singular structure of the two-point function using the covariant point-splitting approach developed by DeWitt and Christensen~\cite{DeWitt:1975ys, Christensen:1976vb, Christensen:1978yd}. In this framework, $\langle T_{\mu \nu} \rangle$ is extracted by evaluating the coincidence limit of a bi-tensor derived from the two-point function, with UV divergences subtracted using a Hadamard parametrix. This subtraction is a local state-independent geometric object that captures the universal UV singular structure of the two-point function. This method cleanly separates the renormalized $\langle T_{\mu \nu} \rangle$ into two finite parts: a state-independent part encoding the trace anomaly, and a state-dependent part determined by the choice of quantum state and boundary conditions.

Wald extended this framework into a powerful axiomatic approach for defining a “reasonable” renormalized stress-energy tensor using Hadamard renormalization~\cite{Wald:1977up, Wald1978, Wald:1978ce}. It ensures the semiclassical Einstein equation remains well-defined and physically consistent. These are known as the \textit{Wald's axioms} for renormalized $\langle T_{\mu \nu} \rangle$:
\begin{itemize}
    \item \textbf{General covariance and conservation}:
    $\langle T_{\mu \nu} \rangle$ must preserve general covariance and therefore be conserved in the classical sense $\nabla^\mu \langle T_{\mu \nu} \rangle=0$. This is the requirement of the Bianchi identity $\nabla^\mu G_{\mu \nu}=0$, for consistency with the semiclassical Einstein equation.
    
    \item \textbf{Local and geometric construction}: It must be locally constructed from the metric, curvature tensors, and their derivatives at point $x$, as well as from the quantum state encoding boundary conditions. This implies that $\langle T_{\mu \nu} \rangle$ must be consistent with the trace anomaly if present.

    \item \textbf{Minkowski normalization}:  In flat spacetime, for the Minkowski vacuum, $\langle T_{\mu \nu} \rangle$ must agree with the usual normal-ordered result (i.e., it should vanish for the vacuum). In other words, Minskowski spacetime must be a solution to the semiclassical Einstein equation.
    \item \textbf{Causality}: The dependence of $\langle T_{\mu \nu} \rangle$ on the quantum state must be smooth and causal. That is, local changes in the state outside the past light cone of $x$ should not affect $\langle T_{\mu \nu} (x) \rangle $.
    \end{itemize}
The Hadamard condition, demanding that the singularity structure of the two-point function mimics that of flat spacetime, satisfies all these axioms and is regarded as the right criterion for physically admissible quantum states in curved backgrounds. These states are called \textit{Hadamard states}. In fact, Wald proved that even after imposing all reasonable axioms, there remains a residual finite, local, and conserved ``ambiguity": 
\begin{itemize}
    \item Any ambiguity in the definition of $\langle T_{\mu \nu} \rangle$ must be expressible as a linear combination of local conserved tensors constructed from the metric, curvature tensors, and their derivatives at a given point $x$.
\end{itemize}
If background matter fields are present, this ambiguity can also involve local terms constructed from those fields and their derivatives.

This residual ambiguity must be fixed by additional physical input, such as symmetry conditions or boundary behavior from the properties of the background spacetime, or matching to known results. It is a manifestation of the local counterterms in \eqref{eq:fulloneloop} from the effective action perspective once we have specified the state, though they in principle come from different origins. Wald’s ambiguity is a reflection of the freedom to choose different finite renormalization schemes, all of which differ by local counterterms. While local counterterms in the effective action are initially introduced to cancel divergences in the $\langle T_{\mu \nu} \rangle$, the freedom to add such terms persists even after renormalization. The residual freedom implies that any two renormalized stress-energy tensors that differ by a local, conserved, and covariant expression are equally valid, and such differences can always be absorbed into the effective action via finite local counterterms.

The above discussion was made for four-dimensional theories. However, Wald's axioms imply that these characteristic features must be reproduced in any ``acceptable" two-dimensional theory that purports to encode the one-loop physics of a higher-dimensional model. See, for example,~\cite{Decanini:2005eg} for the generalization of Hadamard renormalization and Wald’s axioms to arbitrary spacetime dimensions.

Remarks on canonical quantization approaches to critical collapse are provided in Appendix~\ref{sec:appendixD}. While the canonical and path integral formulations should describe the same physics up to renormalization ambiguities, the latter offers a key advantage: it does not require an \textit{a priori} state choice, as we discuss below.

\subsubsection*{The choice of quantum state}

We can now pose the final, and delicate question: how should one specify a physically reasonable quantum state? As argued above, such a state must satisfy the Hadamard condition and ensure that $\langle T_{\mu \nu} \rangle$ obeys Wald's axioms. However, beyond these mathematical constraints, one also expects physically meaningful states to yield stress-energy tensors with clear physical interpretations in the context of the given spacetime.

In the case of a Schwarzschild black hole, it is well known that most Hadamard states yield time-dependent $\langle T_{\mu \nu} \rangle$ due to the lack of global symmetries. However, there are three distinguished Hadamard states that produce stationary and physically interpretable stress-energy tensors~\cite{KAY199149, Fabbri:2005}: the Boulware state~\cite{Boulware1975}, which is static and corresponds to the Minkowski vacuum at infinity, but is singular at the horizon. It describes vacuum polarization outside a static eternal black hole; the Hartle-Hawking state~\cite{Hartle:1976tp, Israel:1976ur}, which is regular everywhere and describes a black hole in thermal equilibrium with a surrounding heat bath; the Unruh state~\cite{Unruh:1976db, Davies:1976ei, Hiscock:1980ze, Hiscock:1981xb}, which is regular on the future horizon and models a black hole formed by collapse, radiating Hawking flux to future infinity.\footnote{Among these, only the Hartle–Hawking state admits a natural Euclidean path integral construction. This state arises from the requirement of regularity on the Euclidean section with compact imaginary time and corresponds to a thermal equilibrium ensemble. In contrast, the Boulware and Unruh states are intrinsically Lorentzian constructs, defined via mode expansions and causal boundary conditions. These states cannot be obtained from the Euclidean path integral since they are either singular at the horizon or lack the periodicity required for the thermal interpretation.}

The black hole case is well-studied; how about a critical collapse spacetime? This is much less understood due to its inherent non-stationary nature, given the lack of a timelike Killing vector.  As briefly discussed in Section~\ref{sec:intro}, confusion often arises when one imposes states mimicking Hawking radiation, an effect expected to be irrelevant, since critical collapse only concerns the first-formed or earliest marginally trapped surface. Instead, we will demonstrate in Sections~\ref{sec:quantum2+1} and \ref{sec:collapse3+1} that regularity of $\langle T_{\mu \nu} \rangle$ singles out a \textit{unique} quantum state. The resulting $\langle T_{\mu \nu} \rangle$ not only satisfies Wald's axioms but also admits a clear physical interpretation: it is asymptotically Minkowskian, meaning that $\langle T_{\mu \nu} \rangle$ vanishes near asymptotic infinities. Hence, the state reduces to the usual Minkowski vacuum corresponding to the asymptotic observers. It reflects vacuum polarization due to the matter field in the critical spacetime, without introducing any artificial Hawking flux. This ensures that the quantum backreaction arises genuinely from the collapsing matter itself.

From a physical perspective, this is the \textit{a priori} expected state for studying critical collapse. Remarkably, this state is not imposed by hand, but rather emerges naturally from the requirement of regularity. The structure is reminiscent of the Boulware state in Schwarzschild spacetime, which is also the only static state that is asymptotically Minkowskian and describes vacuum polarization in the exterior of a static star. 

However, there are important differences. In black hole spacetimes, the Boulware state is typically considered unphysical due to its perturbative divergence at the event horizon. It can only describe the vacuum polarization exterior to a static massive star, whose radius is larger than the Schwarzschild radius such that the physically relevant portion does not contain horizons. This problem is generic for the Boulware state in backgrounds with a horizon.\footnote{But see~\cite{Fabbri:2005zn, Fabbri:2005nt, Ho:2018fwq, Ho:2019pjr, Arrechea:2019jgx, Barcelo:2019eba, Beltran-Palau:2022nec, Wu:2023uyb, Arrechea:2024cnv, Numajiri:2024qgh} for recent developments arguing against this interpretation and showing that the Boulware state naturally leads to horizon-less geometries.} In critical spacetimes, this issue is naturally avoided since no horizon has yet formed. 

Another key distinction lies in how the state is defined. The canonical Boulware state is specified by quantizing a static background (e.g., Schwarzschild) in coordinates adapted to asymptotic flatness, such as Eddington-Finkelstein coordinates, and choosing plane-wave modes aligned with the timelike Killing vector. This leads to a time-independent $\langle T_{\mu \nu} \rangle$ describing pure vacuum polarization. In critical spacetimes, we will likewise find that the two-dimensional $\langle T_{ab} \rangle$ is enforced to be time-independent. However, upon lifting back to the original higher-dimensional setting, a universal time-scaling behavior emerges with respect to the adapted time coordinate $T$, due to the dilaton coupling. The reason is that while the spacetime is not stationary, it is invariant under time rescaling due to self-similarity.

We refer to this unique state in the critical spacetime as a \textit{Boulware-like state} or, more descriptively, an \textit{asymptotically Minkowskian state}. Note that the latter can still be slightly misleading, since the spacetime is not truly asymptotically flat unless it is explicitly junctioned to a different exterior solution. In Sections~\ref{sec:quantum2+1} and \ref{sec:collapse3+1}, we will show that this state choice does not require any Weyl-invariant terms in the one-loop effective action.\footnote{A similar feature is observed in black hole spacetimes~\cite{Balbinot_1999}, though it cannot be understood as simply setting the normal-ordered part $\langle: T_{ab}: \rangle=0$, as emphasized in~\cite{Wu:2023uyb}. Typically, Weyl-invariant terms can violate Wald’s axioms or obstruct solutions to the semiclassical Einstein equation. But of course, legal Weyl-invariant terms can arise from the heat kernel~\cite{Mukhanov_1994, Balbinot_1999, Wu:2023uyb}. In fact, we present counterexamples in Appendix~\ref{sec:WeylGarfinkle} and Appendix~\ref{sec:Hayward} where Weyl-invariant terms are essential for realizing a stationary state, or where the global causal structure precludes the existence of such a state altogether.}

\subsubsection*{Calculating the quantum stress-energy tensor} 
Now we are ready to write down the prescription of computing $\langle T_{ab} \rangle$ from \eqref{eq:fulloneloop}. Since $\Gamma_{\text{anom}}$ is non-local, we can make it local by introducing auxiliary scalar fields $\chi_1$ and $\chi_2$ satisfying~\cite{PhysRevD.52.2239, Wu:2023uyb}
\be \label{eq:chi1}
\Box \chi_1= \lambda_1 R+ \lambda_2 (\nabla \phi)^2,
\ee
\be \label{eq:chi2}
\Box \chi_2=- \mu_1 R-\mu_2 (\nabla \phi)^2,
\ee
where $\lambda_1, \lambda_2, \mu_1$ and $\mu_2$ are arbitrary coefficients. The solutions of $\chi_1$ and $\chi_2$ in the background spacetime therefore encode the choice of state, that is, the boundary conditions. Then the $\Gamma_{\text{anom}}$ can be decomposed into three parts
\bea
\Gamma_{\chi_1}&=& \hbar \int d^2 x \sqrt{-g} \bigg[ \frac{1}{2} (\nabla \chi_1)^2+ \chi_1 (\lambda_1 R + \lambda_2 (\nabla \phi)^2) \bigg],
\\
\Gamma_{\chi_2}&=& \hbar \int d^2 x \sqrt{-g} \bigg[ -\frac{1}{2} (\nabla \chi_2)^2+ \chi_2 (\mu_1 R + \mu_2 (\nabla \phi)^2) \bigg],
\\
\Gamma_{\phi} & =&-\frac{\hbar}{8\pi}\int\sqrt{-g}\phi R,
\eea
where in order to restore $\Gamma_{\text{anom}}$, the following requirements must be satisfied 
\be \la{constraint1}
\lambda_1^2 -\mu_1^2=-\frac{1}{48 \pi}, \quad \lambda_1 \lambda_2-\mu_1 \mu_2=\frac{1}{8 \pi}, \quad \lambda_2^2 -\mu_2^2=0. 
\ee
The last constraint requires that there is no additional Weyl-invariant term that goes like $(\nabla \phi)^2\frac{1}{\Box} (\nabla \phi)^2$ in the action. The set of constraints \er{constraint1} allows us to express the stress-energy tensor in terms of only $\lambda_2$ by the following two sets of solutions
\be \label{eq:sol1}
\{\lambda_1=\frac{1}{16 \pi \lambda_2}-\frac{\lambda_2}{12}, \quad \mu_1=\frac{-1}{16 \pi \lambda_2}-\frac{\lambda_2}{12}, \quad \lambda_2=\mu_2\},
\ee
or
\be \label{eq:sol2}
\{\lambda_1=\frac{1}{16 \pi \lambda_2}-\frac{\lambda_2}{12}, \quad \mu_1=\frac{1}{16 \pi \lambda_2}+\frac{\lambda_2}{12}, \quad \lambda_2=-\mu_2\},
\ee
that would give identical results. The quantum stress-energy tensor in two dimensions would be
\be \label{eq:quantumstress}
\langle T^{(2)}_{ab} \rangle=\langle T^{(\chi_1)}_{ab} \rangle+\langle T^{(\chi_2)}_{ab} \rangle+\langle T^{(\phi)}_{ab} \rangle,
\ee
where
\bea
\langle T^{(\chi_1)}_{ab} \rangle &=& \hbar \bigg[ -\nabla_a \chi_1 \nabla_b \chi_1 +\frac{1}{2} g_{ab} (\nabla \chi_1)^2  + 2\lambda_1 ( \nabla_a \nabla_b \chi_1- g_{ab} \Box \chi_1) 
\no\\
&\quad&
  - 2 \lambda_2 \chi_1 \bigg( \nabla_a \phi \nabla_b \phi-\frac{1}{2} g_{ab} (\nabla \phi)^2 \bigg) \bigg],
\\
\langle T^{(\chi_2)}_{ab} \rangle &=&  \hbar \bigg[ \nabla_a \chi_2 \nabla_b \chi_2 -\frac{1}{2} g_{ab} (\nabla \chi_2)^2  + 2\mu_1 ( \nabla_a \nabla_b \chi_2- g_{ab} \Box \chi_2) 
\no\\
&\quad&
- 2 \mu_2 \chi_2 \bigg( \nabla_a \phi \nabla_b \phi-\frac{1}{2} g_{ab} (\nabla \phi)^2 \bigg) \bigg],
 \\
\langle T^{(\phi)}_{ab} \rangle &=& -\frac{\hbar}{4 \pi} (\nabla_a \nabla_b \phi- g_{ab} \Box \phi).
\eea
One can verify that $\langle T^{(2)}_{ab} \rangle$ restores the trace anomaly \eqref{eq:traceanomaly}. The two-dimensional stress-energy tensor from the $s$-wave sector satisfies simple relations with the higher-dimensional ones $\langle T^{(D)}_{\mu \nu} \rangle$ consistent with the conservation law, as we will see in Sections~\ref{sec:quantum2+1} and \ref{sec:collapse3+1}.

We have carefully laid out the formulation of the one-loop theory to be applied to critical spacetimes in Sections~\ref{sec:quantum2+1} and \ref{sec:collapse3+1}. Before proceeding, we comment on two works that have attempted to incorporate quantum effects in critical spacetimes through the same techniques we are adopting.

\textbf{(i)} The study by Chiba and Siino~\cite{Chiba} is the closest one to ours since the same two-dimensional model and one-loop effective action are employed, but without carefully addressing the choice of the quantum state and local counterterms. It is argued that regularity at the center of critical spacetime demands the disappearance of echoes, and the semiclassical equations do not admit a CSS solution. In contrast, we will demonstrate with explicit examples that there is a unique choice of state in critical spacetime preserving regularity, where the quantum effects act as a growing mode. The semiclassical equations therefore admit solutions that are not exactly CSS but quasi-CSS, with a dynamically relevant scale breaking exact self-similarity.

\textbf{(ii)} A recent work by Moitra~\cite{Moitra:2022umq} has initiated a systematic study of the most general two-dimensional dilaton gravity up to second-order in derivatives compatible with CSS. The gravity sector contains the metric, dilaton $\phi$, and a gauge field
\be
S_{\text{grav}}\propto \int d^2 x \sqrt{-g} A (\phi) [R+\gamma (\nabla \phi)^2+V(\phi) -G(\phi) F_{\mu \nu} F^{\mu \nu}],
\ee
with $A(\phi)$ and $G(\phi)$ being arbitrary dilaton-dependent couplings. $V(\phi)$ is the dilaton potential and $\gamma$ is a constant coefficient. By assuming CSS in the metric and the dilaton, the gauge field term can be absorbed into the potential by a redefinition $V(\phi) \to V_{\text{eff}}$, leading to two different classes of models depending on $A(\phi)$. 

One class is related to Liouville theory and admits a minimally coupled massless scalar field. In this model, one can include one-loop effects through the Polyakov action \eqref{eq:Polyakov} for the matter and obtain closed-form solutions. However, this is a purely two-dimensional theory and does not correspond to the higher-dimensional free scalar field.

The other class is termed ``the stringy model'' and is closely related to ours since it features dilaton-coupled massless scalar $f$, with an  action given by
\be \label{eq:Upamanyu2D}
S\propto \int d^2 x \sqrt{-g} e^{-2 \phi} [R+\gamma(\nabla \phi)^2+V_{\text{eff}} e^{\frac{2 \phi}{\kappa}}- (\nabla f)^2-V_f e^{\frac{2 f}{\lambda}}],
\ee
where $\kappa$ and $\lambda$ are constant coefficients coming from assuming CSS ansatzes for the dilaton and scalar fields, respectively. Here, $V_f$ is an additional scalar field potential. Our theory, given by \eqref{eq:gravityaction} and \eqref{eq:matteraction}, clearly corresponds to a particular choice of these coefficients. However, we emphasize that the general model \eqref{eq:Upamanyu2D} does not necessarily admit a higher-dimensional origin.

It was claimed that a careful analysis of the classical equations of motion leads to no closed-form solutions to \eqref{eq:Upamanyu2D}, which seems to be troublesome if we want to analytically extract quantum effects based on this type of model. In fact, numerous closed-form CSS solutions are known in higher dimensions~\cite{Garfinkle:2000br, Roberts:1989sk, Brady:1994aq, Brady:1994xfa, Hayward:2000ds, Hirschmann:2002bw, Clement:2001ns, Clement:2001ak, Baier:2013gsa, Clement:2014pua, Clement:2014rda}. Upon dimensional reduction, these yield solutions to \eqref{eq:Upamanyu2D}. Some of these geometries exhibit exotic features, and their precise connection to critical collapse is still under debate. The two exceptions are the three-dimensional Garfinkle spacetime and the four-dimensional Roberts spacetime, to be analyzed in Sections~\ref{sec:quantum2+1} and \ref{sec:collapse3+1}, which are both CSS and solve \eqref{eq:Upamanyu2D} in the effective two-dimensional spacetimes. Furthermore, this model motivates us to find a new exact solution to \eqref{eq:Upamanyu2D} parametrized by the one-parameter family $\gamma$ with rich global structures and peculiar quantum properties, which we call the Weyl-Garfinkle spacetime. It is a purely two-dimensional solution that we discuss in Appendix~\ref{sec:WeylGarfinkle}.

Finally, we stress that critical collapse in a purely two-dimensional setting is conceptually obscure since gravity in $1+1$ dimensions lacks transverse spatial volume, and the notion of forming a horizon or concentrating matter to a point becomes ill-defined. There is no physical ``collapse" in the traditional sense, only wave propagation along a line. The two-dimensional model in \eqref{eq:gravityaction} and \eqref{eq:matteraction} is merely an effective description that captures the $s$-wave dynamics of higher-dimensional gravitational collapse. In this reduced framework, the dilaton field encodes the geometry of the transverse sphere, allowing us to study quantum effects within a well-defined and analytically tractable setting.

\section{Quantum critical collapse in $2+1$ dimensions}
\label{sec:quantum2+1}

This section focuses on the semiclassical properties of critical gravitational collapse for the Einstein-scalar system in $2+1$ dimensions, where a fundamental tension arises concerning the role of the cosmological constant.

\subsection{The critical Garfinkle spacetime in $2+1$ dimensions}
\label{sec:classicalGarfinkle}

Critical collapse in $2+1$ dimensions is particularly intriguing for several reasons~\cite{Jalmuzna:2015hoa}. First, the standard argument for mass scaling in Type II collapse, which relies on dimensional analysis, fails in three-dimensional spacetimes. In higher dimensions, the black hole mass exhibits a scaling relation with the maximum curvature in the subcritical regime (see discussion around \eqref{eq:dimensionalrelation}), which can be connected to the amplitude of the collapsing scalar field. However, in $2+1$ dimensions, both the black hole mass and the total energy of the scalar field are dimensionless. Consequently, dimensional analysis provides no guidance for scaling behavior, and the emergence of Type II critical phenomena in this context is surprising. It also makes identifying Type I behavior more subtle, as it usually requires a dynamically relevant scale, such as the quantum corrections that we will explore later.

Second, black holes in $2+1$ dimensions are known to exist classically\footnote{However, this statement is true only in pure gravity. One simple way of seeing this is through brane-world holography, in which one can construct three-dimensional black holes for all values of the cosmological constant \cite{Emparan:1999wa, Emparan:1999fd, Emparan:2002px, 
Arkani-Hamed:2007ryu,  Emparan:2020znc, Emparan:2022ijy}. The key ingredient lies in the new scale generated by brane-world quantum effects, which are enhanced by the large number of fields. This then illustrates that such black holes can exist in non-holographic setups as well; one just needs enough quantum matter.} only in the presence of a negative cosmological constant ($\Lambda < 0$), as exemplified by the Ba\~{n}ados-Teitelboim-Zanelli (BTZ) solution~\cite{Banados:1992wn}. The first numerical study for the appearance of critical behavior in $2+1$ dimensions was indeed done by Pretorius and Choptuik in AdS spacetime for the Einstein-scalar system~\cite{Pretorius:2000yu} (see also~\cite{Husain:2000vm}).\footnote{A consequence of $\Lambda < 0$ is that asymptotic flatness is replaced by asymptotically AdS boundary conditions. For a massless scalar field, the only boundary conditions compatible with the Einstein equations are totally reflecting. Under such conditions, even arbitrarily weak initial data can collapse after undergoing multiple reflections off the AdS boundary, eventually forming a black hole. Nevertheless, the mass of the apparent horizon at the time of its first appearance still obeys the characteristic scaling law.} This result is striking not only because Type II behavior was unexpected, but also because it immediately reveals a conceptual tension: black hole formation requires a negative cosmological constant, and indeed, their solution was shown to asymptotically approach the BTZ geometry exterior to the event horizon, confirming that $\Lambda$ plays a role in the background. However, for the mass scaling characteristic of Type II collapse to occur, the cosmological constant must be dynamically irrelevant.

Finally, the nature of the self-similarity observed adds to the peculiarity. In all higher-dimensional examples of critical collapse in the Einstein-scalar system, the critical solution exhibits DSS with a characteristic echoing period~\cite{Gundlach:2025yje}. In $2+1$ dimensions, the numerical solution is CSS, suggesting that the critical phenomena in three dimensions may belong to a distinct universality class from their higher-dimensional counterparts.

These considerations motivated Garfinkle's analytic construction of a solution that serves as a candidate for the critical spacetime~\cite{Garfinkle:2000br, Garfinkle:2002vn}.  Fortunately, in $2+1$ dimensions, the problem is tractable enough that by imposing CSS on both the metric and the massless scalar field, a closed-form solution can be explicitly constructed. This solution also alleviates the tension discussed above, as it is an exact solution when $\Lambda = 0$. This means that in the regime when $\Lambda$ is dynamically irrelevant and the CSS property emerges, the Pretorius-Choptuik solution may be well approximated by the Garfinkle spacetime. Indeed, to our knowledge, this remains the only known closed-form solution in which the connection to genuine critical collapse is clear~\cite{Jalmuzna:2015hoa}.\footnote{We will discuss the Roberts solution in $3+1$ dimensions in Section~\ref{sec:collapse3+1}, which can also be written in closed form. However, its relation to true critical collapse is more subtle.} 

In the following, we introduce the classical Garfinkle spacetime and discuss issues related to this solution. As we will see, a detailed analysis of its global structure and classical perturbations reveals that it does not satisfy all the criteria for a critical spacetime laid out in Section~\ref{sec:criticalcollapse}. It turns out that puzzles surrounding this solution have to do with the reintroduction of the cosmological constant~\cite{Clement:2001ns, Clement:2001ak, Hirschmann:2002bw, Cavaglia:2004mt, Baier:2013gsa, Clement:2014mja, Clement:2014pua, Clement:2014rda, Jalmuzna:2015hoa}.

The Garfinkle metric is a CSS exact solution to the Einstein-scalar system \er{eq:Einsteinscalar} in $2+1$ dimensions. By further adopting the Christodoulou ansatz, such that the massless scalar field is also CSS~\cite{Christodoulou:1994hg}
\be \la{CSSscalar}
f(T, x)= c T+f(x),
\ee
where $c$ is a constant to be specified later, and the function $f(x)$ depends only on $x$, one can obtain a solution in closed form. An important fact is that introducing the cosmological constant $\Lambda$ would obviously break the scale invariance inherent in the CSS property. However, it is dynamically irrelevant as we approach larger and larger curvature regions, that is, the regime we care about for Type II collapse. Hence, we have an exact solution to the Einstein-scalar field equations only when $\Lambda=0$. 

In the adapted coordinates $(T, x)$, the Garfinkle solution is represented by a one-parameter family $n$ and is given by 
\be \label{eq:Garfinklemetric}
ds^2=  \ell^2 e^{-2 T} \bigg[e^{2 \rho (x, n)} \bigg(dx-\frac{x}{2n} dT \bigg)dT+r^2(x, n) d \theta^2 \bigg],
\ee
where
\be
e^{2\rho (x, n)}=2n \bigg(\frac{1+x^n}{2} \bigg)^{4(1-\frac{1}{2n})}, \quad r(x, n)=\frac{1-x^{2n}}{2},
\ee
and the domain being $T \in (- \infty, \infty)$ and $x \in [0,1]$. The solution can be analytically continued to $x \in [-1, 0)$, where additional subtleties arise that we will address shortly. The scalar field supporting this geometry is
\be
f(T,x)=\sqrt{\frac{2n-1}{2 n}}\bigg[T-2 \ln{\bigg(\frac{1+x^n}{2} \bigg)}\bigg],
\ee
reminiscent of the form \er{CSSscalar}. 

\begin{figure}[t!]
    \centering
    \includegraphics[width=0.4\linewidth]{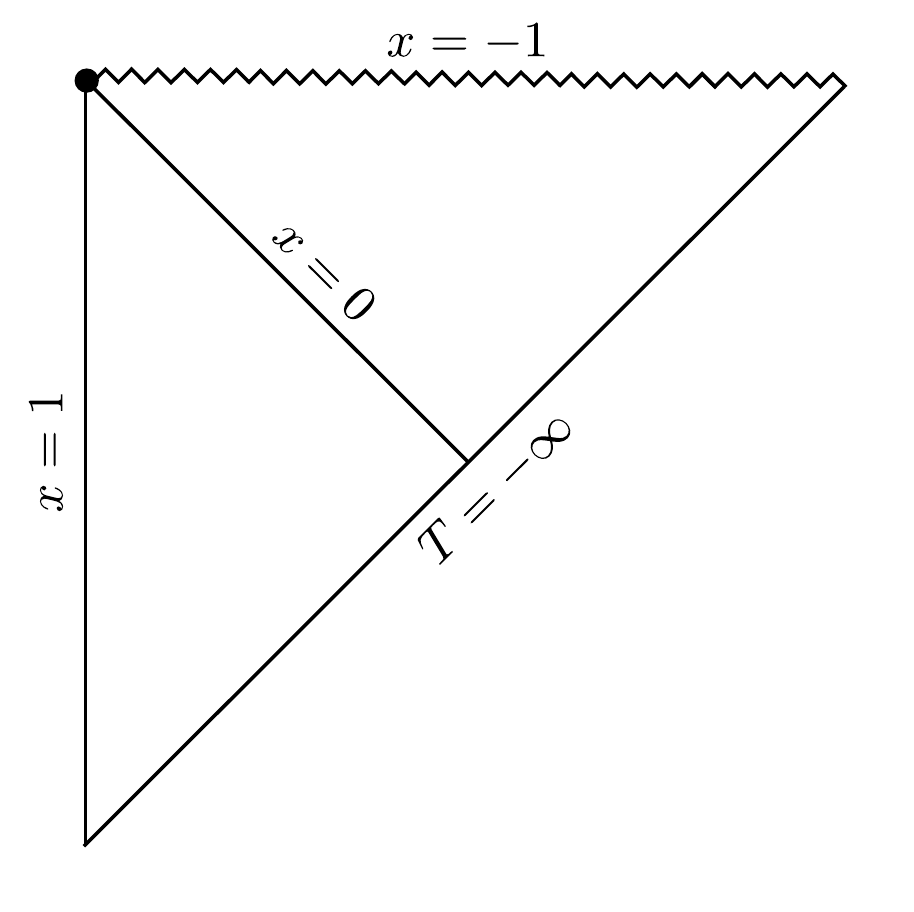}
    \includegraphics[width=0.4\linewidth]{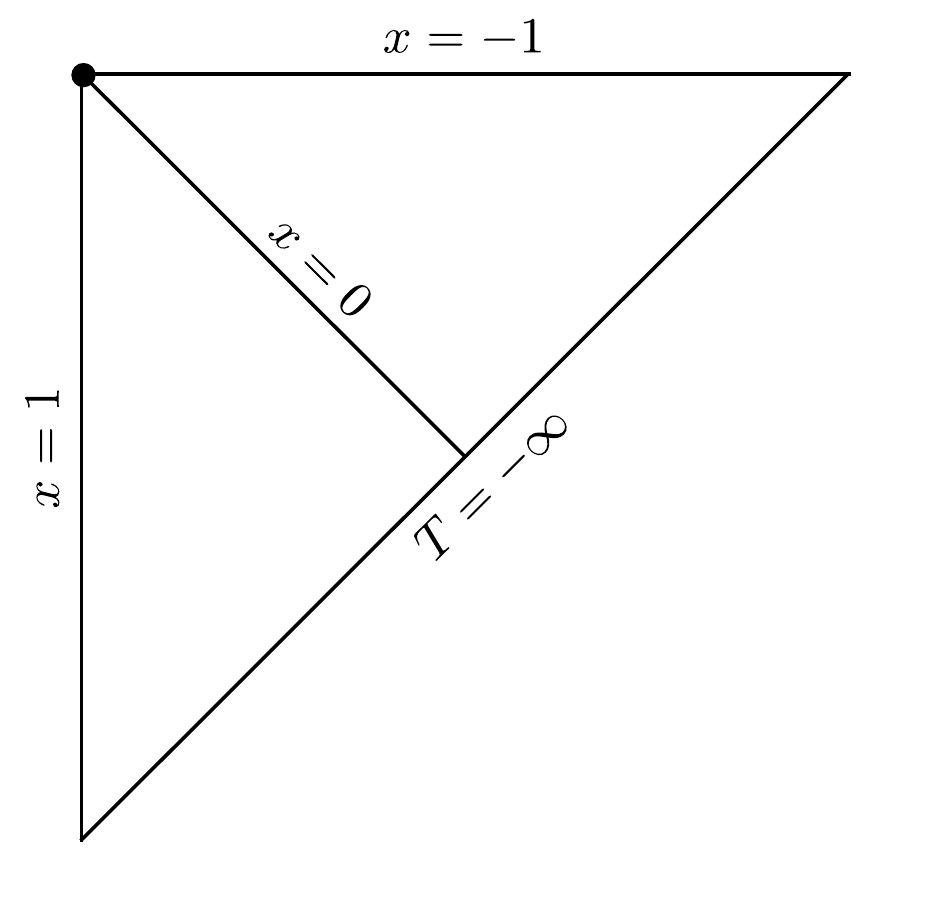}
    \caption{The causal structure of Garfinkle spacetime for odd (left panel) and even $n$ (right panel). The thick black dot represents the kinematical point singularity at $T = \infty$. The primary distinction lies in the analytic continuation beyond the past light cone of the singularity, covering the region $x \in [-1,0)$. As can be seen from the curvature invariants \eqref{eq:GarfinkleRicci}, for odd $n$ there is a spacelike curvature singularity located at $x = -1$. Throughout this paper, we focus exclusively on the interior region, which corresponds to $x \in [0,1]$.}
    \label{fig:garfinklen}
\end{figure}

The Penrose diagrams of the Garfinkle spacetime are depicted in Figure~\ref{fig:garfinklen}. The solution is analytic at the center $x=1$ and the light cone $x=0$ (and anywhere in between) if and only if $n$ is a positive integer. The curvature singularity occurs at $T \to \infty$, which can be confirmed by computing the curvature invariants
\be \label{eq:GarfinkleRicci}
R=\frac{e^{2T}}{\ell^2}\frac{(1-2n)2^{5-\frac{2}{n}}x^{n-1}(x^n+1)^{\frac{2}{n}-6}}{n},
\ee
\be
R^2=R^{\mu \nu}R_{\mu \nu}=\frac{1}{3}R^{\mu \nu \rho \sigma}R_{\mu \nu \rho \sigma},
\ee
which means we only need to examine the Ricci scalar $R$. This is a kinematical singularity as we discussed in Section~\ref{sec:criticalcollapse}. A remark is that for $n=\frac{1}{2}$, which is not the regime under consideration since we require $n \in \mathbb{N}$, the scalar field profile vanishes, and the geometry reduces to Minkowski written in self-similar coordinates. This fact is irrelevant for the classical analysis, but it will serve as a useful consistency check when analyzing quantum corrections in Section~\ref{sec:Garfinkleoneloop}.

The Garfinkle solution is a remarkable surprise and stands out as an excellent candidate for a critical spacetime in $2+1$ dimensions, in part because it is simple enough to be constructed in closed form. This contrasts sharply with the situation in higher dimensions, where the critical solutions for the Einstein-scalar system are inferred numerically. Moreover, the Garfinkle solution is CSS. In higher-dimensional models, the presence of a potential is typically essential to support regular CSS solutions. In $2+1$ dimensions, the Garfinkle solution remains regular even with $\Lambda=0$ and in the absence of any potential.

\FloatBarrier

However, by examining further its global structure and the classical perturbation modes~\cite{Garfinkle:2000br, Garfinkle:2002vn}, the Garfinkle solution cannot be the true critical spacetime for the following reasons:
\begin{itemize}
        \item The surface $x=0$, which corresponds to the past light cone of the naked singularity, is a \textit{marginally outer-trapped surface.} In the spherically symmetric case, the presence of trapped surfaces can be diagnosed by evaluating the scalar~\cite{Frolov:1998} 
        \be
        (\nabla \bar{r})^2=2^{4-\frac{2}{n}}x^{2n-1}(x^n+1)^{\frac{2}{n}-4},
        \ee
        where $\bar{r}=\ell e^{-T}r$. Regions where $(\nabla \bar{r})^2<0$ correspond to trapped surfaces. This condition is satisfied for all $x \in (-1,0)$ with $n \in \mathbb{N}$, indicating that every circle exterior to the light cone lies within the trapped region.

       At $x=0$, we have $(\nabla \bar{r})^2=0$, marking the boundary of the trapped region. Since the lines of constant $x$, which align with the homothetic vector field, are timelike for $x>0$ and spacelike for $x<0$, this confirms that the surface $x=0$ is marginally outer-trapped.

       This is problematic because a genuine critical solution, lying precisely at the threshold between dispersion and black hole formation, should not itself contain trapped surfaces.

        \item Even worse, beyond the light cone there is a spacelike central curvature singularity for every odd $n$ at $x=-1$. This can be confirmed from the Ricci scalar in \eqref{eq:GarfinkleRicci}.

        As discussed in Section~\ref{sec:criticalcollapse}, a legitimate critical solution may allow one to discard regions beyond the future light cone of the naked singularity (i.e., the Cauchy horizon). But analytic continuation beyond the past light cone for $0>x \geq -1$ is relevant and is part of the critical spacetime, since it governs whether the singularity is visible.

        The singularity and horizon structure, therefore, suggest that the Garfinkle geometry is better interpreted as a self-similar black hole solution, rather than as the critical solution itself. 
        
        \item A defining criterion for a critical spacetime, as summarized in Section~\ref{sec:criticalcollapse}, is the presence of a single growing mode that tunes the system toward black hole formation. But a classical linear perturbation indicates that there are additional unstable growing modes~\cite{Garfinkle:2002vn}. Specifically, for a given integer $n$, there are $n-1$ growing modes, as we will briefly review in Section~\ref{sec:GarfinkleHorizon}. The expectation immediately suggests that $n=2$ corresponds to the critical spacetime. Yet, the associated critical exponent does not agree with numerical results. In contrast, the $n=4$ case yields a critical exponent that matches well with the simulations when considering its most dominant growing mode.\footnote{According to~\cite{Pretorius:2000yu}, the first marginally outer-trapped surface exhibits a mass-scaling exponent of $\delta \simeq 0.6$, while~\cite{Husain:2000vm} gave $\delta \simeq 0.81$ for data sufficiently far from criticality. In the latest study~\cite{Jalmuzna:2015hoa}, the numerical result suggests $\delta \simeq 0.68$, which agrees well with the theoretical prediction $\delta \simeq 0.6957$ obtained from the $n=4$ case using the top growing mode. A similar agreement is found for the scaling exponent of the Ricci scalar.} 
\end{itemize}

Resolving these tensions, and more generally, building an analytic bridge for critical collapse in $2+1$ dimensions and numerical results, has been the focus of several studies~\cite{Clement:2001ns, Clement:2001ak, Hirschmann:2002bw, Cavaglia:2004mt, Baier:2013gsa, Clement:2014mja, Clement:2014pua, Clement:2014rda, Jalmuzna:2015hoa}. It turns out that the aforementioned issues in Garfinkle spacetime can be resolved by incorporating the cosmological constant~\cite{Cavaglia:2004mt, Jalmuzna:2015hoa}, indicating that $\Lambda$ plays an important role in $2+1$ dimensions. Let us give an outline below.

As discussed in Section~\ref{sec:criticalcollapse}, when the field equations are not strictly scale-invariant but approach scale invariance asymptotically at small scales, the critical solution can often be approximated by an expansion in powers of the ratio between the solution’s intrinsic length scale and the length scale set by the field equations. It was formally shown in~\cite{Gundlach:1996vv} that, at leading order, this expansion yields a scale-invariant solution.

Indeed, the cosmological constant given by the AdS length scale $\Lambda =-\frac{1}{\ell^2}$ is dynamically irrelevant for Type II self-similar behavior. This is evident in the Garfinkle solution, which requires $\Lambda=0$ to preserve scale invariance. Nevertheless, the cosmological constant can still be treated as a perturbative correction, expressed in powers of  $e^{-T} \propto -\frac{\tilde{u}}{\ell}$ for a suitable coordinate $\tilde{u}$, thereby yielding an approximately CSS or quasi-CSS solution. Let us formally write down the expansion as~\cite{Cavaglia:2004mt}
\be
\tilde{\rho}(T,x)=\tilde{\rho}_0(x) + \sum_{n=1}^\infty e^{-2nT} \rho_n(x),
\ee
\be
r(T,x)=r_0(x)+\sum_{n=1}^\infty e^{-2nT} r_n(x),
\ee
\be
f(T,x)=\sqrt{\frac{2n-1}{2 n}}[T+f_0(x)+\sum_{n=1}^\infty e^{-2nT} f_n(x)],
\ee
where the leading order solution is given by the Garfinkle metric. Here $\tilde{\rho}_0(x)$ is related to $\rho$ defined in \eqref{eq:Garfinklemetric} via $e^{2 \rho}=2n x^{2n-1} e^{2 \tilde{\rho}_0}$. Perturbative corrections can be introduced both inside and outside the light cone. In the interior region, the first-order corrections do not admit closed-form expressions but serve to remove the apparent horizon at $x=0$~\cite{Cavaglia:2004mt}. In the exterior, rather than na\"ively extending the Garfinkle solution into the region $x \in [-1,0)$, one may instead impose a junction condition at the past light cone and match the interior to a distinct exterior solution, consistent with numerical simulations. This exterior solution, obtained both numerically and analytically, is described by an outgoing Vaidya metric and is referred to as the null continuation in~\cite{Jalmuzna:2015hoa}. The first-order corrections in the exterior are given by
\be
\rho_1 (x)=\frac{4^{\frac{1}{n}} n^2 (1-8n)}{16 (1-5n) (1-6n)}x,
\ee
\be
r_1 (x)=\frac{4^{\frac{1}{n}} n^2}{16 (1-6n)}x,
\ee
\be
f_1(x)=-\frac{4^{\frac{1}{n}} n^3}{8 (1-5n)(1-6n)}x,
\ee
all of which vanish at the light cone $x=0$. These $\Lambda$-corrections remove all marginally outer-trapped surfaces in the exterior. Moreover, in the supercritical regime, the classical linear perturbation with a growing mode reintroduces a well-defined apparent horizon in the exterior.\footnote{We will include the classical perturbation inside the light cone that also brings back the apparent horizon in the horizon-tracing analysis in Section~\ref{sec:GarfinkleHorizon}, where it first forms near the original apparent horizon, consistent with~\cite{Jalmuzna:2015hoa}.}

It is conjectured that the true critical solution is well approximated by the $n=4$ Garfinkle solution inside the light cone and its null continuation outside, together with their first-order $\Lambda$-corrections. The issue of the additional growing modes is then resolved from a numerical perspective. By gluing the $\Lambda$-corrected Garfinkle solution inside the light cone to a new exact solution outside, one selects $n=4$ with the top growing mode and eliminates two of the three growing modes. Further numerical studies support this conjecture: the modified $n=4$ Garfinkle solution appears to have only one growing mode when evolved with $\Lambda<0$~\cite{Jalmuzna:2015hoa}.

This analysis shows that $\Lambda$ plays a key role in elevating the Garfinkle solution to a viable critical solution, and it alleviates the tension with the known BTZ black hole geometry in $2+1$ dimensions. However, we emphasize that the precise relationship between the $\Lambda$-corrected Garfinkle solution and the BTZ geometry remains to be fully understood.

With the limitations of the classical Garfinkle solution now properly accounted for, it serves as a useful starting point for investigating how quantum effects may alter the dynamics. Since $\Lambda$ becomes negligible at small scales, it is justified to study quantum corrections within the $\Lambda=0$ approximation.

In the remainder of this section, we focus on the interior of the past light cone, where the central question is whether regular collapsing matter can give rise to a naked singularity. Although the behavior of the apparent horizon and the impact of quantum corrections in the exterior region are also crucial for assessing the visibility of such singularities, we defer a detailed treatment of those aspects for future work.

\subsection{One-loop analysis and the semiclassical Garfinkle spacetime} \label{sec:Garfinkleoneloop}

Having introduced the one-loop effective theory in Section~\ref{sec:onelooptheory}, we are ready to perform an explicit semiclassical analysis of the Garfinkle spacetime. We will find that a regular quantum stress-energy tensor describing the backreaction from the collapsing matter for general $n$ Garfinkle spacetime can be constructed in closed form, thanks to the analytic property of the background geometry.

\begin{figure}
\centering
\includegraphics[width=0.28\textwidth]{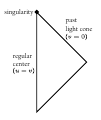}
\caption{The interior region of the Garfinkle spacetime. A consistent quantum stress-energy tensor $\langle T_{\mu \nu} \rangle$ must be regular at both the center $(u=v)$ and the past light cone $(v=0)$ of the singularity.}
\label{Garfinkle_uv}
\end{figure}

Since in the semiclassical calculation we are always perturbing around $\frac{\hbar}{\ell^{D-2}}=\frac{\hbar}{\ell} \ll 1$, here we omit writing both $\hbar$ and $\ell$ explicitly in the intermediate steps, but restore them in the end for the semiclassical Einstein equations. It will be much more convenient to apply the one-loop action in double-null coordinates $(u,v)$, which is related to the adapted coordinates via
\be \label{eq:Garfinkletcoord}
x=\frac{v}{u}, \quad T=-2n \ln{(-u)} \implies u=-e^{\frac{-T}{2n}}, \quad v=ux=-e^{\frac{-T}{2n}} x,
\ee
where we restrict the domain $u \in (-\infty, 0]$ and $ v \in (-\infty, 0]$, yet $|v|\leq|u|$, corresponding to the interior region. Whenever we need to make the physics manifest, we will transform the relevant quantities back to the adapted coordinates $(T, x)$. The metric is then given by~\cite{Garfinkle:2000br, Garfinkle:2002vn, Jalmuzna:2015hoa}
\be
ds^2=-e^{2 \rho} du dv+ r^2 d \theta^2,
\ee
where
\be
e^{2 \rho}=4n^2 \bigg[\frac{(-u)^n+(-v)^n}{2} \bigg]^{4(1-\frac{1}{2n})},\quad r= \frac{1}{2} [(-u)^{2n}-(-v)^{2n}].
\ee
Again with $n \in \mathbb{N}$ for the solution to be analytic at the center $u=v$ and the light cone $v=0$. We express $r$ in this form to highlight that for $n=\frac{1}{2}$, the metric reduces to flat Minkowski spacetime. Note that $u=v=0$ is the curvature singularity; $u=-\infty$ represents the past null infinity $\mathcal{I}^-$; $v=0$ for finite negative $u$ is the past light cone, which, as we mentioned, is also an apparent horizon. See Figure~\ref{Garfinkle_uv} for a Penrose diagram of the interior region. The background scalar field $f$ is then
\be
f=-2\sqrt{\frac{2n-1}{2 n}} \ln{\bigg[\frac{(-u)^n+(-v)^n}{2} \bigg]}.
\ee
In the following, we will not specialize to any particular value of $n$ in the Garfinkle spacetime and work out $\langle T_{\mu \nu} \rangle$ with its full generality. In Section~\ref{sec:GarfinkleHorizon}, we take $n=2$ as a representative example for the horizon-tracing problem, deferring the physically most relevant case $n=4$, along with general lessons for arbitrary $n$, to Appendix~\ref{sec:horizonGarfinkle}.

\FloatBarrier

A consistent quantum backreaction from the matter field in critical collapse must result in a regular $\langle T_{\mu \nu} \rangle$ and regular backreacted spacetime as a solution to the semiclassical Einstein equation \eqref{eq:semiclassicalEinstein}. That is, we will only require the following boundary conditions
\be \label{eq:regularity}
\lim_{u \to v} \langle T_{\mu \nu} \rangle < \infty, \quad \lim_{v \to 0} \langle T_{\mu \nu} \rangle < \infty,
\ee
where all components of $\langle T_{\mu \nu} \rangle$ remain finite in a regular coordinate system as the center $u \to v$ and the light cone $v \to 0$ are approached. We will demonstrate that there is a unique choice of quantum state that is compatible with these regularity conditions in the interior. Surprisingly, such a state will be asymptotically Minkowskian, indicating that it naturally incorporates the physical vacuum polarization from the matter. \textit{A priori}, self-similarity need not be preserved at the quantum level and is therefore not imposed as a requirement for the state. Nonetheless, we will see that a universal homogeneous scaling behavior for $\langle T_{\mu \nu} \rangle$ emerges, leading to a growing perturbation mode, indicating that quantum corrections are indispensable.

The $s$-wave contribution to $\langle T^{(3)}_{\mu \nu} \rangle$ is encoded in the two-dimensional one-loop action \eqref{eq:fulloneloop}. The components of $\langle T^{(3)}_{\mu \nu} \rangle$ are related to the two-dimensional ones $\langle T_{ab} \rangle$ via the following $s$-wave relations compatible with conservation law~\cite{Mukhanov_1994}
\be \label{eq:3Drelations}
\langle T^{(3)}_{ab} \rangle = \frac{\langle T_{ab} \rangle}{2 \pi e^{-2 \phi}}, \quad
\langle T^{(3)}_{\theta \theta} \rangle=-\frac{1}{4 \pi  e^{2 \phi}} \frac{1}{\sqrt{-g_{(2)}}} \frac{\delta \Gamma_{\text{one-loop}}}{\delta \phi}.
\ee
Upon spherical dimensional reduction using the ansatz \eqref{eq:sphericalansatz}, the two-dimensional metric and dilaton are
\be \label{eq:2DGarfinke}
ds^2=-e^{2 \rho} du dv, \quad \phi=-\frac{1}{2}\ln{r}.
\ee
We have to make $\Gamma_{\text{anom}}$ in \eqref{eq:fulloneloop} local first, which can be achieved by solving the auxiliary fields $\chi_1$ and $\chi_2$. Evaluating \eqref{eq:chi1} and \eqref{eq:chi2} in the background \eqref{eq:2DGarfinke}, the solutions to the auxiliary fields can be written as
\be
\chi_1=-4 \lambda_1 \bigg(1-\frac{1}{2n}\bigg)  \ln{[u^n+v^n]}-\frac{\lambda_2}{4} \ln{[u^{2n}-v^{2n}]}+C_1(v)+C_2(u),
\ee
\be
\chi_2=4 \mu_1 \bigg(1-\frac{1}{2n}\bigg)  \ln{[u^n+v^n]}+\frac{\mu_2}{4} \ln{[u^{2n}-v^{2n}]}+C_3(v)+C_4(u),
\ee
where $C_i$ with $i \in \{1,\dots,4 \}$ are functions of the null coordinates accounting for the most general solutions. These functions encode the choice of the quantum state and will therefore be determined by the regularity conditions \eqref{eq:regularity}. Note that from either \eqref{eq:sol1} or \eqref{eq:sol2}, we can express $\lambda_1, \mu_1$, and $\mu_2$ in terms of $\lambda_2$. The auxiliary fields only serve as intermediate steps, and the final physical stress-energy tensor cannot depend on the free parameter $\lambda_2$.

To ensure regularity, we supplement the one-loop action \eqref{eq:fulloneloop} with the following local counterterms
\be \label{eq:counter}
\Gamma_{\text{ct}}=\int d^2 x \sqrt{-g} \alpha_1 \phi R+ \alpha_2 f R+ \alpha_3 (\nabla f)^2.
\ee
Here, the terms involving $\alpha_1$ and $\alpha_2$ cancel divergences at the center, while $\alpha_3$ regulates divergences at the past light cone. These coefficients $\alpha_i$ will be determined uniquely.

The process of determining $\langle T^{(3)}_{\mu \nu} \rangle$ is quite involved though straightforward, and we give a brief sketch here. Regularity of the two-dimensional and three-dimensional trace near the center fixes
\be
\alpha_1=\frac{1}{16 \pi}, \quad \alpha_2=\frac{1}{24 \pi} \sqrt{1-\frac{1}{2n}},
\ee
and implies the constraint among the general functions $C_i(v)$ for $\langle T^{(3)}_{\theta \theta} \rangle$ to be regular at the center
\be \label{eq:cond1}
C_1(v)+C_2(v)-C_3(v)-C_4(v)=\frac{(2n-1) \ln{(2 v^n)}}{4n \pi \lambda_2}.
\ee
In fact, the choice of $\alpha_1$ and $\alpha_2$ makes the two-dimensional trace $\langle T^a{}_a \rangle$ vanish. Imposing regularity for $\langle T^{(3)\mu}{}_\mu \rangle$ and $\langle T^{(3)}_{\theta \theta} \rangle$ near the light cone allows us to set $C_1(v)=C_3(v)=0$ without loss of generality. Equation \eqref{eq:cond1} then reduces to
\be \label{eq:cond2}
C_2(v)-C_4(v)=\frac{(2n-1) \ln{(2 v^n)}}{4n \pi \lambda_2}.
\ee 
The remaining conditions come from the diagonal components $\langle T^{(3)}_{uu} \rangle$ and $\langle T^{(3)}_{vv} \rangle$. Regularity of $\langle T^{(3)}_{vv} \rangle$ at the center gives
\be
\alpha_3=\frac{17n-4}{192 n \pi},
\ee
and our choice $C_1(v)=C_3(v)=0$ would ensure $\langle T^{(3)}_{vv} \rangle$ being regular at the light cone. Substituting $\alpha_3$ into $\langle T^{(3)}_{uu} \rangle$ and requiring it to be regular at the center, we have
\be
C'_2(v)+C'_4(v)=\frac{(1-5 n+6n^2) \lambda_2}{3 (1-4n)v}+K_1 v^{4n-2},  \quad C'_2(v)-C'_4(v)=\frac{2n-1}{ 4\pi \lambda_2 v},
\ee
where $K_1$ is a constant. We can then solve
\be
C_2'(v)=\bigg(\frac{2n-1}{8 \pi \lambda_2}-\frac{(1-5n+6n^2)\lambda_2}{6 (1-4n)} \bigg)\frac{1}{v}+\frac{K_1 v^{4n-2}}{2}, 
\ee
\be
C_4'(v)=\bigg(-\frac{2n-1}{8 \pi \lambda_2}-\frac{(1-5n+6n^2)\lambda_2}{6 (1-4n)} \bigg)\frac{1}{v}+\frac{K_1 v^{4n-2}}{2},
\ee
and then
\be
C_2(v)=\bigg(\frac{2n-1}{8 \pi \lambda_2}-\frac{(1-5n+6n^2)\lambda_2}{6 (1-4n)} \bigg)\ln{v}+\frac{K_1 v^{4n-1}}{2(4n-1)} +K_2,
\ee
\be
C_4(v)=\bigg(-\frac{2n-1}{8 \pi \lambda_2}-\frac{(1-5n+6n^2)\lambda_2}{6 (1-4n)} \bigg)\ln{v}+\frac{K_1 v^{4n-1}}{2(4n-1)} +K_3,
\ee
where $K_2, K_3$ are integration constants. In order to restore \eqref{eq:cond2}, we choose the integration constants to satisfy
\be
K_2-K_3=\frac{(2n-1) \ln{2}}{4 n \pi \lambda_2}.
\ee
We can then completely fix the quantum stress-energy tensor for general $n$ Garfinkle spacetime in the following closed form
\bea
\langle T^{(3)}_{uu}\rangle&=&-\frac{(2n-1)n}{16 \pi^2 u^2 (u^{2n}-v^{2n})^3(u^n+v^n)^2}\bigg[u^{6n} -8 u^{5n} v^n-8u^{4n}v^{2n}+9 u^{2n}v^{4n}
\no\\
&\quad&-4 u^n v^{5n}-2 v^{6n}+12 u^{3n}v^{3n}+8u^{4n} (u^n+v^n)^2 \ln{\bigg(\frac{2u^n}{u^n+v^n}\bigg)}\bigg],
\eea
\bea
\langle T^{(3)}_{vv}\rangle&=&\frac{(2n-1)n}{16 \pi^2 v^2 (u^{2n}-v^{2n})^5}\bigg[-3 u^{6n}v^{2n} +10 u^{5n}v^{3n}-u^{4n}v^{4n} -20u^{3n} v^{5n}+11 u^{2n}v^{6n}
\no\\
&\quad&+10 u^n v^{7n} -7v^{8n} +8(2u^{2n} v^{6n}-v^{8n}-u^{4n} v^{4n}) \ln{\bigg(\frac{2u^n}{u^n+v^n}\bigg)}\bigg],
\eea
\be
\langle T^{(3)}_{uv}\rangle=0,
\ee
\bea
\langle T^{(3)}_{\theta \theta}\rangle&=&- \frac{(2n-1) 4^{-1-\frac{1}{n}}}{n \pi^2 u } \frac{(u^n+v^n)^{-5+\frac{2}{n}}}{u^n-v^n} v^{n-1} \bigg[3 u^{3n}-4 u^{2n} v^n+3u^n v^{2n} -2v^{3n} 
\no\\
&\quad&-8u^{2n} v^n \ln{\bigg(\frac{2u^n}{u^n+v^n}\bigg)}\bigg]
\eea
where the end result is independent of $K_1$ and $\lambda_2$. 

A few remarks are in order. While the two-dimensional trace $\langle T^a{}_a \rangle$ vanishes on-shell due to the inclusion of local counterterms, the three-dimensional trace $\langle T^{(3)\mu}{}_\mu \rangle$ remains non-zero. This does not contradict the well-known result that there is no trace anomaly in odd-dimensional spacetimes, since that statement applies only to conformally coupled matter fields. As a consistency check, we note that when $n = \frac{1}{2}$, all components of the $\langle T^{(3)}_{\mu \nu} \rangle$ vanish identically, in accordance with one of Wald’s axioms for Minkowski normalization discussed in Section~\ref{sec:onelooptheory}. 

For general $n$, the components of $\langle T^{(3)}_{\mu \nu} \rangle$ involve complicated polynomials in $u$ and $v$. Unlike what typically happens for the black hole spacetime, the $uu$- and $vv$-components are not symmetric under the exchange of $u$ and $v$, and the $uv$-component is forced to vanish. These features reflect the nature of the quantum state in Garfinkle spacetime and its adaptation to the $(u, v)$ coordinates, yet the physical information will be clear once we go to the adapted coordinates $(T,x)$.

The quantum stress-energy tensor also exhibits some nice asymptotic behavior that corresponds to an asymptotically Minkowskian or Boulware-like state. One can check that there is no incoming energy from $\mathcal{I}^-$ at $u \to -\infty$ 
\be
\lim_{u \to -\infty} \langle T^{(3)}_{vv} \rangle=0.
\ee
And no outgoing energy at $\mathcal{I}^+$ by taking $v \to \infty$
\be
\lim_{v \to \infty} \langle T^{(3)}_{uu} \rangle=0,
\ee
although we do not have a well-defined notion of the future null infinity in the Garfinkle solution, as the quantum stress-energy tensor may not hold beyond the light cone, where it is argued that the classical solution should be junctioned to an outgoing Vaidya geometry~\cite{Jalmuzna:2015hoa}. Nevertheless, this means that the quantum state reduces to the usual Minkowski vacuum near the asymptotic infinities, and we can interpret the one-loop effects as coming from the vacuum polarization of the matter field itself.

Despite these appealing properties, the expression of $\langle T^{(3)}_{\mu \nu} \rangle$ given here in the double-null coordinates $(u, v)$ is not particularly illuminating. For a clearer interpretation, let us look at its behavior in the adapted coordinates $(T,x)$. With the transformation \eqref{eq:Garfinkletcoord}, we have
\bea
\langle T^{(3)}_{TT} \rangle&=&e^T\frac{ (2n-1)}{64 n \pi^2 (x^{2n}-1)^3} \bigg[1-10x^n+14 x^{2n}-10x^{3n}+5x^{4n} 
\no\\
&\quad&+8(x^{4n}+1) \ln{\bigg(\frac{2}{1+ x^n}\bigg)}\bigg],
\eea
\be
\langle T^{(3)}_{xT} \rangle = e^T\frac{(2n-1)x^{2n-1}}{32 \pi^2 (x^{2n}-1)^3} \bigg[-3+10x^n -7 x^{2n} -8x^{2n} \ln{\bigg(\frac{2}{1+ x^n}\bigg)}\bigg],
\ee
\be
\langle T^{(3)}_{xx} \rangle= e^T \frac{ (2n-1)n x^{2n-2}}{16 \pi^2 (x^{2n}-1)^3} \bigg[3-10 x^n+7 x^{2n}+8x^{2n} \ln{\bigg(\frac{2}{1+ x^n}\bigg)} \bigg],
\ee
\bea
\langle T^{(3)}_{\theta \theta} \rangle&=&-e^T \frac{(2n-1) 4^{-1-\frac{1}{n}} x^{n-1} (x^n+1)^{-5+\frac{2}{n}}}{n \pi^2 (x^n-1)} \bigg[-3+2 x^{3n}-3 x^{2n}+4 x^n
\no\\
&\quad&+8 x^n  \ln{\bigg(\frac{2}{1+ x^n}\bigg) }\bigg].
\eea
Surprisingly, the stress-energy tensor retains a quasi-CSS structure with a homogeneous scaling behavior distinct from the classical background
\be \label{eq:Garfinklestressgen}
\langle T^{(3)}_{\mu \nu} \rangle = e^T F_{\mu \nu}(x,n),
\ee
where we write $F_{\mu \nu} (x, n)$ as a tensor accounting for the parts that only depend on $x$ and $n$. One can verify that all components of $F_{\mu \nu} (x, n)$ are not only regular at both $x=1$ and $x=0$, but also \textit{real-analytic} throughout the domain $x\in [0,1]$ provided $n \in \mathbb{N}$, possessing all the desirable properties discussed above. Most importantly, the Lyapunov exponent is positive, reflecting the curvature singularity at $T \to \infty$, which means the $s$-wave one-loop quantum effects act as a growing mode in Garfinkle spacetime. The exponent is determined kinematically, and it has a clear physical and mathematical origin, which can be understood in terms of the dilaton and the nature of the quantum state, as we will elaborate in Section~\ref{sec:discussion}.

Now, if $\langle T_{\mu \nu} \rangle$ given above makes sense, we must be able to solve the semiclassical Einstein equation in $2+1$ dimensions
\be 
G_{\mu \nu}= T_{\mu \nu}+ \frac{\hbar}{\ell} \langle T_{\mu \nu} \rangle,
\ee
and work out a regular semiclassical Garfinkle spacetime at one-loop. Indeed, we now show that this can be achieved exactly. With the adapted coordinates $(T, x)$, we consider the following metric ansatz
\be
ds^2=\ell^2 e^{-2 T} \bigg[e^{2 \rho} \bigg(dx-\frac{x}{2n} dT \bigg)dT + r^2 d \theta^2\bigg],
\ee
where
\be
e^{2 \rho(T,x)}=F(T,x)=F_0 (x)+ \frac{\hbar}{\ell} F_q(x) e^{\omega_q T},
\ee
\be
r(T,x)=r_0(x)+ \frac{\hbar}{\ell}  r_q (x) e^{\omega_q T}.
\ee
Here $F_0(x)$ and $r_0(x)$ correspond to the classical Garfinkle background as given in \eqref{eq:Garfinklemetric}, and the functions $F_q(x)$ and $r_q(x)$ are to be determined. We solve the equations to $O(\hbar)$ by substituting the ansatz into the semiclassical Einstein equations.\footnote{Note that the right-hand side of~\eqref{eq:semiclassicalEinstein} must be consistently expanded to $O(\hbar)$, including the classical contribution evaluated at the corrected background. This is required by covariance; otherwise, there will not be a self-consistent semiclassical solution.} The zeroth-order equations are satisfied by construction, and at first order, we find $\omega_q = 1$, reflecting the growing quantum mode. 

By subtracting the $xx$-component from the $xT$-component, we get
\be
r'_q+x r''_q=0 \implies r_q=\ln{(x)} c_1+c_2.
\ee
with integration constants $c_1,c_2$. Regularity at $x=0$ demands $c_1=0$. Then substituting this back to either the $xx$- or $xT$-components we can solve
\bea
F_q(x)&=& -\frac{4^{\frac{1}{n}-4}(2n-1) (x^n+1)^{3-\frac{2}{n}} }{\pi^2(x^n-1)}  \bigg\{ (-3+64 \pi^2 c_2) (x^n-1) \bigg[1+
\no\\
&\quad& (1+x^n) \ln{\bigg(\frac{x^n}{1+x^n}\bigg)} \bigg]+4 \ln{\bigg(\frac{2}{1+x^n} \bigg)} \bigg\}+c_3 (1+x^n)^{4-\frac{2}{n}},
\eea
where we have an additional integration constant $c_3$. In order to make $F_q (x)$ regular at $x=0$, we have
\be
c_2=\frac{3}{64 \pi^2}.
\ee
Substituting both $F_q(x)$ and $r_q(x)$ into the $TT$-component, we can fix
\be
c_3=\frac{2^{\frac{2}{n}-5} n^{\frac{2}{n}-4}(1-2n)}{\pi^2}.
\ee
Therefore
\be \label{eq:quantumGarfinkle}
F_q(x)=-\frac{(2n-1)4^{\frac{1}{n}-4} (1+x^n)^{3-\frac{2}{n}}}{\pi^2(x^n-1)} \bigg[x^{2n}-1+4 \ln{\bigg(\frac{2}{1+x^n}\bigg)} \bigg], \quad r_q=\frac{3}{64 \pi^2}.
\ee
The resulting semiclassical geometry is then manifestly real-analytic within the past light cone for all $n \in \mathbb{N}$. One can verify that the full semiclassical Einstein equations, including the $\theta\theta$ component, are satisfied to $O(\hbar)$. The backreaction vanishes when $n=\frac{1}{2}$, as expected. While $r_q$ is a constant, it is still nontrivial for the horizon-tracing analysis that follows.

\subsection{Horizon tracing of semiclassical Garfinkle spacetime} \label{sec:GarfinkleHorizon}

We have demonstrated that quantum effects originating from the $s$-wave sector contribute a universal growing mode in critical collapse, as exemplified by the Garfinkle spacetime. In this section, we incorporate these quantum effects into the standard horizon-tracing and mass-scaling analysis, alongside the classical growing mode, based on linear perturbation theory.

\subsubsection*{Setting up the horizon-tracing problem} 

To study the full linear perturbation problem near criticality, it is essential to account for both the classical and quantum growing modes, which originate from distinct physical mechanisms. The quantum growing mode arises from vacuum polarization effects of the collapsing scalar field, while the classical growing mode reflects deviations from the critical solution as we move away from the critical point. We formulate the horizon-tracing problem by including contributions from both types of growing modes, thereby laying the groundwork for numerical analysis. 

In fact, the problem of analyzing the classical perturbation modes is more delicate than the quantum effects. Fortunately, the classical growing modes in Garfinkle spacetime have already been analyzed in detail~\cite{Garfinkle:2002vn, Jalmuzna:2015hoa}. We will recap these important results.

The idea is that a linear perturbation being regular at the center and the past light cone will fix a general $n$ Garfinkle spacetime to have $n-1$ classical growing modes. Explicitly, for classical growing modes with $e^{\omega_c T}$, we have the following restrictions for the exponent $\omega_c$ with an integer $m$
\be
\omega_c=\frac{m}{2n}, \quad m>1 \text{ and } m=2n-1 \text{ or }m<n,
\ee
this means that we have $n-1$ growing modes given by $m=2,3,...,n-1$ and $m=2n-1$. We consider in particular the cases of $n=2$ and $n=4$
\be
n=2 \implies m=3 \implies \omega_c=\frac{3}{4},
\ee
\be
n=4 \implies m=2, 3, 7 \implies \omega_c=\frac{1}{4}, \frac{3}{8}, \frac{7}{8}.
\ee
These are most relevant for the simple reasons that $n=2$ would have exactly one growing mode leading to black hole formation, satisfying one of the criteria for critical phenomena; while $n=4$ with the top growing mode is conjectured to be the best fit of the numerical analysis, as discussed in Section~\ref{sec:classicalGarfinkle}. It is shown that $n=4$ is a good approximation to a yet unknown true critical solution, with only one growing mode given by $\omega_c=\frac{7}{8}$~\cite{Jalmuzna:2015hoa}. We will nevertheless analyze horizon tracing in both cases, where for $n=4$ we assume there is only one growing mode.\footnote{The $n=4$ and general $n$ horizon tracing problems are discussed in Appendix~\ref{sec:horizonGarfinkle}, where we consistently select the dominant classical growing mode. We find that the quantum growing mode possesses a universal Lyapunov exponent larger than all the classical ones, making this selection more reasonable.}

Under both classical and quantum perturbations, the geometry receives backreaction of the form\footnote{Note that there will also be classical perturbation on the scalar field, which would further contribute to the quantum correction, but they will be of order $\mathcal{O}((p-p^\ast)\hbar)$ or higher.}
\be
e^{2 \rho(T,x)}\equiv F(T, x)=F_0(x)+(p-p^\ast) F_c(x) e^{\omega_c T}+\frac{\hbar}{\ell} F_q(x) e^{\omega_q T},
\ee
\be
r(T, x)=r_0(x)+(p-p^\ast) r_c(x) e^{\omega_c T}+ \frac{\hbar}{\ell} r_q (x) e^{\omega_q T}.
\ee
The leading order terms $F_0(x)$ and $r_0(x)$ are given by the classical Garfinkle spacetime \eqref{eq:Garfinklemetric}. The coefficient $(p-p^\ast)$ characterizes the amplitude that deviates from the classical critical point $p^\ast$, and we will discuss the explicit forms of $F_c$ and $r_c$ shortly. For $n=2$, we have $\omega_c=\frac{3}{4}$, while for $n=4$, we pick the growing mode with $\omega_c=\frac{7}{8}$. The quantum backreactions $F_q$ and $r_q$ were given in \eqref{eq:quantumGarfinkle}, with the exponent $\omega_q=1$. Note that we are always in the interior region $x \in [0,1]$.

Let us now discuss the classical backreactions $F_c$ and $r_c$, which have been worked out explicitly in~\cite{Jalmuzna:2015hoa}. The classical perturbations are introduced via the ansatz
\be
e^{2 \rho}=2n x^{2n-1} e^{2 \tilde{\rho}_0+ 2 e^{\omega_c T} a(x)} \approx 2n x^{2n-1}[ e^{2 \tilde{\rho}_0}+2 a (x) e^{2 \tilde{\rho}_0+\omega_c T}+\cO(a^2 (x) )],
\ee
\be
r=r_0(x)+e^{\omega_c T} b(x),
\ee
where $F_0(x)=2 n x^{2n-1}e^{2 \tilde{\rho}_0}$. By matching with our perturbative ansatz, we can identify
\be
(p-p^\ast)F_c(x)=4 n x^{2n-1} a(x) e^{2 \tilde{\rho}_0}=2 a(x) F_0(x),
\ee
\be
(p-p^\ast) r_c(x)=b(x).
\ee
The functions $a(x)$ and $b(x)$ are given in their notations by\footnote{These functions were not determined completely in earlier studies~\cite{Garfinkle:2000br, Garfinkle:2002vn}. They are now determined by matching with null continuations beyond the light cone, as we discussed in Section~\ref{sec:classicalGarfinkle}.}
\be
a(x)= c_2 \bigg[\frac{C_b(1-n)+C_c n}{2n}+\frac{C_c(1-2n)}{4n}x \bigg],
\ee
\be
b(x)=c_2 \frac{C_c}{2}(1-x),
\ee
where 
\be
C_b \equiv \frac{\Gamma(\frac{1}{2}-\omega_c)}{\sqrt{\pi} \Gamma(1-\omega_c)}, \quad C_c\equiv \frac{\Gamma(\omega_c-\frac{1}{2})}{ \sqrt{\pi} \Gamma(\omega_c)} ,
\ee
are constants with fixed $\omega_c$, and $\omega_c$ must correspond to the top growing mode. $c_2$ is a small parameter in order for the perturbation to make sense and we identify 
\be
c_2= -(p-p^\ast),
\ee
for horizon formation with pure classical perturbation in the interior.

We consider a general quasi-local mass function known as the Hawking mass~\cite{Hawking:1968qt, Hayward:1993ph}, or more specifically, the Misner-Sharp-Hernandez mass in spherically symmetric spacetimes~\cite{Misner:1964je, Hernandez:1966zia}. In three dimensions, it is given by \cite{Jalmuzna:2015hoa}
\be \label{eq:Hawkingmass}
M(T, x) \equiv \frac{\bar{r}^2}{\ell^2}- (\nabla \bar{r})^2,
\ee
where $\bar{r}^2=\ell e^{- T} r (T,x)$. We take $(\nabla \bar{r})^2=0$ as the apparent horizon-tracing condition, where $(\nabla \bar{r})^2<0$ would imply we are in the trapped region.\footnote{The condition $(\nabla \bar r)^2=0$ provides a diffeomorphism-invariant diagnostic for locating the apparent horizon. In double-null coordinates, this condition corresponds to tracing the outgoing marginally trapped surface, characterized by $\partial_v \bar r = 0$. The alternative branch $\partial_u \bar r = 0$ never arises in the class of spacetimes considered in this work, since $\partial_u \bar r \neq 0$ already at the background level.} This simple criterion is valid in the spherically symmetric case.

The classical geometry is set by the scale $\ell e^{-T}$, with the Ricci scalar scaling as $R \propto \tfrac{e^{2T}}{\ell^2}$. To capture interesting dynamics, we focus on mildly positive $T$, which can always be arranged by adjusting $\ell$. While this rescaling is a standard practice in classical analyses, more care is needed once quantum effects are introduced, since we must also track the possible breakdown of the semiclassical approximation. Nevertheless, in our model with a massless scalar field, which is intrinsically scale-free, physical outcomes such as horizon formation remain invariant under any choice of $\ell$, even after including quantum corrections. This is because quantum effects only enter through the perturbative dimensionless ratio $\frac{\hbar}{\ell^{D-2}}$. Numerically, we are therefore free to set $\ell = 1$ for simplicity.

A spoiler is that we find the universal quantum growing mode, arising from vacuum polarization, shifts the classical critical threshold $p^\ast$ to a new quantum threshold $p^\ast_q$, with the difference defined as $\Delta p \equiv p^\ast - p^\ast_q$. Approaching this new threshold $p \to p^\ast_q$, a finite mass gap $M_{\text{gap}}$ emerges. The key question is what is setting the scale of $M_{\text{gap}}$. In Section~\ref{sec:discussion}, we analyze these quantities in more details and show that, while $M_{\text{gap}}$ is indeed independent of $\ell$, it remains sensitive to the background profiles and to the relative strength of the perturbations. 

In geometrized units, we further set $\hbar = 1$ in our numerical analysis, thereby adopting Planck units in which the quantum-gravity curvature scale is of order unity (up to a factor of $\sqrt{8\pi}$). Within this convention, the values of $M_{\text{gap}}$ obtained from the Garfinkle solution (and the Roberts solution in $3+1$ dimensions, discussed in Section~\ref{sec:collapse3+1}) may appear to lie at the Planck scale. This is, however, somewhat artificial: the background curvature in our scale-free model with $\Lambda=0$ is already trans-Planckian. In more realistic settings, the perturbative profiles would be tied to boundary conditions in which the curvature scale is set by the cosmological constant or the size of the Hubble horizon, thereby breaking the exact self-similarity and scale-freeness of the classical background. The broader lessons drawn from the Garfinkle and Roberts solutions still apply, but importantly, the resulting black hole mass gap need not be microscopically small in a way that would invalidate the semiclassical analysis. We will return to this point in Section~\ref{sec:discussion}.

We will also have to justify the validity of linear perturbation theory, where we should examine individual components of the metric. For mildly positive $T$, the linear-order perturbations remain small compared to the background values throughout the domain $x \in [0,1]$.

A particularly nice feature is that the horizon-tracing condition $(\nabla \bar{r})^2$, which determines the location of apparent horizons, is a non-linear function. It has the background $O(1)$ piece independent of $\ell$. However, the quantum correction to $(\nabla \bar{r})^2$ will still go as $\frac{\hbar}{\ell} e^T f_q(x)$, but is a consequence of linearizing $(\nabla \bar{r})^2$. That means since $(\nabla \bar{r})^2$ is a non-linear function of the metric and the areal radius, it can receive perturbations comparable in magnitude to the $O(1)$ piece without implying a breakdown of linear perturbation theory.\footnote{In addition to the horizon-tracing function being non-linear, to fully justify the validity of linear perturbation theory, one will also have to show that quantum effects from higher loops with $O(\hbar^2)+\cdots$ do not get any further enhancement in the Lyapunov exponent as the growing mode. That is, the quantum growing mode remains $e^{\omega_q T}(O(\hbar)+O(\hbar^2)+\cdots)$ schematically with a fixed $\omega_q>0$. We will justify this point in Section~\ref{sec:discussion} once we have clarified the physical origin of the growing mode. We thank Gustavo J. Turiaci for a discussion on this point.} This highlights the extreme sensitivity of apparent horizons to both classical and quantum perturbations, while still preserving the validity of the linear-order analysis.\footnote{Such a sensitivity of the horizon structure was observed already in the $\Lambda$-corrections of Garfinkle spacetime we discussed in Section~\ref{sec:classicalGarfinkle}, where $\Lambda$ is even a dynamically irrelevant scale~\cite{Jalmuzna:2015hoa}.}

A final remark is that, for large enough $T$, linear perturbation analysis for the quantum growing mode will always break down, as for the classical perturbation mode. However, in Section~\ref{sec:discussion}, we will point out that the nature of the quantum growing mode and the fact that quantum corrections from higher loops are suppressed would facilitate the possibility of a full non-linear analysis.

\subsubsection*{Horizon tracing for $n=2$ semiclassical Garfinkle spacetime}

To set the stage for subsequent generalizations, we begin by examining the horizon-tracing problem in the $n=2$ Garfinkle spacetime as a representative example. The $n=4$ and general $n$ cases will be discussed in Appendix~\ref{sec:horizonGarfinkle}.

In $n=2$ case, we have
\be
F_c(x)=-\frac{1}{4}\bigg( 2C_c-C_b-\frac{3C_c}{2}x \bigg)(1+x^2)^3, \quad r_c(x)=-\frac{C_c}{2} (1-x),
\ee
with
\be
C_b \approx -0.763, \quad C_c \approx 1.67,
\ee
and
\be
F_q(x)=-\frac{3(1+x^2)^2[x^4-1+4 \ln{(\frac{2}{1+x^2})}]}{128  \pi^2 (x^2-1)}, \quad r_q(x)=\frac{3}{64 \pi^2}.
\ee
We calculate the horizon-tracing function to linear order in perturbations
\be \label{eq:n=2horizon}
(\nabla \bar{r})^2 \approx f_0(x)+e^{\frac{3T}{4}}(p-p^\ast) f_c(x)+e^T  f_q (x),
\ee
where the $x$-dependent parts are
\be
f_0(x)=\frac{8 x^3}{(1+x^2)^2},
\ee
\be
f_c(x)=-\frac{2 (C_c+2 C_b x^3-3 C_c x^3+3 C_c x^4)}{(x^2+1)^3},
\ee
\be
f_q(x)=\frac{3x^3 [x^4-1+4 \ln{(\frac{2}{1+x^2})}]}{8 \pi^2 (x^2-1) (x^2+1)^4}.
\ee
Note that the classical growing mode $e^{\frac{3}{4}T}$ is subdominant compared to the quantum one with $e^{T}$, and higher-order terms can carry even stronger exponential growth. Therefore, the following numerical analysis is justified only for mildly small $T$ and for data not too far from the critical point. As we will see, it is precisely the interplay between these two modes that gives rise to a new critical point characterized by a finite mass gap.

\begin{figure}[hbt!]
\centering
\includegraphics[width=0.4\textwidth]{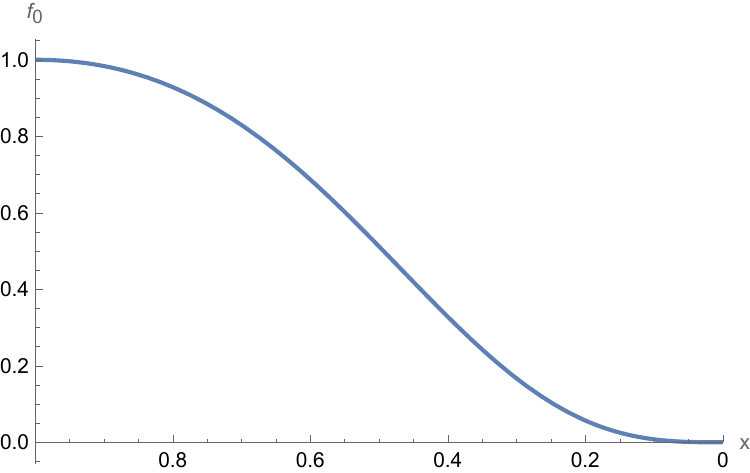}
\includegraphics[width=0.4\textwidth]{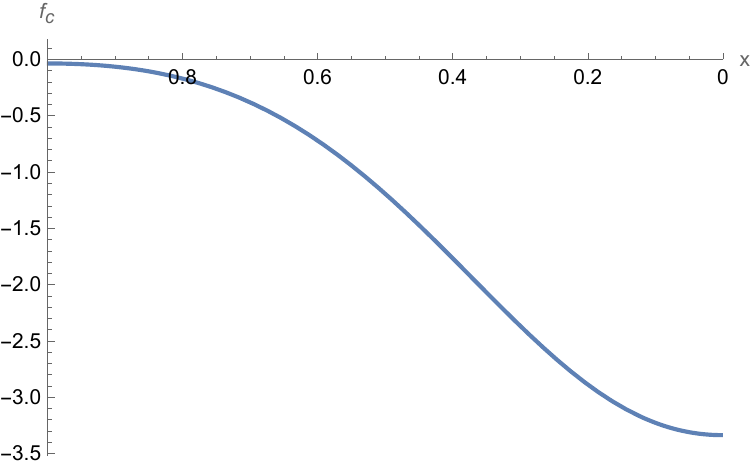}
\includegraphics[width=0.4\textwidth]{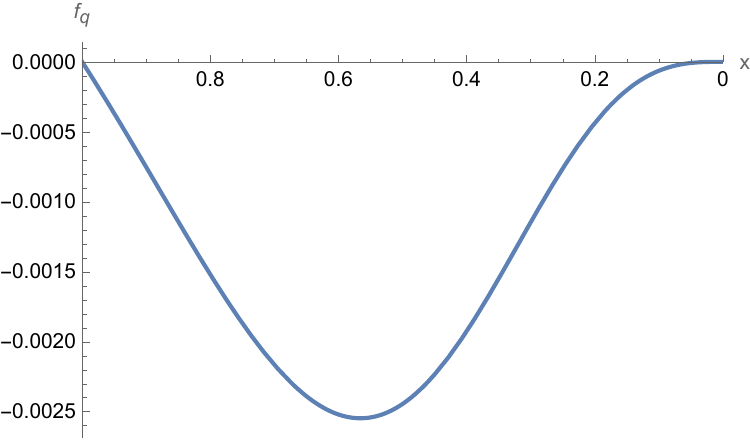}
\caption{The contributions to $(\nabla \bar{r})^2$ in \eqref{eq:n=2horizon} from the functions $f_0(x)$, $f_c(x)$, and $f_q(x)$ in the interior region $x \in [0, 1]$.}
\label{n=2fs}
\end{figure}

\begin{figure}[hbt!]
\centering
\includegraphics[width=0.45\textwidth]{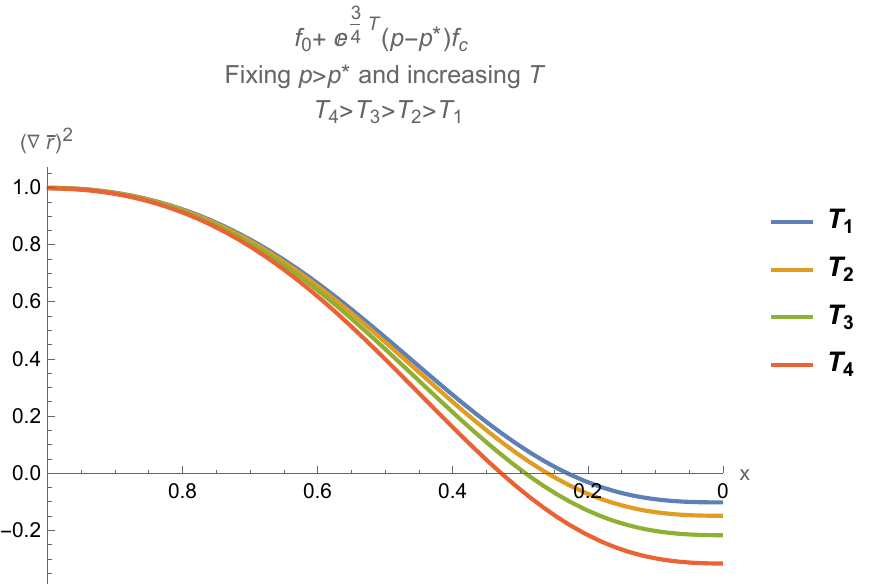}
\includegraphics[width=0.45\textwidth]{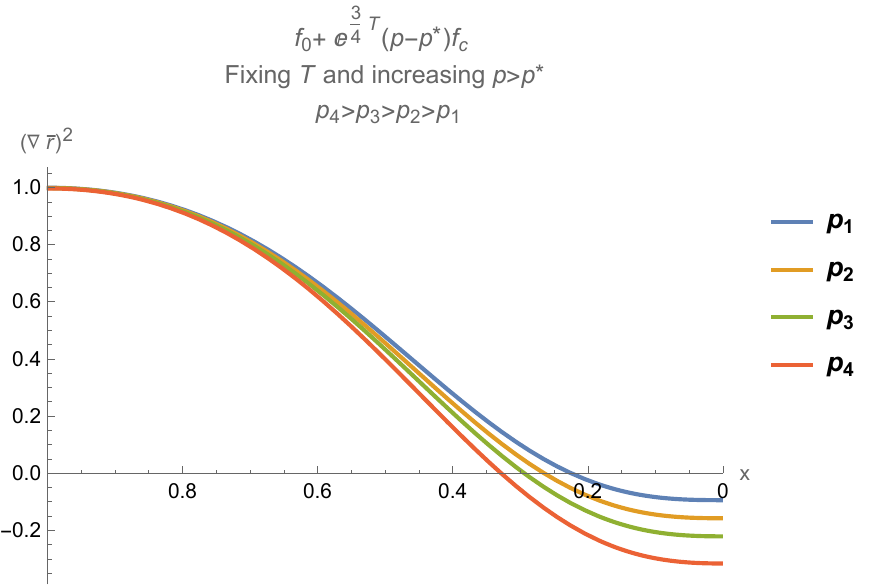}
\caption{When the classical perturbation is turned on, by either fixing $T$ or $(p-p^\ast)$ we see part of the spacetime close to the light cone belongs to the trapped region as $(\nabla \bar{r})^2<0$. We take values with small deviations from criticality $(p - p^\ast) \lesssim 0.01$ and moderately positive values of $T \lesssim 3$.}
\label{n=2f0+fc}
\end{figure}

\paragraph{The structure of the apparent horizons.}Let us examine the behavior of these $x$-dependent functions in Figure~\ref{n=2fs}, where the past light cone ($x = 0$) is always placed on the right side of each panel. In the absence of perturbations, we observe that the leading-order term $f_0$ would cause $(\nabla \bar{r})^2$ to vanish as we approach the past light cone, representing the original apparent horizon in the $\Lambda=0$ Garfinkle solution. When $f_c$ is introduced, it contributes negatively to $(\nabla \bar{r})^2$, leading to the formation of an apparent horizon close to the light cone, as shown in Figure~\ref{n=2f0+fc}. The quantum correction $f_q$ remains regular and vanishes both at the center and near the light cone, yet contributes negatively to $(\nabla \bar{r})^2$.

\begin{figure}[hbt!]
\centering
\includegraphics[width=0.72\textwidth]{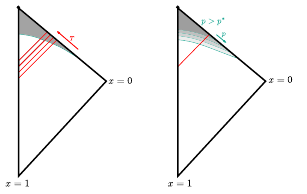}
\caption{In the left panel, we fix a value of $(p - p^\ast) > 0$ and examine a sequence of null slices with $T_i \geq T_{\text{AH}}$. In the right panel, we fix a particular time slice $T_i \geq T_{\text{AH}}$ and vary the amplitude by increasing $(p - p^\ast)$. The apparent horizon profiles shown are schematic; the precise structure can be inferred from the behavior of $(\nabla \bar{r})^2$ in Figure~\ref{n=2f0+fc}.}
\label{horizons}
\end{figure}

As a consistency check, we demonstrate in Figure~\ref{n=2f0+fc} that an apparent horizon forms near the past light cone when only the classical perturbation is included~\cite{Jalmuzna:2015hoa}. One may either fix $T$ and vary $(p - p^\ast)$, or vice versa; the resulting effect is governed by the combination $e^{\frac{3}{4}T}(p - p^\ast)$, which controls the overall amplitude of the growing mode. The physical interpretation differs slightly depending on which parameter is held fixed (see Figure~\ref{horizons} for comparison).

As emphasized in Section~\ref{sec:criticalcollapse}, the formation of the first apparent horizon during the time evolution depends only weakly on the choice of time slicing. We consider a series of null time slices labeled by $T=T_i$. Let $T_{\text{AH}}$ denote the earliest time at which an apparent horizon forms within the light cone $x \in (0,1]$. This construction implies that the resulting trapping horizon is a spacelike surface originating near the light cone and extending into the interior, since by computing $(\nabla \bar{r})^2$ over successive null time slices, we can reconstruct the shape of the apparent horizon. As the condition $(\nabla \bar{r})^2 \leq 0$ characterizes the trapped region, we find that as $T$ increases, a growing portion of each time slice lies within the trapped region.

Moreover, increasing the amplitude $(p - p^\ast)$ causes the entire spacelike trapping horizon to shift toward earlier times, such that its intersection with a fixed $T$-slice moves closer to $x = 1$. Physically, this indicates that the apparent horizon radius increases with $(p - p^\ast)$. Consequently, as the black hole grows larger, more of the spatial slice at any given time becomes trapped.

\begin{figure}[t!]
\centering
\includegraphics[width=0.6\textwidth]{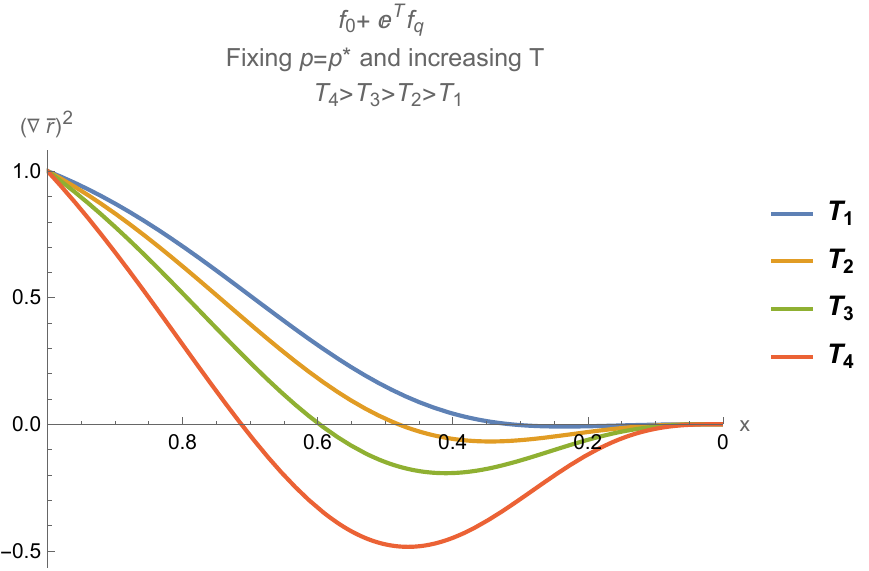}
\caption{We turn off classical perturbation by staying at the classical critical point $p=p^\ast$ but consider different time slices with mildly positive $T \lesssim 6$. The linear perturbation appears to become comparable to the background near $x=0$, this is an artifact of the background metric components vanishing rapidly in this region, merging with the apparent horizon at $x=0$ for $\Lambda=0$ Garfinkle spacetime.}
\label{n=2f0+fq}
\end{figure}

Now let us turn off the classical perturbation by staying at the classical critical point $p = p^\ast$, and we retain only quantum corrections. We examine several null time slices $T_i$ for sufficient yet moderately large $T$ for quantum effects to play a role, while remaining within the regime of linear perturbation. The results are shown in Figure~\ref{n=2f0+fq}. We observe that even the pure critical spacetime, with $p = p^\ast$, develops an apparent horizon at finite $x$ solely due to quantum backreaction. As we increase $T$, progressively larger regions of the spacetime fall into the trapped region where $(\nabla \bar{r})^2 < 0$.

\begin{figure}[t!]
\centering
\includegraphics[width=0.45\textwidth]{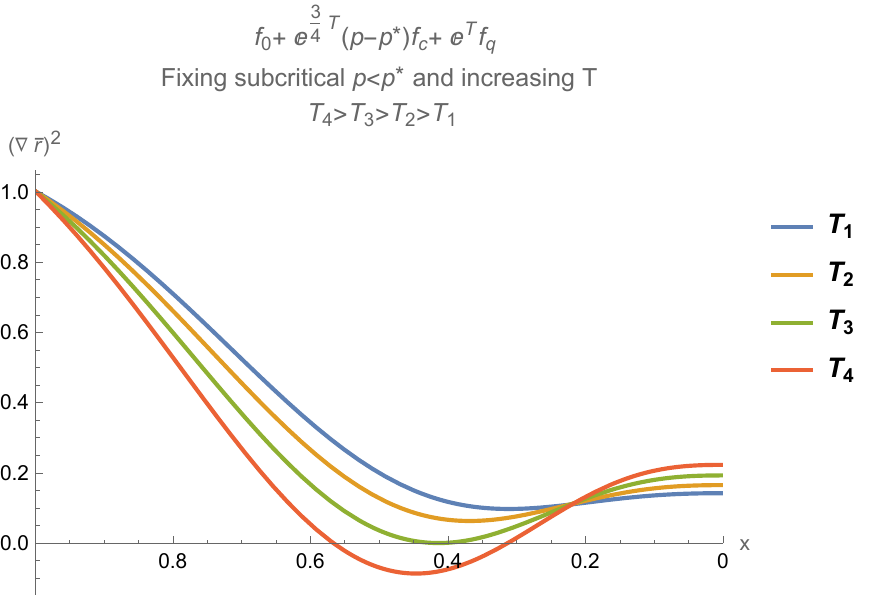}
\includegraphics[width=0.45\textwidth]{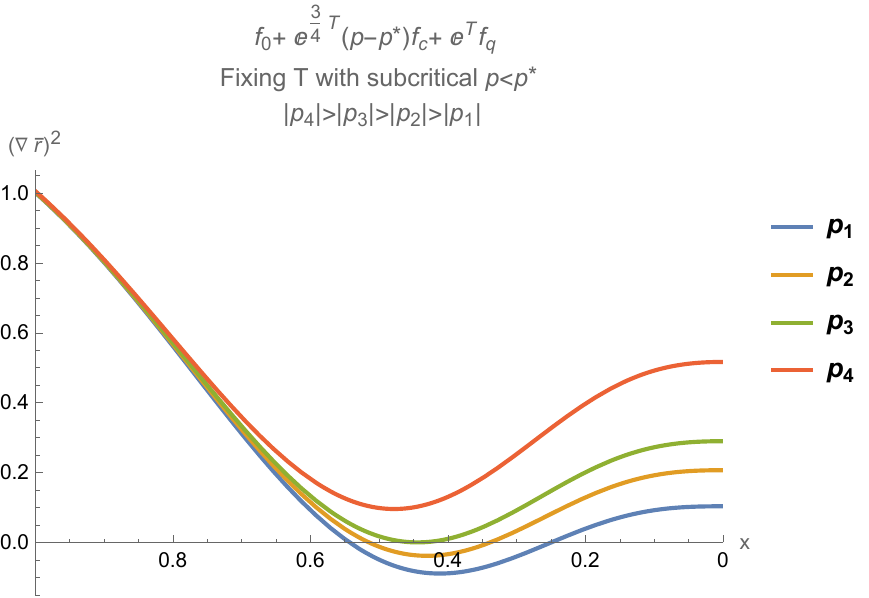}
\caption{On the left-hand side, the horizon-tracing function touches zero at some value $T_3$ that is greater than the previous $T_{\text{AH}}$. This is expected: for subcritical data, black hole formation is more difficult, so one must evolve to later times before quantum effects become strong enough to trigger horizon formation. On the right-hand side, the absence of horizon formation for very subcritical data at a fixed time slice is also consistent with expectations, since the amplitude is too small for either classical or quantum effects to produce a trapped region.}
\label{n=2f0+fc+fq(subc)}
\end{figure}

This demonstrates that quantum effects alone can shield the classical naked singularity and potentially induce horizon formation for certain subcritical configurations (see Figure~\ref{n=2f0+fc+fq(subc)}). However, it is important to note that $f_q$ vanishes at the past light cone, so it does not modify the fact that the light cone itself remains a marginally trapped surface, at least in the absence of a cosmological constant. As we discussed in Section~\ref{sec:classicalGarfinkle}, the past light cone should be matched to an outgoing Vaidya spacetime~\cite{Jalmuzna:2015hoa}, where the quantum backreaction in the exterior could be very different.

\begin{figure}[t!]
\centering
\includegraphics[width=0.6\textwidth]{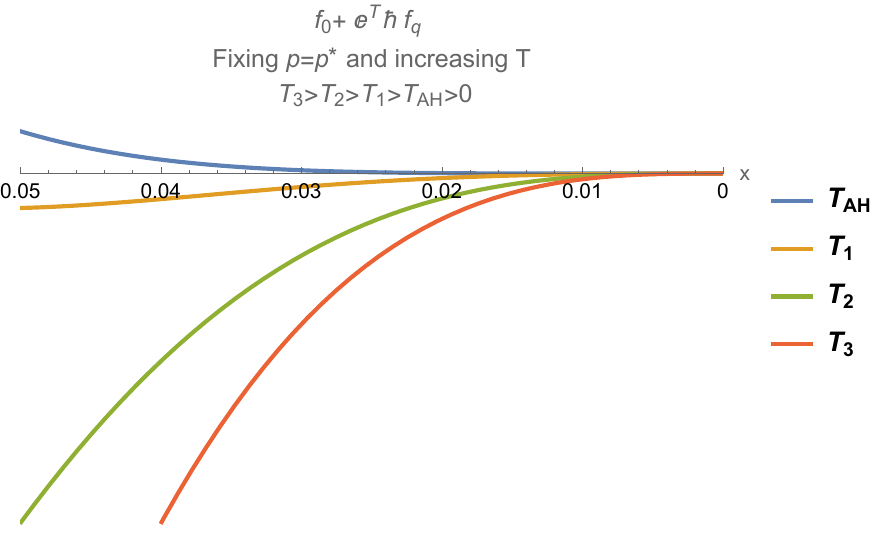}
\caption{We zoom in to the region close to the past light cone where we can identify the onset of horizon formation. Note that for any larger time slices, the curves will not touch zero twice.}
\label{n=2f0+fq(Tslices_zoomin)}
\end{figure}

To ensure consistency, we must examine whether horizon formation arises even with arbitrarily small quantum corrections, i.e., at very early time slices. Recall that $T \in (-\infty, \infty)$, but $e^T$ is always positive. If an apparent horizon were to form at finite $x$ in the limit $T \to -\infty$, it would suggest horizon formation without significant quantum backreaction, leading to a contradiction. To probe this, we focus on early time slices near the onset of horizon formation and zoom in close to the past light cone. We find that at a particular critical time slice, denoted $T = T_{\text{AH}}$ (which is not small, in this case, $T_{\text{AH}} \approx 4.78$), the quantum-corrected critical spacetime gives $(\nabla \bar{r})^2 = 0$ at some $x_{\text{AH}}$ extremely close to the light cone at $x=0$, as shown in Figure~\ref{n=2f0+fq(Tslices_zoomin)}. This is expected given the profile of $f_q$. 

\FloatBarrier 

\FloatBarrier

The appearance of $(\nabla \bar{r})^2 = 0$ at this finite $x_{\text{AH}}$ implies the formation of a marginally trapped surface and thus a non-zero black hole mass given by $M = \bar{r}^2 (T_{\text{AH}}, x_{\text{AH}})$. This confirms that the apparent horizon emerges only after quantum effects accumulate sufficiently, ruling out horizon formation in the asymptotic past, and reinforces the interpretation that the horizon is seeded by quantum backreaction. 

However, we must emphasize that this is not the complete picture without taking into account the classical perturbation. In fact, the role that quantum backreaction plays is to shift the classical critical point, where the earliest onset of the horizon would not be close to the light cone.

\begin{figure}[t!]
\centering
\includegraphics[width=0.45\textwidth]{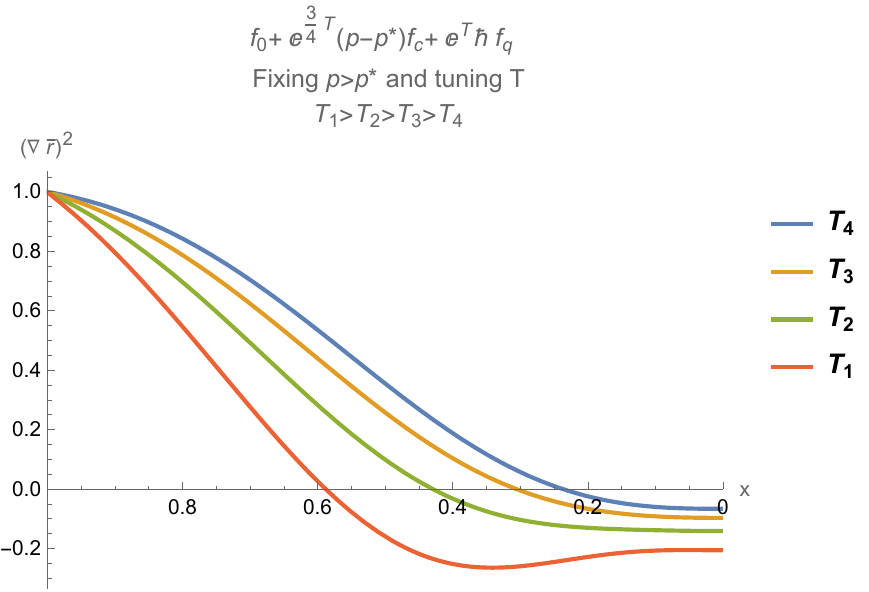}
\includegraphics[width=0.45\textwidth]{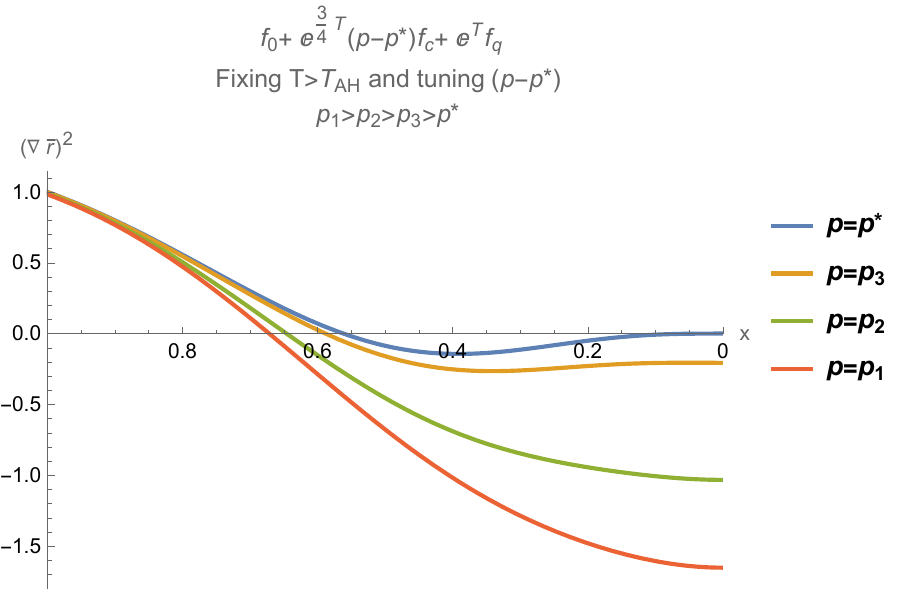}
\caption{Starting from configurations where $T > T_{\text{AH}}$ such that a trapped region has already formed, increasing either the amplitude or considering different time slices would further enhance horizon formation, i.e., more and more regions would be trapped. Here we take $(p-p^\ast) \lesssim 0.008 $, and $T \lesssim 5.5$.}
\label{n=2f0+fc+fq}
\end{figure}
\FloatBarrier

\FloatBarrier 

Now we include both classical and quantum perturbations. The results so far indicate that adding a classical supercritical perturbation simply enhances horizon formation, as illustrated in Figure~\ref{n=2f0+fc+fq}. However, it also suggests a more subtle phenomenon: even certain classically subcritical data with $p< p^\ast$ can lead to horizon formation once quantum effects are included. To explore this, we consider turning on a classical perturbation with $(p-p^\ast)<0$, see Figure~\ref{n=2f0+fc+fq(subc)}. The physical interpretation is that quantum corrections effectively shift the classical critical value to a new quantum-corrected threshold 
$p^\ast_q<p^\ast$. We should therefore reinterpret the dynamics in terms of this shifted new critical value and study the behavior of horizon formation as a function of the amplitude $(p-p^\ast_q)$.

However, a limitation of the above analysis is that we cannot increase $T$ indefinitely, since both classical and quantum perturbations eventually grow large enough to invalidate linear perturbation theory. This motivates the need for a more robust way to characterize the interplay between classical and quantum effects, one that does not rely on independently fine-tuning $T$ and $(p-p_q^\ast)$.

\paragraph{Locations of the EMOTS and mass scaling.}Since quantum effects are governed by a fixed scale set by $\hbar$, the horizon formation condition on a given null slice $T$ is effectively controlled by the amplitude $(p-p^\ast_q)$. In this way, the value of $T$ at which the apparent horizon forms, which we denote as $T_{\text{AH}}$, becomes intimately tied to the amplitude. We assume that for sufficiently large $T$, there exists a surface $T=T_{\text{AH}}$ such that $(p-p^\ast_q)>0$, and the apparent horizon condition $(\nabla \bar{r})^2=0$ is satisfied. On this null surface, the corresponding mass function is given by $M(T,x)=\bar{r}^2(T,x)$, and we refer to the first such surface as the \textit{earliest marginally outer-trapped surface} (EMOTS). While this definition is slicing-dependent, the qualitative trends are expected to be robust.

The key question is how the relative strength between classical and quantum perturbations influences the location $x_{\text{EMOTS}}$. Specifically, we ask whether it is possible to fine-tune the amplitude so that the apparent horizon forms with a single root, that is, the EMOTS just forms without enclosing a trapped region. This defines the quantum-corrected critical value $p^\ast_q$, and the corresponding EMOTS location in $x$ will depend on the value of $T_{\text{AH}}$.

\begin{figure}[t!]
\centering
\includegraphics[width=0.5\textwidth]{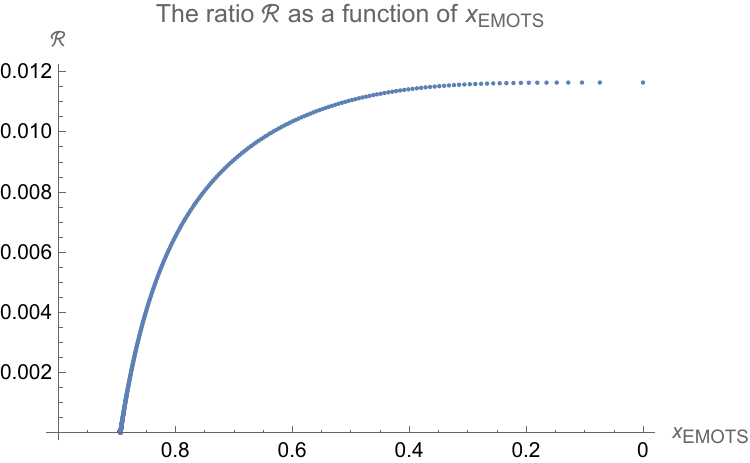}
\caption{We observe that the location of EMOTS in $x$ moves toward the center as the ratio gets smaller and smaller. As the ratio goes to zero, which can be interpreted as approaching the new critical point $p \to p^\ast_q$, corresponding to the minimal mass. On the other hand, increasing the ratio would correspond to approaching the classical critical point $p^\ast$, where horizon formation is moving toward the light cone, consistent with expectations.}
\label{n=2Ratio_x_Qsupercritical}
\end{figure}

To capture this interplay in a more invariant way, we define the following ratio characterizing the relative strength of classical and quantum perturbations ($n=2$)
\be
\mathcal{R} \equiv \frac{e^{\omega_c T }(p-p^\ast_q)}{e^{\omega_q T}  \hbar}=\frac{(p-p^\ast_q)}{e^{\frac{1}{4}T}}.
\ee
This ratio must be small enough in order for the quantum effects to play a role. Our first goal is to understand how the location of $x_{\text{EMOTS}}$, varies with the ratio $\mathcal{R}$. We can in fact plot the whole regime where quantum effects start to kick in until the ratio goes to zero. Specifically, we sample 500 data points of $\mathcal{R}$ versus $x_{\text{EMOTS}}$, with the results presented in Figure~\ref{n=2Ratio_x_Qsupercritical}.

Note that if we continue to increase the ratio beyond the range shown in Figure~\ref{n=2Ratio_x_Qsupercritical}, the quantum correction becomes too small to make any change and reduces to classical analysis. Then, the EMOTS does not make much sense with the null slicing we are adopting, as the apparent horizon would merge with the past light cone, see Figure~\ref{horizons} (we can still identify a trapped region). One will have to incorporate the perturbative $\Lambda$-corrections to lift the apparent horizon at the past light cone, as already demonstrated in~\cite{Jalmuzna:2015hoa}.

\FloatBarrier

\begin{figure}[t!]
\centering
\includegraphics[width=0.5\textwidth]{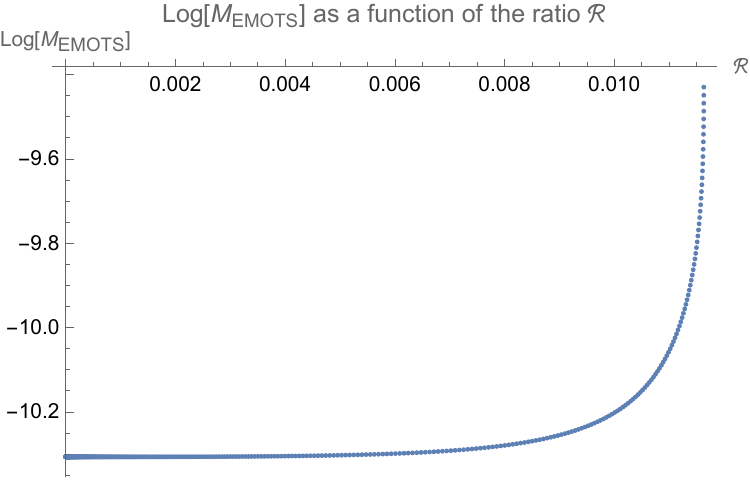}
\caption{The corresponding logarithm of the mass function $M_{\text{EMOTS}}=\bar{r}^2(T_{\text{EMOTS}}, x_{\text{EMOTS}})$ against the ratio $\mathcal{R}$, where we note each point actually occurs at different values of $(T_{\text{EMOTS}}, x_{\text{EMOTS}})$. As we decrease the ratio, the mass function monotonically approaches a constant value, indicating a Type I behavior.}
\label{n=2Mass_Ratio_Qsuper-critical}
\end{figure}
\FloatBarrier

Now we plot the corresponding $M_{\text{EMOTS}}$ evaluated at $x_{\text{EMOTS}}$ in Figure~\ref{n=2Mass_Ratio_Qsuper-critical}. For larger values of the ratios, where classical effects are important, the mass function increases monotonically with the ratio. However, as the ratio decreases and quantum effects begin to take over, the mass saturates to an approximately constant value. This signals the transition to a \textit{quantum-modified Type I behavior with a finite mass gap.} As the ratio is tuned further downward into negative values, corresponding to subcritical data $(p - p^\ast_q) < 0$, no black hole forms, consistent with dispersal.

We have employed the $n=2$ Garfinkle spacetime as the simplest example to provide a detailed analysis of the apparent horizon structure and mass scaling, incorporating both classical and quantum perturbations. As noted, this case possesses exactly one classical growing mode. However, the $n=4$ case with its top growing mode is the most physically relevant as it agrees with the numerical simulation \cite{Pretorius:2000yu, Jalmuzna:2015hoa}. A detailed case study of the $n=4$ Garfinkle spacetime is presented in Appendix~\ref{sec:horizonGarfinkle}, where we find the same qualitative features as in the $n=2$ case. Furthermore, we have carried out a horizon-tracing analysis for general $n$ Garfinkle spacetimes, leading to the surprising observation that the mass gaps for different values of $n$ tend to converge to approximately the same scale, as illustrated in Figure~\ref{nRatio_x_Mass}. 

A model-independent analytic treatment of the threshold shift $\Delta p = p^\ast-p^\ast_q$ and the mass gap $M_{\text{gap}}$ is provided in Section~\ref{sec:discussion}.

\section{Quantum critical collapse in $3+1$ dimensions} \label{sec:collapse3+1}

In this section, we study the semiclassical properties of critical gravitational collapse in the $3+1$ dimensional Einstein-scalar system, focusing on the analytic Roberts solution, which is closely related to the true critical spacetime.

\subsection{The critical Roberts spacetime in $3+1$ dimensions}
\label{sec:classicalRoberts}

Among all cases, the $3+1$-dimensional setting is the most physically relevant, as it corresponds to our universe. In particular, the Einstein-scalar system, a simple yet canonical model, provides a foundational framework for gravitational collapse, as explored in the seminal works of Christodoulou and Choptuik. However, despite significant efforts, an explicit analytic form of the true critical solution remains unknown, except attempts for analytic approximations~\cite{Price:1996sk} and existence proofs~\cite{Reiterer:2012hnr}.\footnote{From a more formal perspective~\cite{Cicortas:2024hpk, Gundlach:2025yje}, the universal solution discovered by Choptuik remains a conjectured solution.} 

The critical solution in this setup is known to exhibit DSS. Unfortunately, DSS solutions are difficult to obtain in closed form, particularly in the Einstein-scalar system. To gain analytic insight, one often considers CSS spacetimes as a simplification. A class of such CSS solutions for the Einstein-scalar system \eqref{eq:Einsteinscalar} was first introduced by Roberts in the context of cosmic censorship~\cite{Roberts:1989sk}, and later rediscovered by Brady and by Oshiro, Nakamura, and Tomimatsu in the study of critical collapse~\cite{Brady:1994xfa, Oshiro:1994hd}. These solutions were the first known analytic examples of CSS scalar collapse in closed form. A complete classification of CSS-type solutions was later provided by Brady~\cite{Brady:1994aq}.\footnote{See~\cite{Frolov:1998zt} for higher-dimensional generalizations of the Roberts solution, where closed-form solutions exist only for $D = 4, 5, 6$. For dS and AdS generalizations, see~\cite{Roberts:2014ogn, Maeda:2015cia}, where these spacetimes admit conformal Killing vectors rather than homothetic ones, and exhibit rich global structures that warrant further investigation. In \cite{Roberts:2017rma}, a canonical quantum gravity perspective is investigated.}

Let us begin by describing the classical properties of the Roberts solution. This is a CSS two-parameter family $(\alpha, \beta)$ solution of the Einstein-scalar system. In the double-null coordinates $(u, v)$, with the domain given by  $v \in [0, \infty)$ and $u \in (-\infty, 0]$, the metric is
\be \label{eq:fullRoberts}
ds^2= -2 e^{2 \rho} du dv+r^2 d \Omega^2,
\ee
with
\be \label{eq:Robertscoeff}
e^{2 \rho}=1, \quad r^2=-\alpha v^2+\beta u^2-uv,
\ee
and the scalar field
\be \label{eq:scalarRoberts}
f=\frac{\sqrt{2}}{2}\ln{\bigg[-\frac{2\alpha v+u(1-\sqrt{1+4 \alpha \beta})}{2 \alpha v+u(1+\sqrt{1+4 \alpha \beta})} \bigg]}+f_0(\alpha, \beta).
\ee
This geometry describes the implosion of scalar radiation from past null infinity $u=-\infty$, with the scalar field switched on at $v=0$. An appropriate normalization of $f_0 (\alpha, \beta)$ can be chosen such that, for instance, $f_{v=0}=0$, allowing the region $v<0$ to be smoothly matched to Minkowski space. We adopt the sign convention for $\alpha$ used in~\cite{Frolov:1997uu}, and the factor of $\sqrt{2}$ in the scalar field reflects the normalization $8 \pi G_N=1$. 

The exact critical spacetime corresponds to setting $\alpha=0$ and $\beta=1$.\footnote{By adjusting $(\alpha, \beta)$~\cite{Brady:1994xfa}, one can obtain spacetimes with a past null singularity, like a time-reversed version of the Roberts spacetime, which could result in a white hole solution. Other choices can also yield self-similar cosmological solutions. Among these, the most intriguing case is $\alpha=\beta=0$, which gives the Hayward solution~\cite{Hayward:2000ds, Clement:2001zd}, a peculiar spacetime featuring a central bifurcate null singularity that also arises in the study of critical collapse. We will examine this solution further in Appendix~\ref{sec:Hayward}.} The most striking property is that, by sticking to $\beta=1$, $\alpha$ plays the role of a one-parameter family of solutions that encodes all three regimes of critical collapse by interpolating from subcritical to supercritical evolutions. The $\alpha=0$ critical spacetime contains a null curvature singularity at $r=0, u=0$, extending all the way to future null infinity $v=\infty$. Outgoing null rays do reach infinity, with no apparent horizon. For $\alpha>0$, a black hole is formed with an apparent horizon surrounding a spacelike $r=0$ curvature singularity, and the apparent horizon merges with the curvature singularity when $\alpha \to 0$; while for $\alpha<0$, the field disperses to future null infinity, leaving behind Minkowski space for $u>0$. See Figures~\ref{alpha-small} and \ref{null-coords}.

Despite such an attractive feature, the Roberts solution cannot represent the true critical spacetime for several key reasons:
\begin{itemize}
\item Unlike the typical naked singularity scenario discussed in Section~\ref{sec:criticalcollapse}, here the curvature singularity is null and extends all the way to future infinity $\mathcal{I}^+$. In this case, no observer can witness the singularity without actually reaching it \cite{Penrose:1978}. An observer may approach arbitrarily high, but not infinite, curvature from the future singularity.

    \item In the supercritical regime $\alpha>0$, the black hole mass grows without bound as $v \to \infty$, and the entire spacetime eventually gets trapped, failing to be asymptotically flat~\cite{Brady:1994xfa, Wang:1996xh}. As a result, one cannot simply read off the power-law scaling from the total mass with marginally non-zero $\alpha$ (which would suggest $\delta=0.5$)~\cite{Brady:1994xfa, Oshiro:1994hd}. 
    \item Although the one-parameter family of solutions labeled by $\alpha$ resembles the usual critical behavior, we can see from the above facts that this interpretation based on interpolating $\alpha$ is misleading. Instead, we should carefully study its linear perturbation modes.
    
    However, it was found that even though exactly self-similar modes vanish, the critical case is not an intermediate attractor described in Section~\ref{sec:criticalcollapse}, but merely a threshold solution, since nearby solutions do not evolve towards it~\cite{Frolov:1997uu}.
    \item Even worse, the classical perturbation spectrum $\omega_c$ is not discrete, but \textit{continuous}. Furthermore, it occupies a region in the \textit{complex} plane~\cite{Frolov:1997uu}:
    \be \label{eq:Robertsmodes}
    \frac{1}{2}< \text{Re}(\omega_c)<1, \quad |\text{Im}(\omega_c)|>\sqrt{\frac{\text{Re}(\omega_c)(2-\text{Re}(\omega_c))}{1-(2\text{Re}(\omega_c))^{-1}}},
    \ee
    The continuous and complex spectrum obviously violates one of the defining criteria in Section~\ref{sec:criticalcollapse} that there should exist exactly one growing mode, and the corresponding Lyapunov exponent should be purely real. 

    \item Even by taking the eigenvalues approaching $\max\{ \text{Re}(\omega_c)\}=1$, it would imply a mass scaling exponent $\delta=1$, which is very far away from Choptuik’s numerically observed value of $\delta=0.37$.

    \item Most importantly, the solution is CSS, while the true critical spacetime observed by Choptuik is DSS.

    \newpage
\begin{figure}[t!]
\centering
\includegraphics[width=0.42\textwidth]{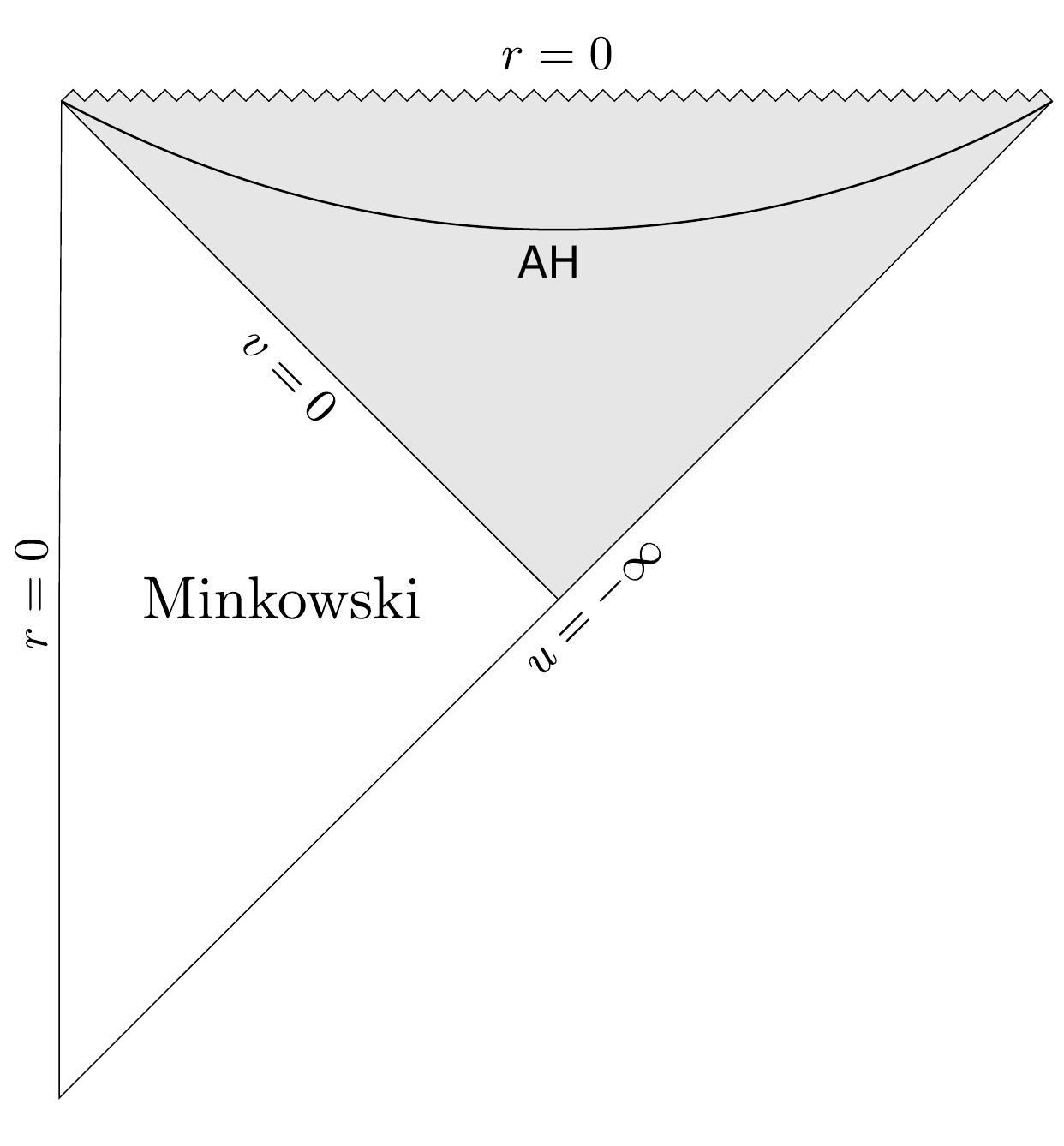}
\includegraphics[width=0.32\textwidth]{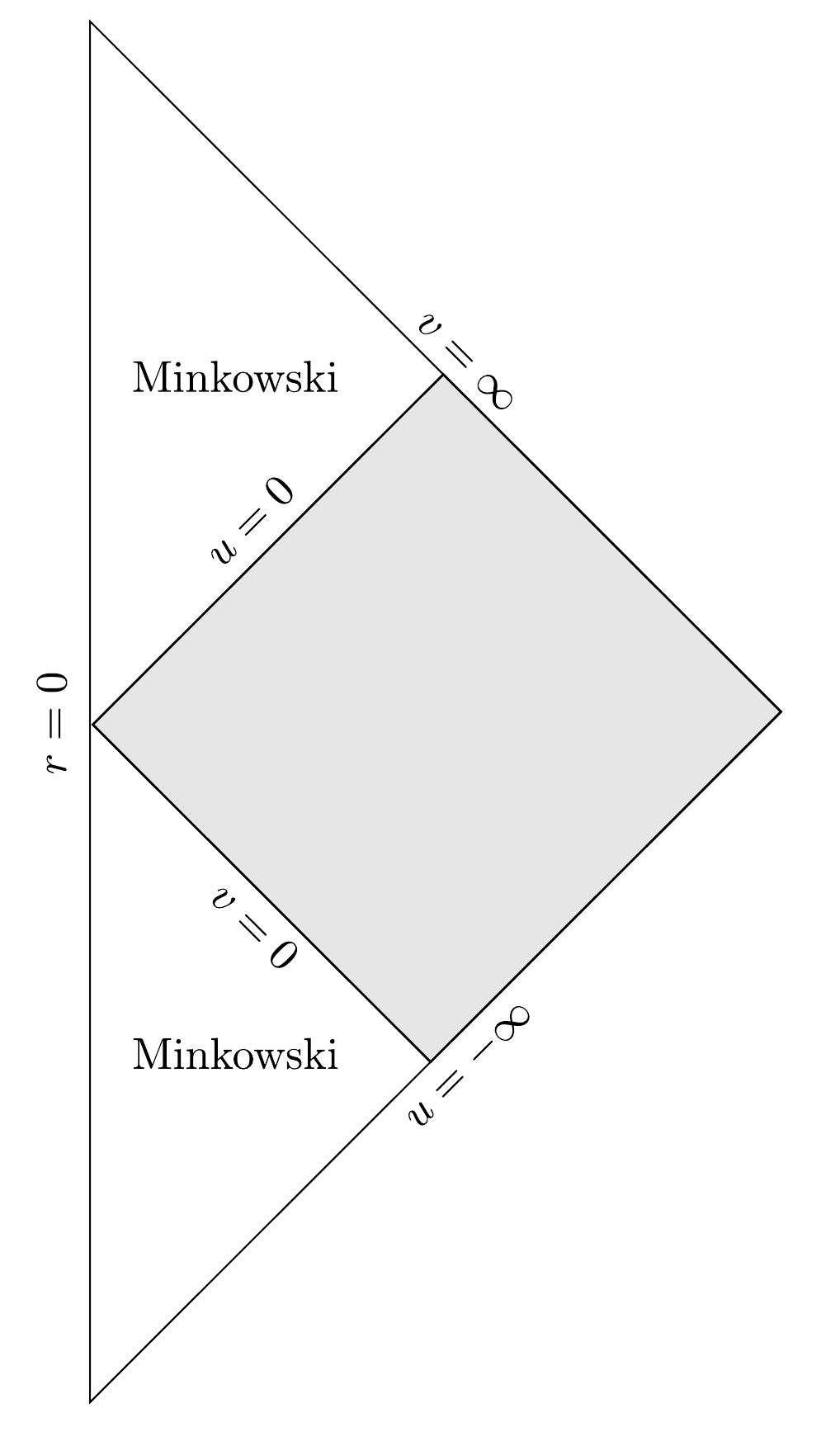}
\caption{The Penrose diagrams for the $\beta=1$ Roberts spacetime are shown for $\alpha>0$ (left panel), featuring an apparent horizon enclosing a spacelike singularity, and for $\alpha<0$ (right panel), corresponding to dispersal without black hole formation. }
\label{alpha-small}
\end{figure}
\FloatBarrier

\begin{figure}[t!]
\centering
\includegraphics[width=0.42\textwidth]{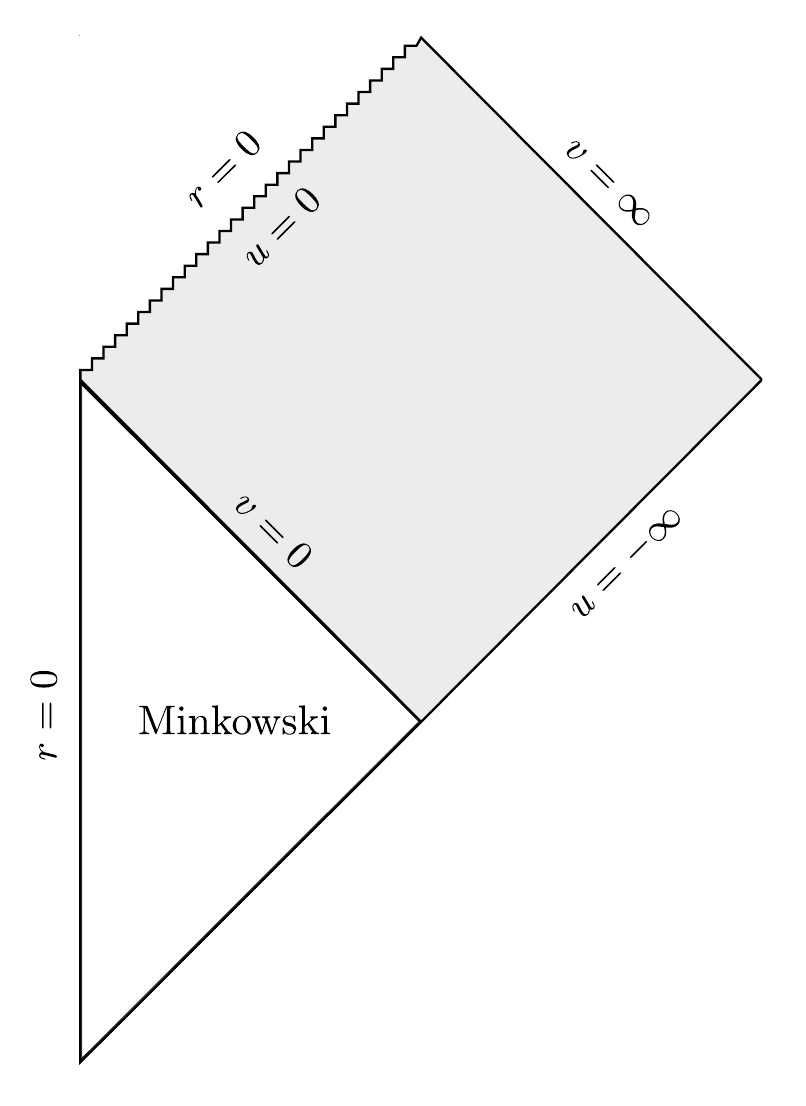}
\includegraphics[width=0.42\textwidth]{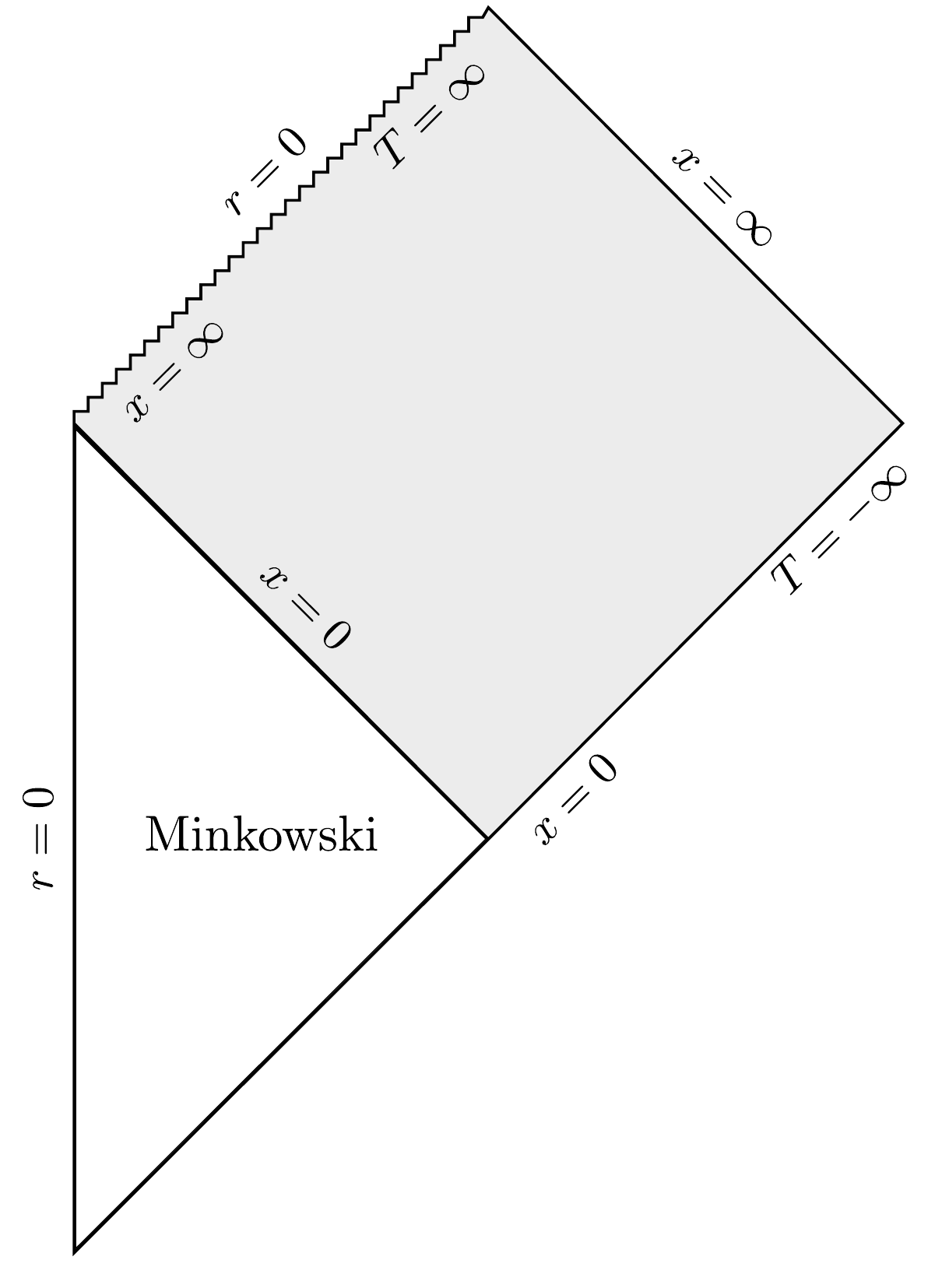}
\caption{The global structure of the exact critical CSS spacetime, corresponding to $\alpha=0$ and $\beta=1$ in the Roberts solution, features a null singularity extending to future infinity. We present this structure in both double-null coordinates $(u,v)$ and adapted coordinates $(T,x)$.}
\label{null-coords}
\end{figure}
\FloatBarrier 
\end{itemize}
It therefore seems that the Roberts solution may not be the most suitable candidate for an analytic study in this context. We shall address these problems before moving on to its semiclassical properties.

The issues of unbounded black hole mass and lack of asymptotic flatness can be addressed by truncating the self-similar solution and matching it to another spacetime, an outgoing Vaidya region, so that the overall geometry becomes asymptotically flat and has finite mass, as demonstrated in~\cite{Wang:1996xh}. The influx is turned off at a finite value $v_0$, where the spacetime is joined to the exterior. This is precisely the construction described in Section~\ref{sec:criticalcollapse}, where the critical solution should be viewed as the interior fill-in region, analogous to the case of the Garfinkle spacetime~\cite{Jalmuzna:2015hoa}.

In fact, this feature justifies our focus on the regime $v<v_0$, with $v_0$ taken to be sufficiently small, where the spacetime is described by the critical Roberts solution. As we will see in Section~\ref{sec:Robertshorizon}, classical perturbations are analytically tractable only for small $v$, and quantum corrections are also expected to contribute only in this regime.

The more severe problem of having classical perturbation modes forming a continuous and complex spectrum is closely related to the distinction between CSS and DSS behaviors, and it turns out to be exactly what we want. It signals a lack of universality in critical phenomena across different incoming wave packets. The complex, oscillatory nature of the growing modes implies that perturbations evolving on the scale-invariant CSS background acquire a scale-dependent structure of the form $e^T \cos(\text{Im}(\omega_c)T)$, where the CSS is dynamically broken into DSS. The associated echoing period, $\Delta =2\pi/\text{Im}(\omega_c) \approx 4.44$ with $\max\{\text{Im}(\omega_c)\}=\sqrt{2}$, closely resembles that observed in Choptuik’s numerical DSS solution ($\Delta \approx 3.44$). The small quantitative discrepancy is unsurprising since we are comparing linear perturbations rather than full nonlinear field configurations, and the similarity in profiles reinforces this interpretation. 

Indeed, the breaking of self-similarity in Roberts spacetime has been studied in detail by Frolov~\cite{Frolov:1999fv}, who showed that a generic growing perturbation causes the Roberts solution to evolve in a universal manner: the original CSS is replaced by a DSS, decaying into the choptuon. Complementary analysis of non-spherical perturbations~\cite{Frolov:1998tq} reveals no growing modes in those sectors, showing that the Roberts solution is unstable only to spherically symmetric perturbations. This resolves the apparent tension: the Roberts solution is not the true intermediate attractor; rather, it decays into the DSS critical spacetime, which plays the role of the universal attractor. In the following, we employ the CSS Roberts spacetime as an analytic approximation to gain insights into quantum effects.

Having addressed the major limitations of the Roberts spacetime, now we set $\beta=1$ throughout, the normalization $f_0$ in the scalar field from \eqref{eq:scalarRoberts} becomes
\be
f_0 (\alpha)=-\frac{\sqrt{2}}{2}\ln{\bigg[\frac{\sqrt{1+4 \alpha }-1}{\sqrt{1+4 \alpha }+1}\bigg]},
\ee
which is nice in the sense that the solution converges to the critical solution as $\alpha \to 0$ from \eqref{eq:Robertscoeff}~\cite{Frolov:1997uu}
\be
r_c^2= u^2-uv, \quad f_c=\frac{\sqrt{2}}{2}\ln{\bigg[1-\frac{v}{u}\bigg]}.
\ee
We introduce the adapted coordinates $(T,x)$~\cite{Frolov:1997uu}
\be \label{eq:adpatedRoberts}
T=-\ln{(-u)}, \quad x=\frac{1}{2}\ln{\bigg[1-\frac{v}{u} \bigg]},
\ee
with the inverse transformation
\be
u=-e^{-T}, \quad v=e^{-T}(e^{2x}-1).
\ee
Then the metric is
\be
ds^2=2 e^{-2T}e^{2 \rho}e^{2x}  [(1-e^{-2x})dT^2-2dT dx]+r^2 d \Omega^2
\ee
and the critical solution is simply
\be
\rho_c=0, \quad r_c=e^{-T}e^x, \quad f_c=\sqrt{2} x.
\ee

\subsection{One-loop analysis and semiclassical Roberts spacetime}
\label{sec:Robertsoneloop}

Now we follow the same procedure established in Sections~\ref{sec:onelooptheory} and \ref{sec:quantum2+1} to study the semiclassical properties of the Roberts spacetime. Again, we first work in the double-null coordinates $(u,v)$. With the spherical dimensional reduction ansatz \eqref{eq:sphericalansatz}, the two-dimensional spacetime is given by the metric and the dilaton field $\phi$
\be
ds^2=-2du dv, \quad \phi=\frac{1}{2}\ln{\bigg( \frac{1}{u^2-uv}\bigg)}. 
\ee
The auxiliary fields from solving \eqref{eq:chi1} and \eqref{eq:chi2} are
\bea
\chi_1&=& \frac{\lambda_2}{4} \bigg\{\ln{(v-u)}\bigg[\ln{\bigg(-\frac{u}{v}\bigg)}-1\bigg]-\frac{1}{2}\bigg[\ln{\bigg(1-\frac{u}{v}\bigg)\bigg]^2-\text{Li}_2\bigg(\frac{v}{v-u}\bigg)} \bigg\}
\no\\
&\quad&+\lambda_2 \gamma \ln{(-u)}+C_1(v)+C_2(u),
\eea
\bea
\chi_2&=& -\frac{\mu_2}{4} \bigg\{\ln{(v-u)}\bigg[\ln{\bigg(-\frac{u}{v}\bigg)}-1\bigg]-\frac{1}{2}\bigg[\ln{\bigg(1-\frac{u}{v}\bigg)\bigg]^2-\text{Li}_2\bigg(\frac{v}{v-u}\bigg)} \bigg\}
\no\\
&\quad&-\mu_2 \gamma \ln{(-u)}+C_3(v)+C_4(u).
\eea
Here we have ensured that the particular solutions to $\chi_1$ and $\chi_2$ are manifestly real in the defined domain of $(u, v)$. A peculiar feature of the reduced Roberts solution is that it has vanishing spacetime curvature, with the null singularity instead encoded in the dilaton field. As a result, the auxiliary fields are independent of the parameters $\lambda_1$ and $\mu_1$. Furthermore, from an \textit{a posteriori} analysis, we find that a one-parameter family of terms proportional to a new parameter $\gamma$ given above, can be added while preserving conservation law and yielding a regular $\langle T^{(4)}_{\mu \nu}\rangle$. These terms can be absorbed into $C_2(u)$ and $C_4(u)$, corresponding to a one-parameter choice of state. However, once quantum backreaction on the geometry is included, regularity at infinity uniquely selects $\gamma=-\frac{3}{4}$.

We need to determine the boundary conditions, i.e., picking a regular choice of state for $\langle T_{ab} \rangle$. In the Roberts solution, we will require regularity at $v=0, \infty$ and $u=-\infty$ (while $u=0$ is the null curvature singularity). Note that $v=0$ is where we turn on the scalar field influx, hence $v<0$ is a Minkowski region. $\langle T_{ab} \rangle$ does not need to vanish at $v = 0$ but must remain regular. In the reduced spacetime, we will allow $\langle T_{ab} \rangle$ to be slowly growing, at most logarithmically as $v \to \infty$, compatible with the conditions in~\cite{Frolov:1997uu, Frolov:1999fv}. In the end, we will only be concerned with the physical four-dimensional $\langle T^{(4)}_{\mu \nu}\rangle$, and ask if it corresponds to a Boulware-like state that vanishes asymptotically. The $s$-wave relations are simply given by~\cite{Mukhanov_1994, Balbinot_1999, Wu:2023uyb}
\be \label{eq:4Dswave}
\langle T^{(4)}_{ab} \rangle = \frac{\langle T^{(2)}_{ab} \rangle}{4 \pi e^{-2 \phi}}, \quad
\langle T^{(4)}_{\theta \theta} \rangle=\frac{\langle T^{(4)}_{\varphi \varphi} \rangle}{\sin^2{\theta}}=-\frac{1}{8 \pi  } \frac{1}{\sqrt{-g_{(2)}}} \frac{\delta \Gamma_{\text{one-loop}}}{\delta \phi}.
\ee
By imposing regularity of $\langle T_{vv} \rangle$ near $v=0$, we find $C_1(v)=C_3(v)$, and 
\be
C_3''(v)=\frac{\lambda_2(1-\ln{v})}{4 v^2}.
\ee
With this choice, $\langle T_{vv} \rangle$ vanishes identically as $u \to -\infty$. Substituting into $\langle T_{uu} \rangle$ and imposing regularity leads to $C_2(u) = C_4(u)$ (while if we pick $C_2(u)=-C_4(u)$, there is no $\lambda_2$ independent results, which is unphysical as $\lambda_2$ is an auxiliary parameter). We then find
\be
C_4''(u)=\frac{\lambda_2 \ln{(-u)}}{4 u^2}.
\ee
Then we obtain a unique quantum stress-energy tensor in the reduced Roberts spacetime 
\be
\langle T_{uu} \rangle=\frac{-(3+4 \gamma)(u-v)^2-uv+(u-v)^2\ln{(\frac{u}{u-v})}}{16 \pi u^2 (u-v)^2},
\ee
\be
\langle T_{vv} \rangle =\frac{uv-2v^2 -(u-v)^2 \ln{(\frac{u}{u-v})}}{16 \pi (u-v)^2 v^2},
\ee
\be
\langle T_{uv} \rangle= -\frac{v}{16 \pi u (u-v)^2}.
\ee
One can verify that the $\langle T_{ab} \rangle$ given above is regular at $v=0$, vanishing at the past null infinity $u=-\infty$, while logarithmically divergent as $v \to \infty$. But this is fine, as we are ultimately interested in four-dimensional quantities, and they are given by
\be
\langle T^{(4)}_{uu} \rangle=\frac{-(3+4 \gamma)(u-v)^2-uv+(u-v)^2\ln{(\frac{u}{u-v})}}{64 \pi^2 u^3 (u-v)^3},
\ee
\be
\langle T^{(4)}_{vv} \rangle=\frac{uv-2v^2 -(u-v)^2 \ln{(\frac{u}{u-v})}}{6 4\pi^2 u v^2 (u-v)^3},
\ee
\be
\langle T_{uv} \rangle= -\frac{v}{64 \pi^2 u^2 (u-v)^3}.
\ee
\be
\langle T^{(4)}_{\theta \theta} \rangle= \frac{\langle T^{(4)}_{\varphi \varphi} \rangle}{\sin^2{\theta}}=0, 
\ee
where the angular components vanish identically. We observe that $\langle T^{(4)}_{\mu \nu} \rangle$ vanishes asymptotically as $v \to \infty$ and $u \to -\infty$, which corresponds to a regular Boulware-like state.

By transforming to the adapted coordinates \eqref{eq:adpatedRoberts}, the quantum stress-energy tensor becomes
\be
\langle T^{(4)}_{TT} \rangle= e^{2 T}\frac{e^{-6 x}(-4+8 e^{2x}- e^{4x}(7+4 \gamma))}{64 \pi^2},
\ee
\be
\langle T^{(4)}_{xT} \rangle=e^{2T}\frac{e^{-4 x}(2-5 e^{2 x}- e^{4x}(2x-3))}{32 \pi^2 (e^{2x}-1)},
\ee
\be
\langle T^{(4)}_{xx} \rangle= e^{2T}\frac{(3-e^{-2x}+2e^{2x}(x-1))}{16 \pi^2 (e^{2x}-1)^2},
\ee
while the angular components remain zero. Similar to the Garfinkle spacetime, we observe that they all have the structure 
\be \label{eq:Robertsstressgen}
\langle T^{(4)}_{\mu \nu} \rangle = e^{2T} F_{\mu \nu}(x),
\ee
indicating that the quantum effects act as a growing mode. The Lyapunov exponent differs from the Garfinkle case, and we will elaborate on this point in Section~\ref{sec:discussion}. The components of $\langle T^{(4)}_{\mu \nu} \rangle$ diverge as $T \to \infty$, corresponding to the approach toward the null singularity. However, they remain completely regular at $x=0$ and vanish asymptotically as $x \to \infty$ and $T \to -\infty$. One can verify in fact all components of $F_{\mu \nu}(x)$ are \textit{real-analytic} in $x \in[0, \infty)$.

The backreaction on the geometry from the semiclassical Einstein equation can be solved. We pick an ansatz similar to~\cite{Frolov:1997uu}
\be
ds^2=2 \ell^2 e^{-2T}e^{2x} [e^{2 \rho_1}(1-e^{-2x})dT^2-2e^{2 \rho_2}dT dx]+r^2 d \Omega^2
\ee
but with 
\be
\rho_1=\rho_c+ \frac{\hbar}{\ell^2} F_q(x) e^{\omega_q T}, 
\ee
\be
\rho_2=\rho_c+ \frac{\hbar}{\ell^2} W_q(x) e^{\omega_q T},
\ee
\be
r =r_c \bigg( 1+\frac{\hbar}{\ell^2} r_q(x)e^{\omega_q T} \bigg).
\ee
In contrast to the Garfinkle case, we need to introduce three independent functions: $F_q, W_q,$, and $r_q$. \textit{A priori}, there is no reason to expect the quantum backreaction can be encoded in a single conformal factor $\rho$. Again, we consistently expand the right-hand side of~\eqref{eq:semiclassicalEinstein}, including the classical piece sitting at the corrected background.

The required conditions for a physical semiclassical spacetime should include vanishing backreaction at $x=0$ (since the past is Minkowski), and regularity as $x \to \infty$. This will pick up a unique value $\gamma=-\frac{3}{4}$ for the one-parameter choice of state. The semiclassical Einstein equation, satisfying the boundary conditions, will determine
\be
F_q(x)=-r_q(x)=\frac{6-6e^{-2x}+\pi^2+12 x^2-12x\ln{(1-e^{2x})}-6 \text{Li}_2(e^{2x})}{768 \pi^2},
\ee
\be
W_q(x)=0.
\ee
This solution satisfies all components of the semiclassical Einstein equation up to order $O(\hbar)$, including the angular components. Although the individual terms in $F_q(x)$ and $r_q(x)$ involve complex-valued functions for $x>0$, their imaginary parts cancel exactly, resulting in a semiclassical spacetime that is real-analytic on $x \in [0, \infty)$ for finite $T$.

As a final remark, we noted that in the two-dimensional reduced Roberts spacetime, we allow $\langle T_{ab} \rangle$ to be at most logarithmically divergent as $v \to \infty$, consistent with the analysis in~\cite{Frolov:1997uu} and the interpretation from the full four-dimensional theory. In this case, $\langle T_{ab} \rangle$ is independent of $T$ when expressed in $(T,x)$ coordinates. While it is indeed possible to choose a quantum state that is strictly regular at $v \to \infty$, such a choice would necessarily introduce explicit $T$-dependence, which diverges as $T \to \infty$.\footnote{As we discuss in detail in Appendix~\ref{sec:appendixD},~\cite{Zahn:2025tnh} reported a discrepancy based on the choice of a self-similar state, identifying an additional linear-$T$ term arising from the state-independent part of $\langle T_{\mu\nu} \rangle$. Here, if one imposes the boundary condition that is strictly regular as $v \to \infty$, a linear-$T$ scaling indeed appears; hence no genuine discrepancy necessarily exists. This, nevertheless, corresponds to a different boundary condition from the one adopted in the linear perturbation analysis~\cite{Frolov:1997uu}.} This implies that no stationary Boulware-like quantum state can remain completely regular throughout the reduced spacetime, unlike in the Garfinkle solution or in black hole spacetimes~\cite{Balbinot_1999, Wu:2023uyb}. 

A similar feature arises in the semiclassical analysis of the Hayward solution, discussed in Appendix~\ref{sec:Hayward}. In that case, one is forced to accept a logarithmically divergent $\langle T_{ab} \rangle$ in the reduced spacetime, reflecting the causal structure characterized by a bifurcate null singularity. As a result, no stationary quantum state can be found.

This can be understood as a consequence of the curvature singularities being null rather than point-like in the reduced spacetime. The two-dimensional stress-energy tensor $\langle T_{ab} \rangle$ must still encode the kinematically divergent behavior near such null singularities. This highlights a subtle but important feature of quantum effects in these backgrounds: a Boulware-like state is not determined solely from self-similarity; in addition, the global causal structure plays a crucial role in determining the appropriate choice of quantum state.

\subsection{Horizon tracing for semiclassical Roberts spacetime}
\label{sec:Robertshorizon}

We now analyze the horizon structure of the semiclassical Roberts spacetime, building on the techniques used in Section~\ref{sec:GarfinkleHorizon}. The starting point is to understand the classical growing modes of the Roberts solution.

\subsubsection*{Classical perturbation modes of the Roberts spacetime}

The classical perturbation spectrum of the Roberts spacetime has been thoroughly studied in~\cite{Frolov:1997uu, Frolov:1999fv}. As reviewed in Section~\ref{sec:classicalRoberts}, these modes fill a continuous region in the complex plane \eqref{eq:Robertsmodes}. Fortunately, this problem is analytically tractable in the regime of interest that we should briefly summarize below.

Following the conventions of~\cite{Frolov:1999fv}, we consider general spherically symmetric metric perturbations of the form
\be
\delta g_{\mu \nu} dx^\mu dx^\nu=k_{uu} du^2+2 k_{uv} du dv+k_{vv}dv^2+r^2 K d \Omega^2,
\ee
alongside scalar field perturbations denoted $\varphi$. To isolate gauge-invariant content under spherical symmetry, one defines the following three gauge-invariant quantities. One for the matter perturbation
\be
f=\frac{K}{2}-\varphi+\frac{1}{2u}\int k_{vv} dv,
\ee
and the others for the metric perturbations
\be
\rho= (r^2 K)_{,uv}+k_{vv}-k_{uv}-\frac{uk_{uu,v}}{2}+\frac{(2u-v) k_{vv,u}}{2},
\ee
\be
\sigma=k_{uv}-\frac{1}{2}\int k_{vv,u} dv-\frac{1}{2}\int k_{uu, v} du.
\ee
The linearized Einstein-scalar system can then be written in terms of these gauge-invariant variables, where
\be
2 u (u-v) f_{,uv}+(2u-v)f_{,v}-u f_{,u}-2 f=0,
\ee
\be
\rho=0, \quad \sigma_{,u}=2f_{,u}+\frac{2f}{u}.
\ee
We are free to switch between different gauges once we have identified the gauge-invariant variables. For instance, we could pick the field gauge with $K=k_{vv}=0$ or the null gauge $k_{uu}=k_{vv}=0$, as discussed in~\cite{Frolov:1999fv}.

We choose initial data on a constant-$u$ slice, equivalent to $T=0$, and apply the junction conditions across the null shell $v = 0$ as boundary conditions. In adapted coordinates $(T, x)$, the main equation to solve becomes
\be
\mathcal{D}f \equiv (1-e^{-2x}) \partial_x^2 f+2 \partial_x \partial_T f+2 \partial_T f-4f=0.
\ee
Due to the scale-invariance of the background, we could reduce the PDE to one dimension by applying a Laplace transform with respect to the scale variable $T$
\be
F(k, x)=\int_{0}^\infty f (T, x) e^{-kT} dT,
\ee
where $k$ is later identified as the spectrum of the classical modes $\omega_c$ in \eqref{eq:Robertsmodes}. The Laplace transform on the operator is
\be
\mathcal{L}_T (\partial_T f)= kF- f(T=0),
\ee
such that the initial condition $f(T=0)$ would be a source term. Hence the Laplace-transformed equation is
\be
\mathcal{D}_k F (k, x)=h(x),
\ee
where $\mathcal{D}_k=\mathcal{L}_T \mathcal{D}$, and now we have an ordinary differential operator, which is algebraic in $k$. Then $h(x)$ encodes information about the initial shape of the wave packet at $T=0$. Introducing $y = e^{2x} = 1 - v/u$ simplifies the equation into a hypergeometric form. The source term becomes
\be
h(y)=-y \frac{d}{dy}f(y, T=0)-\frac{1}{2}f(y, T=0).
\ee
After solving for $F(k, y)$, we perform an inverse Laplace transform to reconstruct $f(T, x)$ and the associated perturbations $\rho$ and $\sigma$.

The scalar wave packet profile can be decomposed into three components depending on the initial and boundary conditions: outgoing, constant, and incoming parts. The outgoing and constant parts would not grow exponentially as $T \to \infty$, hence the most physically interesting case is the incoming part with the conditions
\be
f(T, x=0)=0, \quad f(T=0, x)=f_{I}(x).
\ee
Along with the boundary condition at infinity, one could fix a complex profile of the modes $k$. A formal solution is given in~\cite{Frolov:1999fv}, but it is of little practical use. We could instead rely on the stationary phase method to obtain a late-time asymptotic form of $f(T, y)$ as it is the most physically relevant regime.

Fortunately, there is a useful late-time approximation that can be written in closed form. The dominant contribution at large $T$ arises from the mode $k = 1 + i\sqrt{2}$ identified via the stationary phase approximation. Although Roberts admits a continuum of modes, it is universal in the sense that a single mode would eventually dominate.

We may consider a power-law form for the wave packet near $y=1$, taking $h(\eta) \propto (1-\eta)^\alpha$. However, this leads to a power-law divergence of the perturbation near $v=0$, resulting in a weak null singularity. To avoid this, we can regulate the initial wave profile by cutting off the divergence for sufficiently small values of $y-1$. If this power-law diverging wave, localized near $y=1$, does not backscatter and influence the evolution of the wave packet at larger $y$, then the cutoff can be applied while keeping the rest of the evolution essentially unchanged. This turns out to be feasible, and we instead use the regularized form
\be
h(\eta) \propto (1-\eta)^\alpha e^{\frac{1-\eta}{\lambda}},
\ee
which effectively suppresses $h$ for $y-1 > \lambda$. We are interested in the region outside the initial localization, but still within the small-$y$ limit, satisfying $\lambda \ll y-1 \ll 1$. In this regime, the dominant contribution in the stationary phase approximation arises from the singularities of $F(k, y)$, with the only real singularity being a simple pole at $k=-\alpha$. While $k \in \mathcal{C}$, we retain only the component with the largest real part, $k=1$.

Outside the region of initial localization but still in the small-$y$ regime, the late-time behavior of the solution admits a closed-form approximation
\be
f(T, y) \approx \frac{\Theta(T+\ln{\lambda})}{1+\alpha} Z_{2}(y;-\alpha) e^{-\alpha T}.
\ee
In this expression, we substitute the complex value of $k$ and take the real part of the resulting solution. The explicit form of $Z_{2}(y;-\alpha)=Z_2(y;k)$ will be given below. Here $\Theta$ is defined as
\be
\Theta(x) \equiv e^{-e^{-x}},
\ee
which behaves as a smoothed step function
\be
\Theta(x) \approx 
\begin{cases}
1, & \text{Re}(x) > 0, \\
0, & \text{Re}(x) < 0.
\end{cases}
\ee
This expression implies that the perturbation outside the localized initial region does not feel the influence of the portion near $y-1<\lambda$ until a delayed time $T=-\ln{\lambda}$, when it begins to spread. In other words, the initially localized wave packet for $v< \lambda$ does not backscatter until it reaches the singularity at $u=0$, after which it re-emerges in a narrow band $-u<\lambda$.

The function $Z_2(y; k)$ has been worked out in~\cite{Frolov:1999fv}, and is given in terms of the hypergeometric function
\be
Z_2(y; k)=(1-y)^{c-a-b} {}_2\mathcal{F}_1(c-a,c-b;c+1-a-b;1-y)
\ee
where the parameters are defined by
\be
c=1, \quad a+b=k, \quad ab=\frac{k}{2}-1, \quad a, b=\frac{1}{2}(k \mp \sqrt{k^2-2k+4}).
\ee
Let us now derive the corresponding metric perturbation. We are free to choose any gauge, and we opt for the field gauge with $K=k_{vv}=0$, in which case the scalar perturbation $\varphi$ coincides with the gauge-invariant quantity $f$. The linearized Einstein-scalar field equations take a simple form
\be
f= - \varphi,\quad \rho=-k_{uv}- \frac{u k_{uu,v}}{2}, \quad \sigma=k_{uv}-\frac{1}{2}\int k_{uu,v}du,
\ee
and in this gauge we have
\be
k_{uv}=2f, \quad k_{uu,v}=-\frac{4 f}{u}.
\ee
Let us perform a coordinate transformation from
\be
\delta g_{\mu \nu} dx^\mu dx^\nu=k_{uu} du^2+2k_{uv} du dv,
\ee
to $(T, x)$, where then
\be
\delta g_{\mu \nu} dx^\mu dx^\nu= AdT^2 +2 B dTdx,
\ee
with
\be
A=e^{-2 T}(k_{uu}-2(e^{2x}-1)k_{uv}), \quad B=e^{-2T} (2e^{2x}k_{uv}),
\ee
and now $k_{uu}$ and $k_{uv}$ are functions of $(T, x)$ that are given by
\be \label{eq:kuvkuu}
k_{uv}(T,x)=2 f(T,x), \quad k_{uu,v}(T,x)=\frac{4f(T,x)}{e^{-T}}.
\ee

\subsubsection*{Horizon tracing with classical and quantum perturbations}

Now we are ready to tackle the horizon-tracing problem in the Roberts spacetime. Following the similar approach in Section~\ref{sec:GarfinkleHorizon}, we compute the horizon-tracing function to linear order
\be \label{eq:Robertshorizon}
(\nabla \bar{r})^2 \approx f_0(x)+(p-p^\ast) f_c(T,x)+e^{2T} f_q(x),
\ee
where
\be
f_0(x)=\frac{1}{2}e^{-2x} (1+e^{2x}),
\ee
\be
f_c(T, x)=-\frac{1}{4}(e^{-2x}k_{uu}-2e^{-2x}k_{uv}-2k_{uv}),
\ee
\be
f_q(x)=\frac{e^{-4 x}(1-e^{2x}+2x e^{2x  })}{64 \pi^2 (e^{2x}-1)}.
\ee
Here, $k_{uu}$ and $k_{uv}$ are functions of $(T,x)$ that encode the classical growing mode. Before analyzing the contribution of the classical perturbation, let us first examine the behavior of $f_0(x)$ and $f_q(x)$. 

\begin{figure}[hbt!]
\centering
\includegraphics[width=0.45\textwidth]{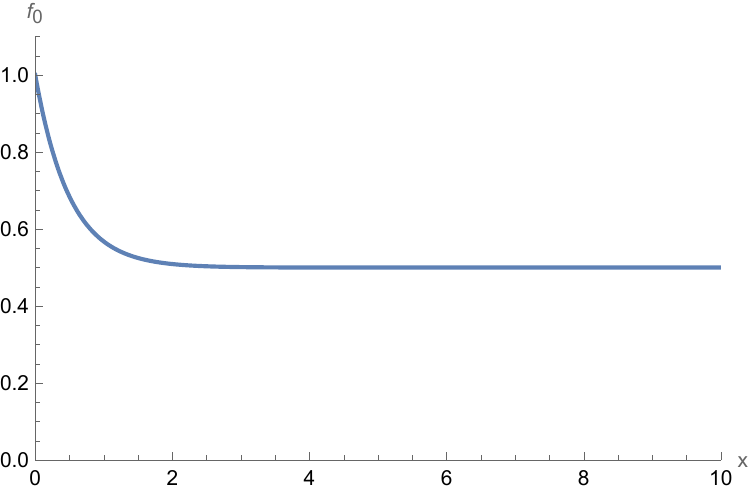}
\includegraphics[width=0.45\textwidth]{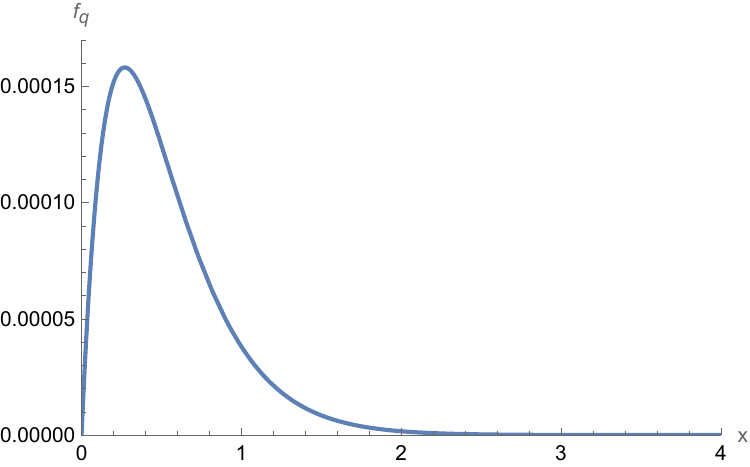}
\caption{The contributions to $(\nabla \bar{r})^2$ in \eqref{eq:Robertshorizon} from the background piece $f_0(x)$ and quantum perturbation $f_q(x)$. Note that the Roberts background admits $x \in [0, \infty )$.}
\label{Robertsf0fq}
\end{figure}

As shown in Figure~\ref{Robertsf0fq}, $f_0(x)$ starts at 1 when $x=0$ and asymptotically approaches $\frac{1}{2}$ as we increase $x$. It never reaches zero, indicating that there is no apparent horizon in the unperturbed critical Roberts spacetime, as expected. 

Interestingly, the quantum correction $f_q(x)$ is nonzero only for a small range of $x$ and vanishes as $x$ increases. This implies that quantum effects only influence the horizon-tracing condition at small values of $x$. Rather than being a drawback, this feature turns out to be highly advantageous for two key reasons. First, as we have discussed, the classical perturbation modes are analytically tractable only in the small-$x$ regime. This allows us to study the subtle interplay between classical and quantum effects precisely where they are both accessible. Second, as discussed in Section~\ref{sec:classicalRoberts}, the Roberts solution should be interpreted as the interior fill-in region, matched to an exterior solution at small enough $x$. This justifies focusing on the small-$x$ region where quantum contributions are significant.

Notably, in contrast to the Garfinkle case, $f_q(x)$ contributes positively at small $x$. As a result, quantum corrections to the critical Roberts spacetime alone do not lead to the formation of an apparent horizon, since the condition $(\nabla \bar{r})^2 <0$ is never satisfied. This might initially suggest that quantum effects are incapable of lifting subcritical data, different from what we observed in the Garfinkle case. However, as we shall see, this conclusion is premature. The intricate balance between the classical and quantum perturbations would still reveal a mass gap, a key signature of Type I critical phenomena.

To proceed, we must examine more closely the contribution to $(\nabla \bar{r})^2$ from the classical perturbation mode $f_c(T,x)$. From \eqref{eq:kuvkuu}, we see that $k_{uu}$ can only be obtained in integral form, as it is governed by the following equation
\be
e^{T+x}[8 f+(\coth{x}-1) \partial_T k_{uu}]-e^{T-x} \partial_x k_{uu}=0.
\ee
The general solution to this equation is
\bea
k_{uu}(T,x)&=&-4\int_1^T dK e^{T-K}(e^{2x}-1)f\bigg[K, \frac{1}{2}\ln{(1+e^{T-K}}(e^{2x}-1))\bigg]
\no\\
&\quad&+  C\bigg(\frac{1}{2}(T+\ln{(e^{2x}-1)})\bigg),
\eea
where $C$ is an arbitrary function of the specific combination $\frac{1}{2}(T+\ln{(e^{2x}-1)})$, to be determined by boundary conditions. Since we could only trust the results within the following regime
\be
\lambda \ll e^{2x}-1 \ll 1, \quad T \leq -\ln{\lambda}.
\ee
This means that when the equality holds, the combination 
\be
s \equiv T+\ln{(e^{2x}-1)} =\ln{\bigg(\frac{e^{2x}-1}{\lambda}\bigg)} \gg 1,
\ee
indicating that even at finite $x$ and within the valid regime of $T$, which is not necessarily the curvature singularity, $s$ could be arbitrarily large. Without loss of generality, the function $C$ could either blow up or vanish as $s$ grows. However, to keep the solution well-behaved, it is physically reasonable to assume that the function $C$ vanishes as $s \to \infty$. Therefore, we can always numerically evaluate the integral form of $k_{uu}$. Note that the classical growing mode scales roughly as $e^{T}$. Compared to the Garfinkle case, the quantum growing mode in the Roberts background is much more dominant than the classical contribution.

Numerically, for $\lambda \ll e^{2x}-1 \ll 1$, we will take $x$ up to $0.1$ and $\lambda \sim 0.005$, which means the classical perturbation remains valid up to $T \approx 5.3$. However, this would already exceed the linear regime. \textit{A posteriori}, we find that quantum effects kick in much earlier, where we only need to stay in mildly positive $T \lesssim 3.3$, which would be perfectly within the validity of linear perturbation. Since we are working with $\lambda \ll e^{2x}-1$, we must impose a lower cutoff in $x$ when $0.005=e^{2x}-1 \implies x \approx 0.0025$. Thus, our analysis will focus on the interval $x \in [0.0025, 0.1]$.

\begin{figure}[hbt!]
\centering
\includegraphics[width=0.45\textwidth]{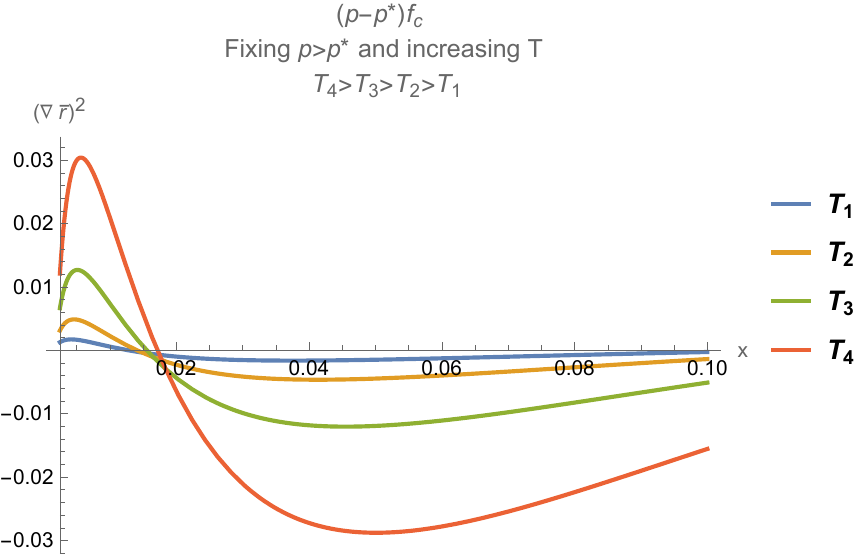}
\includegraphics[width=0.45\textwidth]{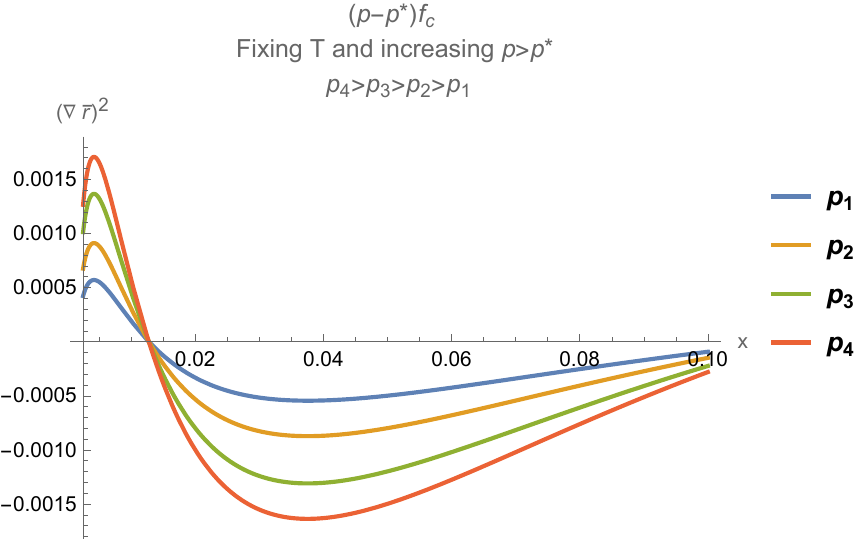}
\caption{We plot the behavior of the classical perturbation mode by staying within $T \lesssim 3.3$ and $(p-p^\ast) \lesssim 0.015$. Even though $f_c(T,x)$ contains subtle $T$-dependence, it is still roughly the combination $e^{T}(p-p^\ast)$ at work.}
\label{Robertsfc}
\end{figure}

\begin{figure}[hbt!]
\centering
\includegraphics[width=0.45\textwidth]{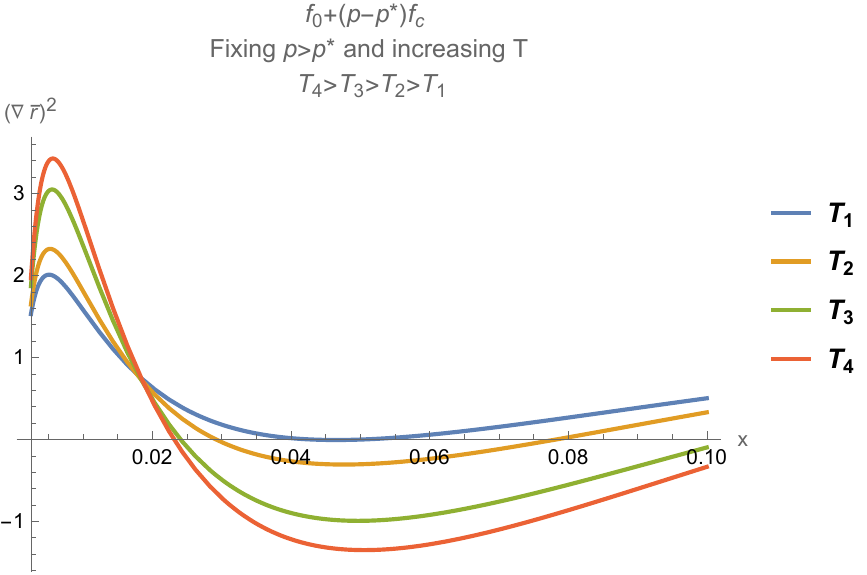}
\includegraphics[width=0.45\textwidth]{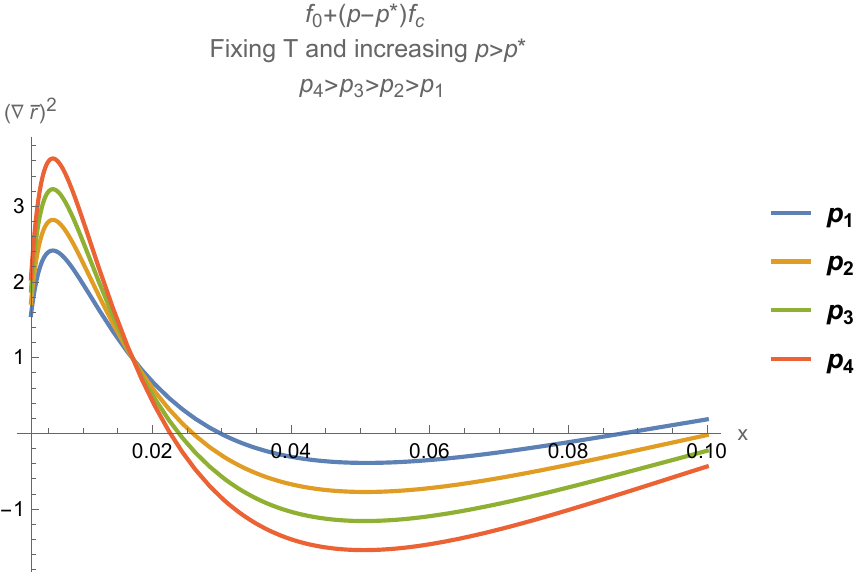}
\caption{Including the background piece with $T \lesssim 3.3$ and mildly large amplitude $(p-p^\ast) \lesssim 1.3$, we can consistently see horizon formation.}
\label{Robertsf0fc}
\end{figure}

We plot $f_c(T, x)$ in Figure~\ref{Robertsfc}, where we see that it is not monotonic and can interpolate between positive and negative values. Combining with the background contribution $f_0(x)$ in Figure~\ref{Robertsf0fc}, we can observe horizon formation for $(\nabla \bar{r})^2<0$, including purely classical growing mode, consistent with the expectation.

\begin{figure}[hbt!]
\centering
\includegraphics[width=0.45\textwidth]{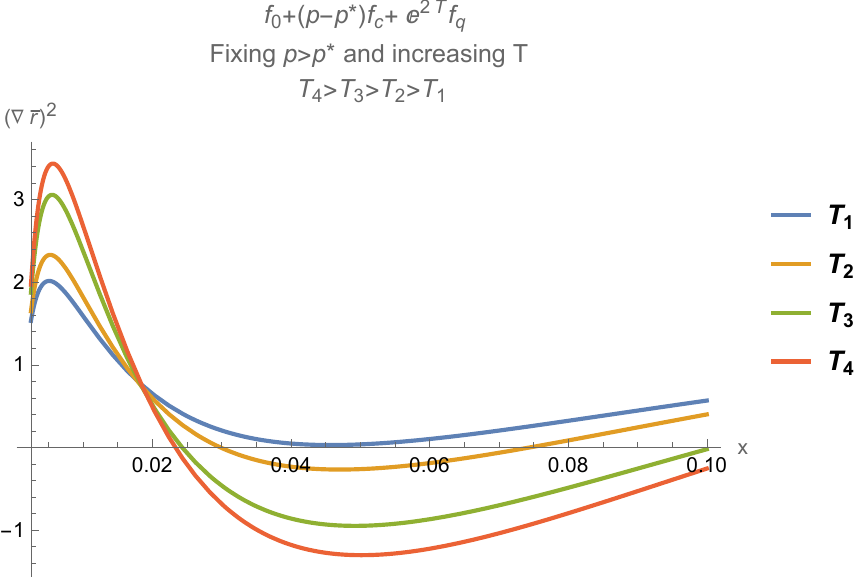}
\includegraphics[width=0.45\textwidth]{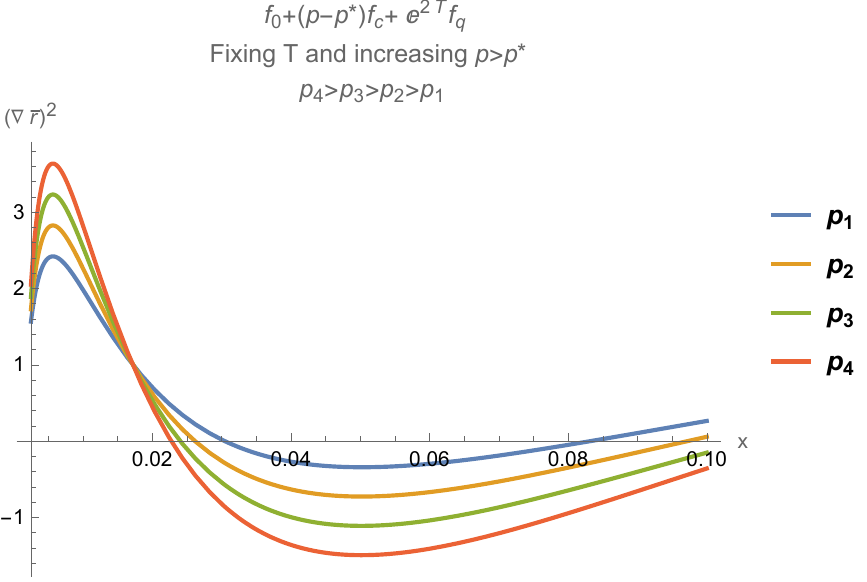}
\caption{Picking the same parameters as in Figure~\ref{Robertsf0fc}, we find adding quantum contribution $f_q$ does not change much of the horizon structure.}
\label{Robertsf0fcfq}
\end{figure}

\begin{figure}[hbt!]
\centering
\includegraphics[width=0.45\textwidth]{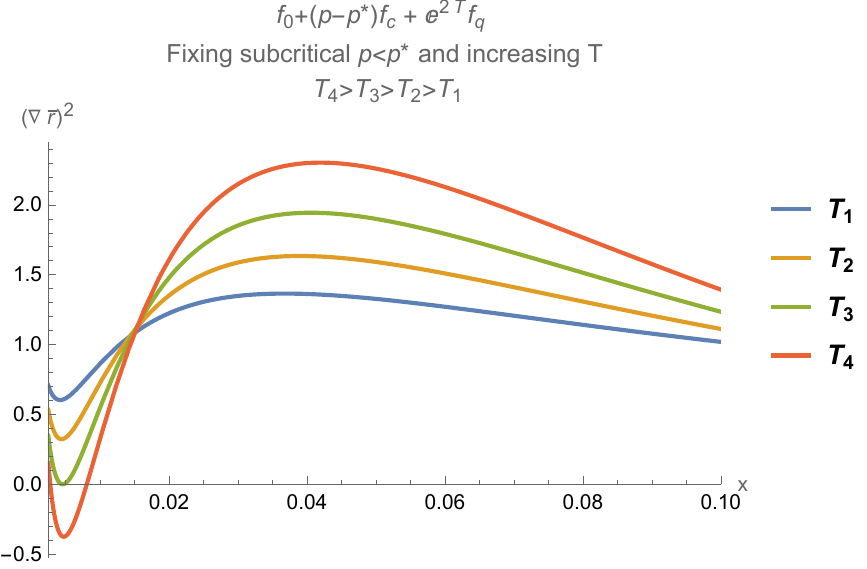}
\includegraphics[width=0.45\textwidth]{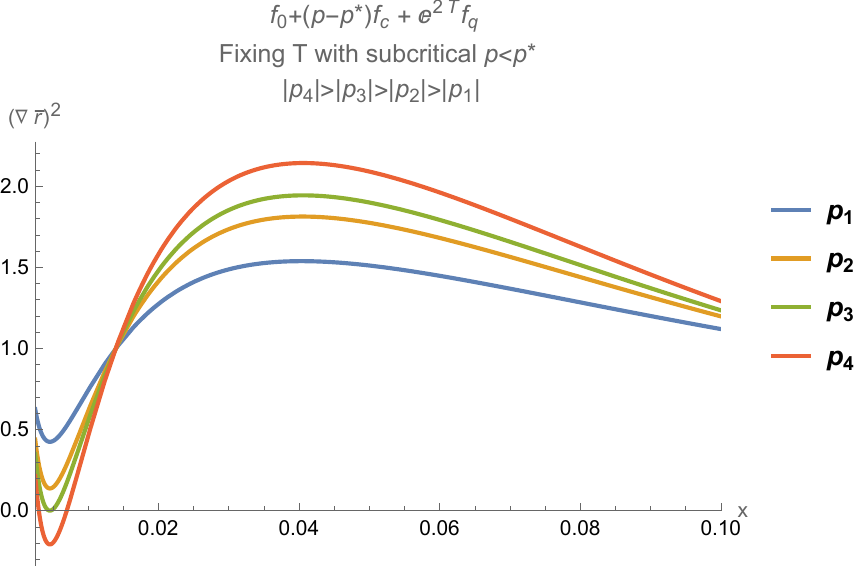}
\caption{subcritical data can be lifted up that result in horizon formation with a single root $(\nabla \bar{r})^2=0$.}
\label{Robertsf0fcfqsub}
\end{figure}

Since $f_q(x)$ contributes positively to $(\nabla \bar{r})^2$, adding it would not cause much difference for the horizon structure, as illustrated in Figure~\ref{Robertsf0fcfq}. However, the non-monotonic behavior of the classical contribution $f_c(T,x)$ indicates that classical effects may counteract quantum corrections for certain subcritical configurations with $p<p^\ast$, as shown in Figure~\ref{Robertsf0fcfqsub}. Remarkably, as in the Garfinkle case, we find that some subcritical data with $p^\ast_q<p^\ast$ can be lifted above the apparent horizon threshold. We therefore reinterpret $p^\ast_q$ as the new critical point in the presence of quantum corrections.

\begin{figure}[hbt!]
\centering
\includegraphics[width=0.45\textwidth]{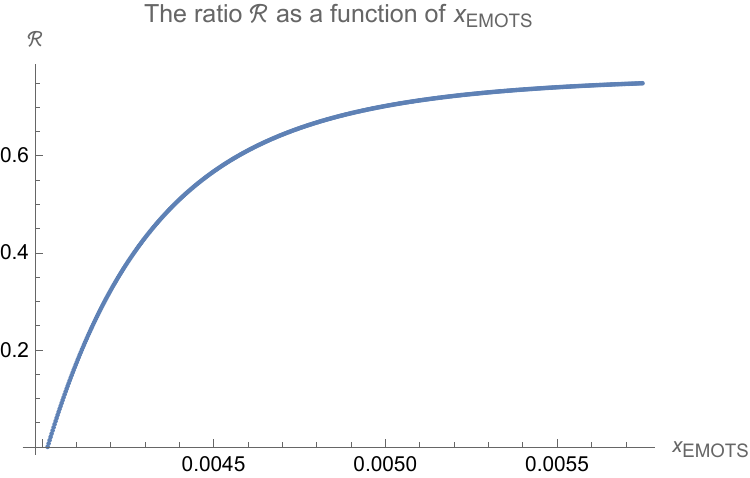}
\includegraphics[width=0.45\textwidth]{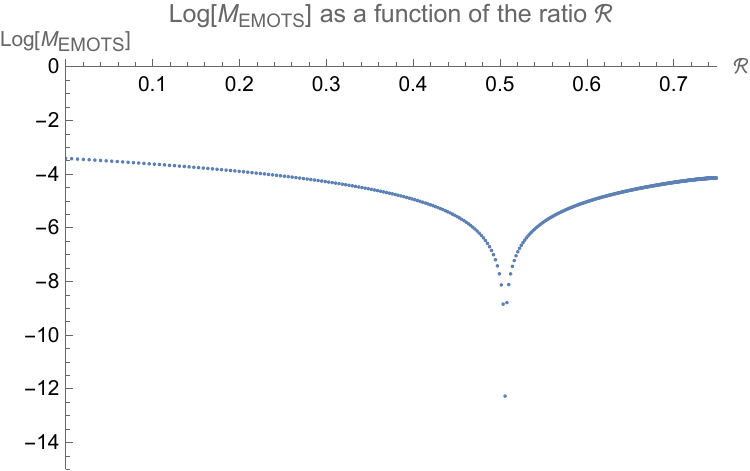}
\caption{We see that $x_{\text{EMOTS}}$ in fact barely move as we decrease the ratio. But quantum corrections cause an abrupt change in the behavior of $M_{\text{EMOTS}}$, in contrast to the Garfinkle case.}
\label{RobertsEMOTS}
\end{figure}

We once again consider the ratio $\mathcal{R}$ between classical and quantum growing modes, defined by taking $\omega_c=\text{max}\{\text{Re}(k)\}$
\be
\mathcal{R}\equiv\frac{(p-p^\ast_q) e^{\omega_c T}}{\hbar e^{\omega_qT}} \approx \frac{(p-p^\ast_q)}{e^T},
\ee
and evaluate it at the time of EMOTS formation. We sample 500 data points and trace both the location $x_{\text{EMOTS}}$ and the mass $M_{\text{EMOTS}}$. It is worth noting that these points occur at slightly different null time slices. To compute $M_{\text{EMOTS}}$, we switch to the null gauge. Note that in four dimensions, the quasi-local mass function is given by
\be
M(T, x) \equiv \frac{\bar{r}}{2}[1-(\nabla \bar{r})^2],
\ee
which means for $M_{\text{EMOTS}}$ with $(\nabla \bar{r})^2=0$, it is proportional to $\bar{r}$. The results are shown in Figure~\ref{RobertsEMOTS}. 

The maximum value of $\mathcal{R}$ corresponds to the point where quantum effects begin to trigger horizon formation. Decreasing $\mathcal{R}$ can be interpreted as approaching the new quantum-shifted critical point $p^\ast_q$. We find that varying $\mathcal{R}$ has minimal impact on the location $x_{\text{EMOTS}}$. However, the behavior of $M_{\text{EMOTS}}$ exhibits a more interesting trend: initially, as $\mathcal{R}$ decreases, $M_{\text{EMOTS}}$ also decreases, consistent with classical expectations. Around $R \sim 0.5$, however, a sudden transition occurs, signaling the onset of significant quantum effects. As $\mathcal{R}$ continues to decrease, quantum corrections become increasingly dominant, ultimately leading to a finite mass gap characteristic of Type I behavior.

\section{Discussion}
\label{sec:discussion}

The phenomenon of critical gravitational collapse provides a unique window into the formation of singularities and the possible breakdown of cosmic censorship in gravity theories. While classical analyses have revealed the existence of naked singularities, the role of quantum effects near criticality remains largely unexplored. Motivated by the limitations of earlier treatments, we develop an anomaly-based one-loop semiclassical framework for analyzing scalar field collapse in Einstein gravity and illustrate its consequences in analytically tractable $2+1$ and $3+1$-dimensional examples that are closely related to true critical solutions. A key feature of our result is the emergence of a unique, Boulware-like quantum state that arises naturally from the regularity of the stress-energy tensor. This state encapsulates only vacuum polarization effects of the collapsing scalar field and excludes artificial fluxes, making it a physically reasonable quantum state for near-critical analysis. Within this framework, we demonstrate that quantum backreaction gives rise to a universal growing mode and a dynamically generated mass gap, suggesting a transition from classical Type II to a quantum-modified Type I behavior. These results provide a concrete mechanism by which quantum effects may shield the singularity and offer a new avenue toward understanding cosmic censorship through semiclassical gravity.

Here, we take the opportunity to clarify key aspects of our findings, and address potential limitations in our analysis.

\subsubsection*{Nature of the quantum growing mode}

We have observed in both $2+1$ and $3+1$ dimensions that the growing modes associated with quantum corrections scale as $e^T$ and $e^{2T}$ in the adapted coordinates $(T,x)$, respectively. This exponential scaling suggests a simple kinematical origin, intuitively attributable to the dilaton field $\phi$, encoding the areal radius in the dimensionally reduced theory under time evolution. This intuition proves to be correct upon closer examination.

Although the quantum stress-energy tensor $\langle T^{(2)}_{ab} \rangle$ appears to be stationary from the two-dimensional perspective, as we are adopting a Boulware-like state in which all components are $T$-independent, this is not the case from the higher-dimensional viewpoint. The $T$-dependence in physical quantities like $\langle T^{(D)}_{\mu \nu} \rangle$ arises not from explicit time dependence in $\langle T^{(2)}_{ab} \rangle$, but rather from the dilaton-dependent mapping to the higher-dimensional quantities.

This becomes evident by inspecting the structure of the dimensional reduction. We first emphasize that with the normalization \eqref{eq:sphericalansatz} of the metric ansatz, the matter sector is always proportional to $e^{-2 \phi} (\nabla f)^2$, as in \eqref{eq:matteraction}. This implies that the dimensional reduction of a minimally coupled free massless scalar field $f$ from any $D$-dimensional theory always results in the same two-dimensional trace anomaly \eqref{eq:traceanomaly} upon quantization. For spherically symmetric spacetimes, the relation between the higher-dimensional stress-energy tensor and its two-dimensional counterpart takes the form
\be
\langle T^{(D)}_{ab} \rangle \propto \frac{\langle T^{(2)}_{ab} \rangle}{ e^{-2 \phi}}, \quad \langle T^{(D)}_{\theta \theta} \rangle \propto  e^{\frac{2 \phi (D-4)}{D-2}} \frac{1}{\sqrt{-g^{(2)}}}\frac{\delta \Gamma^{(2)}_{\text{one-loop}}}{\delta \phi},
\ee
as a straightforward consequence of diffeomorphism invariance and compatibility of conservation laws in the corresponding spacetimes, discussed in Section~\ref{sec:onelooptheory}. Note that all other angular components are proportional to $\langle T^{(D)}_{\theta \theta} \rangle$. Due to our choice of quantum state in the reduced spacetime, both $\langle T^{(2)}_{ab} \rangle$ and $\frac{\delta \Gamma^{(2)}_{\text{one-loop}}}{\delta \phi}$ are independent of $T$, so we only need to examine the remaining multiplicative factors.

From the dilaton profile $\phi=-\frac{D-2}{2} \ln r$, we see that as long as the $D$-dimensional critical spacetime exhibits an overall $e^{-2T}$ scaling, as is true for both CSS and DSS spacetimes, we have 
\be
r^2(x, T)= e^{-2 T} \tilde{r}^2(x) \implies \phi=-\frac{D-2}{2} \ln r =-\frac{D-2}{2} \ln{(e^{-T} \tilde{r}(x))}
\ee
and therefore
\be
e^{-2 \phi} = e^{-(D-2)T} \tilde{r}^{D-2}(x), \quad e^{\frac{2 \phi (D-4)}{D-2}} = e^{(D-4) T} \tilde{r}^{4-D}(x).
\ee
Note that $\sqrt{-g^{(2)}}$ always contributes a factor $e^{-2 T}$, then
\be
\langle T^{(D)}_{ab} \rangle \propto e^{(D-2)T}, \quad \langle T^{(D)}_{\theta \theta} \rangle \propto  e^{(D-2)T},
\ee
corresponding exactly to what we found for $D=3, 4$ in the Garfinkle and Roberts spacetimes.

This shows that the apparent time dependence in higher-dimensional quantities is effectively induced by the time evolution of the areal radius encoded in the dilaton field, even though the two-dimensional effective stress-energy tensor remains stationary. The functional form of the quantum backreaction is thus deeply tied to the self-similar property of the geometry, leading us to conjecture that the $s$-wave quantum growing mode in general dimensions scales universally as $e^{(D-2)T}$, with the exponent determined solely by the spacetime dimensions.\footnote{In the Garfinkle and Roberts spacetimes we studied, the quantum $s$-wave mode consistently dominates the classical growing mode, exhibiting a larger Lyapunov exponent. This need not hold in generic self-similar critical collapse systems. Nevertheless, our conclusions regarding the threshold shift and the mass gap remain unaffected.} This is independent of the detailed matter content, provided that the matter fields are minimally coupled and the spacetime remains approximately self-similar.\footnote{For DSS critical spacetimes, which is generically the case for the Einstein-scalar system in $D \geq 4$, the stress-energy tensor may receive a periodic modulation with a bounded periodic function that should not enhance the net growth. This is further justified given that the existence of a Boulware-like state with its asymptotic Minkowskian property is not tied to self-similarity itself.}

A crucial observation is that higher-loop terms cannot acquire any additional enhancement by factors of $e^{\# T}$. At higher loops, one must still work in a Boulware-like state describing pure vacuum polarization, where the effective two-dimensional stress-energy tensor is stationary.\footnote{There is a caveat. This universal growing behavior is always there, but if $\langle T^{(2)}_{ab} \rangle$ does not admit a stationary choice of state, as we can see in the Appendix~\ref{sec:Hayward}, then in addition to self-similarity, global causal structure could acquire additional time-dependence.} As we just described, the growing mode emerges from the dilaton profile of the corresponding higher-dimensional stress-energy tensor, where the areal radius $r$ entering this relation is always the fixed background radius, even when higher loops are included. The quantum growing mode really has to do with the self-similarity property of the spacetime, rather than the loop counting effects.\footnote{This is valid as we are taking the critical spacetime as the fixed background for the computation of higher-loop effects; however, if we treat the semiclassical Einstein equation in the Hartree-Fock manner, i.e., by iteratively solving the backreaction order by order, then at the next order the new background would feature an apparent horizon and may therefore emit Hawking radiation. This could change the picture, since all the subtleties about the appropriate quantum state in a black hole geometry would then enter. Yet this should not be the main concern for the critical collapse problem, because such effects are still parametrically suppressed (and even exponentially suppressed due to the lack of growing mode associated with such a quantum flux), and mass scaling depends only on the earliest marginally outer-trapped surface. We thank Roberto Emparan for a discussion on this point.} We can then write schematically 
\be \label{eq:higherloops}
\langle T^{(D)}_{\mu \nu} \rangle= e^{(D-2)T} \bigg(\frac{\hbar}{\ell^{D-2}} F_{\mu \nu}(x)+ \frac{\hbar^2}{\ell^{2(D-2)}}\tilde{F}_{\mu \nu}(x)+\cdots \bigg),
\ee
with a universal growing factor $e^{(D-2)T}$ and all higher loop effects in the parentheses being pure functions of $x$ as they must stay in the Boulware-like state. An important implication is that, once this structure is established, it follows that higher loops remain parametrically suppressed compared to the one-loop term. This provides a strong justification for treating the quantum correction on equal footing with the classical contribution when the quantum mode grows to $O(1)$ and thereby probing the genuinely non-linear regime in a semiclassical framework---going beyond linear perturbations while still remaining under theoretical control, which is related to the potential Type I metastable phase that we turn to now.

\subsubsection*{Type I collapse?} 

Our semiclassical analysis suggests that quantum effects play an essential role in critical collapse and, when combined with the classical growing mode, lead to the emergence of a finite black hole mass gap. This behavior is reminiscent of Type I critical collapse observed in various classical systems as discussed in Section~\ref{sec:criticalcollapse}. In this context, \textit{universality} refers to the fact that the black hole mass near the quantum-modified threshold $p \to p^\ast_q$ (defined numerically by the vanishing of the ratio $R \to 0$) approaches a fixed, finite value that is independent of the initial data. Another qualitative similarity is that the black hole mass does not always decrease monotonically with the tuning parameter. While in the $2+1$-dimensional case it does (Figures~\ref{n=2Mass_Ratio_Qsuper-critical}, \ref{n=4Ratio_x_mass}, \ref{nRatio_x_Mass}), the transition in $3+1$ dimensions is more abrupt (Figure~\ref{RobertsEMOTS}). Interestingly, both behaviors are observed in classical Type I systems.

Even at the classical level, Type I behavior is expected when a dynamically relevant scale is introduced into the field equations. This was first demonstrated in the Einstein-Yang-Mills system~\cite{Choptuik:1996yg}, where the gauge coupling introduces a scale. Perhaps the simplest model exhibiting this behavior is a massive scalar field~\cite{Brady:1997fj}, where the scalar mass sets the relevant scale. In our case, quantum effects introduce a scale associated with $\hbar$, and it is a growing mode coming from vacuum polarization that acts analogously to a classical instability. A nontrivial feature, however, is that quantum effects are not inserted by hand; they arise naturally from the unique choice of a Boulware-like state for the collapsing matter, which is always present in every classical gravitational collapse system.

Nevertheless, despite the appearance of a mass gap, our setting differs in important ways from the traditional Type I collapse with classical matter fields. By definition, a Type I collapse involves a metastable soliton phase that precedes black hole formation, with a lifetime obeying Choptuik-type scaling \eqref{eq:lifetime}. This solitonic phase depends on the matter model. For example, boson stars in the case of a massive complex scalar field~\cite{Hawley:2000dt}, or Bartnik-McKinnon solitons in the Einstein-Yang-Mills system~\cite{Bartnik:1988am}. In contrast, it is unclear whether quantum effects can induce an analogous metastable phase with a similar scaling, or what its physical nature would be.

This difficulty is twofold. First, we are quantizing the matter field in the fixed critical background and treating the growing modes, both classical and quantum, as linear perturbations on such a background. Second, incorporating quantum effects into a fully dynamical framework suitable for numerical simulation poses delicate technical challenges. 

To properly study such a soliton-like phase and the associated phase transition would require going beyond linear perturbation theory, into a regime where classical and quantum effects induce $O(1)$ changes to the geometry. This is already true for classical perturbations: once the single growing mode reaches an amplitude of $O(1)$, exponential growth stops, and nonlinearity takes over. We typically define this point as the first appearance of a black hole horizon or the first curvature maximum in subcritical evolution:
\be
|p-p^\ast| e^{\omega_i T_{\text{non-lin}}} \sim O(1).
\ee
The critical exponent should thus be interpreted as determining the time scale at which perturbations cease to be linear. Beyond this point, the evolution departs from the self-similar regime. In this sense, the quantum correction is merely an additional growing mode. In the regime where the ratio $\mathcal{R} \to 0$, it can dominate and produce a mass gap.

Even so, linear perturbation theory remains meaningful if one perturbs around a finite-mass solitonic configuration rather than a scale-invariant one. This soliton breaks scale invariance and lacks self-similarity, with its mass related to the scale introduced. It is unstable in the sense that crossing a threshold in initial data leads to black hole formation. The black hole mass is then approximately the soliton mass.

To probe the non-linear regime, one must recognize an important difference from the purely classical case, where the Einstein equations always apply and a full numerical analysis can in principle be carried out. The subtlety with including quantum effects is that they are intrinsically perturbative in $\hbar$, raising the concern that higher-loop contributions might invalidate a non-linear treatment. However, as shown near \eqref{eq:higherloops}, higher loops are parametrically suppressed, which justifies treating the one-loop correction on the same footing as the classical terms. This opens the possibility of a genuinely non-perturbative, fully non-linear analysis that consistently incorporates quantum effects---a feasible and crucial next step.

This perspective suggests intriguing connections between our semiclassical critical and near-critical solutions and recently proposed horizonless geometries obtained by incorporating quantum effects, either perturbatively or non-perturbatively, and using both analytical and numerical approaches. These include various soliton-like configurations~\cite{Fabbri:2005zn, Fabbri:2005nt, Kawai:2013mda, Kawai:2015uya, Ho:2015fja, Ho:2015vga, Ho:2016acf, Ho:2019qpo, Ho:2017joh, Ho:2017vgi, Ho:2018fwq, Ho:2019pjr, Baccetti:2016lsb, Baccetti:2017ioi, Baccetti:2017oas, Barcelo:2019eba, Arrechea:2019jgx, Beltran-Palau:2022nec, Wu:2023uyb, Arrechea:2024cnv, Numajiri:2024qgh}, while the detailed mechanisms differ. Particularly noteworthy is that a Boulware-like state can support a horizonless geometry, suggesting the potential to extend our analysis beyond linear perturbation theory and fully capture the soliton-like phase. A complete understanding of the phase structure associated with quantum critical collapse remains an important open question.

\subsubsection*{Shift of the critical threshold and the mass gap} 

Having described the origin of the quantum growing modes from vacuum polarization, we now emphasize that such a structure should appear universally in any self-similar critical collapse system. The two key new results introduced by quantum effects are: (i) a shift of the critical threshold $\Delta p= p^\ast-p^\ast_q$ and (ii) the emergence of a universal mass gap $M_{\text{gap}}$ at $p=p^\ast_q$. Several natural questions arise regarding these quantities: What determines the sign of $\Delta p$ (i.e., whether the threshold is raised or lowered)? What scaling behavior do they follow? Are they intrinsic features of critical collapse, or are they special to the toy spacetimes studied here, depending sensitively on the perturbation profiles of a given background? Addressing these questions requires analytic insight beyond the numerical studies presented in Sections~\ref{sec:quantum2+1} and~\ref{sec:collapse3+1}.

The starting point is the quasi-CSS backreaction arising from both classical and quantum perturbations on the critical background. In geometrized units where we stick to $c=1$, $8 \pi G_N=1$, we can assume the areal radius always takes the form
\be
\bar{r}= \ell e^{-T} \bigg[r_0(x)+(p-p^\ast)e^{\omega_c T} r_c(x)+\frac{\hbar}{\ell^{D-2}} e^{\omega_q T} r_q(x) \bigg],
\ee
where we have argued that higher-loop corrections are parametrically small. There is a single growing mode for both the classical and quantum sectors, governed by $\omega_c$ and $\omega_q$, respectively. The quantum Lyapunov exponent $\omega_q = D-2$ is expected to be universal in the $s$-wave sector. 

The corresponding linearized horizon-tracing condition is determined by
\be
(\nabla \bar{r})^2 \approx f_0(x)+ (p-p^\ast)e^{\omega_c T}f_c(x)+\frac{\hbar}{\ell^{D-2}}e^{\omega_q T} f_q(x).
\ee
where, for a general CSS metric ansatz, the functions $f_i(x)$ are determined by the profiles $r_i(x)$ together with the metric. This formulation makes explicit that we are working within linear perturbation theory: perturbations in $(\nabla \bar{r})^2$ may grow to $O(1)$, while perturbations in $\bar{r}$ itself remain small.

The apparent horizon (AH) is defined by the condition $(\nabla \bar{r})^2=0$. The new critical threshold $p^\ast_q$ is the smallest value of $p$ for which there exists $(T, x)$ satisfying this equation. At $p=p^\ast_q$, the AH first appears at some $(T_\ast, x_\ast)$, corresponding to the earliest marginally outer-trapped surface. We define the quantities
\be
\Delta p \equiv p^\ast -p^\ast_q, \quad \epsilon \equiv \frac{\hbar}{\ell^{D-2}} \quad A \equiv e^{\omega_c T_\ast}, \quad B \equiv e^{\omega_q T_\ast}= A^\frac{\omega_q}{\omega_c}.
\ee
At $x=x_\ast$, two equations must hold:
\be \label{eq:firsteq}
(\nabla \bar{r})^2=0 \implies f_0-\Delta p A f_c+\epsilon B f_q=0
\ee
where the minus sign in front of $\Delta p$ appears because we are evaluating at $p=p^\ast_q$. Minimization with respect to $T$ (earliest AH condition) further requires
\be \label{eq:secondeq}
\partial_T (\nabla \bar{r})^2|_{T=T_\ast}=0 \implies -\Delta p \omega_c A f_c+\epsilon \omega_q B f_q=0.
\ee
Here all $f_i(x)$ are evaluated at $x=x_\ast$, so we omit writing this explicitly; they can be treated as numerical numbers at the horizon location.

\paragraph{Threshold shift.} For the threshold shift $\Delta p(x_\ast)$, we eliminate $\Delta p A$ using \eqref{eq:secondeq}. Substituting into \eqref{eq:firsteq} gives
\be
B=\frac{f_0}{\epsilon f_q (\frac{\omega_q}{\omega_c}-1)},
\ee
Since $B = A^{\omega_q/\omega_c}$, we obtain
\be \label{eq:thresholdshift}
\Delta p(x_\ast)=\frac{\omega_q}{\omega_c}\frac{f_q}{f_c}  \bigg[\frac{f_0}{f_q (\frac{\omega_q}{\omega_c}-1)} \bigg]^{1-\frac{\omega_c}{\omega_q}} \epsilon^{\frac{\omega_c}{\omega_q}},
\ee
The earliest AH corresponds to the location
\be
x_\ast= \text{arg} \text{min}_x B(x)=\text{arg} \text{min}_x \frac{f_0(x)}{f_q(x)}.
\ee
\eqref{eq:thresholdshift} shows that the magnitude of the threshold shift depends only on a specific combination of the perturbation profiles $f_i$ evaluated at $x_\ast$. We can further determine the sign of $\Delta p(x_\ast)$, and hence whether quantum effects lower or raise the threshold.

Without loss of generality, we may assume $f_0>0$, since the critical solution by itself cannot contain an apparent horizon. We can also assume $f_c(x_\ast)<0$: for classically supercritical data $p>p^\ast$, black holes must form in the regime of interest. (This does not mean $f_c$ is globally negative, only that it is negative at $x_\ast$.) There are two equivalent ways to determine the sign of the threshold shift:
\begin{itemize}
    \item Directly from the extremality condition. From $\partial_T (\nabla \bar{r})^2|_{T=T\ast}=0$ we obtain
    \be
    \Delta p= \epsilon \frac{\omega_q}{\omega_c} e^{(\omega_q-\omega_c)T_\ast}\frac{f_q}{f_c}.
    \ee
    This implies
    \be
    \text{sgn}(\Delta p)= \text{sgn}\bigg(\frac{f_q}{f_c} \bigg).
    \ee
    Since $f_c(x_\ast)<0$, this reduces to
    \be
    \Delta p>0 \text{ (threshold lowered)} \leftrightarrow f_q (x_\ast)<0,
    \ee
    \be
    \Delta p<0 \text{ (threshold raised)} \leftrightarrow f_q (x_\ast)>0.
    \ee
    The extremality condition holds regardless of the relative size of $\omega_c$ and $\omega_q$, but if $\omega_c>\omega_q$, then at late times we have classical domination instead of quantum domination.
    \item In terms of the ratio of classical to quantum modes. We defined
    \be
    \mathcal{R} \equiv \frac{(p-p^\ast_q) e^{\omega_c T}}{\epsilon e^{\omega_q T}}
    \ee
    in our numerical analysis to characterize the relative strengths of the classical and quantum modes. At $p=p^\ast_q$, we have $\mathcal{R}=0$ by definition. The AH condition can be written as
\be
\epsilon e^{\omega_q T} (\mathcal{R} f_c+f_q)=-f_0,
\ee
since the prefactor $\epsilon e^{\omega_q T}>0$ and we know $f_0>0$, then 
\be
\mathcal{R} f_c+f_q<0.
\ee
For some $x$, define the critical ratio
    \be
    \mathcal{R} >-\frac{f_q}{f_c} \equiv  \mathcal{R}_{\text{crit}}(x)
    \ee
    Then an AH exists iff $\mathcal{R} \geq \mathcal{R}_{\text{crit}}(x)$ for some $x$, which explains the horizon formation even with $\mathcal{R} \neq 0$. Taking the minimum, horizon formation occurs if
    \be
    \mathcal{R} \geq \mathcal{R}_\ast \equiv \text{min}_x \mathcal{R}_{\text{crit}}(x).
    \ee
    If some $x$ has $f_q(x)<0$, then $\mathcal{R}_{\text{crit}}(x)<0$ and hence $\mathcal{R}_\ast \leq 0$. At the shifted threshold $\mathcal{R}=0 \geq \mathcal{R}_\ast$, the inequality is already satisfied, so an AH forms and the threshold is lowered ($\Delta p>0$); If instead $f_q(x)\geq 0$ everywhere, then $\mathcal{R}_{\text{crit}}(x)\geq 0$ and $\mathcal{R}_\ast>0$. In this case, $\mathcal{R}=0$ does not suffice, and one needs $\mathcal{R}>0$, i.e., a larger $p$ than in the classical case. Thus the threshold is raised ($\Delta p<0$).

\end{itemize}

A physical picture is the following. With $f_c<0$, a negative $f_q$ enhances trapping: it drives $(\nabla \bar{r})^2$ further downward, focusing the outgoing null congruence and facilitating horizon formation. This corresponds to a lowered threshold. Heuristically, vacuum polarization contributes additional positive focusing energy to the relevant null component. By contrast, a positive $f_q$ produces defocusing, opposing trapping, which raises the threshold.\footnote{For the Roberts spacetime, the relevant horizon forms at a location where $f_c(x_\ast)>0$, in which case $f_q>0$ corresponds to a lowered threshold. This is indeed what we observed numerically in Section~\ref{sec:collapse3+1}. The general rule, however, remains $\mathrm{sgn}(\Delta p) = \mathrm{sgn}(f_q/f_c)$.}

The relative ordering of $\omega_c$ and $\omega_q$ does not affect the sign of $\Delta p$. It only controls the flow of $\mathcal{R}(T)$. If $\omega_q>\omega_c$ (the quantum mode grows faster), then $\mathcal{R}(T)$ decreases with $T$, driving the late-time dynamics toward the quantum-dominated branch $\mathcal{R}\to 0$. If some $x$ admits $f_q<0$, the global threshold is then attained on this $\mathcal{R}=0$ branch (lowered threshold). If $f_q\geq 0$ everywhere, the $\mathcal{R}=0$ branch is disfavored, and one must instead increase $p$ to reach $\mathcal{R}_\ast>0$, so the threshold is raised.

\paragraph{Mass gap.} The question regarding the $M_{\text{gap}}$ is trickier as the quasi-local Hawking mass would depend on the dimensionality of the spacetime and the presence of a cosmological constant $\Lambda$. Generically at the apparent horizon where $(\nabla \bar{r})^2=0$, we have \cite{Maeda:2006pm, Maeda:2007uu}
\be
M \propto \frac{-2 \Lambda}{(D-1)(D-2)} \bar{r}^{D-1}+\#\bar{r}^{D-3}+\# (D-3)(D-4) \bar{r}^{D-5},
\ee
which highlights $D=3,4$ being special cases. Note that we are working with the scale-free massless scalar field in the strict $\Lambda=0$ critical spacetimes. Hence, this formula does reproduce the quasi-local mass function we used for the $D=3,4$ cases, where for $D=3$ the role of $\Lambda$ is replaced by the arbitrary length scale $\ell^{-2}$.

At the shifted threshold $p=p^\ast_q$, the mass gap is evaluated at $(T_\ast, x_\ast)$, where
\be
\bar{r}_\ast=\ell e^{-T_\ast} \bigg(r_0-\Delta p A r_c+\epsilon B r_q \bigg).
\ee
Using
\be
e^{-T_\ast}= B^{\frac{-1}{\omega_q}}= \bigg[\frac{\epsilon f_q (\frac{\omega_q}{\omega_c}-1)}{f_0} \bigg]^{\frac{1}{\omega_q}}, \quad \Delta p A=\frac{\omega_q}{\omega_c} \epsilon B \frac{f_q}{f_c},
\ee
one finds after straightforward algebra
\bea
\bar{r}_\ast&=& \ell \epsilon^\frac{1}{\omega_q}\bigg[ \frac{f_q}{f_0}\bigg(\frac{\omega_q}{\omega_c}-1 \bigg) \bigg]^{\frac{1}{\omega_q}} \bigg[r_0+\frac{f_0}{\frac{\omega_q}{\omega_c}-1} \bigg(\frac{r_q}{f_q}-\frac{\omega_q}{\omega_c}\frac{r_c}{f_c} \bigg) \bigg]
\no\\
&=& \hbar^{\frac{1}{\omega_q}} \lambda^\frac{1}{\omega_q} S,
\eea
where all functions are evaluated at $x=x_\ast$, and we defined the following numbers
\be
\lambda \equiv \frac{f_q}{f_0}\bigg(\frac{\omega_q}{\omega_c}-1 \bigg), \quad S \equiv \bigg[r_0+\frac{f_0}{\frac{\omega_q}{\omega_c}-1} \bigg(\frac{r_q}{f_q}-\frac{\omega_q}{\omega_c}\frac{r_c}{f_c} \bigg) \bigg],
\ee
and used the fact that $\epsilon=\frac{\hbar}{\ell^{D-2}}$ and $\omega_q=D-2$ for the $\ell$-dependent part. We see clearly that the areal radius associated with the mass gap is manifestly independent of $\ell$. 

\paragraph{Discussion on the scales.} This $\ell$-independence reflects the underlying scale invariance of a massless scalar in a CSS background: no intrinsic length scale is present. Quantum effects, however, do introduce a scale via $\hbar$. The result
\be
\bar{r}_\ast = \# \hbar^{\frac{1}{\omega_q}}
\ee
shows that the mass gap is a genuinely quantum effect. If the coefficients given by $\lambda$ and $S$ happen to be small, the gap may lie near the Planck scale, corresponding to a microscopic black hole where semiclassical analysis itself becomes questionable. In the toy models of Garfinkle and Roberts spacetimes, our numerical evaluations of $\lambda$ and $S$ (see, for examples, Figures~\ref{n=2Mass_Ratio_Qsuper-critical}, \ref{RobertsEMOTS}, \ref{n=4Ratio_x_mass}, \ref{nRatio_x_Mass}) indeed yield very small values, suggesting that the resulting black holes are microscopic and challenge the validity of the semiclassical approximation.

However, it is important to emphasize that this $\ell$-independence and the resulting Planckian scaling are artifacts of the idealized, scale-free toy models we are studying. In the classical backgrounds, the curvature near horizon formation is already trans-Planckian, so the apparent Planck-scale mass gap simply reflects the absence of any other dimensionful parameter.

In realistic gravitational collapse---for example, in primordial black hole formation during radiation domination\footnote{This situation is well modeled by perfect fluid matter~\cite{Evans:1994pj, Maison:1995cc, Neilsen:1998qc}, which in four dimensions behaves like conformal matter. In this setting, the dimensional reduction and anomaly-based methods we adopt should apply, agnostic of the detailed matter Lagrangian. Moreover, the classical critical solution is robustly CSS and exhibits Type II collapse, precisely the category analyzed in this work.}---one must couple the critical solution to an asymptotically FRW background and compare the quasi-local mass to the horizon mass set by the Hubble parameter $H(t)$ \cite{Niemeyer:1997mt}. This coupling introduces an additional length scale, associated with the cosmological curvature, which regulates the system. In that context, the quasi-local mass remains the relevant notion (rather than the ADM mass), and the parameters $\lambda$ and $S$ effectively inherit a dependence on the local curvature scale at which collapse occurs.

In other words, within the semiclassical framework, one cannot tune the artificial scale $\ell$ of the toy models to control the size of the gap. Instead, the physically meaningful scale enters through $\lambda$ and $S$, which in the strict scale-free models are just pure numbers, but in realistic collapse acquire scale-dependence through the asymptotic FRW boundary conditions. This coupling to FRW precisely breaks the scale-free nature of the toy models and justifies treating the semiclassical gap consistently in a separation-of-scales regime (inner CSS region matched to outer FRW region).

For example, in scenarios involving primordial black holes \cite{Niemeyer:1997mt, Carr:2025kdk}, during radiation domination in four dimensions, the energy density and the Hubble horizon mass are (note again we are taking $8 \pi G_N=1$)
\be
\rho=3 H^2, \quad M_{\text{H}}(t)=\frac{4 \pi}{3}\rho H^{-3}=\frac{4 \pi }{ H}.
\ee
The horizon mass sets the natural scale for gravitational collapse at horizon entry, as it characterizes the maximum mass the gravitational collapse from a density fluctuation can assemble within one Hubble volume. It therefore provides a benchmark against which the mass of any forming black hole is compared.

At the shifted threshold $p=p^\ast_q$, the mass gap is $M_{\text{gap}}=\bar{r}_\ast/2 $. Using our previous expression for $\bar{r}_\ast$, we obtain
\be
\frac{M_{\text{gap}}}{ M_{\text{H}}}=\sqrt{\hbar} \sqrt{\lambda} S H(t),
\ee
Thus at each epoch the gap is always a fixed fraction of the horizon mass. Indeed, if $\lambda$ and $S$ are numerically small, this simply reflects the existence of a Planck-scale lower cutoff for critical collapse, which is negligible on cosmological scales.

However, consistency with semiclassical gravity requires that the curvature at the apparent horizon be sub-Planckian:
\be \label{eq:curvaturescale}
R_{\text{AH}} \leq R_{\text{max}} \ll \hbar^{-1}.
\ee
We also know that the following universal relation must hold,
\be
M^2_{\text{gap}} R_{\text{AH}}=K,
\ee
where $K \propto G^{-2}_N$ in general units. In our convention $8 \pi G_N=1$, this simply reduces to a dimensionless constant of order unity, that depends only on the overlap structure of the classical and quantum perturbations at $x_\ast$. It is independent of $\ell, \hbar, \omega_c,$ or $\omega_q$. This follows by noting that the Ricci scalar at the AH can be written as
\be
R_{\text{AH}}=\frac{e^{2 T_\ast}}{\ell^2} \tilde{R}(T_\ast, x_\ast),
\ee
where $\tilde{R}(T_\ast,x_\ast)\equiv C_R$ is a pure number. Substituting the earlier scaling relations gives
\be
R_{\text{AH}}=C_R \lambda^{\frac{-2}{\omega_q}} \hbar^{\frac{-2}{\omega_q}} \ell^{\frac{4}{\omega_q}-2},
\ee
and hence 
\be
M^2_{\text{gap}} R_{\text{AH}}=\frac{1}{4} S^2 C_R  \equiv K.
\ee
It implies that if $R_{\text{AH}}$ is sub-Planckian, then $M_{\text{gap}}$ must be super-Planckian, ensuring that the system remains within the semiclassical regime. To enforce the curvature bound we require 
\be
\lambda \geq \hbar^{-1} \bigg(\frac{C_R}{R_{\text{max}}} \bigg)^{\frac{\omega_q}{2}}\ell^{2- \omega_q}.
\ee
As a consequence, the mass gap has a floor value
\be
M_{\text{gap}}=\sqrt{\frac{K}{R_{\text{max}}}}=\frac{1}{2}S \sqrt{\frac{C_R}{R_{\text{max}}}}.
\ee
Remarkably, all explicit dependence on $\ell, \hbar, \omega_c,$ and $\omega_q$ cancels in this lower bound. The perturbation profiles are hidden in $S$ and $C_R$, but they are just geometric numbers determined by the appropriate magnitudes set by the perturbations, and are independent of the choice of initial data family. This conclusion holds even without knowing the detailed nature of the metastable soliton phase.

For the threshold shift, we can rewrite it in terms of $\eta\equiv \omega_c/\omega_q$ as
\be
\Delta p=\frac{1}{1-\eta} \frac{f_0}{f_c} \lambda^\eta \epsilon^\eta,
\ee
Imposing the curvature bound from the previous section,
\be
\Delta p \geq \frac{1}{1-\eta} \frac{f_0}{f_c} \bigg(\frac{C_R}{R_{\text{max}}} \bigg)^{\frac{\omega_c}{2}} \ell^{-\omega_c}.
\ee
This shows that the threshold shift is bounded away from zero once the curvature is constrained to be sub-Planckian.

For example, to connect with cosmology, we adopt a separation-of-scales ansatz by tying the local AH curvature to the background FRW curvature
\be
R_{\text{AH}}= \kappa_R H^2,
\ee
where $\kappa_R$ parametrizes how many Hubble curvatures reside at the AH. This prescription explicitly breaks scale-freeness and couples the CSS patch to FRW. From the universal relation
\be
M_{\text{gap}}^2 R_{\text{AH}}=\frac{1}{4}S^2 C_R =K, \quad M_{\text{H}} = \frac{4 \pi}{ H},
\ee
we find
\be
M_{\text{gap}}=\sqrt{\frac{K}{R_{\text{AH}}}}=\sqrt{\frac{K}{\kappa_R}}\frac{1}{H} \implies \frac{M_{\text{gap}}}{M_{\text{H}}}=\frac{1}{4 \pi} \sqrt{\frac{K}{\kappa_R}}.
\ee
Thus the gap is a constant fraction of the horizon mass, independent of epoch and independent of $\omega_c,\omega_q$ (though $M_{\text{H}}$ itself evolves with time). The universality class of the matter (e.g., radiation fluid) and the appropriate scales of the perturbations modes then fix $K$, while $\kappa_R$ encodes the separation of scales.

For the threshold shift, the condition $R_{\text{AH}}=\kappa_R H^2$ fixes $\lambda$ to be
\be
\lambda= \bigg(\frac{C_R}{\kappa_R \xi^2} \bigg)^{\frac{\omega_q}{2}}\ell^2,
\ee
where we have parametrized $\ell=\xi H^{-1}$. Substituting back
\be
\Delta p= \frac{1}{1-\eta} \frac{f_0}{f_c} \bigg(\frac{C_R}{\kappa_R} \bigg)^{\frac{\omega_c}{2}} \xi^{-\omega_c},
\ee
again an epoch-independent universal shift.

A useful remark that justifies the above discussion is that the normalization of the growing modes is a gauge choice: one can always rescale the family parameter $p$ (or the quantum amplitude) and simultaneously rescale the mode functions so that $|f_c(x_\ast)|=|f_q(x_\ast)|=1$ at the earliest AH. What enters the horizon condition are only invariant combinations such as $(p-p^\ast) e^{\omega_c T} f_c$ and $\epsilon e^{\omega_q T} f_q$, not the detailed shape or normalization of the mode profiles themselves. Thus the mode shapes carry no intrinsic significance: the magnitudes of the physical quantities like the threshold shift and the mass gap depend only on these invariant products, not on how $O(1)$ factors are distributed between “amplitudes” and “mode shapes.” This makes explicit why universality survives when coupling the CSS region to an FRW background, as it is not sensitive to the detailed shape of the initial fluctuations (e.g., inflationary models), but only to the universality class of the critical solution.

The epoch-independent universal threshold shift and the finite mass gap may have direct observational implications for primordial black hole formation, where the control parameter is the density contrast $\delta \equiv \frac{\rho-\bar{\rho}}{\bar{\rho}}$ relative to the mean energy density $\bar{\rho}$, with a critical threshold $\delta^\ast$. A shift of this threshold would feed exponentially into the black hole formation fraction, while the mass gap truncates the low-mass tail, potentially alleviating fine-tuning issues and easing tensions with observational constraints~\cite{shortpaper}.

\subsubsection*{Quantum effects from higher angular momentum modes}

In this work, we have focused exclusively on the quantum $s$-wave sector, which yields a single growing mode. This simplification neglects the infinite tower of IR non-spherical modes that arise from the spherical harmonic decomposition. As we discussed in Section~\ref{sec:criticalcollapse} from the classical perspective, it is well established, both analytically and numerically, that all non-spherical perturbations decay, and the only known growing mode originates from the $s$-wave. This aligns with intuition, particularly in spherically symmetric critical collapse, and supports the expectation that the dominant quantum effects should emerge from the $s$-wave sector.\footnote{It is also true in black hole evaporation, where the $s$-wave mode dominates the Hawking radiation observed at infinity.}

However, there are a few caveats when considering higher angular modes in the quantum setting. For clarity, it is useful to separate two issues: (i) in what ways higher-$l$ sectors could in principle modify the conclusions, and (ii) why we expect them to remain subleading within the restricted setup adopted in this paper.

Regarding (i), at the level of QFT on a fixed background, higher-$l$ sectors can in principle modify the conclusions in several ways: the late-time IR content for each sector may depend sensitively and delicately on the choice of quantum state and on boundary/regularity conditions; an infinite tower of modes---each requiring nontrivial renormalization and a careful specification of the state---might accumulate through a mode sum (or, if feasible, a controlled resummation); and nonlinear mode coupling can generate angular structure beyond linear order. Moreover, as already seen in the $s$-wave sector, quantum Lyapunov exponents need not coincide with their classical counterparts. These are all legitimate concerns in QFT on a fixed background. Accordingly, we will be explicit below about the scope of our conclusions and the assumptions under which higher-$l$ effects are expected to remain subleading.

However, for (ii), by contrast, once gravity is included, the situation is much more constrained. As discussed in Section~\ref{sec:criticalcollapse}, with appropriate regularity and boundary conditions in the gauge-invariant linear perturbation analysis including the constraint equations (which restrict admissible initial data), all higher-$l$ modes \emph{decay} due to the effective potential barrier in a self-similar, spherically symmetric background \cite{Martin-Garcia:1998zqj, Gundlach:2025yje}. This is a \emph{kinematical} statement at the level of the Einstein equations, independent of whether the matter source is treated classically or semiclassically. Consequently, imposing the same regularity and boundary conditions implies that no choice of Hadamard quantum state can manufacture an IR growing mode absent from the classical spectrum.\footnote{A subtle point is that higher-$l$ growing modes can appear at second order in perturbation theory~\cite{Garfinkle:1998tt}. Yet any backreaction sensitive to angular structure therefore emerges only at a higher order and is not a primary concern within linear perturbation.} Within our two-dimensional reduced framework we further work with a Boulware-like state that is stationary and spherically symmetric, with vanishing asymptotic quantum flux; equivalently, the vacuum is annihilated by all field modes at past infinity so that each mode begins in its ground state. Taken together with compatibility with the constraint equations, these considerations justify that our state corresponds to admissible initial data in the sense relevant for the near-critical evolution.

We therefore adopt the \emph{working assumption} that, under the same regularity and boundary conditions as in the linear perturbation analysis, the dynamically relevant semiclassical content near criticality is captured by the $s$-wave IR growing mode together with the universal UV structure. As discussed in Section~\ref{sec:onelooptheory}, the latter is captured by performing the dimensional reduction carefully, with the trace anomaly given in~\eqref{eq:traceanomaly}. At the same time, we emphasize that a complete treatment of higher angular modes---whether arising from different boundary/regularity choices, from cumulative mode effects, or through higher-order perturbation theory---remains an important open problem. Rather than attempting to tackle these issues in the critical spacetime directly, we close by highlighting a few promising avenues:
\begin{itemize} 
    \item \textbf{Choosing symmetry-respecting quantum states and analyzing low-$l$ or a finite number of angular modes.}
    
    Most existing studies on quantum properties in critical spacetimes have not incorporated higher angular modes. A notable exception is a recent line of work~\cite{Berczi:2020nqy, Guenther:2020kro, Berczi:2021hdh, Hoelbling:2021axl, Varnhorst:2023dew}, which adopts a coherent quantum state that respects spherical symmetry while allowing for nontrivial mode sums with appropriate regularizations. The simulations incorporate a substantial number $(N_l=400)$ of higher-$l$ modes. But they do not infer any kind of self-similarity, making it difficult to determine whether the higher modes grow or decay,

    The resulting deviation in black hole mass scaling due to quantum effects from summing over modes was found to be small. This is perhaps unsurprising, as the quantum sources are treated as part of the initial data that can be fine-tuned, and a coherent state closely resembles the classical limit with minimum uncertainty. Nonetheless, the conclusion is counterintuitive: near criticality, the semiclassical black hole mass is consistently smaller than its classical counterpart, suggesting that quantum effects act effectively as a form of dissipation. However, the physical origin of this dissipative nature remains unclear. It does not naturally align with the standard interpretation of Hawking radiation; whereas one approaches the critical point, an intuitive expectation is that quantum deviations should increase rather than decrease, given that the black hole is becoming smaller and smaller.  
    
    The use of a coherent state allows the Einstein equations to be cleanly separated into classical background and quantum fluctuations, avoiding mixed terms. However, this state is imposed rather than derived from a fundamental quantum formulation. Nevertheless, the findings offer empirical support for the notion that higher angular modes, even if not decaying, are likely subdominant compared to the dominant $s$-wave. These results suggest that focusing on well-behaved quantum states that respect the underlying symmetry, along with analyzing a finite number of low-$l$ modes, remains a promising and tractable strategy.

    \item \textbf{Numerical evaluation via advanced mode-sum techniques.}

    Significant recent progress has been made in numerically computing quantum expectation values via refined mode-sum techniques. Two particularly powerful frameworks are worth highlighting:

    The \textit{extended coordinate method} developed initially by Breen, Ottewill, and Taylor expands the Hadamard form of the two-point function to high orders using cleverly chosen coordinates~\cite{Taylor:2016edd, Taylor:2017sux, Breen:2018ukd, Taylor:2022sly, Arrechea:2024cnv, Breen:2024ggu}. This facilitates singularity subtraction required for renormalization without relying on WKB approximations, though it still uses Euclidean methods. Their recent work~\cite{Arrechea:2024cnv} includes the first direct computation of the renormalized stress-energy tensor in the Boulware state, bypassing inference from the Hartle–Hawking state, marking a significant advancement in methodology.

    The \textit{pragmatic mode-sum prescription} pioneered by Levi and Ori allows for renormalization based on a single symmetry of the background spacetime~\cite{Levi:2015eea, Levi:2016esr, Levi:2016quh, Levi:2016exv, Levi:2016paz}. Different splitting strategies have been developed, including $t$-splitting (stationary spacetimes), angular splitting (spherically symmetric), and azimuthal splitting (axi-symmetric). Remarkably, this method has even been applied successfully to the case of an evaporating Kerr black hole~\cite{Levi:2016paz}. Unlike WKB-based approaches, it can be implemented directly in a Lorentzian setting and is versatile across different quantum states. However, having at least one exact symmetry remains a necessary (though not sufficient) condition for its application.
    
    These methods open the door to reliable, high-precision computations of quantum effects involving higher angular modes. While they have already proven powerful in black hole spacetimes, their applicability to dynamical critical spacetimes and near-critical regimes remains largely unexplored, presenting an exciting frontier.

    \item \textbf{Trace anomaly for non-conformal theories.}

    Much less is known about the fate of trace anomalies in non-conformal theories, with a recent and intriguing development attempting to address this gap~\cite{Casarin:2018odz, Ferrero:2023unz}. In~\cite{Ferrero:2023unz}, the authors propose a new scalar quantity that is scheme-independent and consistent with the conservation of the stress-energy tensor, even in the absence of conformal symmetry. While in general the trace of the quantum stress-energy tensor in non-conformal settings is expected to be state-dependent, the proposed scalar may offer a way to extract universal information from the full quantum theory, potentially from contributions involving all the modes. It could serve as a powerful diagnostic tool for understanding semiclassical effects in critical collapse and beyond.
\end{itemize}

\section{Outlook}
\label{sec:outlook}

In addition to the intriguing questions most relevant to our studies discussed in Section~\ref{sec:discussion}, several directions merit exploration in future work:
\begin{itemize}
    \item \textbf{Semiclassical analysis of exterior-naked singularity regions.}
    
    We have so far focused on the interior fill-in regions, bounded by the self-similarity horizon (the past light cone of the naked singularity, see Figure~\ref{fig:global}). Only within the region is the geometry self-similar. But it is equally important to investigate the semiclassical dynamics in the exterior that yields the nakedness property for the asymptotic observers. By contrast, the exterior ceases to be self-similar and must instead be asymptotically flat (or have other appropriate asymptotics), making the problem fundamentally different. The techniques we employed based on trace anomaly are expected to remain applicable. However, special care must be taken in selecting a physically meaningful quantum state and in imposing suitable junction conditions to ensure global consistency.\footnote{In particular, the unique Boulware-like state, which carries no quantum flux at infinity, is valid within the interior region (so the exterior future infinity should not receive flux sourced from the interior). Any quantum flux, if present, must arise purely from the exterior geometry.}

    Completing the semiclassical picture in the exterior is particularly crucial, since only then can we determine the global apparent horizon and the teleological event horizon. This is not merely a technical refinement but has far-reaching implications. In the interior, vacuum polarization generates a universal quantum growing mode that enforces horizon formation and eliminates naked singularities. If a similar consequence extends to the exterior, then the fate of the Choptuik singularity is directly linked to that of the endpoint of Hawking radiation. In this case, quantum effects push the Cauchy horizon inside the black hole horizon, so that from the outside the geometry resembles an ordinary evaporating black hole. The violation of determinism is then reduced to the same “mild” level already present in the standard evaporation problem, without introducing any new form of information loss paradox \cite{Hawking:1976}.\footnote{We thank Roberto Emparan for emphasizing this point to us.}

    We believe this is a feasible problem at least for toy spacetimes, where a promising starting point is to analyze analytically tractable exterior geometries, such as the Garfinkle-Vaidya and Roberts-Vaidya spacetimes in the exterior~\cite{Jalmuzna:2015hoa, Wang:1996xh}, briefly discussed in Sections~\ref{sec:classicalGarfinkle} and~\ref{sec:classicalRoberts}. 

    \item \textbf{Cosmic censorship and energy condition violations.}
    
    A key insight of our analysis is that quantum effects in the interior fill-in region can dynamically generate an apparent horizon that shields the singularity, effectively enforcing cosmic censorship through quantum backreaction.\footnote{The emergence of a quantum-induced horizon was also argued in the braneworld setup in AdS/CFT \cite{Frassino:2025buh}.} This is not a demonstration of cosmic censorship within the classical theory, but rather one that arises by introducing quantum corrections, which generically violate classical energy conditions. Such violations are essential in allowing new dynamical behaviors, such as horizon formation, that are otherwise forbidden in the classical setting.

    A related question is: all self-similar solutions with naked singularities feature a Cauchy horizon, whose classical stability depends on specific energy conditions being satisfied~\cite{Nolan:2002hr}. This raises a natural question: is the Cauchy horizon stable under quantum perturbations? It is known that Cauchy horizons within black holes are generically unstable due to quantum effects, and in a very universal, state-independent way \cite{Hollands:2019whz, Emparan:2020rnp, Kolanowski:2023hvh, Shahbazi-Moghaddam:2024emr}. Likewise, Cauchy horizons as formed by closed null curves similarly become unstable due to the same quantum effects \cite{Emparan:2021xdy, Emparan:2021yon, Tomasevic:2023ojy}. 
    
    However, Cauchy horizons in critical spacetimes can be fundamentally different from previous cases since critical spacetimes do not have an exact timelike Killing vector. This question is related to the completion of the exterior picture discussed above, where quantum backreaction may push the Cauchy horizon inside the black hole horizon. In that case, its instability would manifest not as a catastrophic breakdown of predictability, but as the more benign cloaking of the Choptuik singularity, tying its fate to the familiar endpoint of black hole evaporation.

    Answering these questions is crucial for understanding the role of quantum effects in determining the spacetime global structure and predictability of semiclassical gravity.

    \item \textbf{Phase structure in the semiclassical regime.}

    Can we construct a meaningful “phase diagram” for critical collapse in the presence of quantum effects?

    This question becomes subtle in the semiclassical regime. Unlike classical gravity, where the Einstein equations define a well-posed system, one cannot simply treat the one-loop corrections to the equations of motion as defining a new self-contained system on equal footing with the classical terms. The quantum corrections are inherently perturbative, and care must be taken in interpreting their effects. In particular, the perturbative nature of the analysis limits our ability to explore possible non-perturbative phases, such as the emergence of a finite-mass solitonic star, as we briefly alluded to. However, as argued in Section~\ref{sec:discussion}, higher-loop contributions are parametrically small, making this a feasible question. Understanding the boundaries and transitions between qualitatively distinct regimes, including dispersal, black hole formation, and potential new semiclassical phases, remains a challenging open problem.

    \item \textbf{Generalizations to other critical collapse scenarios.}

    We believe the general lessons extracted from our study---such as the universal quantum growing mode and the selection of a Boulware-like state---apply to any self-similar critical collapse system. Several promising directions remain for extending the present analysis. One avenue is a more systematic exploration of CSS-type solutions with a broader class of matter fields, including those with nontrivial potentials, as attempted in~\cite{Moitra:2022umq}. This could again be approached using anomaly-based methods. Note that different matter contents can affect the trace anomaly given in \eqref{eq:traceanomaly}~\cite{vanNieuwenhuizen:1999nu, Nojiri:1999br}.

    Second, while we have focused on CSS-type spacetimes in $2+1$ and $3+1$ dimensions (i.e., the Garfinkle and Roberts solutions), many other exact solutions conjectured to model critical collapse exist in spherical symmetry~\cite{Brady:1994aq, Hayward:2000ds, Hirschmann:2002bw, Clement:2001ns, Clement:2001ak, Baier:2013gsa, Clement:2014pua, Clement:2014rda}. Many of these possess unusual global structures, and modeling their semiclassical behaviors would be within reach with the techniques given here.

    Lastly, we have conjectured that quantum $s$-wave effects from a free massless scalar field lead to a universal scaling of the form $e^{(D-2)T}$, depending only on the spacetime dimension $D$. This observation suggests a natural generalization of our analysis to higher-dimensional critical collapse scenarios of the Einstein-scalar system, which, intriguingly, are known to exhibit DSS. We argued in Section~\ref{sec:discussion} that this growth factor can at most acquire a bounded periodic modulation; however, confirming this will likely require dedicated numerical simulations, given the difficulty of treating DSS geometries analytically. On the other hand, a particularly interesting direction is to explore what happens to this mode in the large-$D$ limit \cite{Emparan:2013moa}. The large-$D$ expansion provides a natural separation of scales, tremendously simplifying the equations of motion where different effects dominate at each scale. See, for instance, the recent study of spherically symmetric CSS gravitational collapse of the Einstein-scalar system in the infinite-dimensional limit~\cite{Clark:2025tqi}.

    \item \textbf{Connection to realistic models of gravitational collapse.}
    
    To assess the relevance of our results for physical gravitational collapse, it is important to confront the limitations of using a scalar field as the matter content, see~\cite{Joshi:2008zz, Malafarina:2024qdz} for more details and recent developments. A natural question is whether quantum corrections of the kind considered here can be meaningfully extended to more realistic collapse scenarios.

    One must also go beyond spherical symmetry and CSS~\cite{Gundlach:2025yje}. Recent progress has extended Christodoulou’s framework, demonstrating regularity on the past light cone of the singularity and generalizing to solutions with asymptotically CSS profiles. Notably, naked singularities have been constructed even within the Einstein vacuum equations outside spherical symmetry, where self-similarity is generalized to a \textit{twisted self-similarity}. There has also been progress in constructing exterior regions with DSS profiles that remain smooth on the past light cone, a crucial feature for matching with interior fill-in regions, though in general, it remains an open problem. See~\cite{Rodnianski:2019ylb, Shlapentokh-Rothman:2022byc, Cicortas:2024hpk, Shlapentokh-Rothman:2022uji, Singh:2022uyx, Singh:2024gyx}. Understanding the semiclassical properties of such geometries, including the role of quantum energy conditions and backreaction, presents a rich and largely unexplored area.

    \item \textbf{Pre-Hawking radiation during gravitational collapse.} 
    
    Hawking radiation is typically considered irrelevant to critical collapse, which concerns the apparent horizon when it first forms. Evaporation is expected to kick in only after a trapped region develops. Based on this understanding, we have argued that the natural quantum state in this context is Boulware-like, capturing only the vacuum polarization of the collapsing matter.

    However, this picture may be incomplete. Investigations into the minimal conditions for the existence of Hawking-like radiation led to the surprising results that such radiation does not require a trapped region~\cite{Barcelo:2006uw, Barcelo:2010pj, Barcelo:2010xk}. This so-called \textit{pre-Hawking radiation} has been studied in several effective models that describe Hawking evaporation during gravitational collapse, potentially resulting in horizon-less configurations~\cite{Kawai:2013mda, Kawai:2015uya, Ho:2015fja, Ho:2015vga, Ho:2016acf, Ho:2019qpo, Baccetti:2016lsb, Baccetti:2017ioi, Baccetti:2017oas, Barcelo:2019eba}. However, see~\cite{Chen:2017pkl, Unruh:2018jlu, Arderucio-Costa:2017etb} for a different perspective. Other approaches, including non-perturbative treatments and reinterpretations of the Boulware state, have produced similar configurations~\cite{Fabbri:2005zn, Fabbri:2005nt, Ho:2017joh, Ho:2017vgi, Ho:2018fwq, Ho:2019pjr, Barcelo:2019eba, Arrechea:2019jgx, Beltran-Palau:2022nec, Wu:2023uyb, Arrechea:2024cnv, Numajiri:2024qgh}.

    In addition to the possibility that these configurations may represent the soliton-like phase of the quantum-modified Type I behavior we identified earlier, several important questions arise. Can one select a suitable quantum state for such scenarios and incorporate it into our effective formalism to investigate the role of Hawking-like flux in critical collapse? In particular, existing realizations of pre-Hawking radiation have largely been restricted to thin- or thick-shell collapse, making its implementation in the context of critical collapse especially nontrivial. Furthermore, in these semiclassical models of gravitational collapse that do not lead to black hole formation but instead yield horizonless geometries, is there a notion of critical phenomena associated with these endpoints?

\newpage

    \item \textbf{Primordial black holes.}

    One interesting and compelling direction is to connect critical collapse to observable phenomena. Remarkably, critical collapse is not merely a theoretical construct but has direct relevance to our Universe. In particular, it provides a viable mechanism for the formation of primordial black holes (PBHs) during the radiation-dominated era of the early universe. This possibility was first proposed in~\cite{Niemeyer:1997mt} and has since been further developed~\cite{Yokoyama:1998xd, Niemeyer:1999ak, Jedamzik:1999am, Green:1999xm,  IHawke_2002, Musco:2004ak, Polnarev:2006aa, Musco:2008hv, Kuhnel:2015vtw}.
    
    In such scenarios, overdensities in the early universe can undergo near-critical collapse, leading to the production of PBHs with masses sensitive to the proximity to the critical threshold. If one tracks the evolution of the system sufficiently close to criticality, we expect similar quantum effects we described in this work to kick in. The resulting shift of the critical threshold and the universal mass gap could, in principle, alter the mass spectrum and abundance of PBHs, making this an intriguing arena for exploring the observable consequences of quantum gravitational phenomena. See discussions in Section~\ref{sec:discussion} and our companion Letter \cite{shortpaper}.

    \item \textbf{Holographic stress-energy tensor and non-perturbative backreaction.}

    The trace anomaly method we employed for computing quantum backreaction is robust and analytically tractable. However, it remains intrinsically perturbative and may not capture the full quantum dynamics near criticality. To go beyond this, it is natural to consider holographic methods inspired by the AdS/CFT correspondence, which offers a non-perturbative definition of the quantum stress-energy tensor. Although AdS/CFT is traditionally formulated with a negative cosmological constant $\Lambda<0$ and asymptotically AdS spacetimes, one can treat the appearance of $\Lambda$ as perturbative and quasi-self-similar solutions to the system, as we have seen in Sections~\ref{sec:criticalcollapse} and \ref{sec:quantum2+1}.

    A promising setup involves brane-world models in which a black hole is localized on a brane embedded in a higher-dimensional AdS spacetime~\cite{Emparan:1999wa, Emparan:1999fd, Karch:2000ct, Karch:2000gx, deHaro:2000wj, Emparan:2002px, Emparan:2020znc, Cisterna:2023qhh, Arenas-Henriquez:2023hur, Tian:2024mew}. The brane plays the role of a lower-dimensional universe, and its induced metric satisfies a modified Einstein equation sourced by a holographic stress-energy tensor:
    \be
    G_{ij}+ \cdots= 8 \pi G_d \langle T_{ij} \rangle.
    \ee
    Here, $\langle T_{ij} \rangle$ is the renormalized stress-energy tensor of a strongly coupled large-$N$ CFT, and the ellipsis denotes higher-curvature corrections from integrating out UV degrees of freedom above the cutoff scale. The brane is situated at a finite distance in the AdS bulk, providing a natural IR regulator for the dual field theory.

    In this way, one uses a classical background in AdS$_{d+1}$ bulk to holographically describe $d$-dimensional gravity coupled to a large $N$ CFT. While limited to leading order in $1/N$, it sidesteps the technical challenges of direct loop computations in curved spacetimes. For example, when the bulk geometry is given by the  AdS$_4$ C-metric, these techniques have been used to model quantum-corrected BTZ black holes in AdS$_3$, where one can compare holographic results with those obtained from free conformal scalars. Although the agreement is not perfect due to differences in field content and interactions, the holographic stress-energy tensor generally yields simpler, resummed expressions that remain valid beyond the linearized regime. Moreover, recent work has emphasized that holographic stress-energy tensors, while initially traceless for conformal theories, can acquire trace anomalies when higher-derivative corrections are included in the bulk action. These corrections reflect the fact that the dual theory has an effective UV cutoff, breaking exact conformal invariance. This provides a richer structure for modeling semiclassical backreaction.

    We still need to understand how to realize this kind of holographic setup in the context of critical collapse, and how to properly interpret the holographic stress-energy tensor $\langle T_{ij} \rangle$, given that it arises from a collection of strongly coupled conformal fields rather than from the direct backreaction of collapsing matter. Nevertheless, recent studies such as~\cite{Moitra:2022umq} have shown that self-similarity can be compatible with AdS-like spacetimes that possess timelike boundaries without constant curvature.

One system where one can explore the brane-world critical collapse is the Garfinkle-Vaidya geometry discussed in Section~\ref{sec:classicalGarfinkle}, which naturally incorporates $\Lambda$-corrections and AdS boundary conditions in the exterior region. This provides a concrete setting where holographic methods could be applied to explore non-perturbative quantum effects in near-critical spacetimes.

\item \textbf{Holographic critical collapse in the large-$D$ framework.}

A promising yet underexplored direction is the study of holographic gravitational collapse and threshold behavior in the large-$D$ limit~\cite{Licht:2022rke, Emparan:2023dxm}, where the bulk gravity simplifies considerably and becomes governed by certain effective equations. In this regime, the near-horizon dynamics decouple from the asymptotic region, enabling tractable computations of complex gravitational processes such as black hole formation. This approach has recently been applied to black droplet configurations in AdS braneworlds, where one can model the dynamical collapse of a CFT cloud by launching a Gaussian blob of energy toward a brane. The resulting solution, a bulk black hole that sticks to the brane, describes a black droplet localized near the brane, surrounded by a CFT halo.

A natural question arises: can the large-$D$ holographic framework exhibit critical phenomena analogous to those found in standard gravitational collapse? While the Gaussian blob setup resembles the initial scalar profiles used in classical critical collapse studies, it does not seem to reveal a universal critical behavior. This may not be surprising, as the large-$D$ effective theory does not naturally possess self-similarity, which is typically the symmetry underlying universal critical behaviors. It is worthwhile to understand what happens by further imposing self-similarity in large-$D$ (see~\cite{Clark:2025tqi}) within the holographic context.

\newpage
    
    \item \textbf{Boundary signature of horizon formation and naked singularities.}

    A complementary approach is to investigate how bulk critical phenomena, such as horizon formation, Choptuik scaling, and naked singularities, manifest in the observables of the dual boundary theory. Several calculations have explored how the black hole singularity is encoded, raising the hope that similar observables may be applicable in our context as well~\cite{Fidkowski:2003nf, Festuccia:2005pi, Grinberg:2020fdj, Horowitz:2023ury, Kolanowski:2023hvh, Ceplak:2024bja}.

    Some work has been done in this direction. In particular, recent studies~\cite{Chesler:2019ozd, Emparan:2021ewh} (see also \cite{Dhar:2018pii}) have shown that by imposing self-similar structure, signatures of the classical growing mode and the naked singularity in the bulk are encoded in the holographic stress-energy tensor or boundary one-point function of the dual CFT. These works considered pure gravity in AdS$_5$ or scalar field collapse in general $D$-dimensional AdS spacetimes and demonstrated that the classical growing mode near criticality appears as a diverging factor in the boundary stress-energy tensor at a critical time $t^\ast$, which is interpreted as the moment when the naked singularity becomes causally connected to the boundary.

    This represents a dual description of classical critical collapse, particularly of the growing mode associated with DSS. Another key observable is the boundary one-point function $\langle O_{\varphi} \rangle$ of a linear scalar field $\varphi$, which exhibits a universal divergence $~\frac{1}{t-t^\ast}$. Notably, this divergence appears to be largely independent of the details of the collapse and instead follows from the universal features of AdS spacetimes combined with DSS. In contrast, if the bulk is CSS, as may be the case in certain AdS collapse scenarios, then the one-point function becomes approximately constant near $t^\ast$, which may require analysis of higher-point correlators or alternative bulk models.

    This framework captures the classical growing mode from the boundary perspective, yet it remains unclear whether it can detect additional quantum growing modes, such as those we have identified in our analysis. Furthermore, an open question is whether quantum effects, particularly those that give rise to universal growing modes with well-defined scaling exponents, also imprint themselves on the boundary stress-energy tensor. If the quantum backreaction serves as an independent growing mode in addition to the classical perturbation, it may modify the scaling behavior of boundary observables or produce distinct divergence patterns. One natural expectation is that such effects could be visible in higher-point functions, which would probe beyond the leading one-point structure. This motivates a more systematic holographic study to identify boundary diagnostics of quantum criticality.

    These efforts raise deep conceptual questions. What are the precise boundary criteria for the emergence of critical phenomena? Can quantum growing modes be identified in holographic data? What distinguishes DSS and CSS behaviors from the CFT perspective?

\item \textbf{Other cosmic censorship violations.} 

While the cosmic censorship conjecture has been remarkably resilient, there exist controlled settings exhibiting naked singularities \cite{Emparan:2020vyf}, arising through distinct mechanisms: (i) the endpoint of black hole evaporation \cite{Hawking:1974sw, Hawking:1976}, where semiclassical calculations predict that a black hole shrinks toward arbitrarily small, Planck-scale size; (ii) the Gregory–Laflamme (GL) instability \cite{Gregory:1993vy}, with a higher-dimensional black object undergoing a classical dynamical instability. The canonical example is a black string, which pinches off into a set of smaller black holes, but the pinch-off entails a topology change and is itself a singular event; and (iii) critical gravitational collapse, discussed at length in this work.

We showed that perturbative quantum corrections remove the critical collapse counterexample by enforcing cosmic censorship through a threshold shift and the associated mass gap. Extending the same techniques to (i) and (ii) is far less straightforward. The endpoint of black hole evaporation lies beyond the reach of semiclassical corrections, while the GL instability, though classical, is generically DSS and nonspherical. Our analysis accommodates cases without an exact timelike Killing symmetry, but it still relies on (continuous) self-similarity and spherical symmetry; even DSS solutions lack the analytic control required for treating backreaction, and may have to be done through numerical techniques.

Of course, we can only claim to have resolved the physical naked singularity coming from critical collapse if the size of the black hole induced by quantum effects is larger than the cut-off scale of the theory (say, Planck or string scale). Otherwise, as is the case for our examples of Garfinkle and Roberts, we have uplifted the critical collapse singularity into one resembling the endpoint of black hole evaporation. Another important point is that we have completely neglected the possibility of higher-curvature corrections, which are bound to become important as one approaches the formation of the naked singularity, just as quantum effects were indispensable in the same way.\footnote{See, for example, studies of critical collapse in the context of Gauss–Bonnet, Lovelock, and 
$f(R)$ gravity theories~\cite{Golod:2012yt, Deppe:2012wk, Taves:2013zpb, Baez:2022stg, Zhang:2021nnn}, where the effects on self-similarity and the notion of Type I and Type II systems are found to be nontrivial. Clearly, the role of higher-curvature corrections in critical collapse remains largely unexplored.} Accounting for higher-curvature corrections would also plausibly connect the critical collapse scenario to the other two violations of cosmic censorship, for which we do have an understanding in terms of string theory.

In particular, in the stringy regime, the Horowitz-Polchinski “string star” \cite{Horowitz:1996nw, Horowitz:1997jc} provides a nonsingular endpoint of string-scale black hole evolution and has been applied to smooth the GL pinch-off \cite{Emparan:2024mbp}; worldsheet analyses make the non-perturbative nature explicit \cite{Chen:2021dsw}. Thus, while perturbative effects suffice to resolve (some) critical collapse violations, the evaporation and GL scenarios likely require genuinely non-perturbative quantum gravity.

These settings remain valuable probes of quantum-gravitational phenomena, complementary to critical collapse. Unlike typical black holes, they are accessible to asymptotic observers without the obstruction of macroscopic horizons. Clearly, further work is required to fully understand these important problems.

\end{itemize}

\paragraph{Acknowledgements} We thank David Garfinkle, Viqar Husain, Gabor Kunstatter, Edward Witten, and Jochen Zahn for useful discussions. In particular, we thank Roberto Emparan, Carsten Gundlach, and Gustavo J. Turiaci for reading the draft and providing valuable feedback that clarified various subtleties. CHW also thanks Gustavo J. Turiaci for numerous discussions that significantly improved this work. He further acknowledges an audience member at the 39th Pacific Coast Gravity Meeting, whose suggestion inspired this project, and is deeply indebted to Gary Horowitz and Jiuci Xu for their early collaboration and many insightful discussions. CHW is supported by the University of
Washington and DOE Award DE-SC0011637 and DE-SC0026287. MT is supported by the European Research Council (ERC) under the European Union’s Horizon 2020 research and
innovation programme (grant agreement No 852386). MT is also supported by the Emmy Noether Fellowship program at the Perimeter Institute for Theoretical Physics.

\begin{appendix}

\section{Horizon tracing for $n=4$ and general $n$ Garfinkle spacetime}
\label{sec:horizonGarfinkle}

In this appendix, we present a detailed analysis of the horizon-tracing problem for the physical $n=4$ Garfinkle spacetime, and extract insights for general $n$ by applying the same methods described in Section~\ref{sec:GarfinkleHorizon}.

\subsubsection*{Horizon tracing for $n=4$ semiclassical Garfinkle spacetime}

As discussed, the $n=4$ Garfinkle solution with its top growing mode is the most physically relevant, as it provides the best agreement with numerical simulations~\cite{Pretorius:2000yu, Jalmuzna:2015hoa}. In what follows, we apply the same procedure as in the $n=2$ case, under the assumption that only the top growing mode is present.

\begin{figure}[hbt!]
\centering
\includegraphics[width=0.45\textwidth]{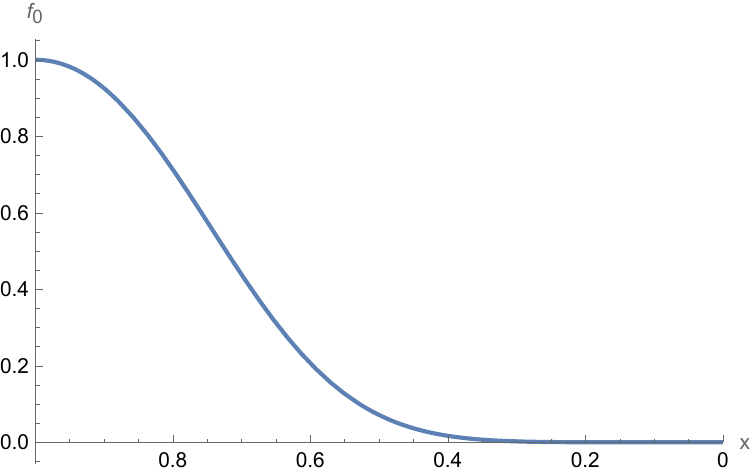}
\includegraphics[width=0.45\textwidth]{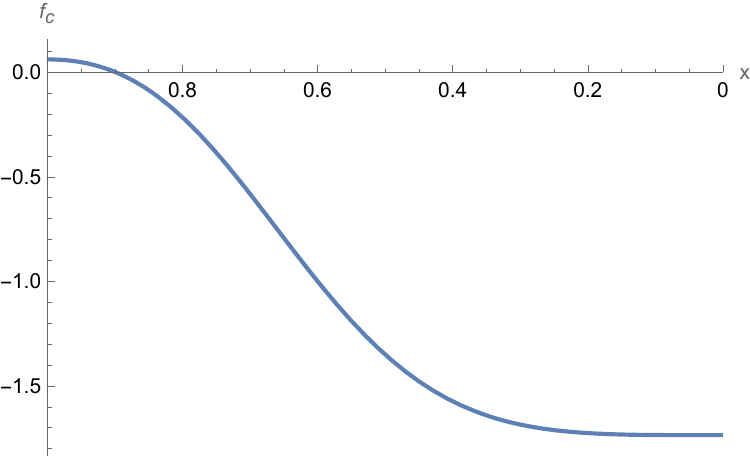}
\includegraphics[width=0.45\textwidth]{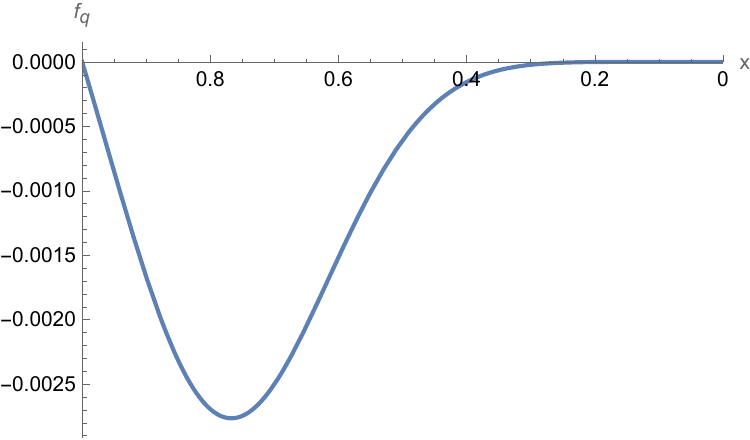}
\caption{Compared with the $n=2$ case shown in Figure~\ref{n=2fs}, we observe that the qualitative features remain similar, but there are some notable differences. For instance, both $f_0$ and $f_q$ decay more rapidly near the light cone, while $f_c$ can become positive near the center.}
\label{n=4fs}
\end{figure}

In the $n=4$ case, we have
\be
F_c(x)=-\frac{1}{\sqrt{2}}\bigg(C_c -\frac{3}{4}C_b-\frac{7}{8}C_c x\bigg)(1+x^4)^\frac{7}{2}, \quad r_c(x)=-\frac{C_c}{2} (1-x),
\ee
with
\be
C_b \approx -0.286, \quad C_c \approx 1.23,
\ee
and
\be
F_q(x)=-\frac{7(1+x^4)^\frac{5}{2}[x^8-1+4 \ln{(\frac{2}{1+x^4})}]}{128 \sqrt{2} \pi^2 (x^4-1)}, \quad r_q(x)=\frac{3}{64 \pi^2}.
\ee
We calculate the horizon-tracing function to linear order
\be
(\nabla \bar{r})^2 \approx f_0(x)+e^{\frac{7T}{8}}(p-p^\ast) f_c(x)+e^T f_q (x),
\ee
where
\be
f_0(x)=\frac{8 \sqrt{2}x^7}{(1+x^4)^\frac{7}{2}},
\ee
\be
f_c(x)=-\frac{\sqrt{2}(C_c+6C_b x^7-7 C_cx^7+7 C_c x^8)}{(x^4+1)^\frac{7}{2}},
\ee
\be
f_q(x)=\frac{7 x^7 [x^8-1+4 \ln{(\frac{2}{1+x^4})}]}{8 \sqrt{2} \pi^2  (x^4-1) (1+x^4)^\frac{9}{2}}.
\ee
The behaviors of these $x$-dependent functions are plotted in Figure~\ref{n=4fs}.
\FloatBarrier 

\begin{figure}[hbt!]
\centering
\includegraphics[width=0.45\textwidth]{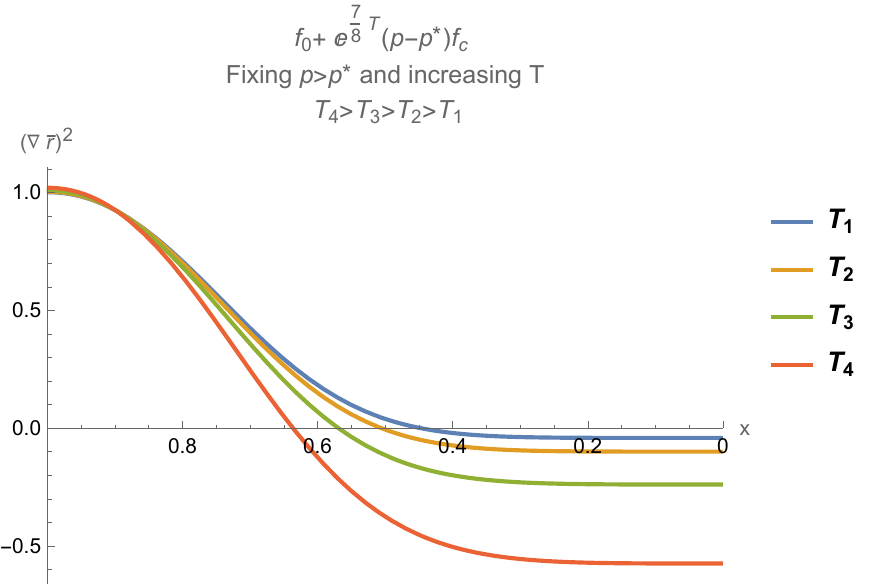}
\includegraphics[width=0.45\textwidth]{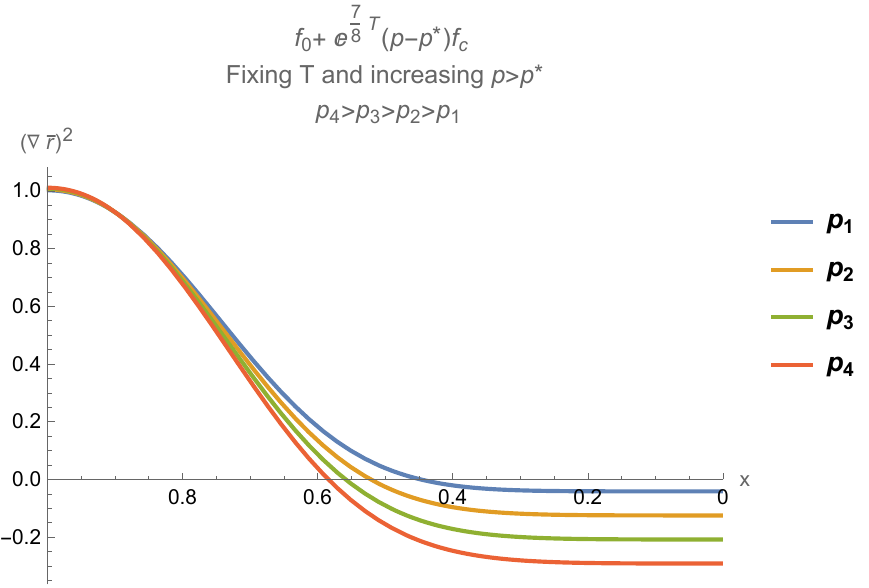}
\caption{When including only the classical perturbation with $T \lesssim 4$ and $(p - p^\ast) \lesssim 0.07$, horizon formation is generally enhanced in the $n=4$ case. This is because the classical growing mode has a larger exponent, $e^{\frac{7}{8}T}$, compared to the $n=2$ case, even though the corresponding profile $f_c(x)$ is smaller.}
\label{n=4f0+fc}
\end{figure}
\FloatBarrier 

\begin{figure}[hbt!]
\centering
\includegraphics[width=0.45\textwidth]{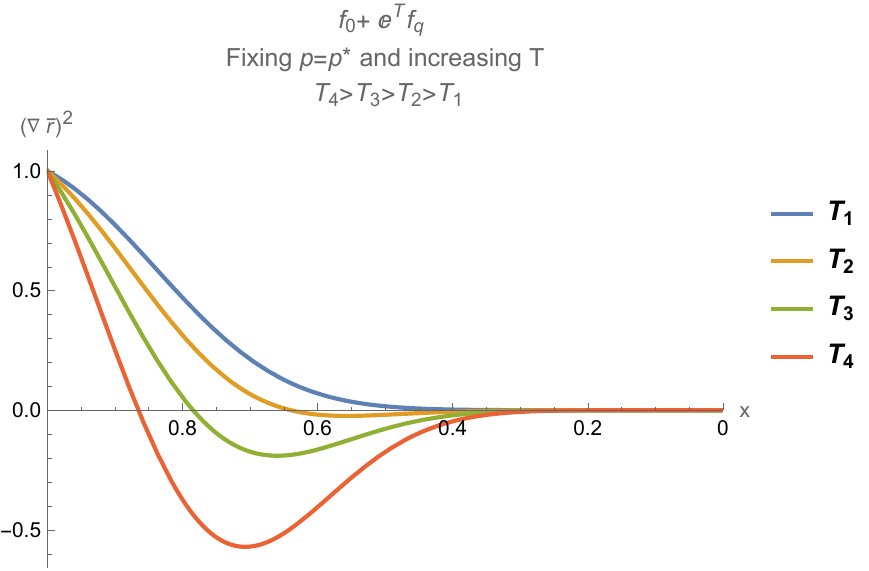}
\caption{Although the quantum perturbation always carries the same exponent, the change in the profile $f_q(x)$ leads to a weaker effect on horizon formation compared to that seen in Figure~\ref{n=2f0+fq}. Nevertheless, it still indicates that certain subcritical data can be lifted above the threshold for critical collapse.}
\label{n=4f0+fq}
\end{figure}
\FloatBarrier

\begin{figure}[hbt!]
\centering
\includegraphics[width=0.45\textwidth]{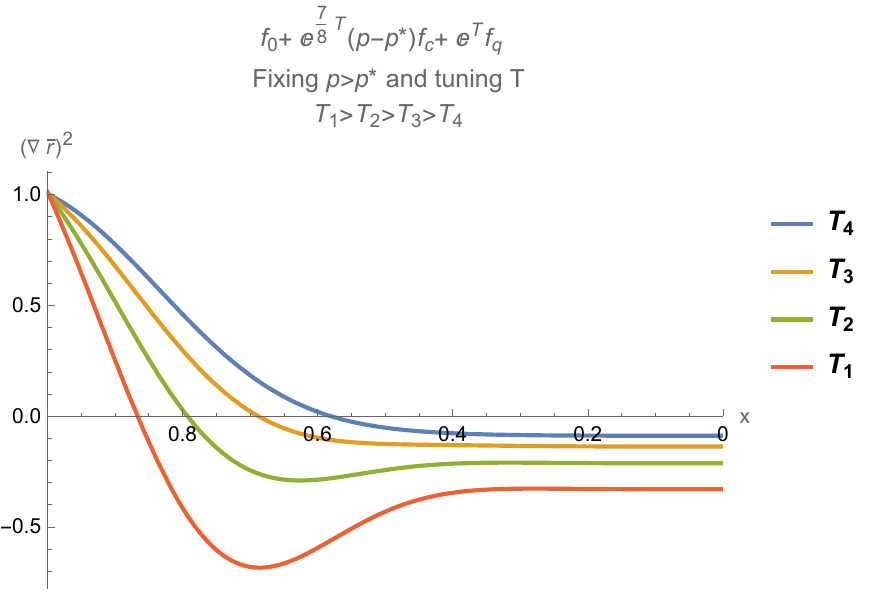}
\includegraphics[width=0.45\textwidth]{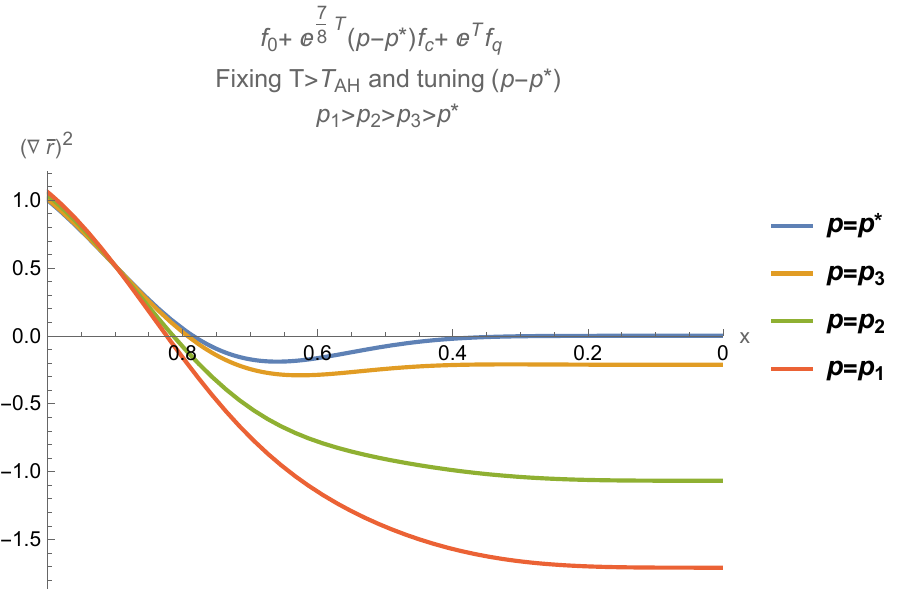}
\caption{Here we consider $T \lesssim 6$ and $(p-p^\ast) \lesssim 0.008$, and again horizon formation is merely enhanced by the inclusion of classical supercritical perturbations.}
\label{n=4f0+fc+fq}
\end{figure}
\FloatBarrier

\begin{figure}[hbt!]
\centering
\includegraphics[width=0.45\textwidth]{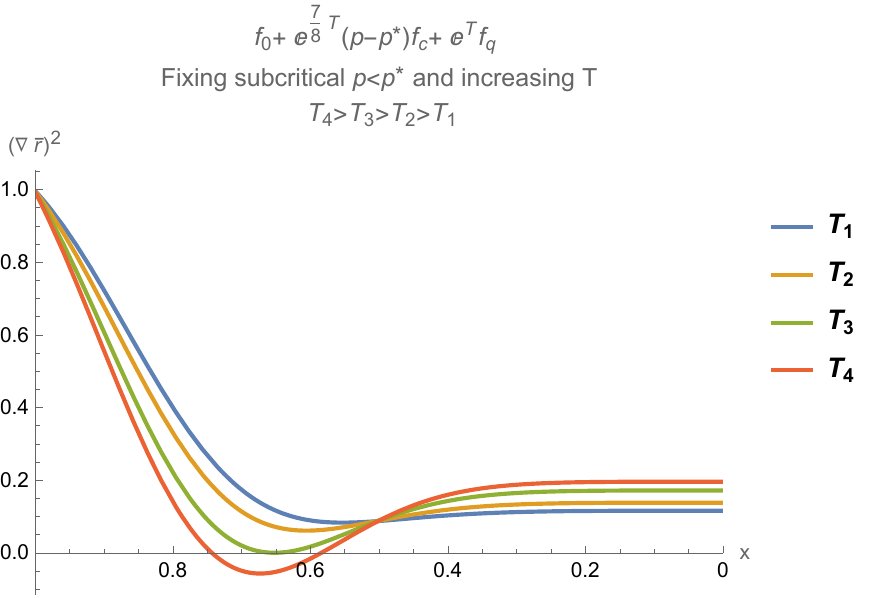}
\includegraphics[width=0.45\textwidth]{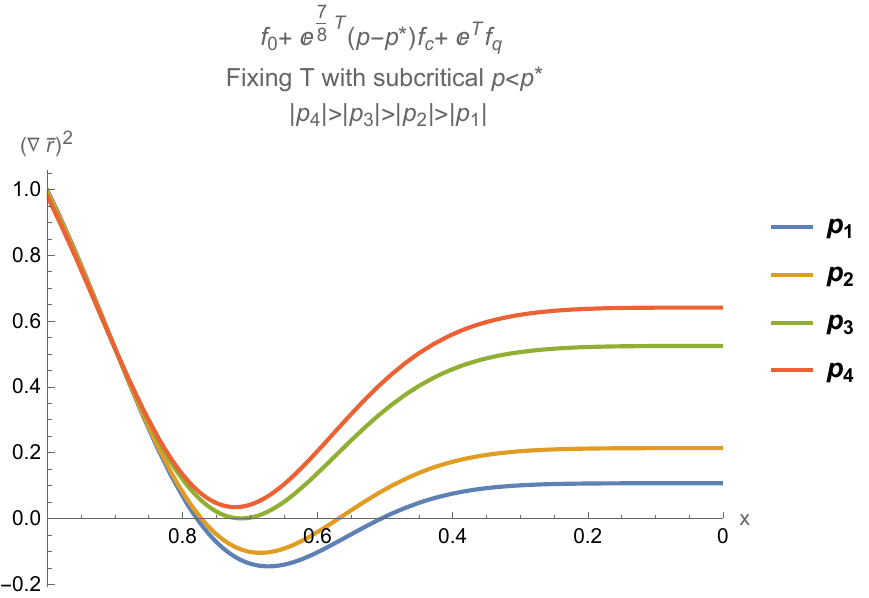}
\caption{Similar to Figure~\ref{n=2f0+fc+fq(subc)}, subcritical data is lifted up. But due to the profile of $f_q$ in the $n=4$ case, the point where $(\nabla \bar{r})^2$ first vanishes is shifted closer to the center.}
\label{n=4f0+fc+fq(subc)}
\end{figure}
\FloatBarrier 

\begin{figure}[hbt!]
\centering
\includegraphics[width=0.45\textwidth]{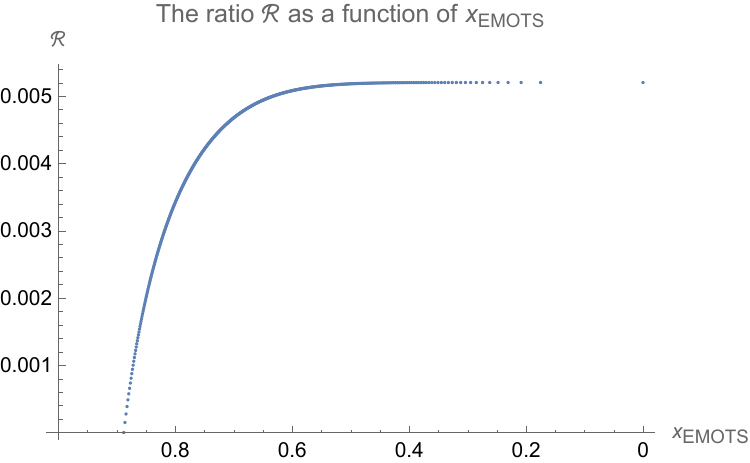}
\includegraphics[width=0.45\textwidth]{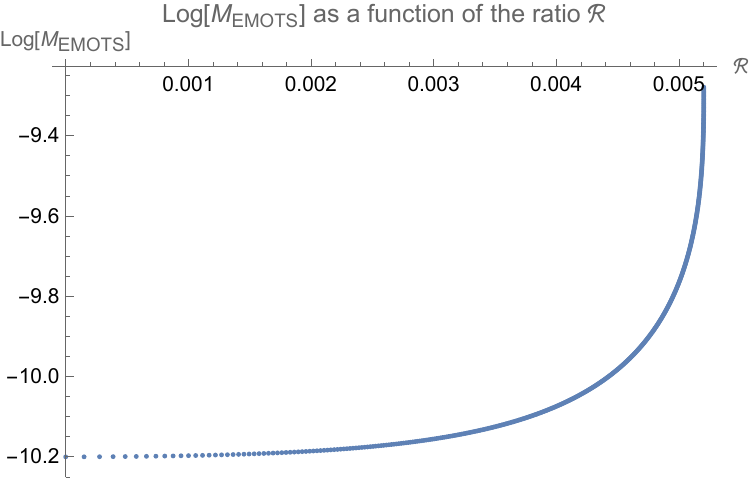}
\caption{We observe qualitatively the same behaviors for $x_{\text{EMOTS}}$ and $M_{\text{EMOTS}}$ as in the $n=2$ case. However, compared with Figure~\ref{n=2Ratio_x_Qsupercritical} and Figure~\ref{n=2Mass_Ratio_Qsuper-critical}, the maximum ratio at which quantum effects begin to play a role has decreased, while the minimum mass gap remains largely unchanged at this scale.}
\label{n=4Ratio_x_mass}
\end{figure}
\FloatBarrier

\subsubsection*{Horizon tracing for general $n$ semiclassical Garfinkle spacetime}

We now study and compare the qualitative and quantitative differences across various values of $n$. For the classical perturbation, we assume that only the top growing mode is present, with an exponent given by $\omega_c = 1 - \frac{1}{2n}$. At sufficiently late times, this mode can be taken as the dominant classical contribution.

\begin{figure}[hbt!]
\centering
\includegraphics[width=0.45\textwidth]{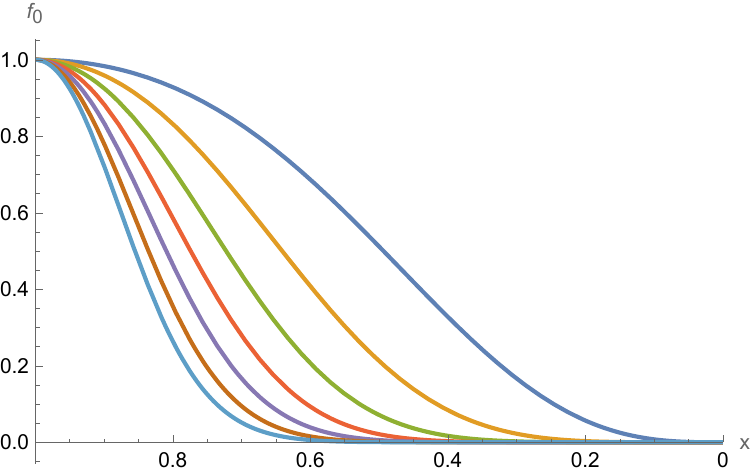}
\includegraphics[width=0.45\textwidth]{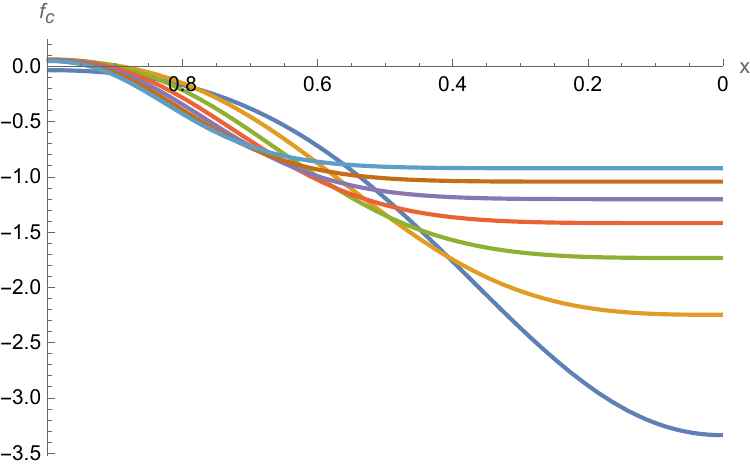}
\includegraphics[width=0.55\textwidth]{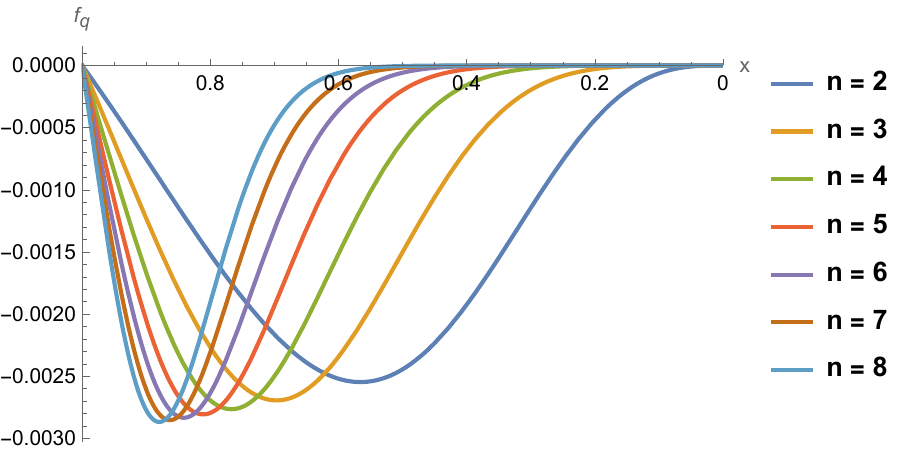}
\caption{We observe consistent trends in all functions for $n=2$ to $n=8$: they become progressively weaker near the past light cone at $x = 0$. However, the maximum amplitude of $f_q$ remains approximately unchanged as $n$ increases.}
\label{nfs(2-8)}
\end{figure}

\begin{figure}[hbt!]
\centering
\includegraphics[width=0.45\textwidth]{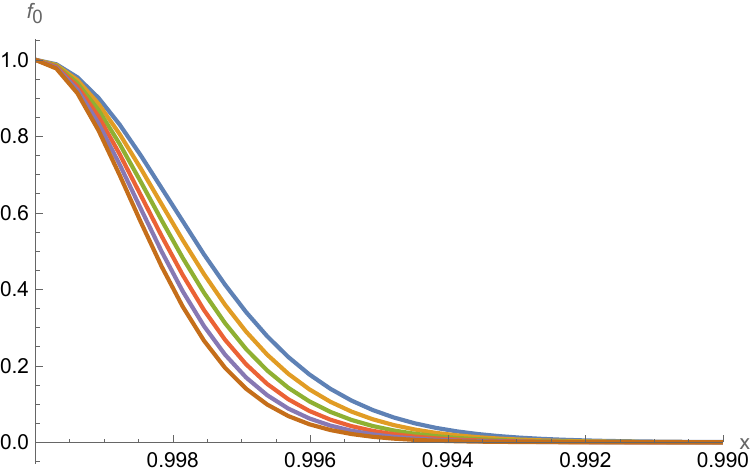}
\includegraphics[width=0.45\textwidth]{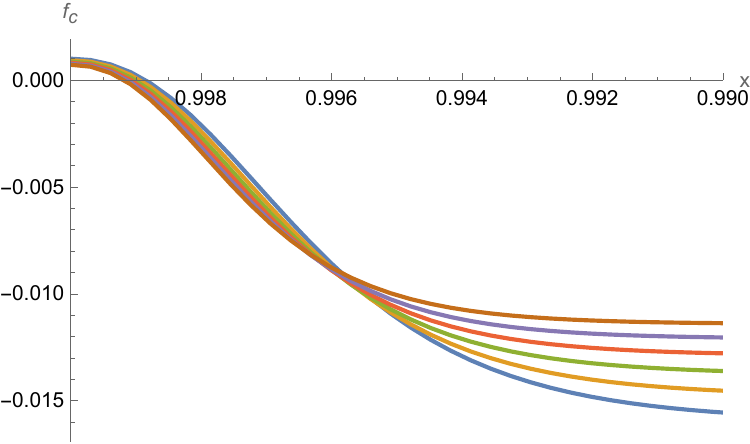}
\includegraphics[width=0.55\textwidth]{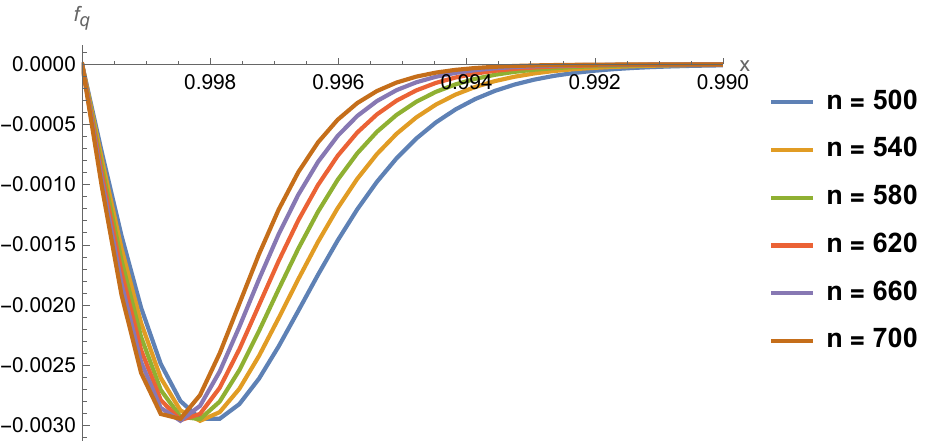}
\caption{For very large values of $n$, we need to zoom in to the region $x \in [0.99, 1]$. The maximum amplitude of $f_c$ drops significantly, while $f_q$ maintains a comparable peak amplitude to the low-$n$ cases, albeit becoming sharply localized near the center at $x = 1$.}
\label{nfs(500-700)}
\end{figure}

For a general $n$ Garfinkle spacetime, we have
\be
F_c(x)=2^{\frac{2}{n}-4} [2(n-1)C_b-C_c x+2n(x-1)C_c](1+x^n)^{4-\frac{2}{n}}, \quad r_c(x)=-\frac{C_c}{2} (1-x),
\ee
\be
F_q(x)=-\frac{(2n-1)4^{\frac{1}{n}-4} (1+x^n)^{3-\frac{2}{n}}}{\pi^2(x^n-1)} \bigg[x^{2n}-1+4 \ln{\bigg(\frac{2}{1+x^n}\bigg)} \bigg], \quad r_q(x)=\frac{3}{64 \pi^2}.
\ee
We calculate the horizon-tracing function
\be
(\nabla \bar{r})^2 \approx f_0(x)+e^{(1-\frac{1}{2n}) T}(p-p^\ast) f_c(x)+e^T \hbar f_q (x),
\ee
where
\be
f_0(x)=4^{2-\frac{1}{n}} x^{2n-1} (x^n+1)^{\frac{2}{n}-4},
\ee
\be
f_c(x)=-\frac{2^{3-\frac{2}{n}} (x^n+1)^{\frac{2}{n}-4}\{C_c x+[2C_b(n-1)+C_c (2n-1)(x-1)]x^{2n} \}}{n x},
\ee
\be
f_q(x)=\frac{(2n-1)2^{1-\frac{2}{n}}x^{2n-1} (x^n+1)^{\frac{2}{n}-5}[x^{2n}-1+4 \ln{(\frac{2}{1+x^n})}]}{n \pi^2 (x^n-1)}.
\ee
The behaviors of these $x$-dependent functions for $n=2 \sim 8$ and for very large $n$ are plotted in Figure~\ref{nfs(2-8)} and Figure~\ref{nfs(500-700)}, respectively.
\FloatBarrier 

\FloatBarrier 

Although the features in Figure~\ref{nfs(500-700)} appear peculiar, it is important to note that all functions vanish in the limit $n \to \infty$ for $x \in [0,1]$:
\be
\lim_{n \to \infty} f_0(x)=0, \quad \lim_{n \to \infty} f_c(x)=0, \quad \lim_{n \to \infty} f_q(x)=0.
\ee
Hence, we turn our attention to the lower-$n$ cases.

\begin{figure}[hbt!]
\centering
\includegraphics[width=0.48\textwidth]{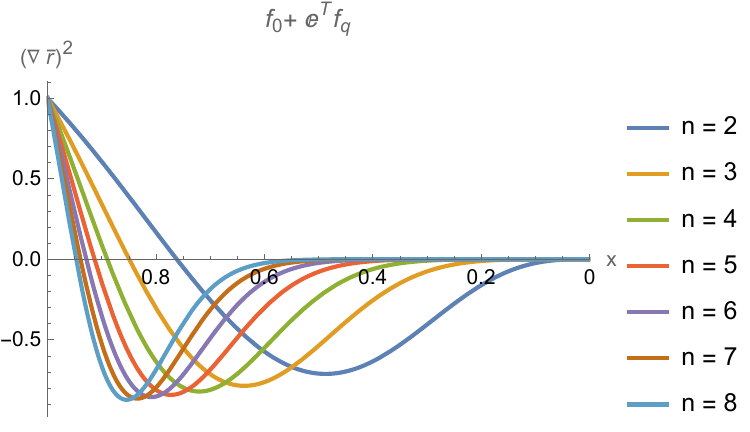}
\includegraphics[width=0.42\textwidth]{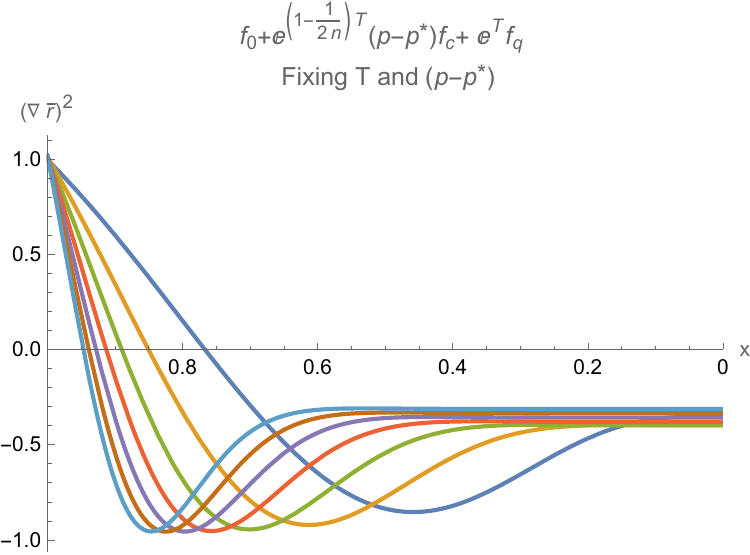}
\caption{In the left panel, we fix a time slice where the trapped region arises from quantum perturbations. The behavior matches the general trend seen in $f_q$ for different values of $n$. In the right panel, we include classical perturbations on top of the quantum effects, which leads to an enhancement of horizon formation, yet preserving the same overall qualitative features.}
\label{nf0+fq}
\end{figure}
\FloatBarrier

\begin{figure}[hbt!]
\centering
\includegraphics[width=0.48\textwidth]{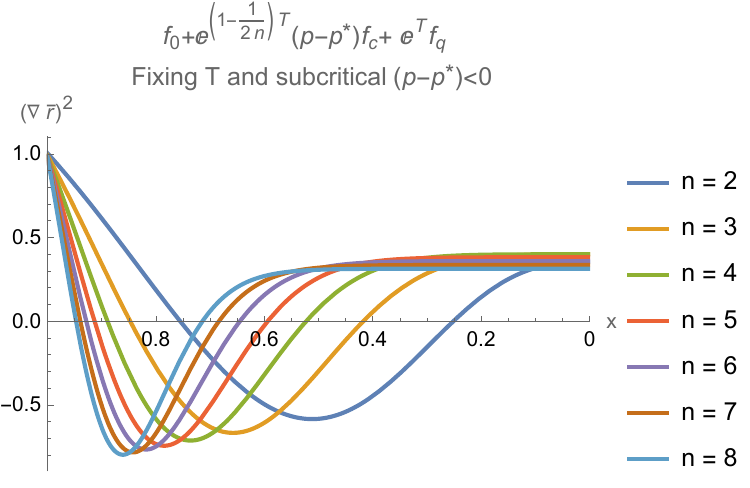}
\includegraphics[width=0.42\textwidth]{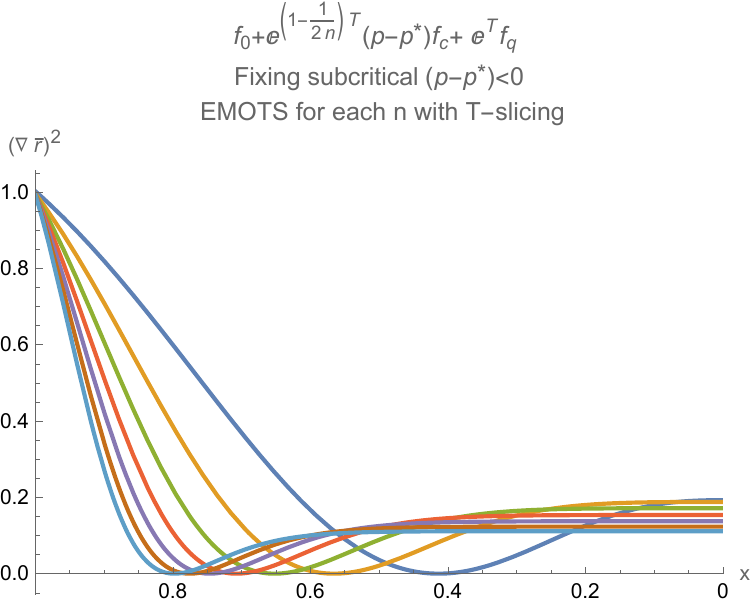}
\caption{Quantum effects lift subcritical data, and the trend agrees with the behaviors of $f_q$. We see that the EMOTS for larger $n$ shifts toward the center.}
\label{nf0+fc+fq(sub)}
\end{figure}
\FloatBarrier

\begin{figure}[hbt!]
\centering
\includegraphics[width=0.48\textwidth]{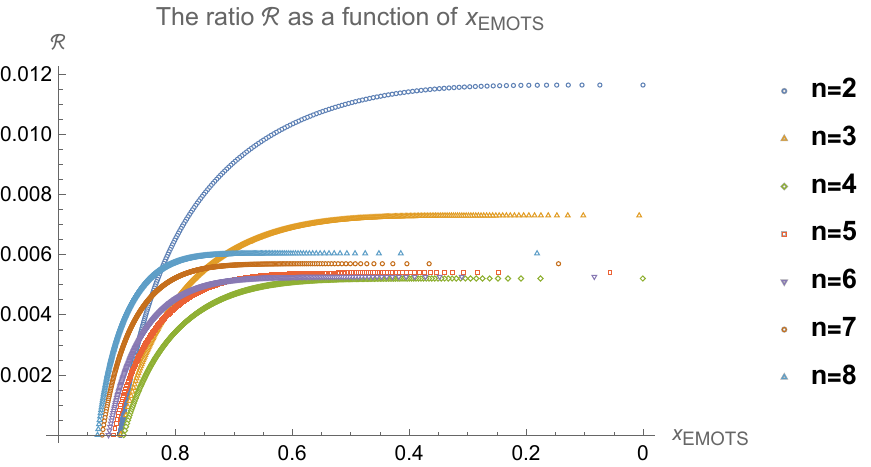}
\includegraphics[width=0.42\textwidth]{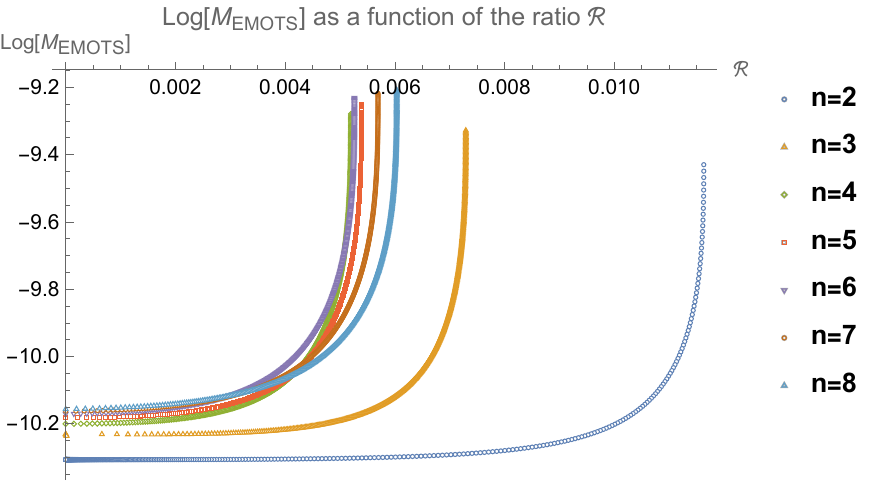}
\caption{We sample 500 data points for each $n$. Among them, $n=2$ is the earliest case where quantum effects begin to play a role. However, for larger $n$, there does not appear to be a consistent trend in the maximum ratio allowed. However, the mass gaps across different $n$ values remain roughly the same.}
\label{nRatio_x_Mass}
\end{figure}
\FloatBarrier 

\section{The Weyl-Garfinkle spacetime}
\label{sec:WeylGarfinkle}

In this appendix, we study a new CSS solution to the most general two-dimensional dilaton gravity theory, obtained by performing a Weyl transformation on the dimensionally reduced Garfinkle spacetime. This defines a two-parameter family of solutions labeled by $(\gamma, n)$. Our goal is to understand how the Weyl transformation affects the global structure of such a self-similar spacetime and the computation of $\langle T_{ab} \rangle$ when the matter sector is not conformal, yet Weyl-invariant. This solution turns out to exhibit rich global properties and reveals previously underexplored features of quantum field theory in curved spacetime.

As discussed at the end of Section~\ref{sec:onelooptheory}, a systematic exploration of the most general two-dimensional dilaton gravity theory up to second order in derivatives, assuming CSS, was carried out in~\cite{Moitra:2022umq}. The relevant action is given by \eqref{eq:Upamanyu2D} that we reproduce here
\be \label{eq:AppUpamanyu2D}
S\propto \int d^2 x \sqrt{-g} e^{-2 \phi} [R+\gamma(\nabla \phi)^2+V_{\text{eff}} e^{\frac{2 \phi}{\kappa}}- (\nabla f)^2-V_f e^{\frac{2 f}{\lambda}}],
\ee
It clearly shows that the couplings of the dilaton to both gravity and matter sectors are generically nontrivial under the requirement of CSS. We previously noted that the effective dilaton gravity model obtained from $D$-dimensional spherical reduction given by \eqref{eq:gravityaction} and \eqref{eq:matteraction}, corresponds to specific choices of the parameters in \eqref{eq:AppUpamanyu2D}. In particular, the simplest case arises for $D=3$, where the dimensionally reduced two-dimensional Garfinkle system \eqref{eq:2DGarfinke} corresponds to a solution with no dilaton kinetic term ($\gamma=0$) and vanishing potential terms ($V_{\text{eff}}=V_f=0$).

However, a well-known practice in dilaton gravity is that a local Weyl transformation can be used to reintroduce a dilaton kinetic term parametrized by $\gamma$~\cite{Mertens:2022irh}, even though it is more commonly employed to eliminate such a term. Since a Weyl transformation can indeed alter the global properties of the spacetime, this remains a nontrivial operation. Furthermore, the origin of the trace anomaly comes from the fact that the path integral measure is not Weyl-invariant. Given the simplicity of the reduced Garfinkle system, we take this opportunity to explore how such a transformation influences the resulting geometry and quantum stress-energy tensor.

Here, we find it clear to work first in the coordinates $(u, w)$ used in~\cite{Garfinkle:2002vn}, in which the reduced Garfinkle system \eqref{eq:2DGarfinke} takes the form
\be
ds^2=2 e^{2\rho} du dw, \quad \phi=-\frac{1}{2} \ln{\bigg[\frac{1}{2}(-u-w^{2n})\bigg]}.
\ee
The coordinates $(u, w)$ are related to the adapted coordinates $(T, x)$ by
\be \label{eq:uwcoordinates}
u=-e^{-T}, \quad w =x e^{-\frac{T}{2n}},
\ee
and again $n \in \mathbb{N}$. We then consider a Weyl transformation through the shift
\be
\rho' = \rho-\frac{\gamma}{8}\ln{(-u-w^{2n})}.
\ee
We note that the $\gamma$-dependent term inherits the same domain of validity as $\phi$. The original Garfinkle solution is smooth for $n \in \mathbb{N}$ with  $u \in (-\infty, -w^{2n}]$ and $w \in [0, 1]$. The metric becomes
\be
ds'^2 = (-u-w^{2n})^{-\frac{\gamma}{4}} (2 e^{2\rho}dudw),
\ee
implying that the Weyl rescaling factor $\omega(x)$ is
\be
ds'^2 =e^{2 \omega(x)}ds^2 \implies e^{2 \omega(x)}=(-u-w^{2n})^{-\frac{\gamma}{4}}.
\ee
We know that under a Weyl transformation, the relevant quantities in the Lagrangian density transform as
\be
\sqrt{-g} \to e^{2 \omega} \sqrt{-g}, \quad R \to e^{-2 \omega} (R-2 \nabla^2 \omega),
\ee
so the gravitational part of the action transforms as (matter sector unchanged since it is Weyl-invariant)
\be
\sqrt{-g} e^{-2 \phi} R \to \sqrt{-g}e^{-2 \phi} (R-2 \nabla^2 \omega).
\ee
One can verify that the additional terms involving the Weyl factor $\omega(x)$ can be canceled by including a dilaton kinetic term $\gamma (\nabla \phi)^2$, given the specific form of the dilaton. However, since a local Weyl transformation can alter the global structure if additional boundary singularities arise, we consider cases $\gamma=0$ and $\gamma \neq 0$ as distinct physical solutions.

We now express everything in the adapted coordinates $(T, x)$ using \eqref{eq:uwcoordinates}
\be
ds^2=e^{(-2+\frac{\gamma}{4} )T}\bigg[\bigg(\frac{1+x^n}{2} \bigg)^{4(1-\frac{1}{2n})} (1-x^{2n})^{-\frac{\gamma}{4}} (2n dx-x dT)dT \bigg], \quad 
\ee
\be
\phi=\frac{1}{2}\bigg[T-\ln{\bigg(\frac{1-x^{2n}}{2}}\bigg) \bigg], \quad f=\sqrt{1-\frac{1}{2n}}\bigg[T-2 \ln{\bigg(\frac{1+x^n}{2} \bigg)}\bigg],
\ee
and one can verify that this configuration is a solution to the following action
\be
S \propto \int d^2 x \sqrt{-g} e^{-2 \phi}[R+\gamma(\nabla \phi)^2-(\nabla f)^2]. 
\ee
This gives a simple, exact two-parameter family of solutions inspired by the Garfinkle spacetime, where $\gamma$ serves as an additional deformation parameter. We refer to this two-dimensional solution, parametrized by $(\gamma, n)$, as the \textit{Weyl-Garfinkle geometry}. Note that the profiles of the dilaton and scalar field remain unchanged compared to the original reduced Garfinkle spacetime. We emphasize that this solution is purely a two-dimensional geometry, not arising from a higher-dimensional reduction, at least not without a corresponding dilaton potential. In particular, the dilaton here does not represent the radius of a higher-dimensional sphere.

In this analysis, we take $\gamma$ to be a general non-negative real parameter, unconstrained by any specific higher-dimensional origin, while avoiding a wrong-sign kinetic term in this frame. Since the dilaton couples directly to the curvature, it also belongs to the gravitational sector. Although our treatment is agnostic to any UV completion, it is worth noting that low-energy effective string theory typically selects $\gamma=4$~\cite{Callan:1985ia, Mandal:1991tz, Witten:1991yr, Moitra:2022umq}.

Let us now examine the global structure of this solution in more detail. We have a two-parameter family of CSS solutions labeled by $(\gamma, n)$, given in closed form, with the domain defined as $T \in (- \infty, \infty)$ and $x \in [-1, 1]$. We still take $n \in \mathbb{N}$ to ensure real analyticity. The metric remains self-similar, with the homothetic scaling exponent modified to be $(-2+\frac{\gamma}{4})$, and the dilaton and scalar field profiles also CSS. To identify the curvature singularities, we compute the curvature invariants explicitly and find the following
\bea
R&=&\frac{e^{(2-\frac{\gamma}{4})T}}{n}2^{4-\frac{2}{n}} x^{n-1}(1+x^n)^{-6+\frac{2}{n}+\frac{\gamma}{4}} (1-x^n)^{-2+\frac{\gamma}{4}} [2 (x^n-1)^2
\no\\
&\quad&-n(4+(\gamma-8)x^n+4 x^{2n})],
\eea
\be
R^2=2 R^{ab}R_{ab}=R^{abcd}R_{abcd}.
\ee
Thus, it suffices to focus on the behavior of the Ricci scalar $R$. 

We first observe that for $n=\frac{1}{2}$, the curvature no longer vanishes, and the geometry does not resemble Minkowski spacetime written in self-similar coordinates. In this setting, $x=1$ represents a timelike boundary, $x=-1$ a spacelike boundary, and $x=0$ a null surface. Dynamical singularities can arise at the boundaries, and while the Penrose diagram resembles the Garfinkle case, for $\gamma>8$ the kinematical singularity shifts to the past instead of the future. Exotic behavior may arise for very large $\gamma$ and $n$, but we do not explore such regimes here.

\begin{figure}[hbt!]
\centering
\includegraphics[width=0.48\textwidth]{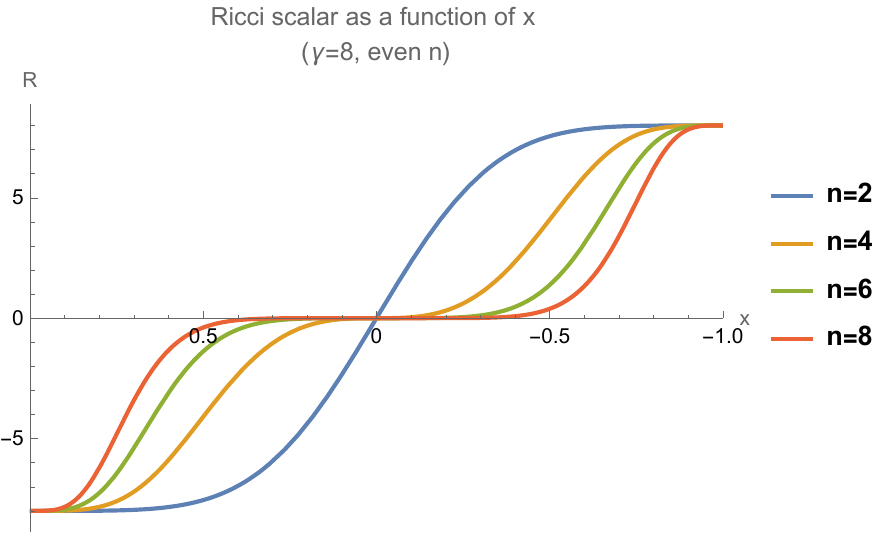}
\includegraphics[width=0.48\textwidth]{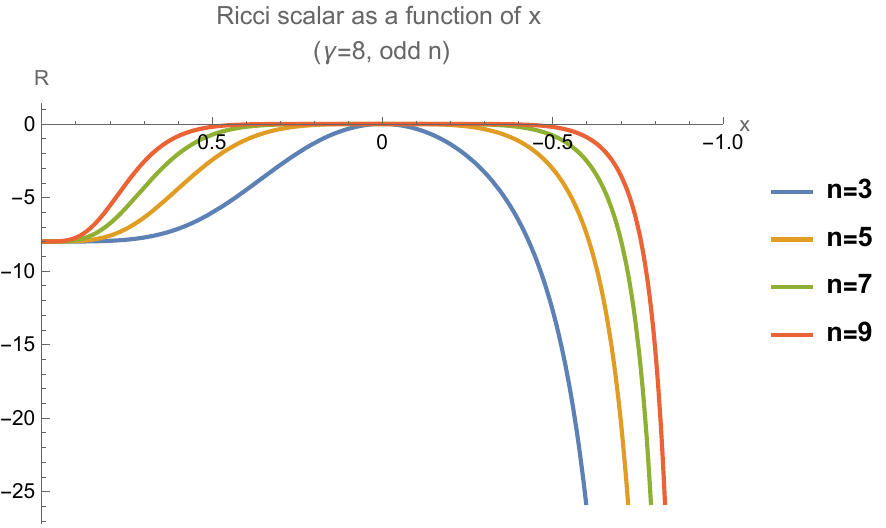}
\caption{From both the metric and the curvature invariants, $\gamma=8$ is obviously a special case, as the self-similar conformal factor is a constant. For even $n$, the spacetime appears everywhere regular, smoothly interpolating from negative to positive curvature without any curvature singularities. For odd $n$, however, a spacelike singularity emerges at $x=-1$.}
\label{Rfunction(gamma=8)}
\end{figure}

\begin{figure}[hbt!]
\centering
\includegraphics[width=0.48\textwidth]{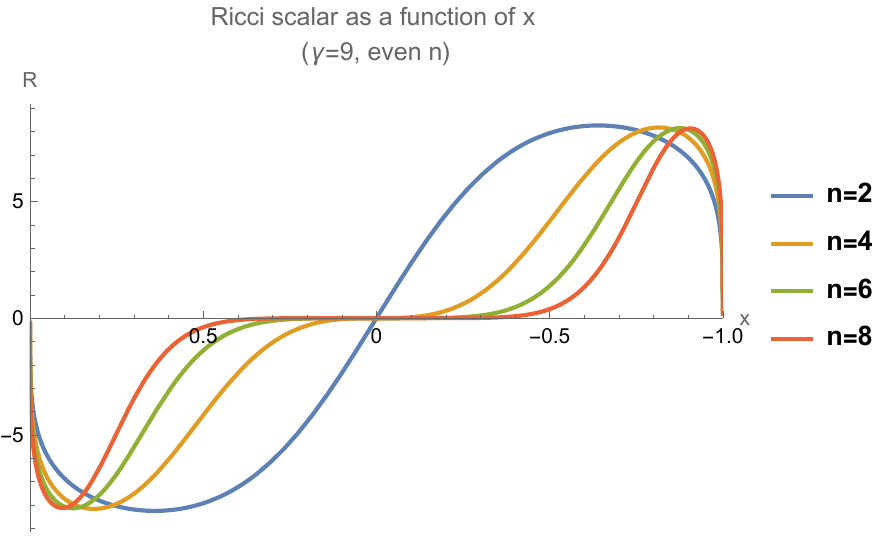}
\includegraphics[width=0.48\textwidth]{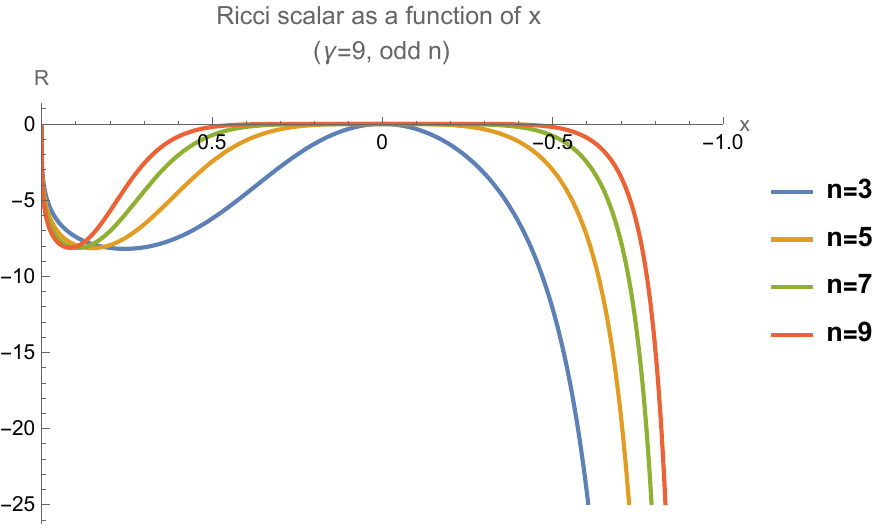}
\caption{For $\gamma>8$, the curvature vanishes near the boundary $x=1$. Yet for $x=-1$, a singularity appears for odd $n$.}
\label{Rfunction(gamma=9)}
\end{figure}

\begin{figure}[hbt!]
\centering
\includegraphics[width=0.48\textwidth]{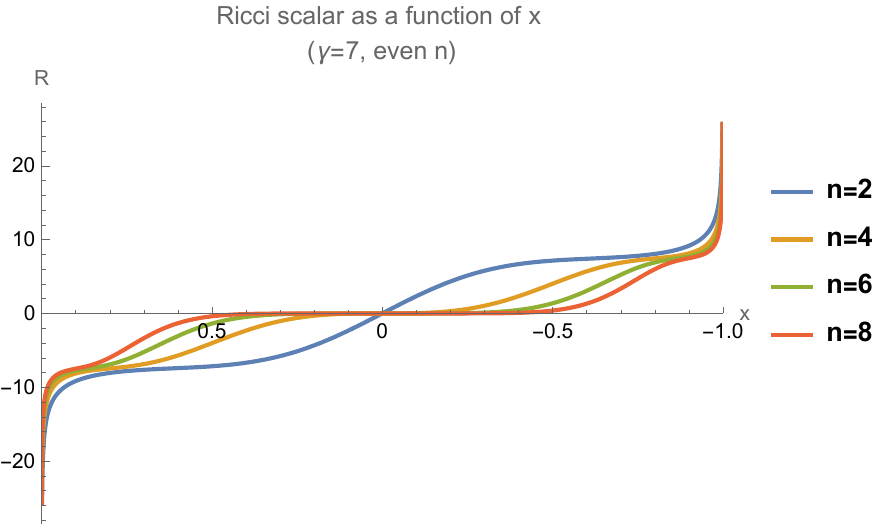}
\includegraphics[width=0.48\textwidth]{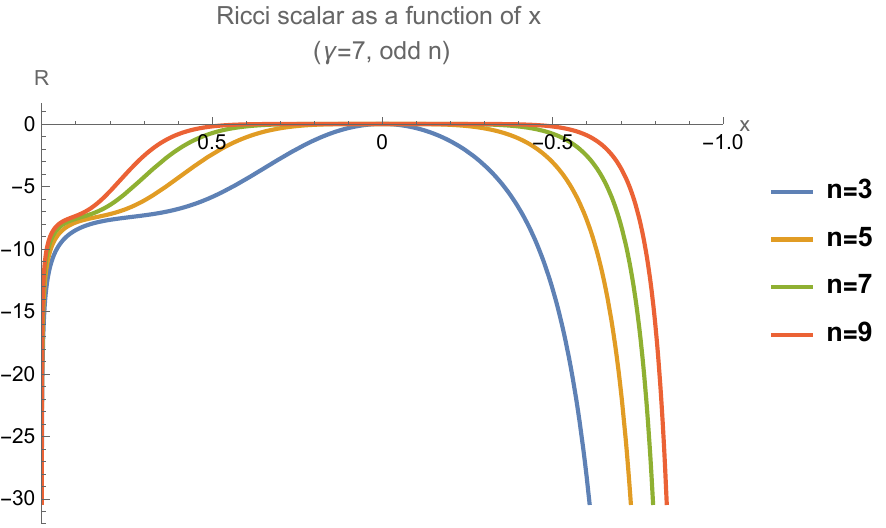}
\caption{For $0<\gamma<8$, singularities develop at both boundaries $x=\pm1$ for both even and odd values of $n$.}
\label{Rfunction(gamma=7)}
\end{figure}

\begin{figure}[hbt!]
\centering
\includegraphics[width=0.48\textwidth]{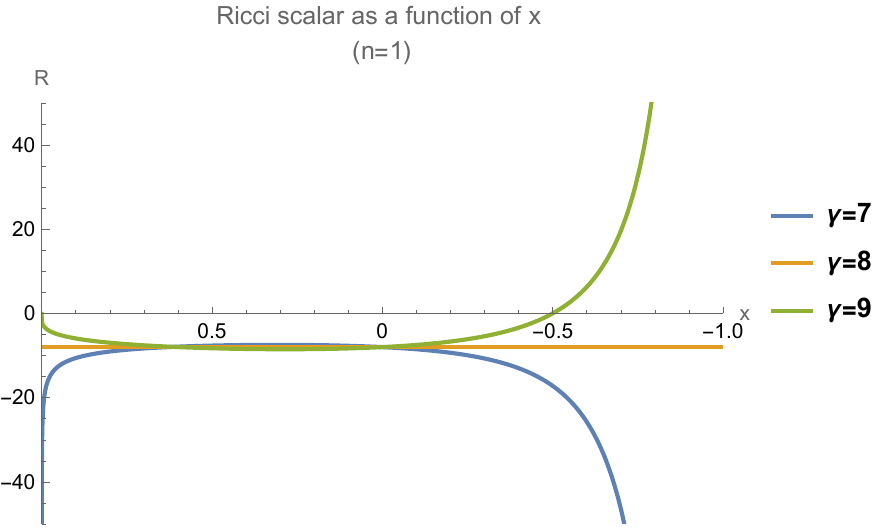}
\caption{The case of $n=1$ is distinct from higher $n$. For the special value $\gamma=8$, the Ricci scalar is constant and negative across the entire slice. Of course, the spacetime is not of constant curvature as the value shown above is evaluated on a fixed time slice. While for $0<\gamma<8$, curvature singularities still develop at both boundaries $x =\pm 1$.}
\label{Rfunction(n=1)}
\end{figure}

To understand the dynamical singularities, we present plots of the Ricci scalar for $x \in [-1,1]$ at fixed finite $T$ in Figures~\ref{Rfunction(gamma=8)},~\ref{Rfunction(gamma=9)},~\ref{Rfunction(gamma=7)}, and~\ref{Rfunction(n=1)}. These correspond to representative values $\gamma=7,8,9$, chosen as integers for simplicity, and capture the essential qualitative behavior. Values below or above the special case $\gamma=8$ yield similar results.

A notable feature is that for $0 < \gamma < 8$,  both endpoints $x=1$ and $x=-1$ develop curvature singularities for all values of $n$. This range of $\gamma$ is physically the most relevant, as only in this regime does the geometry contract as $T$ increases, an expected feature of critical collapse. However, the presence of singularities at both boundaries appears to preclude a direct interpretation of this geometry as a critical solution. But this limitation is not problematic for our purposes, as our goal here is to investigate how a Weyl transformation can alter the global structure and influence the choice of quantum state.

\FloatBarrier 

\FloatBarrier 

\FloatBarrier 

\FloatBarrier

For the apparent horizon, in two-dimensional dilaton gravity, we may compute~\cite{Moitra:2022umq}
\be
(\nabla \phi)^2=e^{(2-\frac{\gamma}{4})T}2^{4-\frac{2}{n}} x^{2n-1}(1+x^n)^{-6+\frac{2}{n}+\frac{\gamma}{4}} (1-x^n )^{-2+\frac{\gamma}{4}},
\ee
where we take $(\nabla \phi)^2=0$ to be the locus of the apparent horizon, while $(\nabla \phi)^2<0$ as the trapping region. At finite $T$ and for $x \in (-1, 1)$, we analyze the sign of this expression. The factors $(1+x^n)^{-6+\frac{2}{n}+\frac{\gamma}{4}}$ and $(1-x^n )^{-2+\frac{\gamma}{4}}$ are always positive in our defined domain. The sign is thus entirely determined by the term $x^{2n-1}$.  Since the power $2n-1$ is odd for all positive integers $n$, then $x^{2n-1}$ is positive if $x>0$ and negative if $x<0$. Therefore, the region $x \in (-1, 0)$ is trapped, and the apparent horizon lies at $x=0$, representing a marginally trapped surface.

For the Weyl-Garfinkle spacetime, we investigate whether a self-consistent renormalized stress-energy tensor corresponding to a stationary Boulware-like quantum state exists in this background. From the analysis of curvature invariants above, we observe that the geometry always exhibits a curvature singularity at $x=-1$ for all values of $(\gamma, n)$. This justifies focusing, as in the original Garfinkle case, on the regular regime $x \in[0, 1]$. To compute the one-loop quantum stress-energy tensor, we again adopt double-null coordinates $(u, v)$ with
\be
x= \frac{v}{u}, \quad T=-2n \ln{(-u)},
\ee
then the metric can be written as
\be
ds^2=-e^{2 A} du dv, \quad e^{2 A}=4n^2 \bigg(\frac{(-u)^n+(-v)^n}{2} \bigg)^{4(1-\frac{1}{2n})} ((-u)^{2n}-(-v)^{2n})^{-\frac{\gamma}{4}}.
\ee
The dilaton and scalar fields take the form
\be
\phi=-\frac{1}{2} \ln{\bigg(\frac{(-u)^{2n}-(-v)^{2n}}{2} \bigg)}, \quad f=-2\sqrt{\frac{2n-1}{2 n}} \ln{\bigg(\frac{(-u)^n+(-v)^n}{2} \bigg)}.
\ee
Then the solution to the auxiliary fields from solving \eqref{eq:chi1} and \eqref{eq:chi2} are
\bea
\chi_1&=&-4 \lambda_1 \bigg(1-\frac{1}{2n}\bigg)  \ln{[u^n+v^n]}+\frac{\lambda_1 \gamma}{4}\ln{[u^{2n}-v^{2n}]}-\frac{\lambda_2}{4} \ln{[u^{2n}-v^{2n}]}
\no\\
&\quad&+C_1(v)+C_2(u),
\eea
\bea
\chi_2&=&4 \mu_1 \bigg(1-\frac{1}{2n}\bigg)  \ln{[u^n+v^n]}-\frac{\mu_1 \gamma}{4}\ln{[u^{2n}-v^{2n}]}+\frac{\mu_2}{4} \ln{[u^{2n}-v^{2n}]}
\no\\
&\quad& +C_3(v)+C_4(u).
\eea
We impose the following boundary conditions to identify a physically reasonable quantum state:
\begin{itemize}
    \item Regularity of $\langle T_{ab} \rangle$ at $u=v$ ($x=1$) and $v=0$ ($x=0$).
    \item Vanishing $\langle T_{ab} \rangle$ as $u \to -\infty$.
\end{itemize}
We begin with the one-loop effective action \eqref{eq:fulloneloop}, as the matter sector remains unchanged under the Weyl transformation. We find that we need to include the following local counterterms for regularization
\be
\Gamma_{\text{ct}}=\int d^2 x \sqrt{-g} \alpha_1 \phi R+\alpha_2 (\nabla \phi)^2,
\ee
and we note this is different from \eqref{eq:counter}. 

To ensure regularity of $\langle T_{uv} \rangle$ at $u=v$, we require
\be
\alpha_1=\frac{6-\gamma}{96 \pi}.
\ee
This component also remains regular and vanishes automatically at $v=0$, and thus imposes no further constraint. However, $\langle T_{uu} \rangle$ is still singular at $u=v$. To cancel the leading divergence, we must choose
\be
\alpha_2=\frac{\gamma (\gamma+6)}{384 \pi}.
\ee
Yet, even after this cancellation, a logarithmic divergence persists at $u=v$ for $\langle T_{uu} \rangle$, which cannot be removed by any local counterterm alone. In fact, it signals the necessity of including a Weyl-invariant non-local term of the form $(\nabla \phi) \frac{1}{\Box} (\nabla \phi)^2$ by relaxing the final constraint in \er{constraint1} to be
\be
\lambda_2^2-\mu_2^2=p,
\ee
where $p$ is a constant. To proceed, without loss of generality, we could set $\lambda_1=0$, $\mu_1=-\frac{1}{4 \sqrt{3}}$ and $\mu_2=\sqrt{\frac{3}{4\pi}}$, then the constant $p$ is fixed to be 
\be
p=\frac{\gamma}{8 \pi},
\ee
which eliminates the logarithmic divergence at $u=v$. This means that a regular quantum state defined for any $\gamma$ is sensitive to this state-dependent term, in contrast to the Garfinkle spacetime. This process will also give a constraint equation for the functions $C_i$ in the auxiliary fields, similar to the Garfinkle case, but we will need additional conditions to solve them. Regularity of $\langle T_{uu} \rangle$ at $v=0$ does not provide further constraints.

A parallel analysis for $\langle T_{vv} \rangle$ at $u=v$  leads to similar conclusions.  However, requiring regularity at $v=0$ and the condition of vanishing flux as $u \to -\infty$ would allow us to set $C_1(v) =C_3(v)=0$. This enables an explicit determination of $C_2(u)$ and $C_4(u)$, leading to a well-defined, regular stress-energy tensor
\bea
\langle T_{uu} \rangle&=&\frac{(2n-1)u^{n-2} }{12 \pi(u^{2n}-v^{2n})^2}\bigg[-2n\bigg(1+3\ln{\bigg(\frac{2 u^n}{u^n+v^n}\bigg)} \bigg) u^{3n}
\no\\
&\quad&+(8n-1)u^{2n} v^n-(7n-2)u^n v^{2n}+(n-1)v^{3n} \bigg],
\eea
\bea
\langle T_{vv} \rangle&=&\frac{(2n-1)}{12 \pi v^2(u^{2n}-v^{2n})^4}\bigg[(n-1)u^{7n}v^n-2(2n-1) u^{6n}v^{2n}+(6n+1)u^{5n}v^{3n}
\no\\
&\quad&+\bigg(3n-4-6n \ln{\bigg(\frac{2u^n}{u^n+v^n}\bigg)}\bigg) u^{4n}v^{4n}-(15n-1)u^{3n}v^{5n}
\no\\
&\quad&+2\bigg(3n+1+6n \ln{\bigg(\frac{2u^n}{u^n+v^n}\bigg)}\bigg)u^{2n}v^{6n}+(8n-1)u^n v^{7n}
\no\\
&\quad&-\bigg(5n+6n \ln{\bigg(\frac{2u^n}{u^n+v^n}\bigg)}\bigg)v^{8n}\bigg],
\eea
\bea
\langle T_{uv} \rangle&=&\frac{(2n-1)n (uv)^{n-1} }{12 \pi (u^{2n}-v^{2n})^4}\bigg[ u^{6n}-2u^{5n}v^n-u^{4n}v^{2n}
\no\\
&\quad&+4u^{3n}v^{3n}-u^{2n}v^{4n}-2u^n v^{5n}+v^{6n} \bigg].
\eea
A surprising feature of this result is that the renormalized stress-energy tensor $\langle T_{ab} \rangle$ is independent of $\gamma$. The expression is not particularly transparent for identifying whether this state resembles a stationary Boulware-like state. Now we transform back to the adapted coordinates $(T, x)$ 
\bea
\langle T_{TT} \rangle &=& -\frac{(2n-1)}{4 8 n^2 \pi (x^{2n}-1)^2} \bigg[(x^n-1) (5nx^{3n}-6nx^{2n}+2 x^{2n}+9nx^n-2x^n-2n)
\no\\
&\quad& +6n(x^{4n}+1) \ln{\frac{2}{1+x^n}}\bigg],
\eea
\bea
\langle T_{xT} \rangle &=&\frac{(2n-1)x^{n-1}}{24 n \pi (x^{2n}-1)^2}\bigg[(x^n-1)(5n x^{2n}-4nx^n+x^n+2n-1) 
\no\\
&\quad&+6nx^{3n}\ln{\bigg(\frac{2}{1+x^n} \bigg)}\bigg],
\eea
\bea
\langle T_{xx} \rangle &=&-\frac{(2n-1)x^{n-2}}{12 \pi (x^{2n}-1)^2}\bigg[(x^n-1)(5n x^{2n}-3nx^n+x^n+n-1) 
\no\\
&\quad&+6nx^{3n}\ln{\bigg(\frac{2}{1+x^n} \bigg)}\bigg],
\eea
which are clearly $T$-independent. Although singularities arise at certain values of $(\gamma, n)$ even within the domain $x \in [0, 1]$, this does not necessarily forbid the construction of a mathematically regular $\langle T_{ab} \rangle$. Regularity at the quantum level does not strictly require the absence of classical singularities, since such singularities are already encoded in the classical stress-energy tensor. However, their presence can restrict the physical interpretation in terms of well-defined Hadamard states.

We discuss several notable features. Remarkably, although the background geometry explicitly depends on the parameter $\gamma$, the derived quantum stress-energy tensor is entirely independent of it. Consequently, the quantum corrections do not smoothly reduce to the original Garfinkle results in the limit $\gamma \to 0$, despite the Weyl-Garfinkle background being a simple Weyl transformation of the Garfinkle geometry. Furthermore, while $\langle T_{ab} \rangle$ vanishes identically for the special value $n=\frac{1}{2}$, the corresponding background remains non-flat if $\gamma \neq 0$. This violates one of Wald’s axioms, which requires agreement with Minkowski space in a suitable flat limit. This subtlety underscores the importance of the state dependence on the global structure, especially given that the Weyl rescaling introduces additional boundary singularities.

This discussion further highlights a subtle point: although the classical matter theory is Weyl-invariant, quantum effects may introduce scheme-dependent ambiguities stemming from the path integral measure and additional local counterterms. Indeed, general statements relating renormalized stress-energy tensors in conformally related spacetimes typically assume conformally coupled matter and special geometric conditions (such as an Einstein metric). In such scenarios, the quantum stress-energy tensors differ by purely local curvature terms derived from the Weyl factor and its derivatives. Since our classical matter theory is only Weyl-invariant, not strictly conformal, the preferred vacuum state might not map straightforwardly under the Weyl transformation. Instead, it could become an excited state or even fail to remain well-defined if horizons, singularities, or boundaries appear or disappear.

\section{The Hayward spacetime}
\label{sec:Hayward}

In this appendix, we examine an extreme example of critical spacetime: the Hayward solution, obtained by setting $\alpha=\beta=0$ in \eqref{eq:fullRoberts}. This solution exhibits several peculiar global properties, as originally discussed in~\cite{Hayward:2000ds, Clement:2001zd}.

\begin{figure}[hbt!]
\centering
\includegraphics[width=0.43\textwidth]{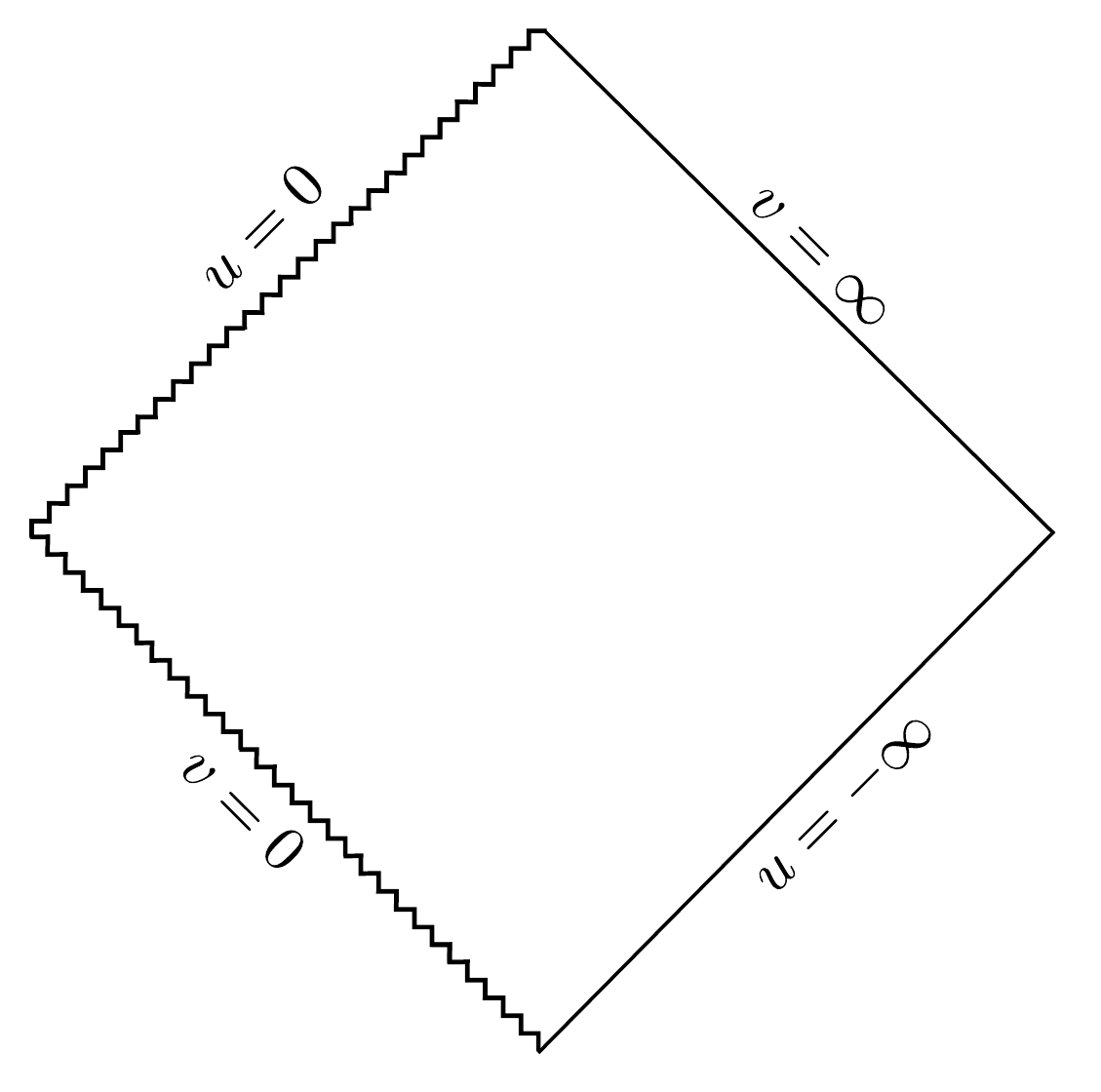}
\caption{The Penrose diagram of the Hayward spacetime, with a central, bifurcate, and null singularity, representing an exotic example of CSS spacetime.}
\label{fig:hayward}
\end{figure}

The background metric and scalar field $f$ in the double-null coodinates $(u, v)$ are given by
\be
ds^2= -2 dudv+r^2 d \Omega^2, \quad r^2=-uv, \quad f=\frac{\sqrt{2}}{2} \ln{\bigg(-\frac{v}{u}\bigg)},
\ee
with the coordinate domain restricted to $u<0$ and $v>0$. By computing the Ricci scalar
\be
R=\frac{1}{uv},
\ee
and similarly, for other curvature invariants, we see there is a central singularity at $uv=0$ ($r=0$). This singularity is bifurcate and null, located at both $u=0$ and $v=0$, representing future and past singularities, respectively. The Penrose diagram is shown in Figure~\ref{fig:hayward}. It is also marginally trapped and massless, reminiscent of the zero-mass central singularity that sits at the threshold between black hole formation and dispersion. The conformal boundary at $r \to \infty$ corresponds to null infinity; however, the spacetime is not asymptotically flat, with the mass unbounded as $r \to \infty$. Much like the Roberts solution, the Hayward spacetime avoids being a black or white hole, as no trapped surfaces form, yet the singularities themselves are marginally trapped. Similarly, no observer can see the future curvature singularity without actually reaching it.
\FloatBarrier 
Spherically symmetric linear perturbations were studied in~\cite{Hayward:2000ds}, revealing two families of dominant modes, labeled $\omega_A$ and $\omega_B$. The latter modes were initially interpreted as signaling a rich phase structure. However, it was later clarified in~\cite{Clement:2001zd} that the $\omega_B$ modes are spurious, artifacts arising from the linearization of a coordinate transformation, and only the $\omega_A$ modes correspond to physical perturbations. These $\omega_A$ modes break CSS through periodic oscillations, reminiscent of DSS behavior. The spectrum of $\omega_A$ modes form two lines in the complex plane that has a fixed real part, with the dominant $\omega_A=1 \pm \sqrt{3} i$, implying an echoing period $\Delta=2\pi/\sqrt{3} \approx 3.62$, which is much closer to Choptuik’s numerically observed value $\Delta \approx 3.44$.

The essential difference from the Roberts solution lies in the presence of a past null singularity at $v=0$, where regularity cannot be imposed. This makes the analysis of $\langle T_{ab} \rangle$ particularly subtle. Despite the geometry being CSS, the expression for $\langle T_{ab} \rangle$ must somehow reflect this global causal structure. Here, we examine how this affects the computation of $\langle T_{ab} \rangle$ in the reduced spacetime. The reduced metric and dilaton are given by
\be
ds^2=-2dudv,\quad \phi=\frac{1}{2}\ln{\bigg(-\frac{1}{uv}\bigg)}.
\ee
The auxiliary fields obtained from solving \eqref{eq:chi1} and \eqref{eq:chi2} are much simpler in this case
\be \label{eq:Haywardchi1}
\chi_1=\frac{1}{4} \lambda_2 \ln{(-u)} \ln{(v)}+C_1(v)+C_2(u),
\ee
\be \label{eq:Haywardchi2}
\chi_2=-\frac{1}{4} \mu_2 \ln{(-u)} \ln{(v)}+C_3(v)+C_4(u).
\ee
Since both $u=0$ and $v=0$ correspond to curvature singularities, we can only impose regularity at $u \to -\infty$ and $v \to \infty$. We then compute the quantum stress-energy tensor $\langle T_{ab} \rangle$ using \eqref{eq:quantumstress}, and substitute either \eqref{eq:sol1} or \eqref{eq:sol2} to express all quantities in terms of $\lambda_2$.

By examining the regularity of $\langle T_{vv} \rangle$ near $u =-\infty$,  we find that in order to cancel the divergence proportional to $\ln{(-u)}$, the only viable choice is $C_2 (u)=-C_4(u)$. If instead $C_2 (u)=C_4(u)$, this would require $C_1(v)=-C_3(v)$, leaving $\langle T_{uu} \rangle$ undetermined with a residual dependence on the arbitrary function $C_4''(u)$. However, no choice of $C_4''(u)$ leads to a stationary result in the adapted coordinates $(T,x)$. 

With the choice $C_2 (u)=-C_4(u)$, consistency then implies $C_1(v)=-C_3(v)$, since taking $C_1(v)=C_3(v)$ would introduce a $\ln{(-u)}$ divergence in $\langle T_{uu} \rangle$. To cancel all potential divergences, we adopt the following ansatz without loss of generality:
\be
C_3(v)=A \ln{(v)}+K_1, \quad C_4(u)= B \ln{(-u)}+K_2,
\ee
where $A, B, K_1,K_2$ are constants. We find that the only choice that eliminates the $\ln{(-u)}$ divergence is 
\be
A=\frac{1}{16 \pi \lambda_2}, \quad B=0.
\ee
To ensure that the result is independent of $\lambda_2$, we may further require either $K_1=-K_2$ or simply set $K_1=K_2=0$. The resulting components of the two-dimensional quantum stress-energy tensor are
\be
\langle T_{uu} \rangle=-\frac{1}{8 \pi u^2},
\ee
\be
\langle T_{uv} \rangle=-\frac{1}{16 \pi uv},
\ee
\be
\langle T_{vv} \rangle=\frac{3\ln{(v)}-7}{4 8 \pi v^2}.
\ee
We see that $\langle T_{ab}\rangle$ captures the expected divergences near the curvature singularities at $u=0$ and $v=0$. Moreover, from $\langle T_{vv} \rangle$, we observe an unavoidable logarithmic divergence as $v \to \infty$, corresponding to future null infinity. This, however, does not pose a problem once we uplift the expression to four dimensions using \eqref{eq:4Dswave}. This is similar to the case of reduced Roberts spacetime, with the key difference being that even if we accept such a logarithmic divergence, the resulting stress-energy tensor in the $(T,x)$ coordinates acquires explicit $T$-dependence from \eqref{eq:adpatedRoberts}:
\be \label{eq:HaywardTT}
\langle T_{TT} \rangle=\frac{-3T+3\ln{(e^{2x}+1)}-19}{48 \pi},
\ee
\be \label{eq:HaywardxT}
\langle T_{xT} \rangle=\frac{e^{2x} (3T-3\ln{(e^{2x}-1)}+10)}{24 \pi (e^{2x}-1)},
\ee
\be \label{eq:Haywardxx}
\langle T_{xx} \rangle=\frac{e^{4x}(-3T+3 \ln{(e^{2x}-1)}-7)}{12 \pi (e^{2x}-1)^2}.
\ee
That is, no stationary choice of quantum state is available in this setting. A similar issue arises in the reduced Roberts spacetime when one does not allow $\langle T_{ab} \rangle$ to be at most logarithmically divergent as $v \to \infty$, as noted at the end of Section~\ref{sec:Robertsoneloop}. 

We find that local counterterms cannot resolve this issue, and introducing additional Weyl-invariant terms typically spoils regularity at $u \to -\infty$. While we cannot entirely rule out the possibility that carefully constructed Weyl-invariant terms or more elaborate counterterms might achieve both goals, such modifications generally compromise the desirable properties of $\langle T_{ab} \rangle$, such as the Wald's axioms, discussed in Section~\ref{sec:onelooptheory}.

This behavior can again be traced to the nature of the curvature singularities in the reduced spacetime: being null rather than point-like, they impose distinct kinematic constraints on the quantum stress-energy tensor $\langle T_{ab} \rangle$. This highlights a subtle but important point: a stationary quantum state cannot be determined solely by demanding compatibility with self-similarity; it is also sensitive to the global causal structure of the spacetime. No choice of quantum state preserves the scale invariance of such a background.

Nevertheless, the quantum state remains Boulware-like when uplifted to four dimensions, as there are no quantum fluxes near infinity. Self-similarity will still introduce an overall $e^{2T}$ dependence, yet global causal structure acquires additional structure, further highlighting the peculiarity of the Hayward spacetime. We emphasize that such time dependence in $\langle T_{ab} \rangle$ is quite generic in curved backgrounds, and does not, by itself, present any issue for physical interpretation. The main complication is practical: solving the semiclassical Einstein equations becomes more challenging, as the resulting PDEs now depend on both $T$ and $x$, likely requiring numerical techniques that we leave to future work.
\\~\

\noindent \textbf{Note added in v2:} We remark on a more ``natural” quantum state in the Hayward spacetime. The effective two-dimensional quantum stress-energy tensor given in~\eqref{eq:HaywardTT}-\eqref{eq:Haywardxx} was obtained by choosing a quantum state (i.e., boundary conditions) compatible with the linear perturbation analysis in the Roberts spacetime discussed in Section~\ref{sec:collapse3+1}. There, we allowed a logarithmic divergence near future null infinity while requiring regularity in the past. This choice naturally aligns with the setup typical in critical collapse, where one prepares smooth initial data in the past and joins the future region to an exterior spacetime with the appropriate asymptotics, as outlined in Section~\ref{sec:criticalcollapse}. Note that when uplifted to four dimensions, no genuine divergences arise.

However, Hayward spacetime is very special. The above stress-energy tensor was derived using the coordinate transformations~\eqref{eq:adpatedRoberts}. If instead we adopt the null coordinates introduced by Hayward~\cite{Hayward:2000ds},
\be
x^\pm=\pm e^{\rho\pm \tau},
\ee
where we still have $u=x^-$ and $v=x^+$, then the metric in $(\tau, \rho)$ coordinates takes the form
\be
ds^2=e^{2 \rho} (-2 d \tau^2+2 d \rho^2+d \Omega^2).
\ee
We then see that $\partial/\partial \rho$ serves as the homothetic Killing vector (corresponding to our notation $T$; the overall scaling is conventional and can be flipped via $\rho \to -\rho$), while $\partial/\partial \tau$ is an exact timelike Killing vector. This makes the Hayward spacetime a special CSS solution in which an exact notion of ``stationarity" is restored.

This observation raises the question of whether there exists a natural quantum state associated with the exact Killing vector, such that the resulting $\langle T_{ab} \rangle$ is $\tau$-independent. Indeed, this is possible if we relax the earlier condition of regularity at past null infinity, i.e., we allow the $\ln(-u)$ divergence to remain uncanceled. In this case, a natural choice of homogeneous solutions $C_i$ in~\eqref{eq:Haywardchi1} and~\eqref{eq:Haywardchi2} would correspond to $C_1(v)=C_3(v)$ and $C_2(u)=C_4(u)$, where we can explicitly solve
\be
C_3 (v)= \frac{1}{8} \lambda_2 \ln{(v)} [\ln{(v)}-8a+2]+ c_1 v+c_2,
\ee
\be
C_4 (u)=\frac{1}{8} \lambda_2 \ln{(-u)} [\ln{(-u)}-8b+2]+ c_3 u+c_4,
\ee
where $c_i$ are integration constants that do not enter $\langle T_{ab} \rangle$. In $(\tau, \rho)$ coordinates, the resulting two-dimensional stress-energy tensor is
\be \label{eq:HaywardstatTT}
\langle T_{\tau \tau} \rangle=\frac{2a+2b-3-2 \rho}{8 \pi}, 
\ee
\be \label{eq:Haywardstatrr}
\langle T_{\rho \rho} \rangle= \frac{2a+2b-1-2 \rho}{8 \pi}, 
\ee
\be \label{eq:Haywardstatrt}
\langle T_{\rho \tau} \rangle=\frac{a-b}{4 \pi}.
\ee
By further requiring the absence of spurious quantum flux, we can set $a = b$, which ensures $\langle T_{\rho \tau} \rangle = 0$. The stress-energy tensor is then stationary with respect to the exact timelike Killing vector $\partial/\partial \tau$, while necessarily retaining a linear dependence on $\rho$ (or equivalently $T$). Upon uplifting to four dimensions, one gets an extra overall $e^{2T}$ Lyapunov scaling. A similar structure was identified in~\cite{Zahn:2025tnh} by inputting a self-similar state, as we discuss in Appendix~\ref{sec:appendixD}.

Finally, we emphasize that throughout this paper, “stationarity” refers to invariance under self-similar time translations, since typical critical spacetimes lack an exact timelike Killing vector. We use this notion to describe the two-dimensional effective stress tensor $\langle T_{ab} \rangle$. In contrast, the higher-dimensional $\langle T^{(D)}_{\mu\nu} \rangle$ always carries an overall factor of $e^{(D-2)T}$, and is therefore “non-stationary” while still respecting self-similarity. Under appropriate conventions and conditions discussed in Section~\ref{sec:criticalcollapse}, the self-similar time $T$ can be regarded as the natural time coordinate describing the flow toward the future singularity. Indeed, the Garfinkle and Roberts solutions admit no such freedom of a natural quantum state, distinguishing them from the special Hayward case considered here.

\section{Remarks on approaches based on canonical quantization}
\label{sec:appendixD}

In this appendix, we offer remarks on two closely related works~\cite{Brady:1998fh, Zahn:2025tnh} that address quantum effects in critical spacetimes through canonical quantization. These studies are \emph{among the more systematic} analyses outlined in Section~\ref{sec:intro} and provide a complementary first-principles perspective compared to the path integral framework adopted in this paper. The general lessons drawn from them suggest an agreement with our calculation of the quantum Lyapunov exponent $\omega_q=D-2$. However, in the non-stationary critical collapse setting the conclusions can depend sensitively on assumptions about the state and boundary conditions (where self-similar states are employed~\cite{Brady:1998fh, Zahn:2025tnh}), and, to our knowledge, a complete evaluation of $\langle T_{\mu \nu} \rangle$ under physically specified regularity/flux conditions across the relevant domains has not been carried out yet.

On the other hand, before comparing frameworks, we emphasize an important scope point: our anomaly-based dimensional reduction defines an effective $s$-wave semiclassical description with state fixed by regularity criteria in the reduced theory (which is not necessarily self-similar). We do not claim that there exists a four-dimensional (or more generally any $D>2$) Hadamard state whose full $\langle T_{\mu \nu} \rangle$ reproduces, upon projection, the expectation values obtained in our reduced effective action calculation. Establishing such an uplift is a separate and subtle question, and we view this as a natural and potentially feasible open problem within the Hadamard framework.

Our goal here is to make explicit which inputs are required in each framework and how these relate to the semiclassical critical collapse.

\textbf{(i)} Brady and Ottewill~\cite{Brady:1998fh} approached the problem from a four-dimensional perspective using the trace anomaly to model quantum effects in critical collapse and reported qualitatively similar features to ours, such as the appearance of a mass gap. They also extracted a quantum Lyapunov exponent $\omega_q = 2$ through a kinematical analysis, which coincides with our result for the four-dimensional model presented in Section~\ref{sec:collapse3+1}.\footnote{Intriguingly, despite the limitations of the Page approximation adopted in this work, which we discuss below, the quantum Lyapunov exponent arising from the Weyl transformation appears to be robust and independent of the specific quantization details. In Section~\ref{sec:discussion}, we have demonstrated that this robustness has a clear physical origin. On the other hand, it is also argued that quantum effects generate a non-universal mass gap~\cite{Brady:1998fh}; contrary to this expectation, we have shown in Section~\ref{sec:discussion} that a universal notion of the mass gap in fact emerges. We thank Gustavo J. Turiaci for a discussion on this point.} From the viewpoint of semiclassical critical collapse, the construction highlights several points where additional clarifications are particularly useful.

First, the trace anomaly used in their work pertains to conformally coupled matter in four dimensions, rather than to a minimally coupled massless scalar field, the central model in the seminal works by Christodoulou and Choptuik. These should be viewed as physically distinct systems.

Second, the computation relies on the so-called Page approximation~\cite{Page:1982fm, Brown:1986jy}, in which one evaluates $\langle T_{\mu \nu} \rangle$ by mapping to a conformally related spacetime. However, this method requires that (i) the physical spacetime be static, (ii) the Weyl rescaling factor be time-independent, and (iii) the background metric be an Einstein metric with vanishing classical stress-energy tensor, so that all curvature arises from quantum backreaction. These assumptions are tailored to a stationary, effectively vacuum setting, whereas a generic critical collapse spacetime is dynamical and supported by nontrivial classical matter; accordingly, extra care is needed when translating conclusions between the two contexts. In the absence of nontrivial classical matter supporting the background, it is natural to ask how the underlying geometry is specified for the quantum calculation. Alternatively, if one views the $\langle T_{\mu \nu} \rangle$ as a perturbative correction around a prescribed classical critical spacetime, then the treatment does not attempt to model the coupled backreaction of quantum effects with the collapsing matter dynamics, and is closer in spirit to approaches that parameterize an effective quantum flux as an additional input.

Lastly, the Page approximation cannot consistently describe this perturbative process either, as it assumes a specific thermal equilibrium structure incompatible with critical collapse. In four dimensions, $\langle T_{\mu \nu} \rangle$ is much more sensitive to the choice of quantum state due to a larger number of degrees of freedom. Since their analysis relies on conformal mapping, the state dependence is encoded in the choice of state in the conformally related spacetime.\footnote{In fact, \emph{non-trivial non-local effects can arise} for non-conformal matter in the Weyl-transformed spacetime. See Appendix~\ref{sec:WeylGarfinkle} for an example indicating that the choice of state before and after Weyl transformations will not just be a local function of the Weyl factor. Rather, the relation is highly non-local, effectively mixing different quantum excitations in the two frames.} The authors do not specify the quantum state beyond assuming it is time-independent, i.e., respects Killing symmetry.\footnote{The state is implicitly assumed to respect self-similarity in the physical spacetime; equivalently, this assumption is encoded by taking the state on the conformally related stationary spacetime to be stationary. We thank Jochen Zahn for pointing this out to us.} In fact, if the requirements above are met, it would correspond to a thermal state akin to the Hartle-Hawking vacuum, and would require additional justification in a critical collapse setting. In contrast, we have demonstrated that the physically motivated quantum state is Boulware-like: Minkowskian both in the asymptotic past and future and contains no incoming or outgoing flux. 

\textbf{(ii)} After the publication of our work,~\cite{Zahn:2025tnh} presented a complementary analysis and discussed subtleties related to the dimensional reduction anomaly and the role of the two-dimensional trace anomaly. We have clarified the reduction anomaly in Section~\ref{sec:onelooptheory}; likewise, the trace anomaly~\eqref{eq:traceanomaly} is fixed by universal UV physics, where we have properly included the contribution from the dilaton. 

By contrast,~\cite{Zahn:2025tnh} extends Brady-Ottewill~\cite{Brady:1998fh} using Hadamard renormalization without restricting to conformal coupling and without the Page approximation. In particular, two nontrivial CSS geometries were analyzed---Roberts and Hayward spacetimes---also treated here (Section~\ref{sec:collapse3+1} and Appendix~\ref{sec:Hayward}), and differences were noted. In the following, we explain that the differences arise from differing assumptions, and we spell out the conditions under which the construction of~\cite{Zahn:2025tnh} applies.

The approach adopts a self-similar state and motivates this choice by arguing that ``generic and unavoidable" quantum effects are expected to respect the spacetime’s homothety. This motivation is intuitive but does not by itself uniquely fix the state: generic self-similar critical spacetimes lack an exact timelike Killing vector, and the self-similar flow is not an isometry; hence, self-similarity serves as a constraint rather than providing, by itself, a constructive prescription for constructing modes. 

Therefore, a canonical choice of quantum state is highly non-trivial, and importing results proven only in stationary or asymptotically stationary geometries calls for additional justification. Typically, a classical property need not automatically persist at the quantum level: fine-tuning of classical initial data and a self-similar background do not by themselves determine a unique quantum state. The selection of a state should be guided by physical criteria applied to the quantum state itself, and not solely by classical symmetries or initial conditions.\footnote{In our case, self-similarity is not imposed as part of the physical condition defining the state; a self-similar pattern for the stress-energy tensor, $\langle T_{\mu \nu} \rangle \propto e^{(D-2)T}F_{\mu \nu}(x^i)$, emerges naturally under reasonable regularity assumptions, which then singles out a physically meaningful quantum state.}

The construction in~\cite{Zahn:2025tnh} warrants a more detailed discussion and clarification that we discuss in detail below. Assuming the self-similarity of the two-point function, the four-dimensional renormalized stress-energy tensor in adapted coordinates $(T, x^i)$ must take the form 
\be \label{eq:selfsimilartensor}
\langle T_{\mu \nu} \rangle (T, x^i)= e^{2T} \bigg({\langle T_{\mu \nu} \rangle} (x^i)+\frac{T}{4 \pi}  {V_{\mu \nu}} (x^i) \bigg),
\ee
where ${\langle T_{\mu \nu} \rangle} (x^i)$ is the state-dependent part and ${V_{\mu \nu}} (x^i)$ is local and state-independent, with the linear-$T$ factor arising from the logarithmic UV structure of the Hadamard parametrix. In odd spacetime dimensions, the logarithmic term (hence the $T$-piece) is absent and the splitting becomes ambiguous; in even dimensions, $V_{\mu \nu}$ is purely geometric. The structure given in~\eqref{eq:selfsimilartensor} is a necessary but not sufficient condition given the assumption of a self-similar two-point function, as it probes essentially the coincidence limit of the two-point function. A similar pattern already appears in Brady-Ottewill~\cite{Brady:1998fh}, where the compatibility of the two-point function underlying self-similar structure is taken as a working assumption, though implicitly encoded in the way the state is specified on a conformally related problem.

The author then proceeds to analyze matter modes from the scalar wave equation, constructs the symplectic form of solutions, and argues that the corresponding two-point function is self-similar on both Hayward and Roberts, without explicitly evaluating the state-dependent piece ${\langle T_{\mu \nu} \rangle} (x^i)$. On that basis, the author claims a discrepancy with our results, suggesting that we miss the linear-$T$ term $T V_{\mu \nu}$ that should arise from the UV structure. However, in Hayward space under physical boundary conditions, we do obtain the same structure (see~\eqref{eq:HaywardTT}-\eqref{eq:Haywardxx} and~\eqref{eq:HaywardstatTT}-\eqref{eq:Haywardstatrt}). A difference appears only for Roberts, where our result~\eqref{eq:Robertsstressgen} does not exhibit a linear-$T$ term. As we have explained in Section~\ref{sec:Robertsoneloop}, Roberts admits two boundary-condition choices: the one compatible with linear perturbation analysis yields \eqref{eq:Robertsstressgen} (no linear-$T$ term), while the other choice does generically produce a linear-$T$ contribution, i.e., there is not necessarily a discrepancy.\footnote{In the three-dimensional Garfinkle spacetime, regularity uniquely fixes the stress-energy tensor~\eqref{eq:Garfinklestressgen} with no linear-$T$ term. Since $V_{\mu \nu}$ vanishes in odd dimensions, there is likewise no discrepancy.}

At the level of QFT on a fixed background, the argument in~\cite{Zahn:2025tnh} relies on three ingredients:\footnote{With the implicit assumption of a quasi-free state (vanishing one-point function). While any nonzero one-point function can be absorbed into the classical background scalar as it is never $O(\hbar)$, one must then re-check that the new background scalar still solves the same geometry and preserves self-similarity.}
\begin{itemize}
    \item (i) the chosen state is Hadamard (so point-splitting renormalization is well-defined and convergent);
    \item (ii) a homothety-covariance property of the anti-symmetric part of the two-point function ($i \Delta (x;x')$ from the canonical commutation relation, CCR), understood as a property of the \emph{full} boundary-value problem, which controls the universal UV structure that yields the $ e^{2T} T V_{\mu \nu}$ term;\footnote{On a fixed, globally hyperbolic background with no additional boundary conditions, $\Delta$ is uniquely determined by the difference of the retarded and advanced Green’s functions, which is indeed homothety covariant in self-similar backgrounds like Hayward and Roberts. In critical collapse applications, however, one typically imposes further \emph{physical} conditions---most notably regularity and/or vanishing-flux requirements at similarity horizons and the center. In that case, the relevant object is the entire boundary-value problem, and homothety-covariance of $\Delta$ requires checking that the homothety maps this full set of inputs into itself. Since homothety is not an isometry, this invariance is not guaranteed \textit{a priori}.}
    \item (iii) one restricts to self-similar Hadamard states so that $e^{2T} \langle T_{\mu \nu} \rangle(x^i)$ holds (determined by the symmetric part of the two-point function). 
\end{itemize}
The discussion in~\cite{Zahn:2025tnh} could benefit from a clearer separation of these logically distinct requirements; as presented, the argument appears to treat some of them as additional inputs rather than derived consequences.

For (i) (Hadamard): in Hayward, Hadamard follows readily, as Hayward is a special CSS spacetime that also admits a timelike Killing vector, restoring the usual stationary framework (Appendix~\ref{sec:Hayward}). In Roberts, by contrast, establishing Hadamard is nontrivial because there is no timelike Killing vector and no general theorem that guarantees it. One can nevertheless obtain a Hadamard in/out state by exploiting the fact that Roberts asymptotically approaches Hayward: in the large-$r$ limit, the Klein-Gordon operator differs from its Hayward counterpart by terms that decay exponentially with all derivatives. Under this control, the asymptotically static machinery applies~\cite{Gerard:2016bbj}: the scattering map is tame, so pulling back the Hayward vacuum yields a Hadamard state on Roberts. Crucially, this hinges on special features of the Roberts-Hayward pair and does not automatically extend to generic self-similar critical spacetimes. The caveat is that in those broader settings, symplectic normalization or formal mode sums are insufficient to certify Hadamard without additional microlocal/regularity input.

For (ii) and (iii), it is helpful to keep conceptually distinct the geometric property of $i\Delta$ from symmetry requirements imposed on the state. Since $\Delta$ transforms covariantly under the homothety, one then infers (iii) by invoking positivity: the symmetric part would be ``bounded from below" by the anti-symmetric part and therefore must scale at least as strongly, so the full two-point function inherits the same weight, yielding the $e^{2T}$ scaling (up to finite/anomalous terms). On this basis, the structure in~\eqref{eq:selfsimilartensor} is presented as generic, with the suggestion that it is implausible for reasonable states to behave otherwise.

However, the step from (ii) to (iii) requires additional input beyond CCR and positivity. First, positivity is a statement about the quadratic form defined by the symmetric part of the two-point function; it is not a pointwise/distributional inequality tying the symmetric kernel to the commutator, and it does not fix a homothetic weight for the symmetric part. Second, CCR and positivity do not enforce homothety-covariance of the state: starting from any Hadamard state $W$, one can form $W'=W+\epsilon S$ with $S$ a smooth, real, symmetric bisolution\footnote{Note that not every smooth $S$ will do: one needs a controlled $S$---e.g., its quadratic form is bounded with respect to the $W$-induced norm, or $W'$ is obtained via a small Bogoliubov transform. For the non-implication we are highlighting, however, it suffices to exhibit some nearby Hadamard state $W'$ that generically breaks homothety invariance while preserving positivity. Crucially, we do not require an unbounded $S$ (nor do we aim) to cancel the geometric $e^{2T}T V_{\mu \nu}$ in the argument presented here.} or that is ``not" homothety-covariant; for small $|\epsilon|$, $W'$ remains Hadamard and positive, while $i \Delta$ (and its homothety scaling) is unchanged. Thus, (iii) should therefore be viewed as an additional restriction on the class of states under consideration

This caveat is even sharper if we move to the semiclassical Einstein equation. In semiclassical critical collapse, (ii) should be evaluated within the coupled Einstein-matter system, where boundary data encode physical initial conditions and the geometry (hence $\Delta$) is not fixed \textit{a priori}. Homothety-covariance of $\Delta$ then represents an additional symmetry of the full coupled system, not merely of the Klein-Gordon symbol. This also clarifies why, in our formulation, alternative boundary conditions---especially those compatible with linear perturbation theory---need not preserve the specific UV structure leading to the linear-$T$ term proposed in~\cite{Zahn:2025tnh}, even though the underlying renormalized UV structure remains intact. This underscores the intrinsic boundary-condition sensitivity of such constructions.

Moreover, the construction for~\eqref{eq:selfsimilartensor} would suggest a universal $e^{2T}$ growth for all the angular $l$-modes. It is argued that since the commutator $i \Delta$ contains all $l$-modes and transforms with homothetic weight $e^{2T}$, the CCR would enforce an $e^{2T}$ type growth in the $l>0$ sector of the state. However, this does not follow, as the homothety-covariance of $\Delta$ is a global constraint on the full bidistribution, not a per-mode law: mode decompositions are non-unique, the homothety generator need not act diagonally in any chosen basis since it is not an isometry, and the Ward-type constraint can be realized by the integral over modes together with the transformed spectral measure/coefficient kernel, without any single mode growing exponentially.

It is also incompatible with linear perturbation analysis in the context of critical collapse. Linear metric perturbations are decomposed into spherical harmonics and analyzed without imposing spherical symmetry. The result is robust: at most a single unstable $s$-wave exists; all $l>0$ sectors are stable/decaying. This linear analysis is kinematic, it does not depend on whether the source is classical or quantum. If the quantum state carries non-decaying or growing $l>0$ content in self-similar time, then $\delta \langle T_{\mu \nu} \rangle$ provides a growing source in those $l$ sectors, unless one imposes non-generic orthogonality or solvability conditions, which are not guaranteed by the CCR, positivity, or regularity.\footnote{There is no direct contradiction as~\cite{Zahn:2025tnh} is constructing a formal fixed-background state with scalar-only modes. But in applications to semiclassical critical collapse, one would additionally need to check consistency with the semiclassical Einstein equations, which may rule out certain states. Scalar fluctuations also source the metric at linear order when the background scalar field is nontrivial. Therefore, a meaningful scalar-only perturbation must still satisfy the constraints and boundary conditions of the full coupled system, so that it can serve as valid initial data for both field and geometry. Meanwhile, state-independent counterterms (including anomalies) are purely geometric and, on a spherically symmetric background, contribute only to the $l=0$ sector; they cannot cancel or absorb $l>0$ state structure. Thus, any persisting $l>0$ growth in the fixed-background state is incompatible with both linear perturbation theory and the semiclassical Einstein equation under the standard boundary conditions.}

A common rebuttal is that ``higher-$l$ quantum fluctuations still yield a spherically symmetric $\langle T_{\mu \nu} \rangle$, hence they only source the $s$-wave geometry." However, this is a modeling choice as it rests on extra symmetry assumptions: the semiclassical Einstein equation sources the geometry through the mean field; and for it to remain purely $s$-wave, one must assume the quantum state is exactly rotationally invariant. It also overlooks a methodological mismatch. The no-growth result for $l>0$ in critical collapse was derived sector by sector without assuming spherical symmetry of perturbations. Consistency with that framework therefore does not require the state to be rotationally invariant. Invoking ``only $s$-wave fluctuations” to neutralize growing higher-$l$ quantum content effectively changes the problem being solved.

Finally, let us address a claim made in~\cite{Zahn:2025tnh} that, because the state-independent part of $\langle T_{\mu \nu} \rangle$ contains a term that grows linearly with the self-similar time $T$ (via the local tensor $V_{\mu \nu}$), one ``does not'' need to compute the state-dependent part: at ``large-$T$” the state-independent piece will dominate the backreaction anyway. This, by itself, does not provide a sufficient basis for omitting the state-dependent contribution. Large-$T$ is precisely where semiclassical and linearized approximations are expected to break down. In that regime, neither linear response nor QFT on a fixed background is under control, and any claim about which contribution ``dominates" becomes difficult to interpret reliably. The relevant comparison must be made within the controlled window, i.e., at finite $T$ where linear response and Hadamard renormalization remain valid, and where one must derive explicit bounds on their relative magnitudes before drawing conclusions (see Section~\ref{sec:GarfinkleHorizon} for a detailed discussion, including why one must introduce a classical perturbation mode to compete with the quantum growing mode at finite $T$, even when the quantum mode ``exponentially'' dominates the classical one).

At finite $T$, once a renormalization prescription is fixed, the split between state-dependent and state-independent parts acquires operational meaning. Their relative size then becomes a genuine dynamical question: there is no theorem that the local $T V_{\mu \nu}$ term must dominate, nor any microlocal or positivity argument that bounds the state-dependent contribution by it. The magnitude and time-dependence of the state-dependent part must be explicitly computed for the chosen state and boundary conditions. In fact, with the author’s own proposed $l>0$ mode content, it can be of comparable or even larger order, precisely where stability is most sensitive, making its evaluation indispensable. Yet obtaining this part through explicit mode summation, especially for $l>0$, is typically the hardest part of the problem and requires uniform sector-by-sector bounds demonstrating that the state-dependent term remains small on the domain (see Section~\ref{sec:discussion} for further discussion of higher angular modes).

Furthermore, bypassing the state-dependent part raises an important question: it becomes unclear whether the construction could inadvertently produce an apparent Hawking-like flux even before horizon formation---a point we previously highlighted in earlier approaches. Moreover, this issue is particularly subtle in the absence of an exact timelike Killing vector, where the standard notion of a ground state ceases to exist, making it far from obvious how such quantum state should be physically interpreted.

Taken together, the set of conditions invoked to secure (i)–(iii), together with possible subtleties regarding semiclassical consistency under standard boundary conditions, suggest that the proposal of~\cite{Zahn:2025tnh} would benefit from further clarification and development. Moreover, if one adopts these conditions as part of the state definition, then the value $\omega_q=D-2$ is effectively built into the construction rather than obtained as an output, so additional discussion is needed to clarify in what sense the quantum Lyapunov exponent is being derived within that framework.

Regardless, the discussion also prompts a natural open question: are the physically relevant quantum states we determine from first principles in Garfinkle, Roberts, and Hayward actually self-similar? However, the structure of $\langle T_{\mu \nu} \rangle$ implied by self-similarity is at best a necessary condition; it is not sufficient. In our approach, the states are selected by requiring the full covariant $\langle T_{\mu \nu} \rangle$ to be regular on the appropriate domain, i.e., by enforcing cancellations of singular structures in both the state-dependent and state-independent sectors, exactly as in standard constructions on black hole spacetimes. By contrast, Ref.~\cite{Zahn:2025tnh} focuses primarily on the state-independent (universal) structure and does not, as presented, fully specify the physical regularity/flux conditions or the complete state-dependent contribution. As a result, it remains to be checked in that framework whether the full covariant $\langle T_{\mu \nu} \rangle$ is regular on the relevant domain, and correspondingly a direct term-by-term comparison with our results depends on how these additional inputs are fixed.

Overall, this discussion illustrates that several conceptual points about quantum effects in critical collapse remain subtle, and it also highlights a practical advantage of the path integral formulation: the choice of state is fixed by explicit physical criteria, and the calculation is carried out directly at the level of the semiclassical Einstein equation.

\end{appendix}

%%%%%%%%%%%%%%%%%%%%%%%%%%%%%%%%%%%%%%%%%%%%%%%%%%
\addcontentsline{toc}{section}{References}
\bibliographystyle{JHEP}
\bibliography{bibliography}

\end{document}